%% file: main.tex
\documentclass[twocolumn,epjc3]{svjour3}
\input{Preamble.tex}
\begin{document}

\input{Afteramble.tex}

\title{Measurements of neutrino oscillation parameters from the T2K experiment using $3.6\times10^{21}$ protons on target}

\input{Authorlist_converted}

\date{Received: 07-Mar-2023 / Accepted: 10-Jul-2023}
\onecolumn
\maketitle

\input{Abstract.tex}
\twocolumn

\section{Introduction}
\label{sec:introduction}
\input{Introduction.tex}

\section{The T2K experiment}
\label{sec:t2k}
\input{T2K.tex}

\section{Updates from previous analysis}
\label{sec:updates}
\input{Updates.tex}

\section{Neutrino flux model}
\label{sec:flux}
\input{Flux.tex}

\section{Neutrino interaction model}
\label{sec:interactionModel}
\input{Interaction.tex}

\section{Near-detector analysis}
\label{sec:nd_fit}
\input{NDfit.tex}

\section{Far-detector selection}
\label{sec:sk}
\input{SK.tex}

\section{Oscillation analysis}
\label{sec:oa_results}
\input{OA.tex}

\section{Simulated data studies}
\label{sec:fakeData}
\input{Fakedata.tex}

\section{Conclusions}
\label{sec:conclusions}
\input{Conclusions.tex}

\input{Acknowledgements.tex}

\input{Appendix.tex}

\bibliographystyle{utphys}
\bibliography{Bib.bib,T2K.bib,Interaction.bib,OA.bib,Fakedata.bib}

\end{document}

%% file: Afteramble.tex

\renewcommand{\tableautorefname}{Tab.}
\renewcommand{\figureautorefname}{Fig.}
\renewcommand{\equationautorefname}{Eq.}
\renewcommand{\sectionautorefname}{Sec.}
\renewcommand{\subsectionautorefname}{Sec.}
\renewcommand{\subsubsectionautorefname}{Sec.}

\renewcommand{\appendixautorefname}{App.}

%% file: Authorlist_converted.tex
\institute{ University Autonoma Madrid, Department of Theoretical Physics, 28049 Madrid, Spain \label{INSTHD} \and\pagebreak[0] University of Bern, Albert Einstein Center for Fundamental Physics, Laboratory for High Energy Physics (LHEP), Bern, Switzerland \label{INSTEE} \and\pagebreak[0] Boston University, Department of Physics, Boston, Massachusetts, U.S.A. \label{INSTFE} \and\pagebreak[0] University of California, Irvine, Department of Physics and Astronomy, Irvine, California, U.S.A. \label{INSTGA} \and\pagebreak[0] IRFU, CEA, Universit\'e Paris-Saclay, F-91191 Gif-sur-Yvette, France \label{INSTI} \and\pagebreak[0] University of Colorado at Boulder, Department of Physics, Boulder, Colorado, U.S.A. \label{INSTGB} \and\pagebreak[0] Colorado State University, Department of Physics, Fort Collins, Colorado, U.S.A. \label{INSTFG} \and\pagebreak[0] Duke University, Department of Physics, Durham, North Carolina, U.S.A. \label{INSTFH} \and\pagebreak[0] E\"{o}tv\"{o}s Lor\'{a}nd University, Department of Atomic Physics, Budapest, Hungary \label{INSTJA} \and\pagebreak[0] ETH Zurich, Institute for Particle Physics and Astrophysics, Zurich, Switzerland \label{INSTEF} \and\pagebreak[0] CERN European Organization for Nuclear Research, CH-1211 Gen\'eve 23, Switzerland \label{INSTIE} \and\pagebreak[0] University of Geneva, Section de Physique, DPNC, Geneva, Switzerland \label{INSTEG} \and\pagebreak[0] University of Glasgow, School of Physics and Astronomy, Glasgow, United Kingdom \label{INSTHJ} \and\pagebreak[0] H. Niewodniczanski Institute of Nuclear Physics PAN, Cracow, Poland \label{INSTDG} \and\pagebreak[0] High Energy Accelerator Research Organization (KEK), Tsukuba, Ibaraki, Japan \label{INSTCB} \and\pagebreak[0] University of Houston, Department of Physics, Houston, Texas, U.S.A. \label{INSTIB} \and\pagebreak[0] Institut de Fisica d'Altes Energies (IFAE) - The Barcelona Institute of Science and Technology, Campus UAB, Bellaterra (Barcelona) Spain \label{INSTED} \and\pagebreak[0] Institut f\"ur Physik, Johannes Gutenberg-Universit\"at Mainz, Staudingerweg 7, 55128 Mainz, Germany \label{INSTJC} \and\pagebreak[0] IFIC (CSIC \& University of Valencia), Valencia, Spain \label{INSTEC} \and\pagebreak[0] Institute For Interdisciplinary Research in Science and Education (IFIRSE), ICISE, Quy Nhon, Vietnam \label{INSTHH} \and\pagebreak[0] Imperial College London, Department of Physics, London, United Kingdom \label{INSTEI} \and\pagebreak[0] INFN Sezione di Bari and Universit\`a e Politecnico di Bari, Dipartimento Interuniversitario di Fisica, Bari, Italy \label{INSTGF} \and\pagebreak[0] INFN Sezione di Napoli and Universit\`a di Napoli, Dipartimento di Fisica, Napoli, Italy \label{INSTBE} \and\pagebreak[0] INFN Sezione di Padova and Universit\`a di Padova, Dipartimento di Fisica, Padova, Italy \label{INSTBF} \and\pagebreak[0] INFN Sezione di Roma and Universit\`a di Roma ``La Sapienza'', Roma, Italy \label{INSTBD} \and\pagebreak[0] Institute for Nuclear Research of the Russian Academy of Sciences, Moscow, Russia \label{INSTEB} \and\pagebreak[0] International Centre of Physics, Institute of Physics (IOP), Vietnam Academy of Science and Technology (VAST), 10 Dao Tan, Ba Dinh, Hanoi, Vietnam \label{INSTHI} \and\pagebreak[0] ILANCE, CNRS – University of Tokyo International Research Laboratory, Kashiwa, Chiba 277-8582, Japan \label{INSTJD} \and\pagebreak[0] Kavli Institute for the Physics and Mathematics of the Universe (WPI), The University of Tokyo Institutes for Advanced Study, University of Tokyo, Kashiwa, Chiba, Japan \label{INSTHA} \and\pagebreak[0] Keio University, Department of Physics, Kanagawa, Japan \label{INSTID} \and\pagebreak[0] King's College London, Department of Physics, Strand, London WC2R 2LS, United Kingdom \label{INSTIF} \and\pagebreak[0] Kobe University, Kobe, Japan \label{INSTCC} \and\pagebreak[0] Kyoto University, Department of Physics, Kyoto, Japan \label{INSTCD} \and\pagebreak[0] Lancaster University, Physics Department, Lancaster, United Kingdom \label{INSTEJ} \and\pagebreak[0] Lawrence Berkeley National Laboratory, Berkeley, CA 94720, USA \label{INSTII} \and\pagebreak[0] Ecole Polytechnique, IN2P3-CNRS, Laboratoire Leprince-Ringuet, Palaiseau, France \label{INSTBA} \and\pagebreak[0] University of Liverpool, Department of Physics, Liverpool, United Kingdom \label{INSTFC} \and\pagebreak[0] Louisiana State University, Department of Physics and Astronomy, Baton Rouge, Louisiana, U.S.A. \label{INSTFI} \and\pagebreak[0] Joint Institute for Nuclear Research, Dubna, Moscow Region, Russia \label{INSTIH} \and\pagebreak[0] Michigan State University, Department of Physics and Astronomy,  East Lansing, Michigan, U.S.A. \label{INSTHB} \and\pagebreak[0] Miyagi University of Education, Department of Physics, Sendai, Japan \label{INSTCE} \and\pagebreak[0] National Centre for Nuclear Research, Warsaw, Poland \label{INSTDF} \and\pagebreak[0] State University of New York at Stony Brook, Department of Physics and Astronomy, Stony Brook, New York, U.S.A. \label{INSTFJ} \and\pagebreak[0] Okayama University, Department of Physics, Okayama, Japan \label{INSTGJ} \and\pagebreak[0] Osaka Metropolitan University, Department of Physics, Osaka, Japan \label{INSTCF} \and\pagebreak[0] Oxford University, Department of Physics, Oxford, United Kingdom \label{INSTGG} \and\pagebreak[0] University of Pennsylvania, Department of Physics and Astronomy,  Philadelphia, PA, 19104, USA. \label{INSTIC} \and\pagebreak[0] University of Pittsburgh, Department of Physics and Astronomy, Pittsburgh, Pennsylvania, U.S.A. \label{INSTGC} \and\pagebreak[0] Queen Mary University of London, School of Physics and Astronomy, London, United Kingdom \label{INSTFA} \and\pagebreak[0] University of Regina, Department of Physics, Regina, Saskatchewan, Canada \label{INSTE} \and\pagebreak[0] University of Rochester, Department of Physics and Astronomy, Rochester, New York, U.S.A. \label{INSTGD} \and\pagebreak[0] Royal Holloway University of London, Department of Physics, Egham, Surrey, United Kingdom \label{INSTHC} \and\pagebreak[0] RWTH Aachen University, III. Physikalisches Institut, Aachen, Germany \label{INSTBC} \and\pagebreak[0] Departamento de F\'isica At\'omica, Molecular y Nuclear, Universidad de Sevilla, 41080 Sevilla, Spain \label{INSTJB} \and\pagebreak[0] University of Sheffield, Department of Physics and Astronomy, Sheffield, United Kingdom \label{INSTFB} \and\pagebreak[0] University of Silesia, Institute of Physics, Katowice, Poland \label{INSTDI} \and\pagebreak[0] Sorbonne Universit\'e, Universit\'e Paris Diderot, CNRS/IN2P3, Laboratoire de Physique Nucl\'eaire et de Hautes Energies (LPNHE), Paris, France \label{INSTBB} \and\pagebreak[0] STFC, Rutherford Appleton Laboratory, Harwell Oxford,  and  Daresbury Laboratory, Warrington, United Kingdom \label{INSTEH} \and\pagebreak[0] University of Tokyo, Department of Physics, Tokyo, Japan \label{INSTCH} \and\pagebreak[0] University of Tokyo, Institute for Cosmic Ray Research, Kamioka Observatory, Kamioka, Japan \label{INSTBJ} \and\pagebreak[0] University of Tokyo, Institute for Cosmic Ray Research, Research Center for Cosmic Neutrinos, Kashiwa, Japan \label{INSTCG} \and\pagebreak[0] Tokyo Institute of Technology, Department of Physics, Tokyo, Japan \label{INSTHF} \and\pagebreak[0] Tokyo Metropolitan University, Department of Physics, Tokyo, Japan \label{INSTGI} \and\pagebreak[0] Tokyo University of Science, Faculty of Science and Technology, Department of Physics, Noda, Chiba, Japan \label{INSTHG} \and\pagebreak[0] University of Toronto, Department of Physics, Toronto, Ontario, Canada \label{INSTF} \and\pagebreak[0] TRIUMF, Vancouver, British Columbia, Canada \label{INSTB} \and\pagebreak[0] University of Warsaw, Faculty of Physics, Warsaw, Poland \label{INSTDJ} \and\pagebreak[0] Warsaw University of Technology, Institute of Radioelectronics and Multimedia Technology, Warsaw, Poland \label{INSTDH} \and\pagebreak[0] Tohoku University, Faculty of Science, Department of Physics, Miyagi, Japan \label{INSTIJ} \and\pagebreak[0] University of Warwick, Department of Physics, Coventry, United Kingdom \label{INSTFD} \and\pagebreak[0] University of Winnipeg, Department of Physics, Winnipeg, Manitoba, Canada \label{INSTGH} \and\pagebreak[0] Wroclaw University, Faculty of Physics and Astronomy, Wroclaw, Poland \label{INSTEA} \and\pagebreak[0] Yokohama National University, Department of Physics, Yokohama, Japan \label{INSTHE} \and\pagebreak[0] York University, Department of Physics and Astronomy, Toronto, Ontario, Canada \label{INSTH}}
\thankstext{thanks0}{also at Universit\'e Paris-Saclay} \thankstext{thanks1}{also at INFN-Laboratori Nazionali di Legnaro} \thankstext{thanks2}{also at J-PARC, Tokai, Japan} \thankstext{thanks3}{affiliated member at Kavli IPMU (WPI), the University of Tokyo, Japan} \thankstext{thanks4}{also at Moscow Institute of Physics and Technology (MIPT), Moscow region, Russia and National Research Nuclear University "MEPhI", Moscow, Russia} \thankstext{thanks5}{also at IPSA-DRII, France} \thankstext{thanks6}{also at the Graduate University of Science and Technology, Vietnam Academy of Science and Technology} \thankstext{thanks7}{also at JINR, Dubna, Russia} \thankstext{thanks8}{also at Nambu Yoichiro Institute of Theoretical and Experimental Physics (NITEP)} \thankstext{thanks9}{also at BMCC/CUNY, Science Department, New York, New York, U.S.A.}
\author{The T2K Collaboration: K.\,Abe\thanksref{INSTBJ}, N.\,Akhlaq\thanksref{INSTFA}, R.\,Akutsu\thanksref{INSTCB}, A.\,Ali\thanksref{INSTGH,INSTB}, S.\,Alonso Monsalve\thanksref{INSTEF}, C.\,Alt\thanksref{INSTEF}, C.\,Andreopoulos\thanksref{INSTFC}, M.\,Antonova\thanksref{INSTEC}, S.\,Aoki\thanksref{INSTCC}, T.\,Arihara\thanksref{INSTGI}
% ...
}

%% file: Abstract.tex
\begin{abstract}
The T2K experiment presents new measurements of neutrino oscillation parameters using $19.7(16.3)\times10^{20}$ protons on target (POT) in (anti-)neutrino mode at the far detector (FD). Compared to the previous analysis, an additional $4.7\times10^{20}$ POT neutrino data was collected at the FD. Significant improvements were made to the analysis methodology, with the near-detector analysis introducing new selections and using more than double the data. Additionally, this is the first T2K oscillation analysis to use NA61/SHINE data on a replica of the T2K target to tune the neutrino flux model, and the neutrino interaction model was improved to include new nuclear effects and calculations. Frequentist and Bayesian analyses are presented, including results on $\sin^2\theta_{13}$ and the impact of priors on the $\delta_\mathrm{CP}$ measurement. Both analyses prefer the normal mass ordering and upper octant of $\sin^2\theta_{23}$ with a nearly maximally CP-violating phase. Assuming the normal ordering and using the constraint on $\sin^2\theta_{13}$ from reactors, $\sin^2\theta_{23}=0.561^{+0.021}_{-0.032}$ using Feldman--Cousins corrected intervals, and $\Delta{}m^2_{32}=2.494_{-0.058}^{+0.041}\times10^{-3}~\mathrm{eV^2}$ using constant $\Delta\chi^{2}$ intervals. The CP-violating phase is constrained to $\delta_\mathrm{CP}=-1.97_{-0.70}^{+0.97}$ using Feldman--Cousins corrected intervals, and $\delta_\mathrm{CP}=0,\pi$ is excluded at more than 90\% confidence level. A Jarlskog invariant of zero is excluded at more than $2\sigma$ credible level using a flat prior in $\delta_\mathrm{CP}$, and just below $2\sigma$ using a flat prior in $\sin\delta_\mathrm{CP}$. When the external constraint on $\sin^2\theta_{13}$ is removed, $\sin^2\theta_{13}=28.0^{+2.8}_{-6.5}\times10^{-3}$, in agreement with measurements from reactor experiments. These results are consistent with previous T2K analyses.
\end{abstract}

%% file: Introduction.tex
The Tokai to Kamioka (T2K) experiment produces a beam of predominantly muon neutrinos by impinging protons from an accelerator onto a target, using magnetic horns to direct the outgoing collision products which thereafter decay into the neutrinos that form the beam. A suite of near detectors, 280 m downstream of the production target, characterise the neutrinos before long-baseline oscillations take effect, and a far detector, $295~\text{km}$ away, measures the long-baseline oscillations. This paper first introduces the neutrino oscillation formalism in \autoref{sec:introduction} and summarises the T2K experiment in \autoref{sec:t2k}. \autoref{sec:updates} outlines the updates to the previous analysis~\cite{Abe:2021gky,T2K:2019bcf}, with the systematic uncertainties presented in detail in \autoref{sec:flux} for the neutrino flux, and in \autoref{sec:interactionModel} for the neutrino interaction model. The analysis of near-detector data, which constrains the majority of the systematic uncertainties in the oscillation analysis, is described in \autoref{sec:nd_fit}. The far-detector selections are described in \autoref{sec:sk}, and the new constraints on the oscillation parameters are presented in \autoref{sec:oa_results}. \autoref{sec:fakeData} summarises the simulated data studies, which act to increase the uncertainty on the oscillation parameters by studying the impact of alternative interaction models. The results are summarised in \autoref{sec:conclusions}, and the data release, amongst other supplementary material, is provided in the appendices.

The observation of neutrino survival probabilities changing as a function of both flavour and distance travelled was established in the late 1990s by Super-Kamiokande (SK)~\cite{Fukuda:1998mi}. Their measurements of neutrinos produced by cosmic rays in the atmosphere found that muon neutrinos disappeared after travelling through the Earth, whereas electron neutrinos did not.
A few years later, the Sudbury Neutrino Observatory (SNO) found evidence that neutrino flavour change was responsible for the measured deficit of electron neutrinos compared to what was predicted from the Sun~\cite{Ahmad:2002jz}.
Neutrino flavour changing was also confirmed using artificial sources of neutrinos in the long-baseline reactor experiment KamLAND~\cite{Eguchi:2002dm} which measured the disappearance of \nueb, and accelerator experiments K2K~\cite{Ahn:2006zza} and MINOS~\cite{Adamson:2014vgd} which measured the disappearance of \numu and \numub. These experiments additionally characterised the oscillation curve in the ratio of the distance travelled over the neutrino energy, $L/E$, which governs the oscillation probability. 
The results can be summarised in a framework with three active neutrinos, where at least two neutrinos have non-zero mass.  
The flavour and mass eigenstates of the neutrinos, $\ket{\nu_l}$ and $\ket{\nu_i}$ respectively, are separate and can be related by a $3\times3$ unitary mixing matrix $U$ as $\ket{\nu_l}=U\ket{\nu_i}$. The mixing matrix is the Pontecorvo--Maki--Nakagawa--Sakata (PMNS) matrix, which can be \allowbreak parametrised by three mixing angles, \thsol, \thatm, \thint, and a CP-violating phase, \deltacp~\cite{Pontecorvo:1967fh,Maki:1962mu}. The probabilities for neutrino flavour oscillations can then be expressed as functions of these mixing angles and the mass-squared differences, $\Delta m^2_{ij}=m^2_i-m^2_j$ where $m_i$ is the mass of the $i$th neutrino mass eigenstate. The $m_2>m_1$ ordering was established by measurements of solar neutrinos across multiple experiments~\cite{Super-Kamiokande:2002ujc}. The ordering of the remaining mass states is unknown, with $m_3>m_2>m_1$ referred to as the normal ordering (NO), and $m_2>m_1>m_3$ as the inverted ordering (IO). This analysis uses the Particle Data Group (PDG)~\cite{Tanabashi:2018oca} convention for the order of the mixing matrices, $U=U_{23} \otimes U_{13} \otimes U_{12}$.

The results from SK, SNO, and KamLAND showed that both \thatm and \thsol were non-zero. 
The last mixing angle, \thint, was indicated to be non-zero through T2K's $2.5\sigma$ measurement of $\numu \rightarrow \nue$~\cite{Abe:2011sj}. It was later precisely measured by short-baseline experiments Daya Bay~\cite{An:2012eh}, RENO~\cite{Ahn:2012nd}, and Double Chooz~\cite{Abe:2011fz}, observing the disappearance of \nueb from nuclear reactors. The long-baseline accelerator experiments T2K and \nova subsequently observed the appearance of \nue in a \numu beam at high significance~\cite{Abe:2013hdq,Adamson:2016tbq}, and \nova observed \nueb appearance in a \numub beam at $4.4\sigma$~\cite{NOvA:2019cyt}. The non-zero \thint mixing angle implies that a measurement of \deltacp is possible in long-baseline accelerator-based experiments by measuring the appearance of \nue and \nueb in \numu and \numub beams, respectively.

On its way to the FD, the beam passes through matter and the presence of electrons modifies the oscillation probabilities as compared to those in vacuum. Namely, charged-current elastic scattering on electrons is possible for \nue and \nueb (hereafter referred to as \nueany), but not for the other flavours~\cite{Wolfenstein:1977ue,Mikheev:1986wj}. 
The sign of the matter effect differs for neutrinos and anti-neutrinos, and the magnitude is a function of the density of electrons in the path of the neutrinos, $n_e$, the weak interaction coupling strength, $G_F$, and the neutrino energy, $E$. The matter effect in the sun was central to measuring $m_2>m_1$.
The probability for \nueany appearance as a function of neutrino energy, $E$, and baseline, $L$, including a first-order approximation of the matter effects, is~\cite{Arafune:1997hd}
\begin{eqnarray}
P\left(\numuany \rightarrow \nueany\right) & \approx & \sin^2 \theta_{23} \frac{\sin^2 2 \theta_{13}}{(A-1)^2} \sin^2 [(A-1) \Delta_{31}] \nonumber \\
&\let\scriptstyle\textstyle \substack{- \\ (+)} & \alpha \frac{J_0 \sin  \delta_{CP}}{A (1-A)} \sin \Delta_{31} \sin (A\Delta_{31})\sin[ (1-A) \Delta_{31}] \nonumber \\
&+& \alpha \frac{J_0 \cos  \delta_{CP}}{A (1-A)} \cos \Delta_{31} \sin (A\Delta_{31})\sin[ (1-A) \Delta_{31}] \nonumber \\
&+& \alpha^2 \cos^2 \theta_{23} \frac{\sin^2 2 \theta_{12}}{A^2} \sin^2 ( A\Delta_{31} )
\label{eq:psurv_nue}
\end{eqnarray}
where
\begin{eqnarray*}
 \alpha &       =& \Delta m^2_{21} / \Delta m^2_{31} \\
 \Delta_{ij}&   =& \Delta m^2_{ij} L / 4 E\\
  A&            =&(-) 2\sqrt 2 G_F n_e E/ \Delta m^2_{31}\\
  J_0 &         =& \sin 2 \theta_{12}\sin 2 \theta_{13}\sin 2 \theta_{23}\cos  \theta_{13}.
\end{eqnarray*}
The first term in \autoref{eq:psurv_nue} is proportional to
\ssqthtwothree, which renders the \nueany appearance sensitive to whether \thatm is above or below $\pi/4$, referred to as the octant of \thatm. This in turn determines whether the $\nu_3$ mass eigenstate has a larger admixture of \numu or \nutau. The term containing \sindcp in \autoref{eq:psurv_nue} has the opposite sign for neutrinos and anti-neutrinos, and allows for CP symmetry violation if \deltacp is different from 0 or $\pi$.
The term containing \cosdcp does not violate CP symmetry, but can change the shape of the \nueany appearance energy spectrum, and is important for precisely measuring \deltacp. In T2K, the term proportional to \sindcp can change the appearance probability by as much as $\pm30\%$ given the current knowledge of the other mixing angles. $J=J_0\sindcp$ is referred to as the Jarlskog invariant~\cite{PhysRevLett.55.1039, JARLSKOG2005323} and is a basis-independent measure of the CP-violation. This analysis presents T2K's constraints on \dmsqtwothree, \ssqthtwothree, \ssqthonethree, \deltacp, $J$, and the mass ordering.

%% file: T2K.tex
To measure \deltacp and the other oscillation parameters, T2K uses a beamline that produces predominantly muon-flavoured neutrinos or anti-neutrinos with a peak energy of $E_\nu\approx0.6~\text{GeV}$, and has been alternating between neutrino and anti-neutrino configurations since 2014.
A suite of near detectors (NDs), approximately 280 m from the beam production target, characterise T2K's neutrino beam before long-baseline oscillations become likely. The far detector (FD) is 295 km away and measures the appearance of \nueany and the disappearance of \numuany in the \numuany-dominated beam. The rate and directional stability of the neutrino beam are measured by the on-axis neutrino ND, INGRID. The second ND, ND280, and the FD, SK, are $2.5\degree$ off-axis with respect to the upstream proton beam that impinges on the neutrino production target. By being placed off-axis, the detectors sample a narrower neutrino energy distribution, peaking near the maximum of the \nueany appearance spectrum.

\subsection{Beamline}
The T2K neutrino beam is produced at the Japan Proton Accelerator Research Complex (J-PARC) in Tokai, Ibaraki, by a high-intensity proton beam, incident on a production target~\cite{Abe:2011ks}. At J-PARC, $\text{H}^-$ ions from an ion source are accelerated to an energy of 400~MeV in a linear accelerator. Charge-stripping foils convert the beam to $\text{H}^+$ at injection into the rapid-cycling synchrotron, which accelerates the proton beam to 3~GeV. These protons are then injected into the main ring (MR) synchrotron, where they are accelerated to 30~GeV. The proton beam from the MR consists of eight bunches with width $\sim80~\text{ns}$ ($3\sigma$), referred to as a ``spill'', produced every 2.48~s.
\begin{figure*}[htbp]
    \centering
    \includegraphics[width=\linewidth]{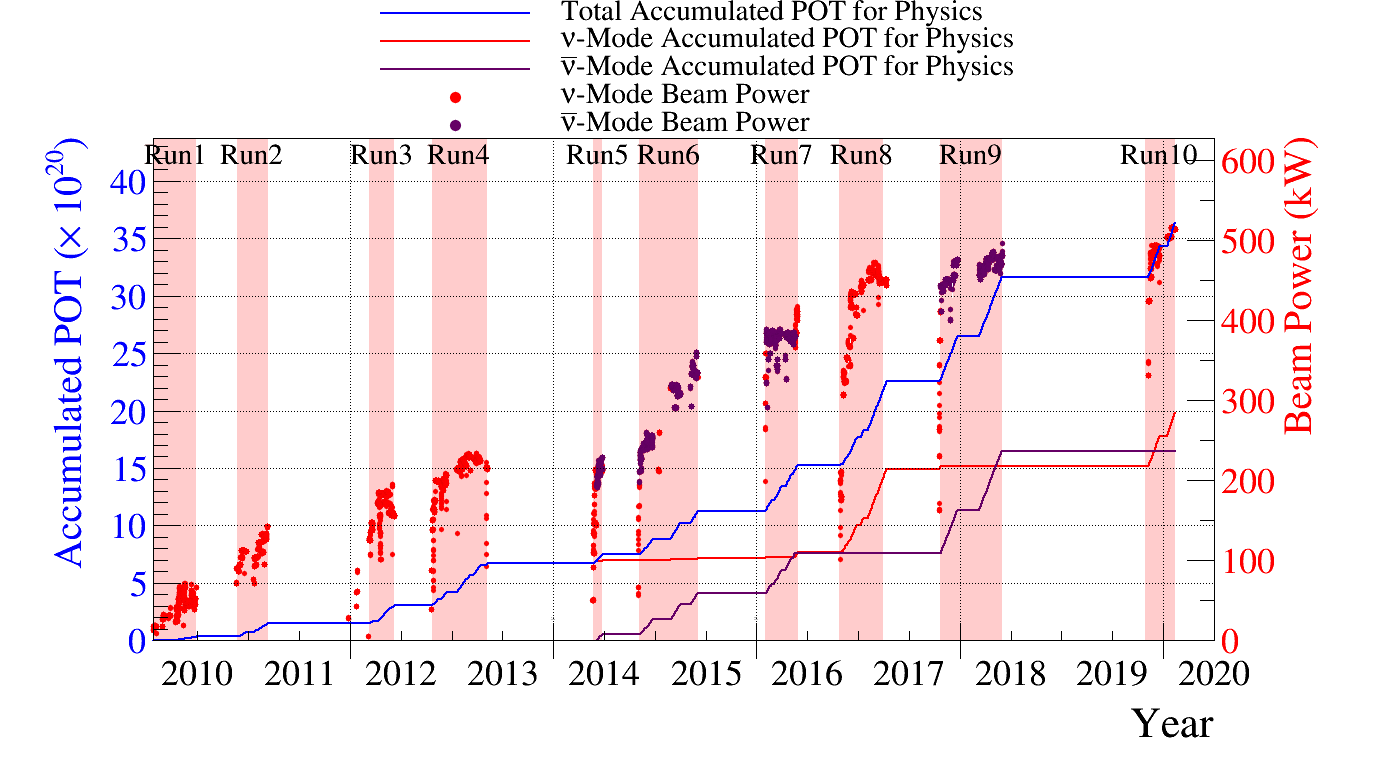}
    \caption{The protons on target (POT) delivered to T2K by the MR over time, with the beam intensity overlaid. The ND280 analysis uses runs 2 to 9, and the INGRID and FD analyses use runs 1 to 10, with run-by-run POT listed in \autoref{tab:pot_2020}.}
    \label{fig:POT}
\end{figure*}

The protons are extracted from the MR to the neutrino beamline, which consists of a series of normal- and super-conducting magnets that are used to bend the proton beam in the direction of the FD, and to focus the beam onto the neutrino production target. The proton beam power, as well as the position, angle, and size of the proton beam at the target, are precisely measured by a series of proton beam monitors~\cite{Abe:2011ks,BHADRA201345} installed along the neutrino beamline.

The 30~GeV protons strike a 91.4~cm-long monolithic graphite target installed in the first of three electromagnetic focusing horns. Outgoing charged pions and kaons are focused by these horns, which have been operating at a current of $\pm250~\text{kA}$ for nearly the full T2K run to date. The polarity of the horns can be set to focus either positively or negatively charged outgoing particles, and a 96~m-long decay volume is located directly downstream of the focusing system. Positively charged pions decay into positively charged muons and muon neutrinos, whilst negatively charged pions decay into negatively charged muons and muon anti-neutrinos. The former is referred to as \fhcalt and the latter as \rhcalt. Kaon and muon decays are the primary contributors to the \nue contamination in the \numu-dominated beam. 

A beam dump is situated at the end of the decay volume and absorbs surviving hadrons. A muon monitor downstream of the beam dump, MUMON~\cite{SUZUKI2012453}, measures the intensity and profile of muons that have more than 5 GeV of energy. This measurement is used as a proxy for stability of the associated neutrino beam.
The predicted neutrino fluxes and uncertainties are described in detail in \autoref{sec:flux}. 

The MR proton beam power has reached a maximum of 515~kW, and the protons on target (POT) and power history are shown in \autoref{fig:POT}. Scheduled upgrades will increase the beam power to 1.3~MW and operate the focusing horns at $\pm$320~kA current. This will significantly increase the POT per run cycle and provide more neutrinos at the ND and FD per POT. It will also reduce the \numub and \numu backgrounds in \fhcalt and \rhcalt~\cite{T2K:2019eao,Oyama:2020kev}, respectively, referred to as the wrong-sign component.

\subsection{Near detectors}
Two NDs are used directly in the oscillation analysis: the on-axis INGRID, and the off-axis ND280. Both detectors are housed in the same pit underground, with the centres of ND280 and INGRID approximately 24~m and 33~m, respectively, below the surface.
\begin{figure}[htbp]
    \centering
    \includegraphics[width=0.3\textwidth]{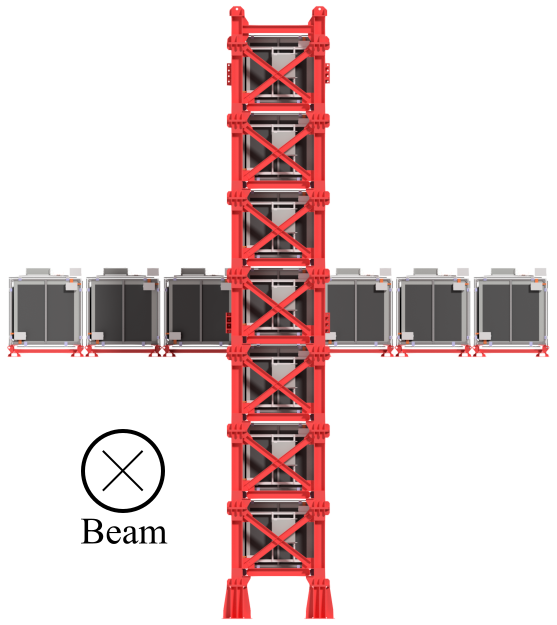}
    \caption{The INGRID on-axis ND, used to measure the neutrino beam profile and rate~\cite{Abe:2011xv}. The beam direction is shown as into the paper.}
    \label{fig:ingrid}
\end{figure}

INGRID~\cite{Abe:2011xv} is designed to measure the profile and stability of the neutrino beam. It samples the beam spill-by-spill with a transverse cross section of $10\times10~\text{m}^2$ with 14 identical modules arranged as a cross, as shown in \autoref{fig:ingrid}. Each of the modules alternates iron target plates of 6.5 cm thickness with tracking scintillator planes of 1 cm thickness, for a total of 9 iron plates and 11 scintillator planes, and is surrounded by scintillator planes acting as vetoes. A module exposes a $1.24\times1.24~\text{m}^2$ area facing the beam, and provides a 7.1~t target mass. INGRID measures the beam direction with an accuracy higher than 0.4~mrad, within the required precision of $\pm1~\text{mrad}$ for the oscillation analysis.
\begin{figure}[htbp]
    \centering
    \includegraphics[width=0.4\textwidth]{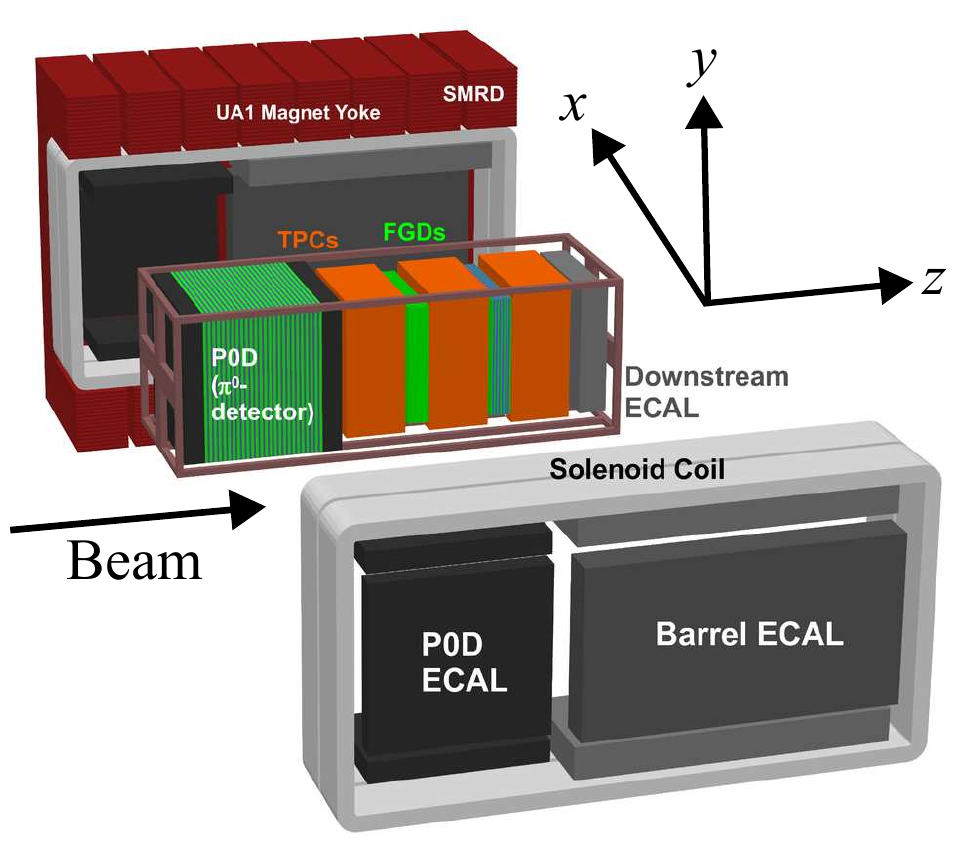}
    \caption{The ND280 off-axis ND, used to measure the neutrino flux and interactions before long-baseline oscillations~\cite{Abe:2011ks}. The detector coordinates and beam direction are superimposed, with the sub-detectors are labelled accordingly.}
    \label{fig:nd280}
\end{figure}

ND280, hereafter referred to as the ND, is used to constrain the uncertainties on the neutrino flux and interactions in the analysis. It is a magnetised detector consisting of different sub-detectors as shown in \autoref{fig:nd280}. The ND measures $5.6~\text{m} \times 6.1~\text{m} \times 7.6~\text{m}$ (width $\times$ height $\times$ length) around its outer edges including the magnet with the coordinate convention being $z$ pointing along the nominal neutrino beam axis, with $x$ and $y$ being the horizontal and vertical directions, respectively.
The refurbished magnet from the UA1~\cite{Astbury:1978uia,UA1:1980ndt} and NOMAD~\cite{NOMAD:1997pcg} experiments at CERN provides a magnetic field of 0.2~T, and the magnet yoke is instrumented with layers of plastic scintillator called the Side Muon Range Detector (SMRD)~\cite{Aoki:2012mf}. 
Inside the magnet enclosure there is an electromagnetic calorimeter (ECal)~\cite{Allan:2013ofa} surrounding the inner detector, which is used to distinguish track-like and shower-like objects, and is made of alternating layers of plastic scintillator and lead. 

The inner detector region houses the $\pi^0$ detector (\poddet)~\cite{Assylbekov:2011sh} in the upstream portion, which is made of alternating layers of water bags, brass sheets, and triangular $x-y$ scintillator planes. The water bags can be filled with either water or air. The \poddet has its own ECal modules upstream and downstream of the water target region, made from alternating scintillator planes and lead sheets.
The \poddet, ECal and SMRD also act as vetoes for interactions originating outside the detector, e.g. cosmic-ray muons and neutrino interactions in the sand upstream of the detector hall.
Downstream in the direction of the FD, there are two Fine-Grained Detectors (FGDs)~\cite{Amaudruz:2012esa}, which are each sandwiched by Time Projection Chambers (TPCs)~\cite{Abgrall:2010hi}. These sub-detectors are together referred to as the ``tracker''.
The most upstream FGD (FGD1) is made of 15 polystyrene scintillator modules. One module is $186.4~\text{cm}\times186.4~\text{cm}\times2.02~\text{cm}$ and consists of two scintillator layers oriented in $x$ and $y$, with each layer containing 192 9.6 mm wide square bars approximately 2 m long, which are read out at one end. The second FGD (FGD2) contains six passive water modules, each sandwiched by polystyrene scintillator modules identical to those in FGD1. The TPCs use a $\text{Ar}:\text{CF}_4:i\text{C}_4\text{H}_{10}$ gas mixture in a 95:3:2 concentration, and have a space point resolution of approximately 1 mm.

This analysis selects interactions occurring in either FGD, using the FGDs and TPCs for track reconstruction and particle identification. The selection is detailed in \autoref{subsec:nd_sel}. The FGDs are capable of tracking charged particles, performing particle identification, and calculating momentum-by-range for contained particles.
The TPCs are three-dimensional trackers which measure momentum through the curvature of the tracks in the magnetic field, with a resolution of $\frac{\delta p_\perp}{p_\perp} \sim 0.1 p_\perp$, where $p_\perp$ is the momentum perpendicular to the magnetic field. The TPCs also provide excellent particle identification. 

\begin{table}[htbp]
	\centering
	\begin{tabular}{ l c c | c | c c }
    \hline
    \hline
		Run     &  Run    & Run  & Beam      & \multicolumn{2}{c}{POT $(\times10^{19})$} \\
		number   &  start  & end  & mode & ND & FD \\
		\hline
		1 & Jan. 2010 & Jun. 2010 & $\nu$ & ---     & 3.26 \\
		2 & Nov. 2010 & Mar. 2011 & $\nu$ & 7.93    & 11.22 \\
		3 & Mar. 2012 & Jun. 2012 & $\nu$ & 15.81   & 15.99 \\
		4 & Oct. 2012 & May 2013  & $\nu$ & 34.26   & 35.97 \\
		5 & May 2014  & Jun. 2014 & $\overline{\nu}$ & 4.35    & 5.12 \\
		  &           &           & $\nu$ & ---     & 2.44 \\
		6 & Oct. 2014 & Jun. 2015 & $\overline{\nu}$ & 34.09   & 35.46 \\
		  &           &           & $\nu$ & ---     & 1.92 \\
		7 & Feb. 2016 & May 2016  & $\overline{\nu}$ & 24.38   & 34.98 \\
		  &           &           & $\nu$ & ---     & 4.84 \\
		8 & Oct. 2016 & Apr. 2017 & $\nu$ & 57.31   & 71.69 \\
		9 & Oct. 2017 & May 2018  & $\overline{\nu}$ & 20.54   & 87.88 \\
		  &           &           & $\nu$ & ---     & 2.04 \\
		10& Oct. 2019 & Feb. 2020 & $\nu$& ---     & 47.26 \\
		\hline
		\multicolumn{3}{c|}{Total} & $\nu$ & 115.31 & 196.64 \\
		\multicolumn{3}{c|}{Total} & $\overline{\nu}$ & 83.36  & 163.46 \\
		\hline
		\multicolumn{3}{c|}{Total} & $\nu+\overline{\nu}$ & 198.67 & 360.10 \\
	\hline
	\hline
	\end{tabular}
	\caption{Collected protons-on-target (POT) for each T2K run included in the analysis of T2K data at the ND and FD. The recorded POT at INGRID closely follows that of the FD.}
	\label{tab:pot_2020}
\end{table}

\subsection{Far detector}
The Super-Kamiokande (SK) detector~\cite{Abe:2011ks,Fukuda:2002uc} is the far detector (FD) for T2K. SK is a large water Cherenkov detector located 295.3~km from the neutrino production target with a 2.7 km water-equivalent overburden.
It is filled with 50~kt of ultrapure water that is optically separated into an inner detector, ID, which forms the primary target for neutrino interactions, and an outer detector, OD, which serves to veto external backgrounds. 

The ID is instrumented with 11,129 inward-facing photomultiplier tubes (PMTs) with 20-inch diameter, providing a total photocathode coverage of 40\%. The OD is instrumented with 1,885 8-inch outward-facing PMTs, which are connected to wavelength shifting plates and are attached to the same stainless steel structure that houses the ID PMTs. 
The structure is offset 2~m from the wall of the OD and there is a 55~cm dead region between the ID and OD surfaces.

Charged particles are detected by their Cherenkov ring pattern, and events are classified by the number of primary rings, the ring pattern of each ring, and the number of time-delayed electron rings consistent with a muon decay, hereafter referred to as ``Michel electrons''. This analysis selects single-ring (``1R'') events, where the ring is either electron-like (\re) or muon-like (\rmu), with a selection-dependent cut on the number of delayed Michel electrons (``\de''). The FD selections are detailed in \autoref{sec:sk}.

The data used in this analysis were taken over two different periods of the SK detector operations and span the years 2010--2020, during what is referred to as the SK-IV period.
Of the $36.01\times10^{20}$ POT reported here, $31.29\times10^{20}$ (runs 1--9) were collected in 2010--2018.
In June 2018, SK detector operations were stopped for refurbishment in preparation for the gadolinium (Gd) loading of the water target for the SK-Gd project~\cite{Super-Kamiokande:2021the,Beacom:2003nk}.
During this work the detector surfaces were cleaned to remove rust and other impurities, detector walls were repaired to fix minor leaks, and failed PMTs were replaced in the ID and OD. This SK detector period is referred to as SK-V.

SK-V resumed data taking in January 2019 with ultrapure water and collected $4.73\times10^{20}$ POT during October 2019--February 2020 (run 10). These data were collected entirely in \fhcalt, resulting in a total of $19.66\times10^{20}$ and $16.34\times10^{20}$ POT available for analysis in the \numu and \numub modes, respectively. For a detailed breakdown of the POT in each run period, consult \autoref{tab:pot_2020}. Gadolinium loading commenced in July 2020, and this analysis does not include such data.

%% file: Updates.tex
This section provides an overview of the improvements to T2K's previously published oscillation analysis~\cite{Abe:2021gky,T2K:2019bcf}, which are detailed in the subsequent sections. 

\begin{itemize}
    \item{\textbf{Data at the FD:} The data at INGRID and the FD increased by $4.73\times10^{20}$ POT (+33\%) in \fhcalt, increasing the overall amount by 15\%, detailed in~\autoref{sec:sk}.}
    
    \item{\textbf{Data at the ND:} The data at the ND increased by $5.73\times10^{20}$ POT (+99\%) in \fhcalt, and by $4.48\times10^{20}$ POT (+116\%) in \rhcalt, increasing the overall amount by 106\%, detailed in~\autoref{sec:nd_fit}.}

    \item{\textbf{Selections at the ND:} The increased data allowed for refining the \rhcalt selections and re-binning all existing selections, improving the constraints on the systematic uncertainties from the ND in the oscillation analysis, detailed in~\autoref{subsec:nd_sel}.}
    
    \item{\textbf{FD reprocessing:} An updated model for the dark rate and gain drift in the PMTs had a slight impact on the reconstruction and the number of observed data events. The processing introduced one more \rhcalt electron-like event, and three fewer \rhcalt muon-like events, and had no overall effect on the \fhcalt samples, detailed in ~\autoref{sec:sk}.}
    
    \item{\textbf{Neutrino flux model:} The neutrino flux was constrained using charged pion production data on a replica of the T2K production target from NA61/SHINE~\cite{Abgrall:2016jif}. Data on a thin target~\cite{Abgrall:2015hmv} was also used when appropriate. This reduced the flux uncertainties before the ND analysis from $\sim9\%$ down to $\sim5\%$ in the neutrino flux peak, detailed in~\autoref{sec:flux}.}
    
    \item{\textbf{Neutrino interaction model:} 
    Several changes to the neutrino interaction model were made. The largest changes were switching to a more sophisticated spectral-function based nuclear model~\cite{Benhar:1994hw} for charged-current quasi-elastic (CCQE) interactions, introducing an additional uncertainty due to nuclear effects in the four-momentum transferred to the nucleus ($Q^2$), and adding an uncertainty for the nucleon removal energy. The nuclear-cascade model for pions was tuned to external data~\cite{PinzonGuerra:2018rju}, and the FD parametrisation was constrained by the fit to ND data, whereas it was previously allowed to vary separately. The interaction model for pions re-scattering within the detector at the ND and FD were unified, and is identical to the pion final-state interaction model, detailed in~\autoref{sec:interactionModel}. However, constraints of re-scattering within the ND were not propagated to re-scattering at the FD, as the uncertainties were kept uncorrelated.}
\end{itemize}

%% file: Flux.tex
This is the first T2K oscillation analysis to use hadron production measurements made on a replica of the T2K target by the NA61/SHINE experiment at CERN~\cite{Abgrall:2016jif}. The method for predicting the neutrino flux and propagating the associated uncertainties remains the same as in  previous results~\cite{Abe:2021gky,T2K:2019bcf,Abe:2012av}. FLUKA~2011.2x~\cite{Bohlen:2014buj,Ferrari:2005zk} is used to simulate interactions inside the target. The outgoing particles from the target, which later decay to neutrinos, are tracked through the horn field using the GEANT3-based JNUBEAM package~\cite{T2K:2012bge}.

The prediction for pions exiting the target's surface are tuned to $\pi^+$ and $\pi^-$ yields measured by the NA61/SHINE experiment, using data collected in 2009 with a replica of the T2K production target~\cite{Abgrall:2016jif}. Pions that leave the target and are within the phase space covered by the replica target data, which is about 90\% of the neutrinos at the flux peak, are given a weight
\begin{equation}
w(p,\theta,z,i) = \frac{\mathrm dn^\mathrm{NA61}(p,\theta,z,i)}{\mathrm dn^\mathrm{MC}(p,\theta,z,i)}
\end{equation}
where $\mathrm dn$ is the POT-normalised differential yield for data (``NA61'') and simulation Monte-Carlo (``MC''), with exiting momentum $p$, polar angle $\theta$, and longitudinal position $z$ along the target for an exiting particle of type $i = \{\pi^+,\pi^-\}$. For the particles leaving the target, no additional tuning weight is applied for any of the interactions or trajectories inside the target. 
Simulations for particles that are not covered by the replica target data, and interactions occurring outside the target, are tuned to NA61/SHINE data on $\pi^\pm$, $K^\pm$, $K^0_s$, $\Lambda$, and $p$ yields from a thin target taken in 2009~\cite{Abgrall:2015hmv}, and other external measurements, applying the same method as previous T2K analyses~\cite{Abe:2012av}. The percentage of hadronic interactions which are tuned by external data is shown in \autoref{tab:fluxtune}. 

\begin{table}[htbp]
    \centering
    \begin{tabular}{l | c c c c}
    \hline
    \hline
    & \numu & \numub & \nue & \nueb \\ 
    \hline
    \fhcalt & 96.5\% & 87.6\% & 90.5\% & 77.8\% \\
    \rhcalt & 87.8\% & 96.2\% & 78.3\% & 91.1\% \\
    \hline
    \hline
    \end{tabular}
    \caption{Percentage of hadronic interactions in the target and downstream beam line for which external measurements are used in the tuning or uncertainty evaluation. The interactions are weighted by their contribution to the neutrino flux at the FD, separated into different horn focusing modes and neutrino flavours.}
    \label{tab:fluxtune}
\end{table}

In the previous thin-target tuning, a large uncertainty on the cross section of proton production was assigned. In the replica-target based tuning, this uncertainty is no longer necessary for particles covered by the replica target data, because the exiting particle yields can be tuned directly without referring to the interaction history inside the target. The uncertainties from NA61/SHINE are then incorporated with the uncertainties associated with the proton beam profile and out-of-target interactions to give the total uncertainty.  

\begin{figure}[htbp]
\includegraphics[width=0.49\textwidth,trim=0mm 18mm 10mm 0mm, clip]{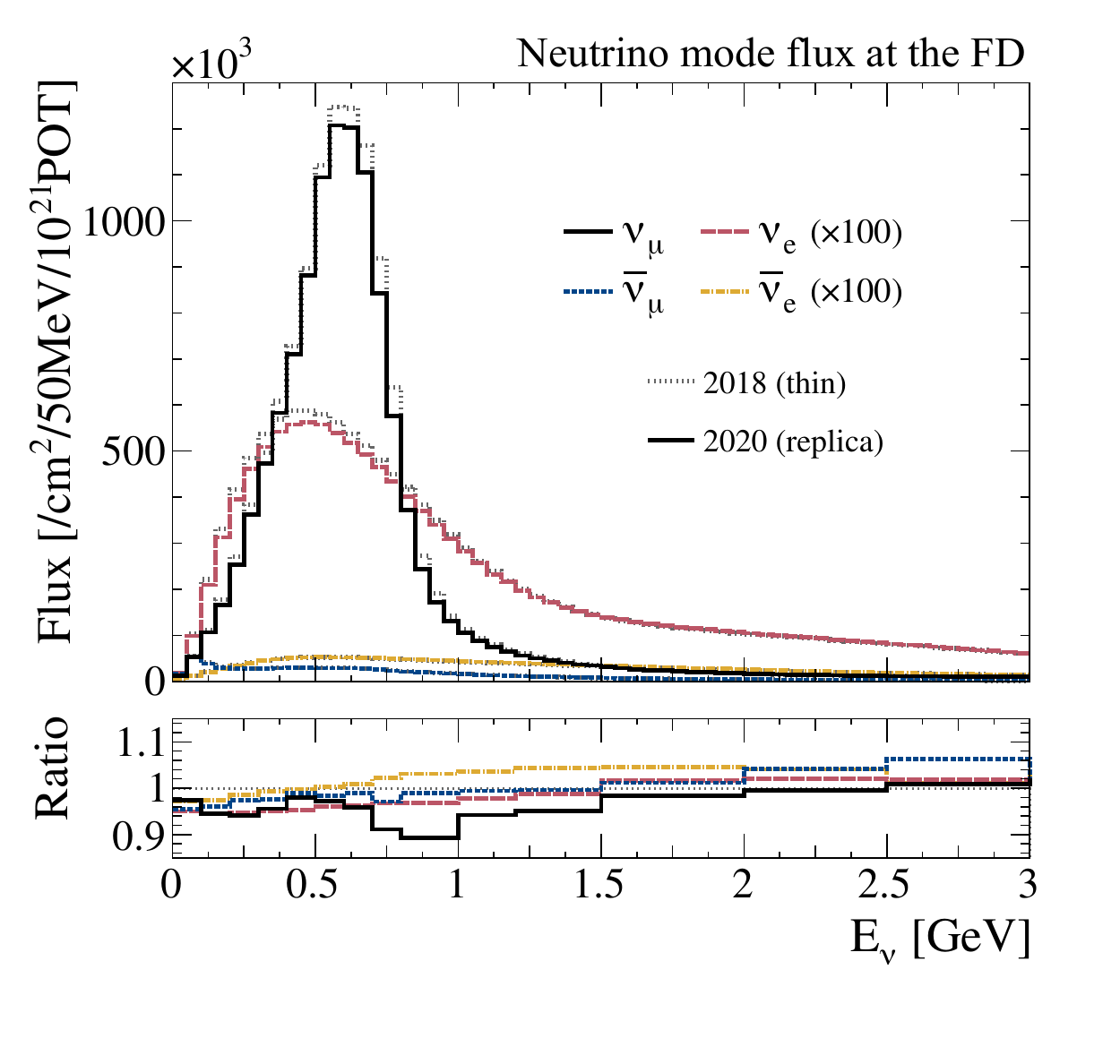}
\includegraphics[width=0.49\textwidth,trim=0mm 18mm 10mm 0mm, clip]{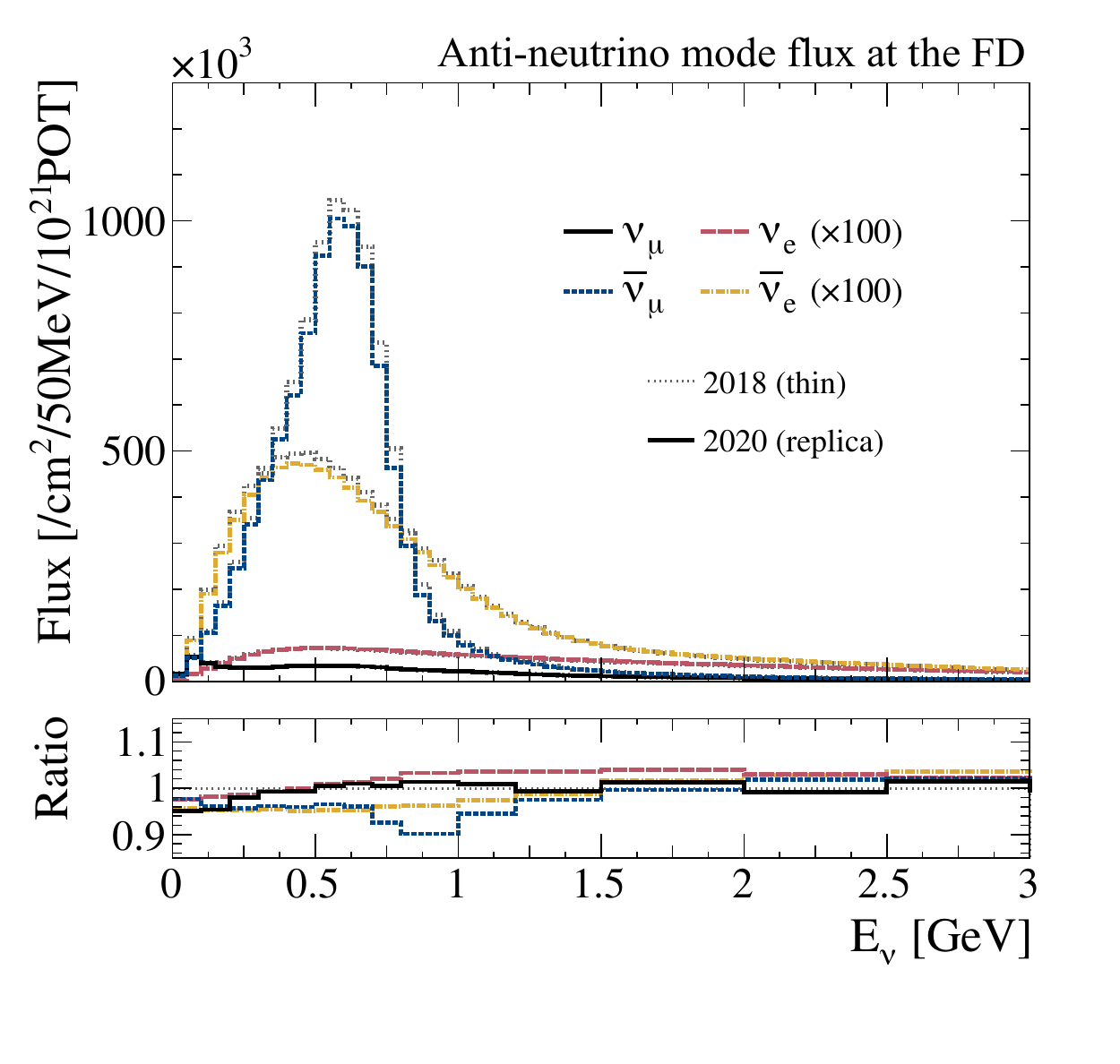}
\caption{The predicted unoscillated neutrino fluxes at the FD in \fhcalt (top) and \rhcalt (bottom). The \nue and \nueb components are scaled by $\times100$. The solid lines show the predictions after tuning to NA61/SHINE data on the T2K replica target, and the dotted grey lines show the predictions in the previous T2K analysis~\cite{Abe:2021gky,T2K:2019bcf}, tuned to thin target hadron production data. The bottom inset shows the ratio of the flux from the replica target tuning to the flux from the thin target tuning.
}
\label{fig:flux:comp}
\end{figure}

\begin{figure}[htbp]
\includegraphics[width=0.49\textwidth,trim=0mm 5mm 0mm 0mm, clip]{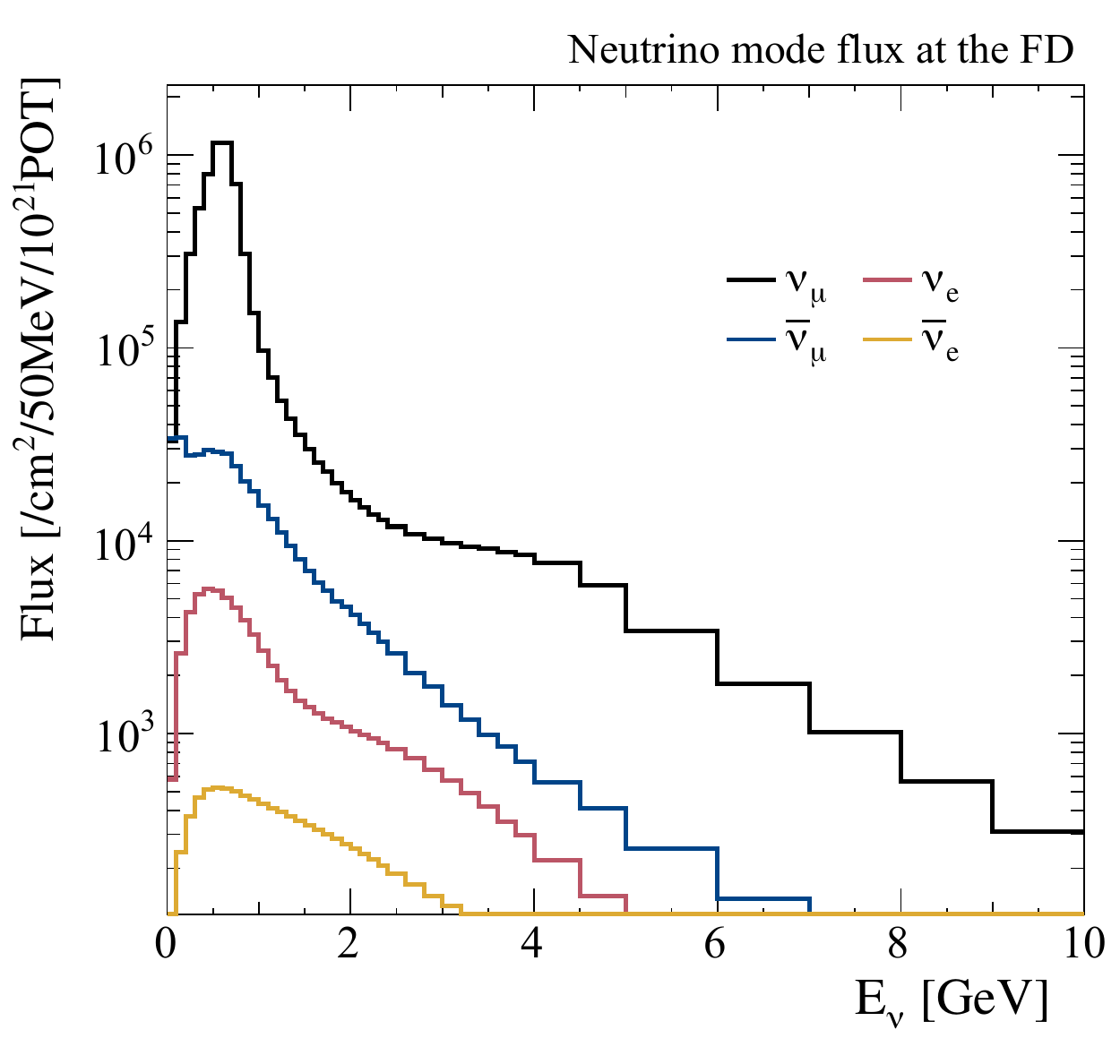}
\includegraphics[width=0.49\textwidth,trim=0mm 5mm 0mm 0mm, clip]{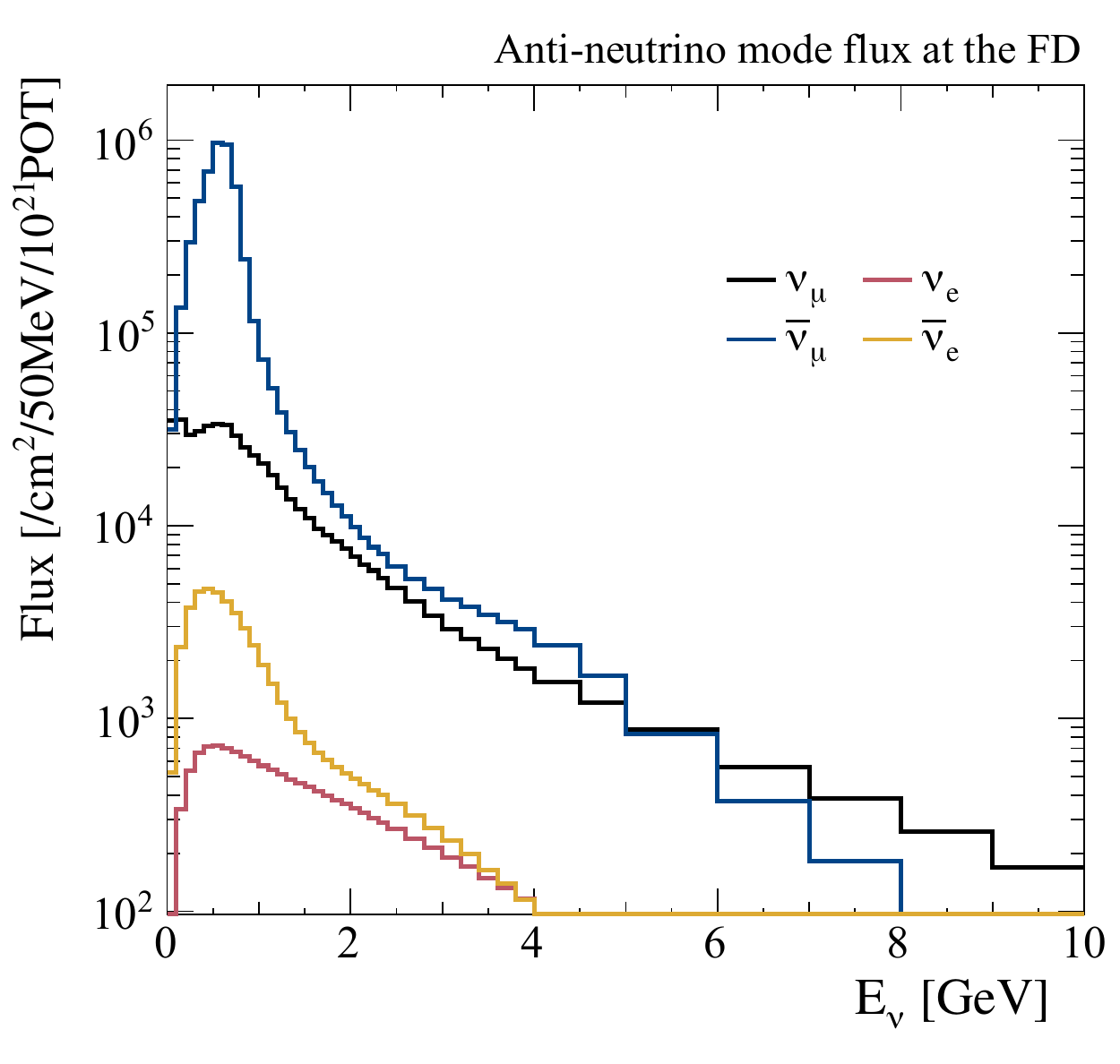}
\caption{The predicted unoscillated neutrino fluxes at the FD in \fhcalt (top) and \rhcalt (bottom) in logarithmic scale with an extended $E_\nu$ range, after the tuning to NA61/SHINE data on the T2K replica target.
}
\label{fig:flux:log}
\end{figure}

\begin{figure}[htbp]
\includegraphics[width=0.49\textwidth]{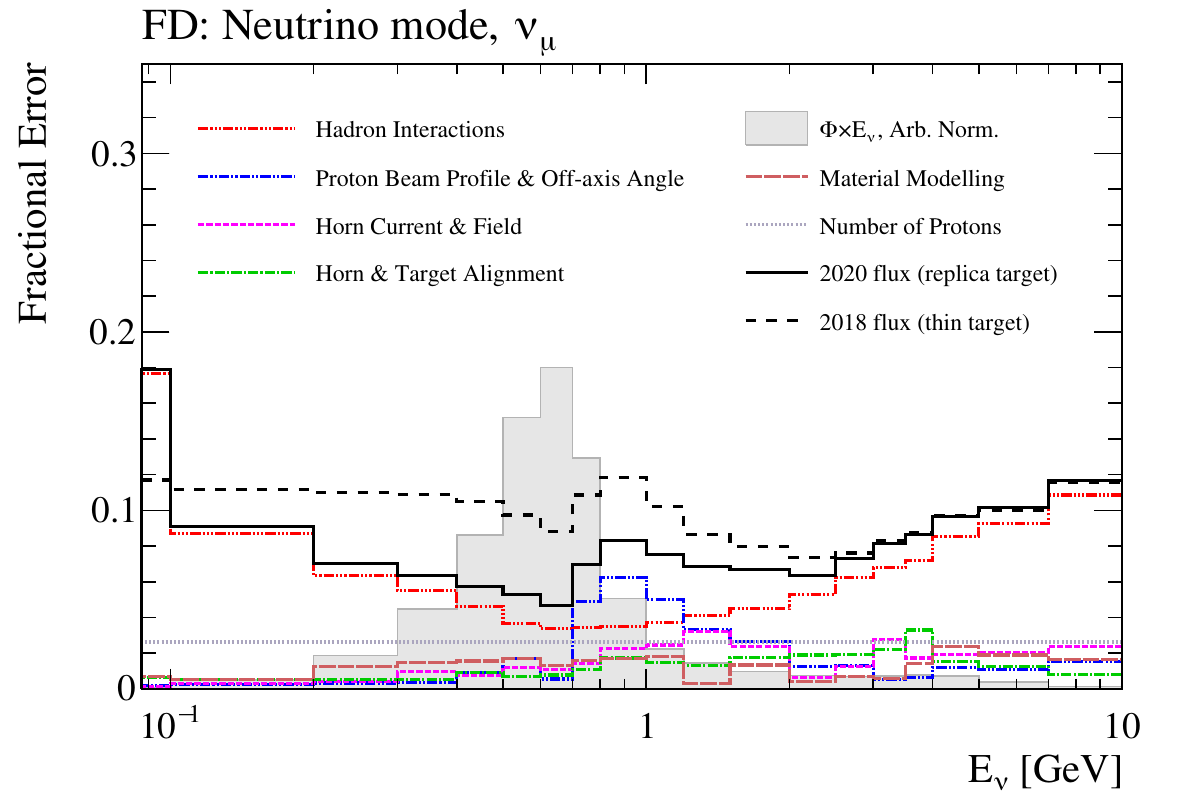}
\includegraphics[width=0.49\textwidth]{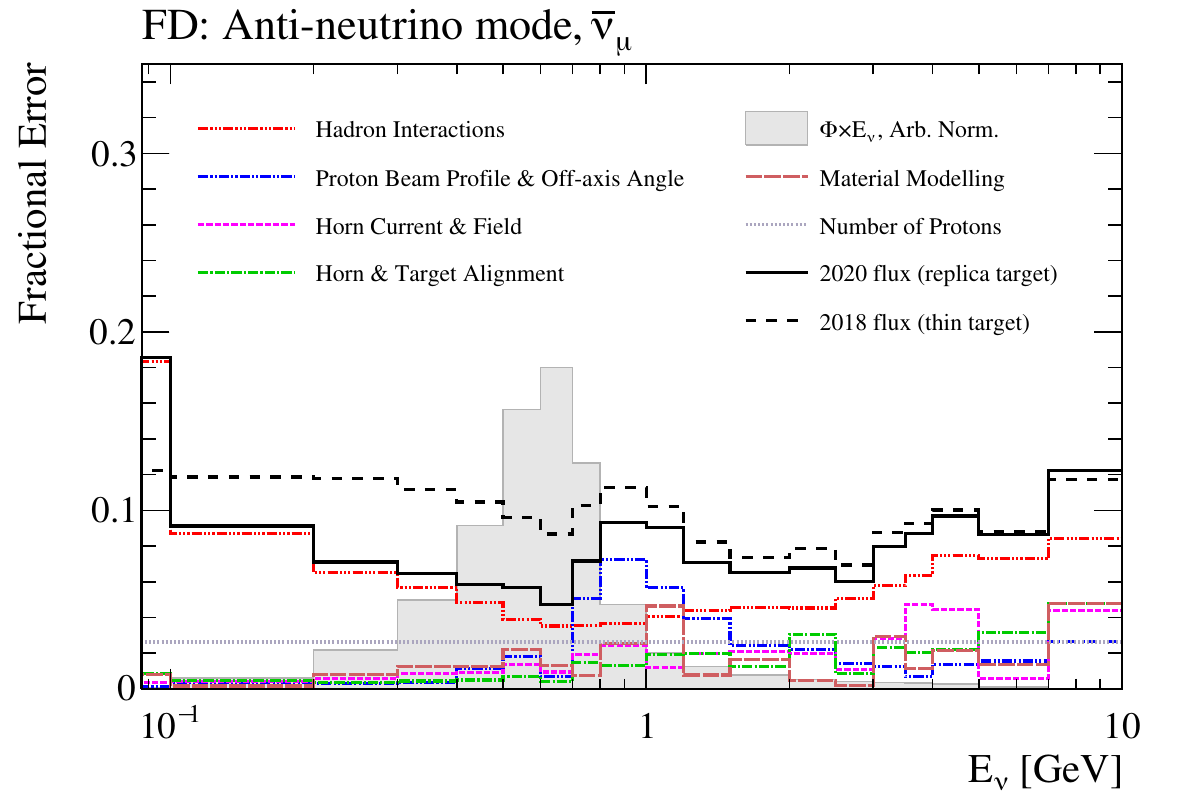}
\includegraphics[width=0.49\textwidth]{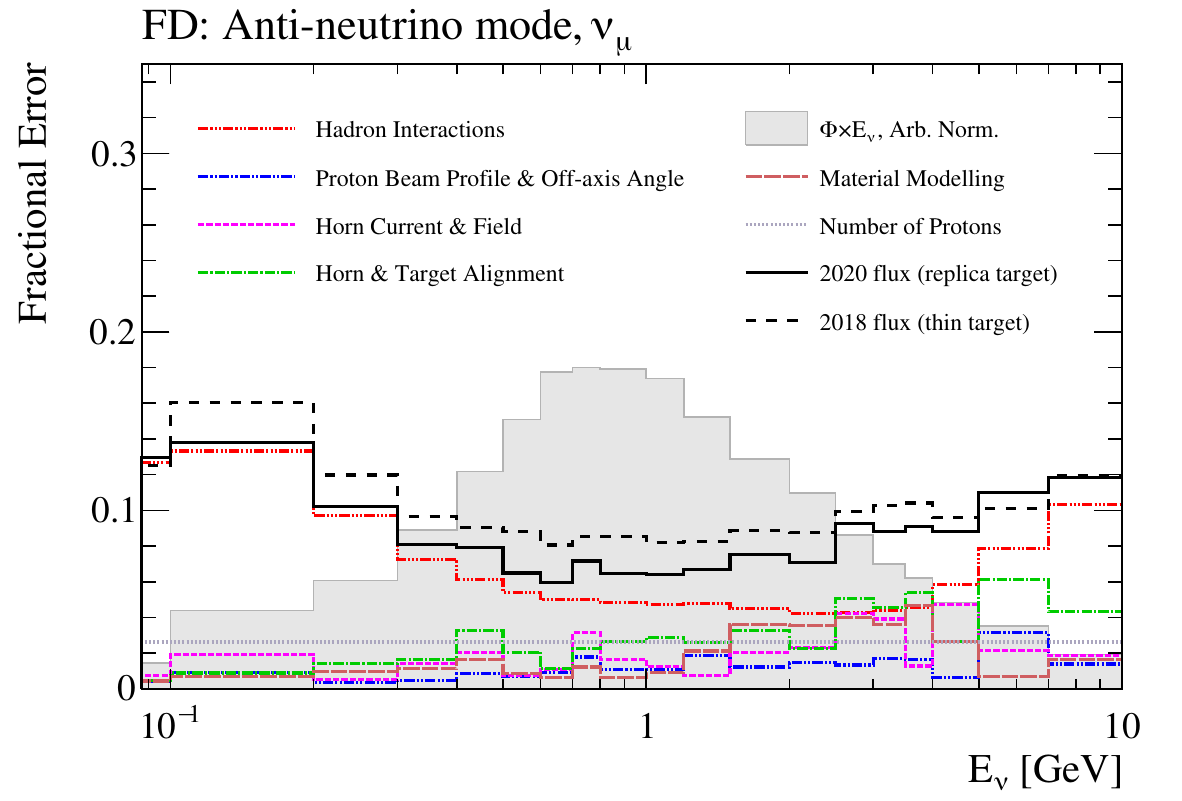}
\caption{Uncertainty on the right-sign flux in \fhcalt (top) and right- (middle) and wrong-sign (bottom) fluxes in \rhcalt, broken down by the sources of uncertainty. The solid black line shows the total flux uncertainty in this analysis, and the dashed black line shows the total uncertainty for the previous T2K analysis~\cite{Abe:2021gky,T2K:2019bcf}, which used NA61/SHINE thin target data. The grey shaded region shows the shape of the neutrino flux.}
\label{fig:flux:uncertainty}
\end{figure}

For the unconstrained interactions not covered by thin- or replica-target data, a systematic uncertainty is calculated by dividing the kinematic phase space parametrised by Feynman-$x_\mathrm{F}$ and transverse momentum, $p_\mathrm{T}$, into six regions. A 50\% fully correlated normalisation uncertainty and a 50\% shape uncertainty uncorrelated between the regions is assigned. The size of the uncertainty is motivated by comparing the hadron interaction models in FLUKA 2011.2c~\cite{Ferrari:2005zk,Bohlen:2014buj} and the GEANT 4.10.03~\cite{GEANT4:2002zbu} \texttt{FTFP\_BERT} and \texttt{FTF\_BIC} physics lists.

The predicted flux distributions are provided in Ref.~\cite{megan_friend_2021_5734307} and are shown for the FD in \autoref{fig:flux:comp}. The largest difference compared to the previous neutrino flux prediction is the reduction of the \numu component in \fhcalt, and the \numub component in \rhcalt (``right-sign''), by 5--10\% around the flux peak. 
Due to the large uncertainty on the hadron interactions in the previous tuning, this difference was covered by the flux uncertainties. To more clearly see wrong-sign and background contributions, the predicted neutrino flux spectra are also shown in logarithmic scale and for a wider range of energies in \autoref{fig:flux:log}.

Overall, tuning with the NA61/SHINE 2009 replica target data reduces the uncertainty from 9\% to 5\% near the flux peak, as shown in \autoref{fig:flux:uncertainty}. In future T2K analyses, outgoing kaons will also be tuned using NA61/SHINE T2K replica target data from 2010, published in 2019~\cite{Abgrall:2019tap}. This will reduce the flux uncertainty at higher energies to $\sim5\%$. 
With a reduced uncertainty contribution from hadron production errors, uncertainties coming from other sources are now dominant in some energy regions.  In particular, uncertainties on the proton beam profile and neutrino beam off-axis angle significantly contribute to the uncertainty on the high-energy edge of the flux peak, since the width of the energy spectrum is directly affected by shifts in the off-axis angle.

%% file: Interaction.tex
Measurements of neutrino oscillations at T2K rely on comparing the neutrino interaction rates at the ND and the FD as a function of the incoming neutrino energy and flavour. 
These are determined from the observed products of neutrinos interacting with the nuclei inside the detectors, which requires a model to translate what is observed in the detector to information about the neutrino that interacted. 
Neutrino interaction uncertainties impact the oscillation analysis by changing the expected rate of neutrino interactions, altering the accuracy of the neutrino energy reconstruction, and complicating the extrapolation of model constraints from the ND to the FD. More details can be found in references~\cite{Abe:2021gky,Alvarez-Ruso:2017oui,Mosel:2016cwa,Katori:2016yel}.

The neutrino interaction model has been significantly improved since the last analysis~\cite{Abe:2021gky}. This section first provides an overview of the components of the model and then discusses the associated uncertainties and their parametrisations. As briefly mentioned in~\autoref{sec:t2k} and detailed further in~\autoref{subsec:nd_sel} and \ref{sec:sk}, this analysis selects charged-current (CC) neutrino interaction events and has no dedicated neutral-current (NC) selections. The oscillation analysis at the FD specifically selects single-ring events and the model focuses on the treatment of such interactions. In these interactions, CCQE and 2p2h are the main contributors and are discussed next. Neutrino interactions in which a single pion is produced and the pion is missed---either due to its kinematics or by it being absorbed in the nuclear medium---are also an important contributor.

\subsection{Base interaction model}
\label{subsec:baseIntModel}
Simulations of neutrino interactions are performed with version 5.4.0 of the NEUT neutrino-nucleus interaction event generator~\cite{Hayato:2021heg,Hayato:2009zz,HAYATO2002171}. NEUT takes inputs from a variety of theoretical models for separate neutrino interaction channels. The total cross sections for each channel as a function of neutrino energy, overlaid on the T2K oscillated and unoscillated muon neutrino fluxes, are shown in~\autoref{fig:xsecAndFlux}. An overview of the channels most relevant to this analysis is presented below. 

\begin{figure}[htbp]
    \centering
    \includegraphics[width=0.49\textwidth]{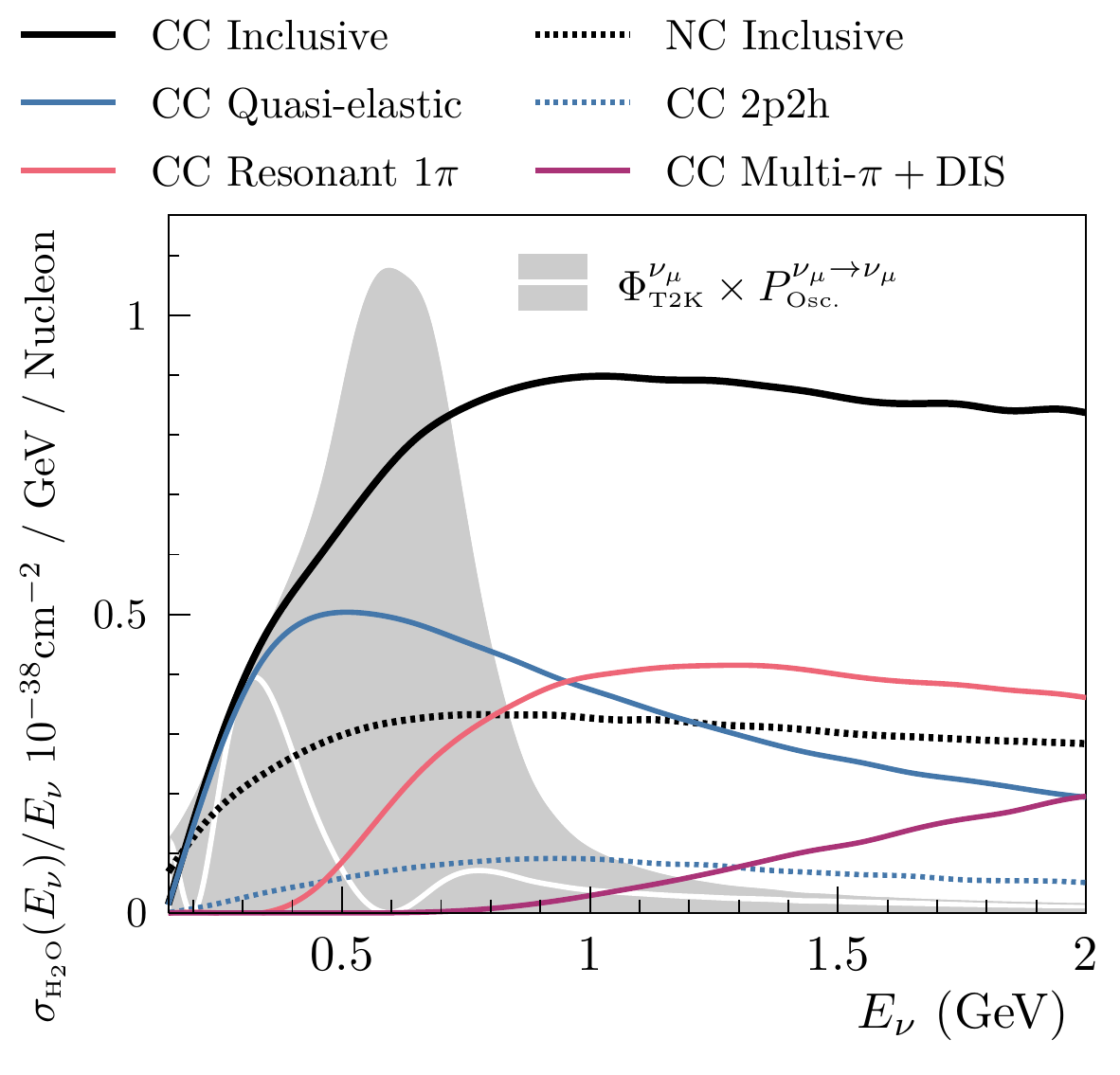}
    \caption{Neutrino cross sections for muon neutrinos interacting on a water target in NEUT, broken down by interaction mode as a function of neutrino energy. The predictions have been modified from their default to reflect the input model used in the oscillation analysis. The surviving muon neutrino flux as seen by the FD is shown with a white line, and the unoscillated muon neutrino flux as seen by the ND is shown as the grey shaded region. The figure is adapted from Ref.~\cite{Hayato:2021heg}.}
    \label{fig:xsecAndFlux}
\end{figure}

\subsubsection{1p1h}
\begin{figure*}[htbp]
    \centering
    \includegraphics[width=0.49\textwidth]{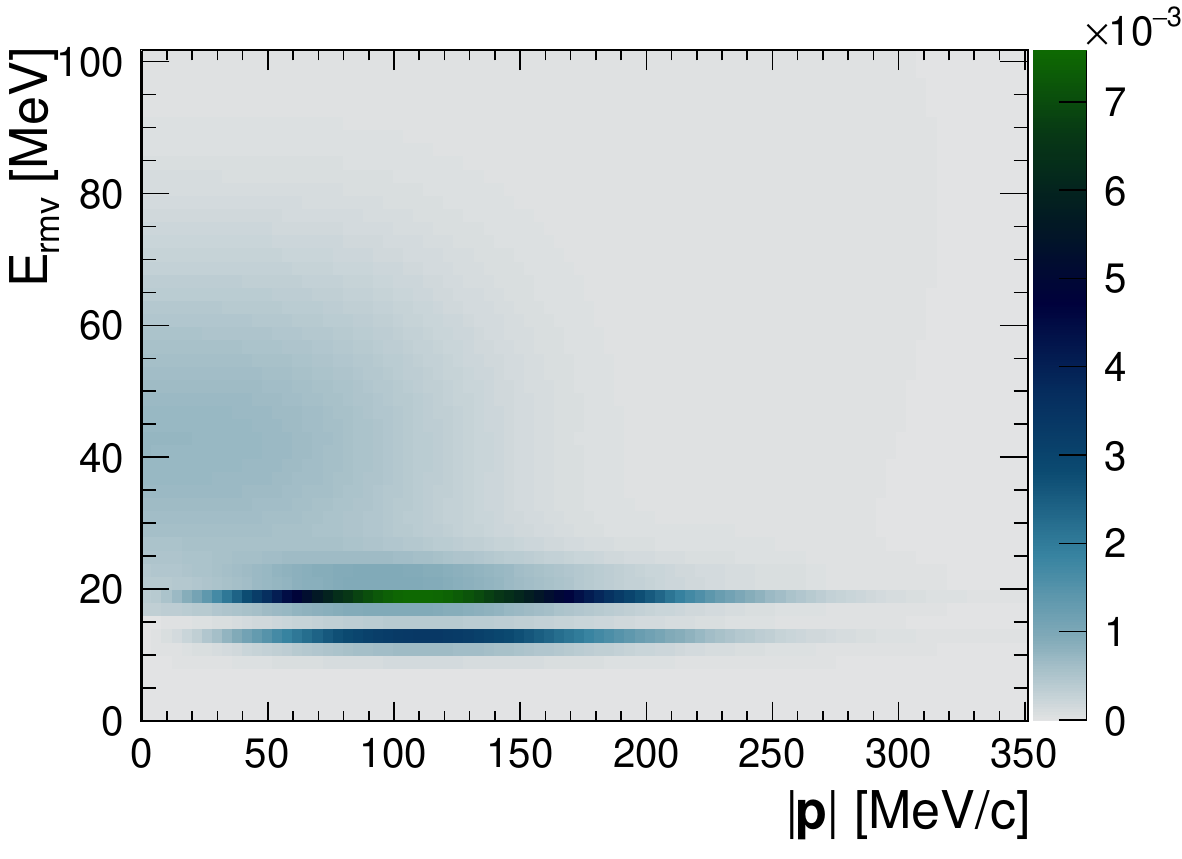}
    \includegraphics[width=0.49\textwidth]{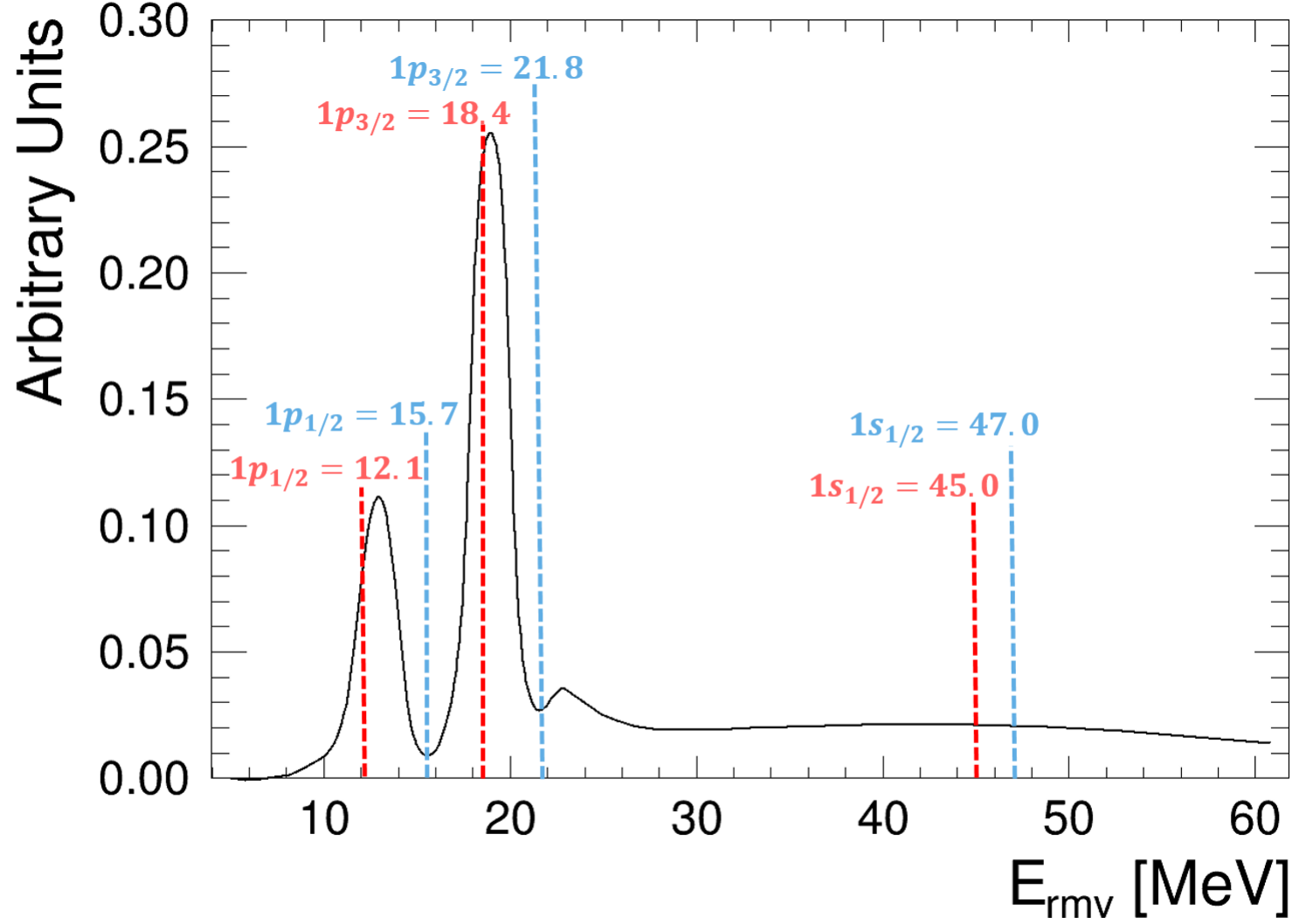}
    \caption{The two-dimensional probability density distribution for the spectral function for oxygen in NEUT~\cite{Benhar:1994hw} (left), and the projection onto the removal energy axis (right). On the left, the darker colour represents a higher probability of finding an initial-state nucleon with a particular removal energy and momentum. The two sharp p-shells at $E_{rmv}\sim12~\text{MeV}$ and $E_{rmv}\sim18~\text{MeV}$, and the larger diffuse s-shell at $E_{rmv}\sim20-65~\text{MeV}$ and $|\textbf{p}|<100~\text{MeV/c}$, are visible. The predictions for the shell positions from another model~\cite{Bodek:2018lmc} are overlaid on the right with dashed lines, for protons (red) and neutrons (blue). The energy in MeV is labelled for each prediction.}
    \label{fig:sf2DO}
\end{figure*}

One-particle one-hole (1p1h) interactions describe charged-current quasi-elastic (CCQE) and neutral-current elastic (NCE) neutrino interactions in which a single nucleon from inside a target nucleus is ejected.
CCQE interactions, which usually produce single-ring electron-like or muon-like events, are the dominant contributor to the FD event samples, making up roughly 70\% of the \rmu selection.
In NEUT, 1p1h interactions are modelled according to the scheme presented in Ref.~\cite{Hayato:2021heg,Benhar:1994hw}, sometimes referred to as the ``Benhar Spectral Function'' model. This approach relies on the plane wave impulse approximation to factorise the 1p1h cross-section calculation into an expression containing a single-nucleon factor alongside a spectral function (SF). The SF is a two-dimensional distribution describing the probability of finding a nucleon with momentum, $|\mathbf{p}|$, and removal energy, $E_{rmv}$, which corresponds to the energy required to remove the nucleon from the nuclear potential. This formalism provides a realistic description of the nuclear ground state and is built largely from exclusive measurements of 1p1h interactions in electron scattering, with additional theory-based contributions to describe the role of initial-state correlations between neighbouring nucleons. As an example, the two-dimensional SF for oxygen is shown in~\autoref{fig:sf2DO}, which exhibits the shell structure of the nucleus.

The single-nucleon component of the 1p1h cross section uses the BBBA05~\cite{BRADFORD2006127} description for the vector part of the nucleon form factors, and a simple dipole form for the axial part. The nucleon axial mass parameter appearing in the form factor, $M_A^{QE}$, is constrained using bubble chamber measurements of neutrino interactions on light nuclear targets, as detailed later in~\autoref{subsec:intModelUncert}.

\subsubsection{2p2h}
In two-particle two-hole (2p2h) interactions, a neutrino interacts with a correlated pair of nucleons, ejecting both from the nucleus. Although this is not a dominant process at T2K, it usually produces single-ring electron-like or muon-like events in the FD---making up about 12\% of the \rmu selection at the FD---and is therefore important to the oscillation analysis.
As T2K's neutrino energy estimator is based on the assumption that the interaction was CCQE, applying it to 2p2h events causes a natural bias. Thus it is crucial that the relative contribution of 2p2h events to the selections, and the bias they cause to the neutrino energy estimator, are well modelled.  
NEUT describes the charged-current 2p2h cross section and outgoing lepton kinematics with the Nieves \etal model~\cite{Nieves:2011yp}. In this model, the 2p2h cross section peaks in two distinct regions of momentum and energy transfer, referred to as ``$\Delta$'' and ``non-$\Delta$'' excitation regions, which each cause distinctly different biases in neutrino energy reconstruction~\cite{Abe:2021gky}. Neutral-current 2p2h interactions are not simulated in NEUT. Their inclusion would have a negligible impact on the oscillation analysis as such interactions would make a small contribution to an already small NC background, which is prescribed large uncertainties.

\subsubsection{Single-pion production}
Single-pion production (SPP) processes are the dominant contributor for the T2K FD sample that requires a single electron-like ring with one delayed decay electron (referred to as \rede in \autoref{sec:sk}). The events also contribute to the other event samples when the pion is not observed due to interactions in the detector or the nucleus, or due to reconstruction inefficiencies. 
SPP at T2K stems mostly from the neutrino-induced excitation of an initial-state nucleon to a baryon resonance that decays into a pion and a nucleon, and makes up about 13\% of the \rmu selection.
These processes are described in NEUT by the Rein--Sehgal (RS) model~\cite{Rein:1980wg} in the outgoing hadronic mass region $W<2.0~\text{GeV}$, with additional improvements to the nucleon axial form factors~\cite{Graczyk:2014dpa,Graczyk:2007bc} and the inclusion of the final-state lepton mass in the calculation~\cite{Berger:2007rq,Graczyk:2007xk,Kuzmin:2003ji}. 
Whilst $\Delta(1232)$ excitations are the dominant contributors to the SPP cross section, a total of 18 baryonic resonances are included in addition to a non-resonant process in the mixed isospin channels. Interference between the resonances is incorporated, but not between the resonant and non-resonant components. The initial-state model for SPP interactions in NEUT is a simple relativistic Fermi gas.

Coherent scattering off nuclei also contributes to the SPP cross section, especially at low four-momentum transfer. In this analysis, NEUT models coherent interactions with the Berger--Sehgal model~\cite{Berger:2008xs}, updated from the RS model~\cite{Rein:1982pf}, and includes Rein's model of diffractive pion production~\cite{Rein:1986cd}.

\subsubsection{Deep inelastic scattering}
Deep inelastic scattering (DIS) describes neutrino interactions with the quark constituents of nucleons. It is a sub-dominant process in T2K's oscillation analysis due to the neutrino energy and the single-ring event selections at the FD. The cross section in NEUT is calculated using the GRV98~\cite{GRV98} Parton Distribution Functions (PDFs), which describe the probability to find a quark of a given type with a given value of the Bjorken scaling variables, $x$ and $y$, inside the target nucleon. Bodek--Yang (BY) modifications~\cite{Bodek:2003wc, Bodek:2005de} are made to extend the validity of this approach to the relatively low four-momentum transfers, $Q^2\lesssim1.5~\text{GeV}^2$, typical for interactions at T2K.

In NEUT, the modelling of DIS processes begins for interactions where the hadronic invariant mass $W>1.3~\text{GeV}$. To avoid double counting the aforementioned non-resonant single-pion production, only DIS interactions that produce more than one pion in the final state are considered. The generation of the hadronic state is split depending on $W$: for interactions with $W>2~\text{GeV}$ PYTHIA 5.72~\cite{SJOSTRAND199474} is used, whilst for $W<2~\text{GeV}$ a custom model interpolating between the $\Delta(1232)$ and DIS interactions is employed, described in Sec.V C of Ref.~\cite{Aliaga:2020rqb}. 

\subsubsection{Final-state interactions}
\label{sec:fsi}
The simulated neutrino interaction events produce an outgoing hadronic system at the interaction vertex inside the nucleus, in addition to the outgoing lepton. These hadrons can undergo final-state interactions (FSI) in the nuclear medium. In NEUT, pion FSI are described using the semi-classical intranuclear cascade model by Salcedo and Oset~\cite{Salcedo:1987md,Oset:1987re}, tuned to modern $\pi-A$ scattering data~\cite{PinzonGuerra:2018rju}. Nucleon FSI are described in an analogous cascade model~\cite{Hayato:2009zz}. Within the cascade, the outgoing hadrons are individually stepped through the remnant nucleus where they can elastically scatter, be re-absorbed, undergo charge-exchange processes, and/or emit additional hadrons which are also stepped through the cascade. Amongst other effects, such cascades allow for SPP events to have no observable pions in the final state after FSI, and for 1p1h interactions to appear as pion production interactions. 

\subsubsection{Coulomb corrections}
Following a charged-current neutrino interaction, the electrostatic interaction between the remnant nucleus and the outgoing charged lepton can cause a small shift in the lepton's momentum. The size of this Coulomb correction has been determined from the analysis of electron scattering data~\cite{Gueye:1999mm} and is implemented as a small nucleus and lepton-flavour dependent shift in the momentum of the outgoing lepton. The size of this shift is $-3.6~\text{MeV}$ ($+2.6~\text{MeV}$) for outgoing $\mu^-$ ($\mu^+$) from a carbon target, and $-4.3~\text{MeV}$ ($+3.3~\text{MeV}$) for outgoing $\mu^-$ ($\mu^+$) from an oxygen target.

\subsection{Uncertainty parametrisation}
\label{subsec:intModelUncert}
Mismodelling of neutrino interactions can bias the measurements of oscillation parameters---for instance attributing an increase in single-ring events to an increase in 2p2h interactions instead of CCQE interactions. It is crucial to evaluate the impact that plausible variations of NEUT's interaction model can have on the neutrino oscillation analysis. This section describes the chosen parametrisation of such variations and the corresponding parameters' uncertainties. When possible, theory-driven uncertainties are used, but in many cases this offers insufficient freedom to describe available data, and additional empirically driven parameters are required. To cover the caveats of such an approach, and to consider plausible model variations not included in the  model parametrisation, a variety of simulated data studies are performed. These are detailed in \autoref{sec:interactionModel_fds}, and applied to the oscillation analysis in \autoref{sec:fakeData} and \autoref{app:appendix_fakedata}.

\subsubsection{1p1h uncertainties}
The 1p1h uncertainty model is split into three categories: removal energy related to the initial state described by the SF, the neutrino-nucleon interaction, and \emph{ad hoc} freedoms in $Q^2$ from nuclear effects, amongst others, inspired by external data. The central values and uncertainties are summarised in~\autoref{tab:1p1hparams}.

\paragraph{\textbf{Removal energy:}}
A mismodelling of nucleon removal energy would directly bias the reconstructed neutrino energy, which would subsequently bias the extraction of the neutrino oscillation parameters, notably $\Delta m^2$. This was identified as a leading source of uncertainty in a simulated-data study in the last T2K oscillation analysis~\cite{Abe:2021gky,T2K:2019bcf}. In this analysis, a more reliable modelling of removal energy with accompanying uncertainties was developed.

Unlike the simplistic Fermi-gas models used in the previous iterations of T2K's neutrino oscillation analyses, the SF model does not have a single fixed value for the nuclear binding energy that can be varied as a parameter. Instead, the SF removal energy distribution, extracted largely from exclusive electron scattering data, reflects the shell structure of the nucleus, shown earlier in \autoref{fig:sf2DO}. The positions of the removal energy peaks, used as an input to the SF model, are measured with a resolution of $2-6~\text{MeV}$~\cite{Dutta:2003yt} and lower~\cite{Leuschner:1994zz}. 
Measurements of the peak positions for carbon differ by up to 2 MeV for the s-shell and 6 MeV for the p-shell~\cite{Bodek:2018lmc}. The relative strength of each peak also has an uncertainty of up to 10\% for carbon~\cite{Huberts_1990,Benhar:1994hw}. To extract a SF from ($e,e^\prime p$) data, the impact of nuclear effects such as FSI must be incorporated, and an uncertainty of 5~MeV in this correction is applied~\cite{Bodek:2018lmc}.
In view of these uncertainties, a global removal energy shift uncertainty of 6~MeV is included in the analysis alongside a 3~MeV uncertainty on the difference between the carbon and oxygen removal energies. Further uncertainties are accounted for by the introduction of parameters that allow freedom as a function of $Q^2$, described in more detail below. 

The construction of the SF from ($e,e^\prime p$) data, and the associated uncertainties, can only be directly applied to modelling 1p1h neutrino interactions with initial-state protons, i.e. anti-neutrino CCQE interactions. The SF for initial-state neutrons cannot be directly constrained in the same way and the implementation in NEUT assumes that protons and neutrons have the same removal energy distributions. However, as can be seen in ~\autoref{fig:sf2DO}, nuclear shell models predict that this is not the case. Calculations suggest that proton and neutron ground states differ in their removal energy by $1-4~\text{MeV}$, depending on the shell and target~\cite{Bodek:2018lmc}. For the sharper p-shells, where an energy shift is more important relative to the width of the shell, the offset between the SF and the model calculations for neutrons is around 4~MeV for oxygen and 2~MeV for carbon. To account for this, the central value removal energies of the SF for neutrino interactions are shifted by these amounts, and an uncertainty of 4~MeV is applied on the difference between neutrino and anti-neutrino removal energies.

The removal energy shifts are encoded in four parameters depending on whether they affect initial-state protons (\nub CCQE interactions) or neutrons ($\nu\xspace$ CCQE interactions), and whether the target is carbon or oxygen: $\Delta E_{rmv}^{\nu O}$, $\Delta E_{rmv}^{\overline{\nu}O}$, $\Delta E_{rmv}^{\nu C}$, $\Delta E_{rmv}^{\overline{\nu}C}$. The removal energy parameters shift a CCQE event's outgoing lepton momentum and depends on the event's lepton kinematics, neutrino energy, and neutrino flavour.

\paragraph{\textbf{``Low $Q^2$'' parameters:}}
NEUT's cross section for charged-current interactions leaving no mesons in the final state (CC0$\pi$) interactions must be suppressed at low $Q^2$ to match recent measurements from \minerva~\cite{Ruterbories:2018gub,Rodrigues:2015hik} and T2K~\cite{Abe:2020jbf,Abe:2020uub}. This is often applied as a suppression of the CCQE cross section via the inclusion of a nuclear screening effect using the Random Phase Approximation (RPA)~\cite{Nieves:2011yp}. However, such effects are not included in the SF CCQE model used in this analysis. Since the SF model is built largely on the impulse approximation---which is expected to break down at low momentum transfers $\lesssim 400~\text{MeV}/c$~\cite{Katori:2016yel}---extra uncertainties are added in the region where discrepancies with measurements are observed. 

The low $Q^2$ suppression is implemented as five parameters which alter the normalisation of the CCQE cross section in a particular $Q^2$ range. The parameters span $Q^2=\{0,0.25\}~\text{GeV}^2$ and are split into sub-ranges of $0.05~\text{GeV}^2$. Since the origin of this low $Q^2$ suppression in SF predictions is poorly understood, these parameters do not have an external constraint. Whilst this free parametrisation is effective at facilitating a ND-driven modification to the CCQE cross section, the lack of a theoretical basis limits the model's overall predictive power. Several simulated data studies are therefore discussed in \autoref{sec:interactionModel_fds} to evaluate the bias from this technique in the extraction of neutrino oscillation parameters.

\paragraph{\textbf{$M_A^{QE}$ and ``high $Q^2$'' parameters:}}
The nucleon axial mass, $M_A^{QE}$, is tuned to neutrino-deuterium scattering data in NUISANCE~\cite{Stowell:2016jfr}. CCQE cross-section data from ANL~\cite{Miller:1982qi,Barish:1977qk}, BNL~\cite{Baker:1981su}, BEBC~\cite{Allasia:1990uy}, and FNAL~\cite{Kitagaki:1983px} is used, and deuterium nuclear effects~\cite{Singh:1971md} and flux uncertainties for ANL and BNL are included. The central value and its uncertainty are adjusted and inflated to cover the result and previous global fit results~\cite{Bernard:2001rs}, giving $M_A^{QE}=1.03\pm0.06~\text{GeV}$.

Uncertainties on the higher $Q^2>0.25~\text{GeV}^2$ predictions of the SF model are driven by the axial component of the neutrino-nucleon interaction,
where the dipole model may be inadequate~\cite{Bhattacharya:2011ah}. An additional three ``high $Q^2$'' parameters are added to allow an \emph{ad hoc} freedom, with the goal of lessening the extent to which $M_A^{QE}$ is used as an effective parameter to correct for deviations from the dipole model. The $Q^2$ ranges and uncertainties of the new high $Q^2$ parameters are based on comparisons of the $Q^2$ shape of the dipole and z-expansion models~\cite{Bhattacharya:2011ah}.

\begin{table}[htbp]
    \centering
    \begin{tabular}{c|c|c}
    \hline
    \hline
         Parameter & Central Value & Uncertainty \\
         \hline
         $M_A^{QE}$ (GeV)         & 1.03 & 0.06   \\
         \hline
         $\Delta E_{rmv}^{\nu O}$ (MeV)      & +4   & 6 \\
         $\Delta E_{rmv}^{\overline{\nu}O}$ (MeV) & 0    & 6 \\
         $\Delta E_{rmv}^{\nu C}$ (MeV)      & +2   & 6 \\
         $\Delta E_{rmv}^{\overline{\nu}C}$ (MeV) & 0    & 6 \\
         \hline
         $0.00 < Q^2 < 0.05$ & 1.00 & ---  \\
         $0.05 < Q^2 < 0.10$ & 1.00 & ---  \\
         $0.10 < Q^2 < 0.15$ & 1.00 & ---  \\
         $0.15 < Q^2 < 0.20$ & 1.00 & ---  \\
         $0.20 < Q^2 < 0.25$ & 1.00 & ---  \\
         $0.25 < Q^2 < 0.50$ & 1.00 & 0.11    \\
         $0.50 < Q^2 < 1.00$ & 1.00 & 0.18    \\
         $Q^2 > 1.00$        & 1.00 & 0.40    \\
         
    \hline
    \hline
    \end{tabular}
    \caption{The parameters included in the 1p1h uncertainty model with their values and uncertainties before the ND analysis. The uncertainties for the removal energy parameters are around their central value and contain the carbon-oxygen and $\nu$-$\nub$ correlations described in the text. The first five $Q^2$ parameters are not externally constrained before the analysis, and are free to vary between 0 and $\infty$. The units of the $Q^2$ ranges are $\text{GeV}^2$.}
    \label{tab:1p1hparams}
\end{table}

\subsubsection{2p2h uncertainties}
The uncertainties related to 2p2h interactions are similar to those in T2K's previous oscillation analysis~\cite{Abe:2021gky,T2K:2019bcf}. Parameters altering the 2p2h normalisation independently for neutrinos and anti-neutrinos, and for carbon and oxygen interactions, are used. The 2p2h normalisations are unconstrained, and the carbon-oxygen scaling parameter has a 20\% prior uncertainty. A separate shape uncertainty is also applied, which allows shifts in the $\Delta$ and non-$\Delta$ contributions in the energy and momentum transfer to the nucleus, $(q_0,|\textbf{q}|)$, of the Nieves model, also separated for carbon and oxygen interactions.

\begin{figure}[htbp]
    \centering
    \includegraphics[width=0.49\textwidth, trim=0mm 7mm 18mm 12mm, clip]{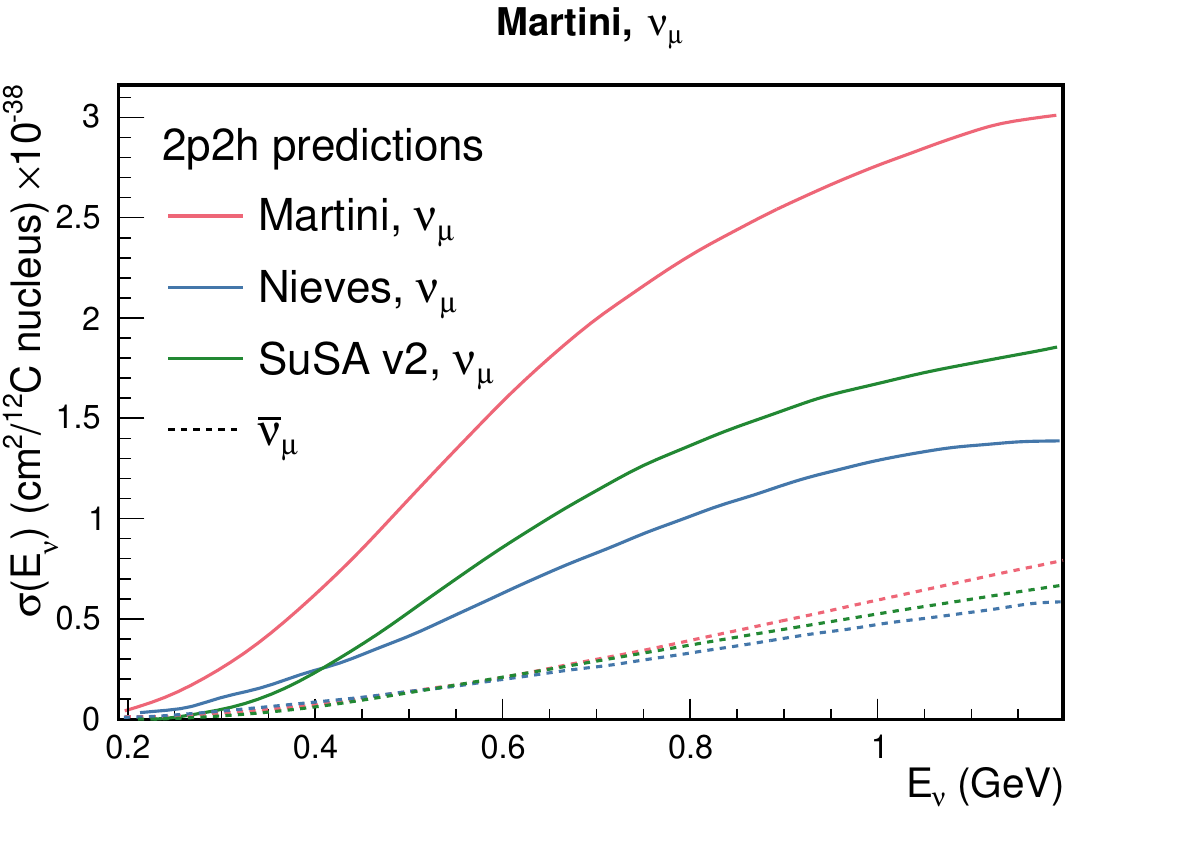}
    \caption{Cross-section predictions for \numu (solid) and \numub (dashed) 2p2h interactions on $^{12}\textrm{C}$ from Martini \etal~\cite{Martini:2009uj}, Nieves \etal~\cite{Nieves:2011yp}, and SuSA v2~\cite{Megias:2016lke,Simo:2016ikv}.}
    \label{fig:int_nieves_susa_martini}
\end{figure}

This analysis also includes additional new uncertainties that reflect the shape of the energy dependence of 2p2h using three different plausible models of the process, also studied by T2K cross-section analyses~\cite{T2K:2020sbd,T2K:2020jav}. The uncertainties span the maximal difference in 2p2h predictions from Martini et. al.~\cite{Martini:2009uj}, Nieves \etal~\cite{Nieves:2011yp}, and SuSAv2~\cite{Megias:2016lke,Simo:2016ikv}, shown in \autoref{fig:int_nieves_susa_martini}. Four parameters are added which control the shape of the energy dependence of 2p2h below and above $E_\nu=600~\text{MeV}$, and are separately applied to neutrino and anti-neutrino events.

\subsubsection{Single-pion production uncertainties}
The uncertainty treatment for SPP remains almost identical to previous T2K analyses~\cite{Abe:2021gky, T2K:2019bcf, Abe:2017vif}. There are three central parameters in the modified RS model: the resonant axial mass, $M_A^{RES}$; the value of the axial form factor at zero transferred four-momentum, $C_5^{A}(Q^2=0)$; and the normalisation of the $I_{1/2}$ non-resonant component. As for $M_A^{QE}$, the parameters have been tuned to deuterium bubble chamber data using NUISANCE~\cite{Stowell:2016jfr}, selecting SPP data from ANL~\cite{ANL_CC1pi,ANL_NC1pi} and BNL~\cite{BNL_CC1pi,BNL_CC1pi_isospin}, including corrected data~\cite{ANL_BNL_corr}. The uncertainties are inflated so that the model adequately describes the SPP cross section in different hadronic mass regions from ANL and BNL, and SPP cross-section measurements on nuclear targets from \miniboone~\cite{AguilarArevalo:2010bm,AguilarArevalo:2009ww,AguilarArevalo:2010xt} and \minerva~\cite{MIN_CC1pip,MIN_CC1pi0,MIN_CC1pi0_2,MIN_pion_2016}.

A new parameter was introduced for anti-neutrino interactions producing low momentum pions, which constitute a background for the single-ring \rhcalt samples. This extra freedom is added through an $I_{1/2}$ non-resonant normalisation parameter that affects both \numub and \nueb single pion interactions with $p_\pi<200~\text{MeV}/c$ in the Rein--Sehgal model. The parameter is not constrained by the ND and has an uncertainty of 100\%.

Normalisation parameters on the CC and NC coherent cross sections are included separately, and each is assigned an uncorrelated 30\% uncertainty based on comparisons to \minerva data~\cite{Higuera:2014azj}. The uncertainty on coherent scattering is fully correlated between carbon and oxygen. 

\subsubsection{Deep inelastic scattering uncertainties}
DIS interactions make a small contribution to the samples in this oscillation analysis due to T2K's neutrino energy. Nevertheless, uncertainties that  cover variations in muon kinematics from CC DIS interactions are needed for the ND fit, whose selections contain some multi-$\pi$ events, and have been significantly updated from previous analyses~\cite{Abe:2017vif}. 

As discussed in~\autoref{subsec:baseIntModel}, NEUT uses PDFs with BY corrections to calculate the DIS cross section. The uncertainty in the BY corrections is parametrised as a fraction of the difference between using the GRV98 PDFs with and without the BY corrections. At $Q^2>1.5~\text{GeV}^2$ the impact is marginal, but in the peak region at lower $Q^2$ the impact is large, altering the predicted cross section by $\sim40\%$. This parameter is split for $W < 2~\text{GeV}$ (multi-$\pi$) and $W>2~\text{GeV}$ (DIS) interactions. 

Another parameter is introduced to modify the generation of the hadronic state for $W < 2~\text{GeV}$ DIS interactions, which uses a custom model~\cite{Hayato:2021heg} to choose the particle multiplicities in an event. This parameter accounts for the differences between the custom model and the AGKY model~\cite{Yang:2009zx} used in the GENIE event generator~\cite{GENIE}.

Two normalisation uncertainties are also included, motivated by comparing the NEUT CC-inclusive cross section to the world average of measurements at higher neutrino energies~\cite{Tanabashi:2018oca}. The uncertainties are $3.5\%$ for neutrino interactions and $6.5\%$ for anti-neutrino interactions, and the two are uncorrelated.

\subsubsection{Final-state interactions uncertainties}
\label{sec:int_fsi}
The NEUT pion cascade model has been tuned to better match external $\pi-A$ scattering data~\cite{PinzonGuerra:2016uae}. The tuning procedure constrains the probability for different interaction processes to occur in the pion cascade (e.g. pion absorption or charge exchange), and is notably more robust than previous parametrisations. The constraints on the pion FSI cascade from the ND analysis are propagated to the FD in this analysis, which was not done before. Furthermore, the simulations at the ND and the FD now use a consistent model for pions from the interaction vertex propagating through the nucleus (``pion final-state interactions''), and for pions propagating through the detector (``pion secondary interactions''), mentioned later in \autoref{sec:nd_syst}. The ND constraint on the FSI parameters is only used to constrain the FD modelling of FSI and not the FD modelling of secondary interactions.

\subsubsection{Other uncertainties}
Additional uncertainties are applied to processes with small contributions to the analysis. As in previous analyses, the NC$1\gamma$ production cross section has a 100\% normalisation uncertainty. The NC elastic, NC resonant kaon and eta production, and NC DIS interactions are grouped together and referred to as ``NC other'' interactions, which have a 30\% normalisation uncertainty that is uncorrelated at ND and FD.
There is one uncertainty controlling the normalisation of the electron neutrino cross section, and another controlling the electron anti-neutrino cross section. The uncertainties are composed of two parts: one 2\% uncorrelated part and one 2\% anti-correlated part, which connects the two parameters~\cite{Day:2012gb}. The parameters only affect electron (anti-)neutrino interactions, and have no effect on the other neutrino flavours.
The total cross sections of CC resonant single-photon production, CC resonant kaon production, CC resonant eta production, and CC diffractive pion production are controlled by a single new parameter referred to as ``CC misc'', which is a 100\% normalisation uncertainty, and such interactions are not affected by other model parameters. Two new parameters are included to account for Coulomb corrections~\cite{Kim:2009zzq,Engel:1997fy}. They control the normalisation of the (anti-)neutrino cross section for $E_\nu=0.4-0.6~\text{GeV}$ with a 2\%(1\%) uncertainty, and are 100\% anti-correlated.

\subsection{Simulated data studies}
\label{sec:interactionModel_fds}
The systematic uncertainties in the analysis are constructed to account for known uncertainties in neutrino interaction physics, but can not possibly cover every model scenario. For instance, cross-section measurements from T2K and other experiments have shown that no single 1p1h model describes the kinematic phase space in T2K and MINERvA~\cite{Abe:2018pwo,Lu:2018stk,Dolan:2018zye,T2K:2020jav,T2K:2020sbd,Ruterbories:2018gub,MINERvA:2022mnw}.
In addition, the ND analysis, presented later in \autoref{sec:nd_fit}, may compensate for cross-section mis-modelling by varying the flux parameters instead of the cross-section parameters, leading to good agreement with the observed event spectrum in lepton kinematics. However, the fitted model may scale the effect incorrectly in other important physics variables, e.g. $E_\nu$. 
It is therefore crucial to test whether the uncertainty model is flexible enough to capture variations under alternative cross-section models which are not directly implemented in the default uncertainty model, and whether the subsequent extrapolation of model constraints to the FD has an effect on constraining the oscillation parameters.

Some of the simulated data sets are similar to those presented in T2K's previous analyses~\cite{Abe:2021gky,T2K:2019bcf}. The studies are updated due to the significant changes in the uncertainty model and ND analysis. The alternative models and tunes are selected to cover a number of interaction types and effects, listed next.

\paragraph{\textbf{\CCzeropi simulated data sets:}}The dominant \CCzeropi samples at the ND and the single-ring samples at the FD are designed to select CCQE-like events. The larger statistics in these samples requires testing for a range of alternative models, and the robustness of the neutrino interaction model.
    
\begin{itemize}
    \item \textit{Non-CC-Quasi-Elastic (non-CCQE) contributions}---Before the fit to data, the prediction of the \CCzeropi selection at the ND is underestimated by $0-20\%$, depending on the outgoing lepton kinematics. Projecting the data and prediction onto the reconstructed four-momentum transfer, $Q^2_{rec}$, defined as the $Q^2$ calculated for a CCQE interaction on a stationary nucleon, and with a binding energy $E_b$,
    \begin{align}
        Q^2_{rec} &= 2E_\nu^{rec}\left(E_\mu - |\vec{p}_\mu|\cos\theta_\mu\right) - m^2_\mu 
    \label{eq:q2rec} \\
        E_\nu^{rec} &= \frac{1}{2} \frac{m^2_\mu+(m_{n}^{eff})^ 2-m^2_p -2E_\mu m_n^{eff}}{E_\mu - |\vec{p_{\mu}}|\cos\theta_\mu - m_n^{eff}} 
    \label{eq:enurec} \\
        m_n^{eff} &=m_n-E_b \nonumber
    \end{align}
    the discrepancy is less than 5\% at $Q^2_{rec}<0.1~\text{GeV}^2$ and approximately 20\% for higher $Q^2_{rec}$. The CCQE cross section is modified after the fit to ND data to account for the difference. This simulated data tests the hypothesis that the underestimation of data is actually due to non-CCQE contributions, and does so by scaling up their predictions instead of the CCQE components. The study is given in detail in \autoref{app:appendix_fakedata}.
        
    \item \textit{Alternative CCQE form factors}---The baseline model used in this analysis assumes a dipole parametrisation of the nucleon form factor. There are other form factor models, of which the 3-component (an extension of Ref.~\cite{Adamuscin:2007fk}) and z-expansion\cite{Bhattacharya:2011ah} formalisms were tested. The effect is largely expected to be covered by the $Q^2$-related freedoms of the cross-section model.
        
    \item \textit{Multi-nucleon (2p2h) model}---The Nieves \etal model~\cite{Nieves:2011yp} was used to describe 2p2h interactions in this analysis. An alternative 2p2h model from Martini \etal~\cite{Martini:2009uj} was tested in the simulated data studies, because its 2p2h cross section is larger and evolves differently in $E_\nu$ for neutrinos and anti-neutrinos, shown earlier in~\autoref{fig:int_nieves_susa_martini}. Modelling the 2p2h spectrum is important in the ND to FD extrapolation, as it is one of the main sources of bias in the reconstructed neutrino energy spectrum of CCQE-like samples. The SuSAv2 model~\cite{Megias:2016lke,Simo:2016ikv}, also shown earlier in \autoref{fig:int_nieves_susa_martini}, is a less extreme variation compared to the Martini model, so was not included.

    \item \textit{Removal energy}---The nuclear removal energy in the relativistic Fermi gas (RFG) model~\cite{Smith:1972xh} was the largest contributor to uncertainty in the previous T2K analysis~\cite{Abe:2021gky,T2K:2019bcf}. This analysis' spectral function (SF) model~\cite{Benhar:1994hw}, mentioned earlier in \autoref{subsec:baseIntModel}, introduced an improved parametrisation for the removal energy uncertainty, and simulated data sets were developed to study its impact.
\end{itemize}

\paragraph{\textbf{CC1$\pi$ simulated data sets:}}Single-pion events are a background for the single-ring selections at the FD and contribute to the bias in reconstructed neutrino energy. Additionally, the \rede sample at the FD specifically targets single-pion events, which motivates the need to have a robust uncertainty model of these interactions. Three simulated data sets were produced:
    
\begin{itemize}
    \item \textit{ND data-driven pion momentum modification}---The \rede selection at the FD tags low momentum pions below Cherenkov threshold by the presence of a delayed Michel electron. The ND analysis in \autoref{sec:nd_fit} uses selections based on muon kinematics and pion tagging to constrain the uncertainties, and does not study the pion kinematics directly. As such, single-pion events may be modelled well in muon kinematics and poorly in pion kinematics. A data-driven simulated data set was created by studying the \CConepi selections at the ND, using the model that was fit to ND data in lepton kinematics. The model was used to predict the reconstructed pion momentum spectrum, $p_{\pi}^{reco}$, in the single-pion ND selections, which was compared to the data in the $p_{\pi}^{reco}<200~\text{MeV}/c$ region. The number of events was underestimated by $\sim20\%$, which was applied as an overall normalisation to the simulation of all single-pion events at the FD that had a pion with generated (true) momentum below $200~\text{MeV}/c$. This is the only simulated data set that was not applied at the ND, and tested only at the FD.
        
    \item \textit{\minerva pion suppression}---A low-$Q^2$ suppression of the single-pion production cross section in GENIE~\cite{GENIE} was needed to consistently describe neutrino interactions on plastic scintillator (CH) from \minerva and bubble chamber data on nucleons~\cite{Stowell:2019zsh}. The function parametrising this discrepancy was used to create simulated data at both the ND and the FD and the study is presented in detail in \autoref{app:appendix_fakedata}.

    \item \textit{Pion secondary interactions}---This analysis introduced a new model for pions rescattering in the ND. The GEANT4 model~\cite{GEANT4:2002zbu} was replaced with NEUT's Salcedo--Oset model~\cite{Salcedo:1987md,Oset:1987re} which was tuned to $\pi-A$ scattering data~\cite{PinzonGuerra:2018rju}. A hybrid simulated data set which blended features of the two models was used in the ND analysis to study the impact of choosing one model over the other.
\end{itemize}

A summary of the simulated data studies is presented in \autoref{sec:fakeData} after the analysis sections, and the simulated data studies are detailed in \autoref{app:appendix_fakedata}.

%% file: NDfit.tex
The high statistics data at the ND are used to constrain many of the neutrino flux and neutrino-nucleus interaction models present in the neutrino oscillation analysis. Sampling the unoscillated neutrinos at a high rate and tuning the prediction to the ND data allows for significant reduction of the uncertainties of the FD prediction. The ND analysis targets \CCzeropi events as these are the signal at the FD, and additionally constrains the background contributions such as \CConepi and CC multi-$\pi$ events. Separation of \numu and \numub events is possible due to the magnetised sign-selecting ND, and there are \numu selections in the \rhcalt which constrain the wrong-sign background.

As in previous T2K oscillation analyses~\cite{Abe:2021gky,T2K:2019bcf}, two complementary likelihood sampling methods are used and are cross-validated. One is based on Markov Chain Monte Carlo (MCMC) methods~\cite{metropolis,hastings}, and the other is based on minimising a test-statistic through gradient-descent methods in Minuit~\cite{James:1975dr}. 
The MCMC analysis is inherently Bayesian, and has the ability to run a simultaneous ND+FD analysis whose results are presented in \autoref{sec:oa:bayesian}. The gradient-descent analysis instead fits the systematic uncertainties in the simulation to find the global minimum of the test statistic that best describes the data at the ND, discussed in \autoref{sec:nd:likelihood}. The central value and covariance matrix of the systematic uncertainties around that best-fit point is then propagated to the FD.
The MCMC framework directly implements the removal energy shift parameters described in \autoref{subsec:intModelUncert} which allows for discrete event migrations between bins, whereas the gradient-descent framework smooths the effect by an effective binned treatment to avoid discontinuous likelihoods. The MCMC analysis also implements a non-uniform rectangular binning, meaning the binning in the $x$ variable is not uniform in the $y$ variable, allowing the events to be binned finer and more effectively, generally leading to improved sensitivity to the systematic uncertainties. The gradient-descent framework instead uses a uniform rectangular binning. 
The effect of these differences is tested at the FD by propagating the results from the gradient-descent framework, which assumes correlated Gaussian parameter constraints, and comparing to propagating the constraints from the steps in the MCMC, which is detailed in \autoref{sec:oa_cross_fitter}. 
This section shows the results from the gradient-descent based analysis.

This analysis of ND data uses $19.867\times10^{20}$ POT, with $11.531\times10^{20}$ collected in \fhcalt and $8.336\times10^{20}$ collected in \rhcalt, as listed earlier in \autoref{tab:pot_2020}. This is an overall POT increase of 106\% compared to the previous analysis.

\subsection{ND selections}
\label{subsec:nd_sel}
The doubling of \rhcalt data in the ND allowed for a refining of the anti-neutrino selections. Additionally, the \rhcalt beam samples now match the \fhcalt beam samples in the separation of events by their reconstructed pion multiplicity. Previous analyses only split the \rhcalt selections into events with a single muon-like track (CC-1Track) and events with a single muon-like track with at least one charged or neutral pion candidate (CC-NTrack).

The events are categorised into 18 samples, split into nine equivalent FGD1 and FGD2 samples to separate neutrino interactions on plastic scintillator (FGD1), and plastic scintillator and water (FGD2). The samples first require a reconstructed muon to be present. They are then split by the sign of the muon candidate---which implies the identity of the incoming neutrino---classifying events as \numu events in \fhcalt, \numub events in \rhcalt, and \numu events in \rhcalt. Each of these charged-current inclusive selections are separated into three reconstructed topologies based on the number of reconstructed charged pions. An event with no reconstructed pions is classified as \CCzeropi; an event with a single charged pion with opposite charge to the muon is \CConepi; and an event with any other number of charged pions (e.g. $1\mu^{-}2\pi^+$ or $1\mu^{-}1\pi^-$ in \fhcalt), or at least one neutral pion, is classified as \CCother. There is no requirement on the number of proton tracks and there is no dedicated \nue or \nueb selection.

The pion tagging in the \numu selections is the same as in previous T2K 
analyses~\cite{Abe:2021gky,T2K:2019bcf}. A pion is tagged by either a pion-like track in the TPC, a pion-like track contained in the FGD, or an isolated delayed Michel electron in an FGD. In the FGD and TPC tagging, the pion candidate is required to share its vertex with the muon candidate, and for the Michel tag it is required to be in the same FGD as the candidate vertex.
For the anti-neutrino selections, TPC and FGD pion-like tracks are identified similarly to the neutrino selections, whilst the Michel tag can only identify positively charged pions since negatively charged pions are more likely to be absorbed. For \fhcalt selections, Michel-tagged pions dominate for $p_\pi<175~\text{MeV}/c$, TPC-tagged pions dominate when $p_\pi>250~\text{MeV}/c$, and the FGD-contained pions make up 30\% of all pion tags when $100~\text{MeV}/c<p_\pi<250~\text{MeV}/c$. There are virtually no Michel-tagged or FGD-contained pions when $p_\pi>400~\text{MeV}/c$. Combining the tags, the selection has about $25\%$ charged pion tagging efficiency when $p_\pi < 300 ~\text{MeV}/c$, increasing roughly linearly to $\sim50\%$ at $p_\pi=1~\text{GeV}/c$. Neutral pions are tagged by identifying a displaced $e^\pm$ candidate in the TPC, indicating the presence of a photon conversion.

\begin{table}[htbp]
    \centering
    \begin{tabular}{c|c|c||c|c}
        \hline
        \hline
        Selection & Topology & Target & Eff. (\%) & Pur. (\%) \\
        \hline
        \multirow{6}{*}{\numu in \fhcalt} 
        &\multirow{2}{*}{0$\pi$}      
            & FGD1 & 48.0 & 71.3 \\
            && FGD2 & 48.0 & 68.2 \\
            \cline{2-5}
        &\multirow{2}{*}{$1\pi^+$}      
            & FGD1 & 29.0 & 52.5 \\
            && FGD2 & 24.0 & 51.3 \\
            \cline{2-5}
        &\multirow{2}{*}{Other}      
            & FGD1 & 30.0 & 71.4 \\
            && FGD2 & 30.0 & 71.2 \\
        \hline
        \multirow{6}{*}{\numub in \rhcalt}
            &\multirow{2}{*}{0$\pi$}      
            & FGD1 & 70.0 & 74.5 \\
            && FGD2 & 69.0 & 72.7 \\
            \cline{2-5}
        &\multirow{2}{*}{1$\pi^-$}      
            & FGD1 & 19.3 & 45.4 \\
            && FGD2 & 17.2 & 41.0 \\
            \cline{2-5}
        &\multirow{2}{*}{Other}      
            & FGD1 & 26.5 & 26.3 \\
            && FGD2 & 25.2 & 26.0 \\
        \hline
        \multirow{6}{*}{\numu in \rhcalt}
            &\multirow{2}{*}{0$\pi$}      
            & FGD1 & 60.3 & 55.9 \\
            && FGD2 & 60.3 & 52.8 \\
            \cline{2-5}
        &\multirow{2}{*}{1$\pi^+$}      
            & FGD1 & 30.3 & 44.4 \\
            && FGD2 & 26.0 & 44.8 \\
            \cline{2-5}
        &\multirow{2}{*}{Other}      
            & FGD1 & 27.4 & 68.3 \\
            && FGD2 & 27.1 & 69.5 \\
    \hline
    \hline
    \end{tabular}
    \caption{Efficiencies and purities for each of the selections at the ND in this analysis, including wrong-sign background components.
    The efficiency is defined as the number of events that have a reconstructed selection that matches the true selection, divided by the total number of events with that same true selection.
    The purity is defined as the number of events with the desired selection divided by the total number of events in the selection.}
    \label{tab:nd_sel_eff_pur}
\end{table}

The efficiencies and purities are determined from reconstructed simulated events, and are provided in \autoref{tab:nd_sel_eff_pur}, which shows similar performance for the two FGDs. FGD2 has worse Michel tagging and FGD-contained track reconstruction than FGD1 due to the passive water layers, resulting in a lower efficiency for \CConepi selections. The purity for \CCzeropi selections for \fhcalt and \rhcalt is above 70\%, and $\sim55\%$ for the \numu in \rhcalt due to the wrong-sign neutrino flux having a longer tail, which makes multi-particle final states more likely. 
The \numub \CCzeropi efficiency is higher than \numu \CCzeropi due to \numub CCQE interactions usually producing a neutron in lieu of the proton from \numu CCQE interactions. In \numu CCQE interactions, the outgoing proton may produce a clear track in the detector, which has a probability of being mis-tagged for a $\pi^+$ (or $\mu^+$ for \numub selections), and so enters another selection; this is very unlikely when the outgoing particle is a neutron. Furthermore, \numub interactions generally produce a larger proportion of forward-going events, where the ND has better acceptance.

The \numub \CCother selections' low purities compared to the \numu in \fhcalt and \numu in \rhcalt equivalents stem from the larger wrong-sign background that, for the reasons stated earlier, produces multiple pions which may be wrongly selected as the $\mu^+$ candidate. In addition, the muon candidate in \rhcalt can be incorrectly assigned as a high momentum proton around $p\sim1~\text{GeV}/c$, where the energy loss in the TPC for a proton is similar to that of a muon. This track confusion seldom happens in the \numu selections, since it selects a negatively charged track. A $\pi^-$ is rarely selected as the $\mu^-$ in \numu selections since it requires a higher energy multi-$\pi$ event or final-state interactions of a hadron from the primary interaction.

Generally, the mis-identification of the muon candidate is largest at low momentum, when it does not leave a long enough track to reliably assess the degree of bending in the ND's magnetic field. Almost all wrong-sign muons, pions and electrons selected as the muon candidate occupy this region. In the case of mis-identification, the muon candidate is otherwise a pion with same charge due to their similar energy loss in the FGDs and TPCs. Using the combined FGD+TPC detector system, there is a 94\%, 86\%, and 77\% probability that the muon candidate is a muon in the \CCzeropi, \CConepi and \CCother selections, respectively. 

\subsection{ND related uncertainties}
\label{sec:nd_syst}
Dedicated control samples have been developed to evaluate the response of the ND and to quantify systematic uncertainties~\cite{Abe:2015awa}. These uncertainties include the modelling of pion and proton secondary interactions in the detector, particle mis-identification probabilities in the TPCs and FGDs, magnetic field distortions, momentum resolutions and scales, efficiencies related to clustering, tracking and track matching, Michel-tagging efficiencies, pile-up, FGD mass, out of fiducial volume (OOFV) background events, and sand muon backgrounds. Sand muon backgrounds enter the selections when neutrinos from a beam spill interact in the sand surrounding the ND pit, creating a muon that enters the ND. These uncertainties can migrate events into or out of selections and change the reconstructed particles' kinematics. The uncertainties can either be efficiency-like (dependent on a particle's kinematics) or normalisation-like (independent of a particle's kinematics).

This analysis is the first to use NEUT's semi-classical Salcedo--Oset cascade model~\cite{Salcedo:1987md,Oset:1987re}, mentioned in \autoref{sec:fsi}, for pion secondary interactions in the detector, where previous analyses used GEANT4~\cite{GEANT4:2002zbu}. The model was tuned to external $\pi-A$ scattering data~\cite{PinzonGuerra:2018rju}, and was found to agree better with data and be more consistent across the interaction channels and pion energy ranges compared to GEANT4.
Additionally, T2K now uses the same model for pion final-state and secondary interactions in both the ND and the FD. The ND constraint on pion final-state interactions is propagated to the FD, whereas the constraint on the secondary interactions is not.
\begin{table}[htbp]
    \centering
    \begin{tabular}{c|c|c||c}
        \hline
        \hline
        Selection & Topology & Target & Uncertainty (\%)\\
        \hline
        \multirow{6}{*}{\numu in \fhcalt} 
        &\multirow{2}{*}{0$\pi$}      
            & FGD1 & 1.20 \\
            && FGD2 & 1.40 \\
            \cline{2-4}
        &\multirow{2}{*}{$1\pi^+$}      
            & FGD1 & 2.65 \\
            && FGD2 & 2.57 \\
            \cline{2-4}
        &\multirow{2}{*}{Other}      
            & FGD1 & 2.33 \\
            && FGD2 & 2.19 \\
        \hline
        \multirow{6}{*}{\numub in \rhcalt}
            &\multirow{2}{*}{0$\pi$}      
            & FGD1 & 1.96 \\
            && FGD2 & 2.08 \\
            \cline{2-4}
        &\multirow{2}{*}{1$\pi^-$}      
            & FGD1 & 4.04 \\
            && FGD2 & 3.63 \\
            \cline{2-4}
        &\multirow{2}{*}{Other}      
            & FGD1 & 3.61 \\
            && FGD2 & 3.23 \\
        \hline
        \multirow{6}{*}{\numu in \rhcalt}
            &\multirow{2}{*}{0$\pi$}      
            & FGD1 & 1.61 \\
            && FGD2 & 1.76 \\
            \cline{2-4}
        &\multirow{2}{*}{1$\pi^+$}      
            & FGD1 & 3.00 \\
            && FGD2 & 2.72 \\
            \cline{2-4}
        &\multirow{2}{*}{Other}      
            & FGD1 & 2.35 \\
            && FGD2 & 2.35 \\
    \hline
    \hline
    \end{tabular}
    \caption{Uncertainties on the total number of events in the ND analysis from detector uncertainties only, broken down by selection.}
    \label{tab:nd_syst}
\end{table}

The uncertainties from the detector uncertainties are presented in \autoref{tab:nd_syst}, and are $1.2-2.1\%$ for the \CCzeropi selections, and $2.5-4.0\%$ for the \CConepi and \CCother selections. The secondary interaction uncertainty for pions contribute $70-95\%$ of the total detector-related uncertainties, depending on the selection. For reference, the statistical uncertainty on the number of events in the ND selections, presented later in \autoref{tab:nd_event_rates}, is $0.5-1.3\%$ for the \fhcalt selections, and $1.1-3.9\%$ for the \rhcalt selections.

\subsection{Defining the likelihood}
\label{sec:nd:likelihood}
Each selection is binned in the reconstructed muon momentum, \pmu, and the cosine of the muon angle with respect to the detector $z$-axis, $\cos{\theta_{\mu}}$, which nearly lines up with the average neutrino direction\footnote{The average neutrino direction in the ND coordinates is $\hat{r}=\left(-0.0128224, -0.0249785, 0.999586\right)$, where the coordinate system shown in ~\autoref{fig:nd280}, with the $z$-axis defined as the side of ND280 which is most parallel to the neutrino beam, and the $y$-axis is defined as the vertical.}. 
The ND likelihood is constructed by calculating the $-2\ln\mathcal{L}_\textrm{total}$ of the data and simulation (MC) across all bins in all samples at each set of the parameter values. The systematic uncertainties in the models for the ND response, neutrino interactions, and neutrino flux, detailed in previous sections, are encoded via a Gaussian penalty term, which includes the covariances between the systematic uncertainties, shown in \autoref{eq:prior}. The treatment of statistical uncertainties in the simulation has been updated~\cite{Barlow:1993dm,Conway:2011in} and was validated against a complementary approach~\cite{Arguelles:2019izp} and the previously used method. The total likelihood is defined as
\begin{equation}
    \label{eq:nd_llh}
    \mathcal{L}_\textrm{total} = \mathcal{L}_\textrm{stat} \times \mathcal{L}_\textrm{MC stat} \times \mathcal{L}_\textrm{syst}
\end{equation}
where $\mathcal{L}_\textrm{stat}$ is the statistical likelihood, $\mathcal{L}_\textrm{MC stat}$ is the MC statistical uncertainty likelihood, and $\mathcal{L}_\textrm{syst}$ is the likelihood of the systematic uncertainties. The frequentist analysis maximises $\mathcal{L}_\textrm{total}$ by finding the minimum of $-2\ln\mathcal{L}_\textrm{total}$, and the Bayesian analysis samples the $-2\ln\mathcal{L}_\textrm{total}$ around the minimum in proportion to the posterior probability. The first two terms in \autoref{eq:nd_llh} are linked, as the statistical uncertainty on the MC affects the number of MC events. The two statistical contributions read,
\begin{equation}
\begin{split}
-2\ln\mathcal{L}_\textrm{stat} - 2\ln\mathcal{L}_\textrm{MC stat} &=  \\
   2 \sum^{\text{samples}}_{i} \sum^{\text{bins}}_{j} \Bigg[ \bigg( N_{\text{MC}}-N_{\text{Data}} \Bigg. \bigg.\Bigg. \bigg. &+N_{\text{Data}}\ln{\frac{N_{\text{Data}}}{N_{\text{MC}}}} \bigg)+\frac{\left(\beta_j-1\right)^{2}}{2\sigma^{2}_{\beta_j}} \Bigg]
\end{split}
\label{eq:barlowbeeston_total}
\end{equation}
where in each bin $j$ of sample $i$, $N_\text{Data}$ ($N_\text{MC}$) is the number of events in data (MC), $\beta_j$ scales the unweighted MC events, and $\sigma_{\beta_j}$ is a measure of the MC statistical uncertainty. The systematic uncertainties are parametrised as correlated Gaussian penalties,
\begin{equation}
    -2\ln\mathcal{L}_\textrm{syst} = \left(\vec{x}-\vec{\mu}\right)^T \mathbf{V}^{-1} \left(\vec{x}-\vec{\mu}\right)
    \label{eq:prior}
\end{equation}
where $\vec{x}$ ($\vec{\mu}$) are the values of the systematic uncertainty parameters during (before) the fit, and $\mathbf{V}$ is their covariance matrix. The ND constrains the flux uncertainty at the FD through such a covariance matrix. The low-momentum \numub SPP, neutrino energy-dependent 2p2h, NC other, NC$1\gamma$, and \nueany/\numuany parameters are barely constrained by the ND analysis, so their constraints are not propagated to the FD in the frequentist analysis. In the simultaneous ND+FD Bayesian analysis, both detectors are used to constrain these parameters.

\subsection{Results of the ND analysis}
\label{sec:nd_results}
\begin{figure}[htbp]
    \centering
    \includegraphics[width=0.49\textwidth, trim=0mm 7mm 0mm 10mm, clip]{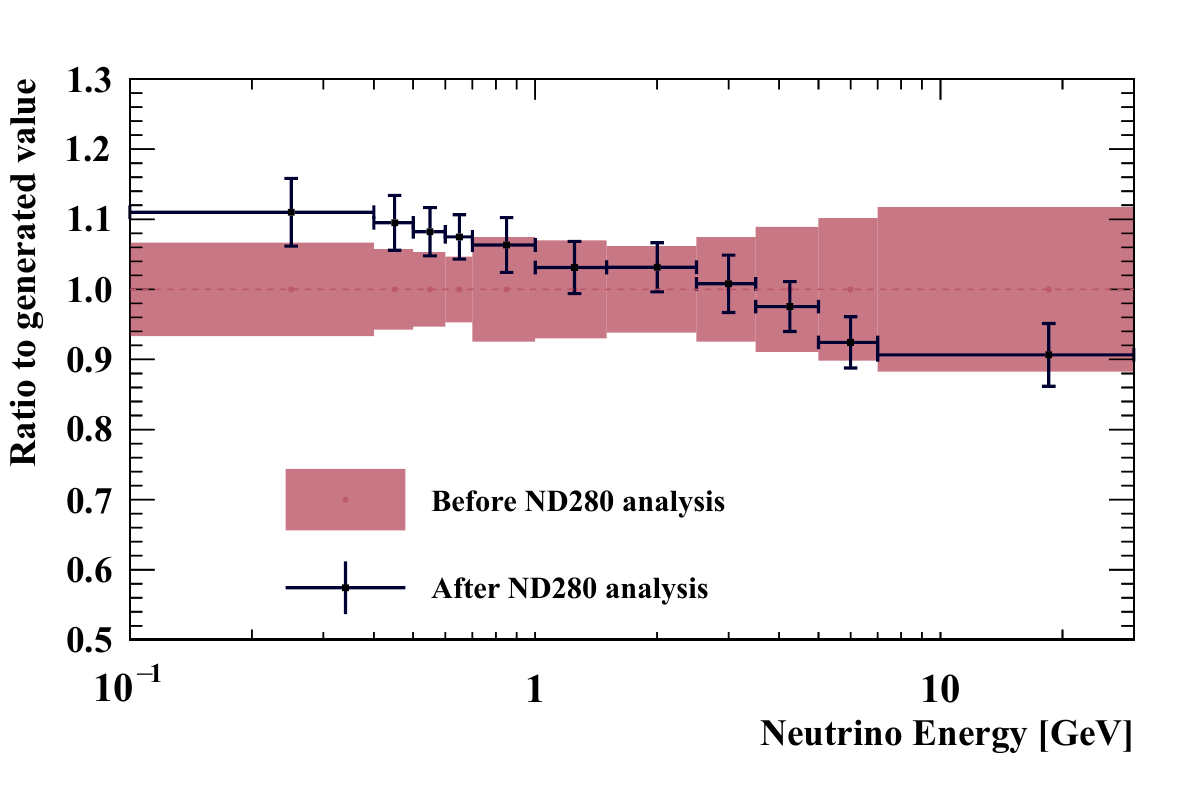}
    \caption{Constraints on the \fhcalt $\nu_{\mu}$ flux uncertainty parameters at the FD from the fit to ND data (black points, black lines), overlaid on the input uncertainty (red band).}
    \label{fig:NDFluxParams}
\end{figure}

\begin{figure*}[htbp]
    \centering
    \includegraphics[width=0.8\textwidth]{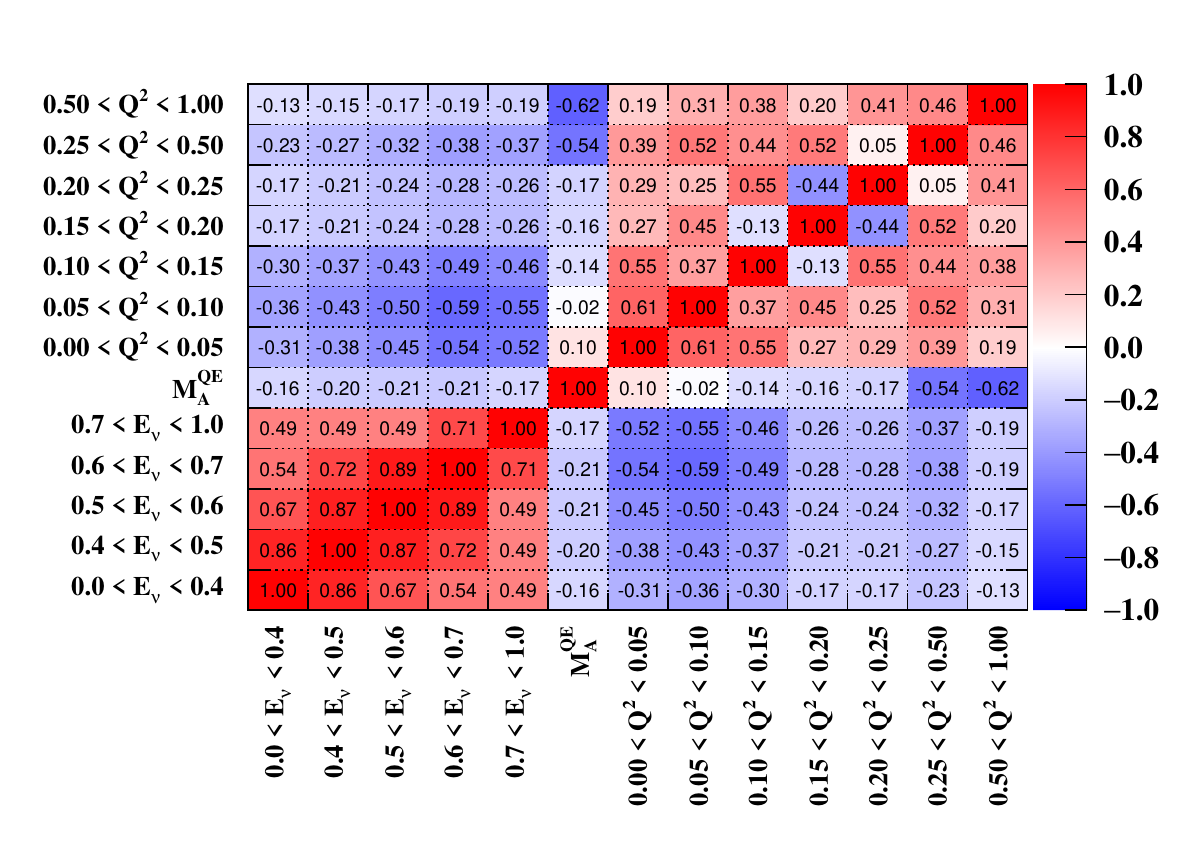}
    \caption{Correlations between selected \fhcalt $\nu_{\mu}$ FD flux and CCQE cross-section parameters. The flux and $Q^2$ normalisation parameters' ranges are in units of~GeV. The strong anti-correlations between the flux and cross-section parameters significantly reduce the uncertainties on the predictions at the FD.}
    \label{fig:corr_matrix}
\end{figure*}

The ND analysis sees large shape changes in the \fhcalt \numu flux parameters with roughly 10\% enhancement at low $E_{\nu}$ and 10\% suppression at high $E_{\nu}$, as shown in \autoref{fig:NDFluxParams}. The neutrino flux parameters have strong correlations with each other and with some cross-section parameters, such as $M_A^{QE}$ and the $Q^2$ parameters, shown in \autoref{fig:corr_matrix}. Moving the flux parameters by this amount incurs a penalty of $-2\ln\mathcal{L_\textrm{flux}}/N_\text{dof}\sim1$ for this variation in flux parameters due to the large correlations, confirmed by $p$-value studies in \autoref{sec:ndfit_pvalue}.

\begin{figure}[htbp]
    \centering
    \includegraphics[width=0.49\textwidth]{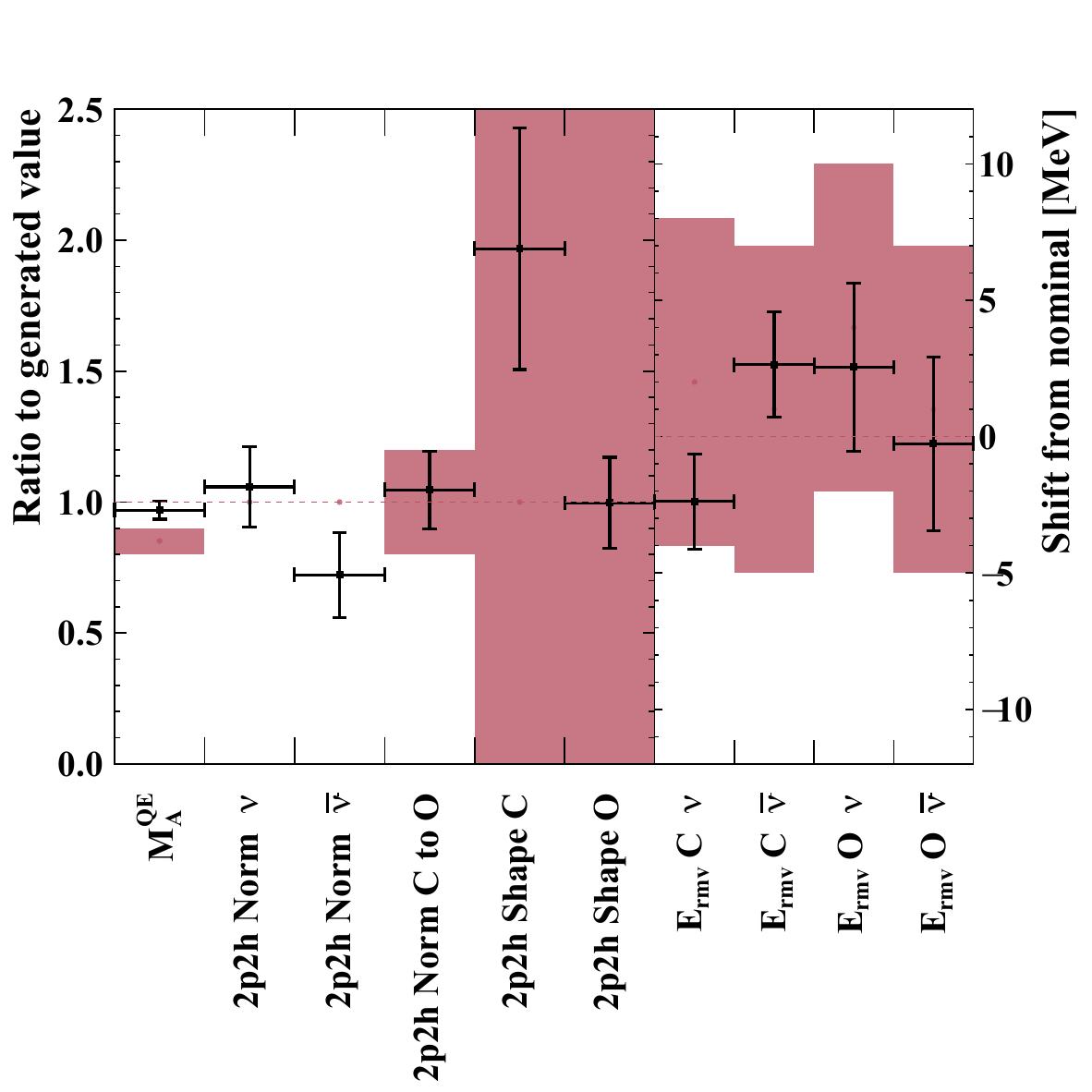}
    \caption{Constraints on the \CCzeropi parameters, excluding the CCQE $Q^2$ parameters, from the fit to ND data (black points, black lines), overlaid on the input uncertainty (red band). The parameters on the left-hand side of the figure are presented as a ratio to the generated value in NEUT, and the right side shows the removal energy parameters, $E_{rmv}$, with shifts in units of MeV. 
    CCQE interactions are generated in NEUT with $M_A^{\textrm{QE}}=1.21~\mathrm{GeV}$, but a pre-fit value of $1.03~\mathrm{GeV}$ was used after analysis of CCQE bubble chamber data. The absence of an uncertainty band reflects that the parameter was not constrained by external inputs before the analysis of ND data.}
    \label{fig:NDCC0piParams}
\end{figure}

\begin{figure}[htbp]
    \centering
        \includegraphics[width=0.49\textwidth,trim=0mm 7mm 0mm 0mm, clip]{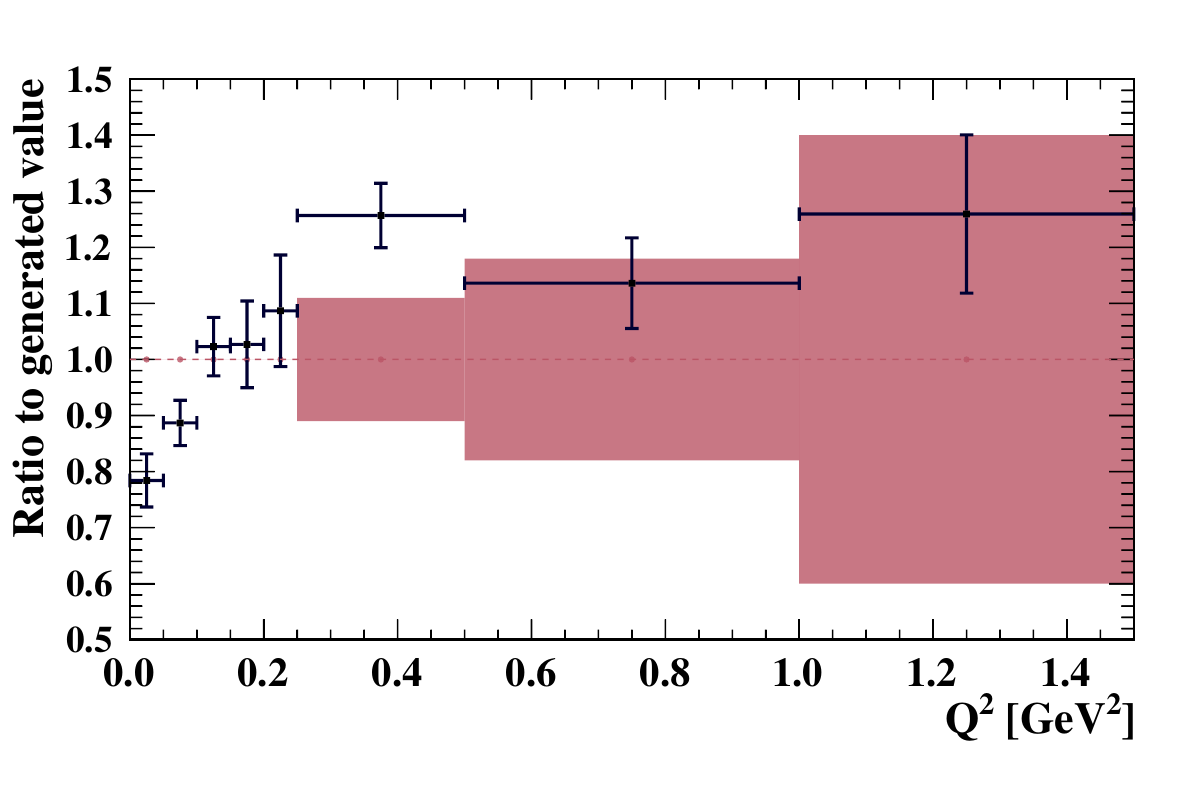}
    \caption{Constraints on the CCQE $Q^2$ parameters as a function of $Q^2$ from the fit to ND data (black points, black lines), overlaid on the input uncertainty (red band). The absence of an uncertainty band reflects that the parameter was not constrained by external inputs before the analysis of ND data.}
    \label{fig:NDQ2Params}
\end{figure}

\autoref{fig:NDCC0piParams} and \autoref{fig:NDQ2Params} show the \CCzeropi cross-section parameters after the fit. Despite the external constraint on $M_A^{QE}$, the data prefers a larger value of $M_A^{QE}=1.16~\text{GeV}$. A complementary fit, changing the uncertainty on $M_A^{QE}$ to be unconstrained instead of informed by bubble chamber data, had little impact on the ND analysis and the predictions at the FD; hence the constraint on $M_A^{QE}$ is primarily driven by the ND data. The 2p2h normalisation is different for neutrinos and anti-neutrinos, which are both constrained to $\sim15\%$ uncertainty, with the 2p2h normalisation for neutrinos consistent with the prediction from Nieves \etal The 2p2h normalisation for carbon and oxygen is consistent with 1, although the shape parameter for oxygen agrees with the Nieves model, whereas the carbon parameter is pulled to be more $\Delta$-like, differing by $\sim1\sigma$. The removal energy parameters are within their uncertainties before the fit, and are compatible for the carbon, oxygen, neutrino and anti-neutrino parameters.

The CCQE $Q^2$ parameters are shown in \autoref{fig:NDQ2Params}, where there is a suppression at low $Q^2$ until $0.2~\text{GeV}^2$, consistent with other cross-section data mentioned in \autoref{subsec:intModelUncert}. At higher $Q^2$ the data prefers an enhancement of $20-30\%$. The $Q^2$ parameters have strong anti-correlations with the flux parameters, as shown in \autoref{fig:corr_matrix}, and studies with fixed values of the $Q^2$ parameters showed that the flux parameters compensate for differences in $Q^2$ for CCQE events, a testament to the parameters' correlations.

The 2p2h normalisation has been given the freedom to independently vary for neutrino and anti-neutrinos, and differences in 2p2h neutrino and anti-neutrino parameters may reflect a more general mismodelling of \CCzeropi interactions. This may 
allow deficiencies in the anti-neutrino CCQE model to be absorbed in the 2p2h normalisation parameters. Similarly, $M_A^{QE}$ and the CCQE $Q^2$ normalisation parameters may be absorbing effects from a different axial form factor parametrisation, which may evolve differently as a function of other variables, e.g. $E_\nu$, as mentioned in \autoref{sec:interactionModel_fds}. Both of these effects, amongst others, are studied through simulated data studies in \autoref{sec:fakeData} and \autoref{app:appendix_fakedata}. The full parameter set with their values before and after the analysis of ND data is provided in \autoref{app:appendix_NDFitValues}.

The MCMC and gradient-descent analyses differed in the treatment of the removal energy uncertainty. 
The MCMC allows for discrete movement of events between bins, which may produce multi-modal posterior probability distributions (output constraint) of the removal energy parameters. The smoothed binned implementation in the gradient-descent framework prevents this from disrupting the ability to find the maximum likelihood, whilst still capturing the overall physics behaviour of the removal energy uncertainty.
The impact of this and other differences between the analyses, such as the non-uniform rectangular binning scheme, were addressed by separately propagating the covariance matrix from the gradient-descent framework and the parameter variations sampled by the MCMC to the oscillation analysis in \autoref{sec:oa_results}.

In general, the constraints on the parameters and the impact of the ND analysis agrees with the expected sensitivity. Furthermore, compatible results are found between the MCMC and the gradient-descent analyses in the central value estimates, uncertainties, and correlations of the parameters, leading to consistent sample predictions at the ND and the FD.

\subsection{ND predictions}
The aforementioned selections in the data and simulation are compared before and after fitting to data, using the constraints on the systematic uncertainties from \autoref{sec:nd_results}. \autoref{tab:nd_event_rates} shows the number of events in each selection, where the agreement between the post-fit prediction and the data is notably improved compared to that of the pre-fit prediction, especially for the \CCzeropi events, which comprise the main signal at the FD. There is a consistent rise across all \CCzeropi selections and a small suppression of \fhcalt $1\pi^+$ events, improving agreement with the data. This causes the smaller \rhcalt $1\pi^-$ prediction to also be suppressed, since they share parameters in the interaction model, with the neutrino flux and detector uncertainties being more loosely correlated, connected only through their input covariance matrices.

\begin{table}[htbp]
    \centering
    
\resizebox{.5\textwidth}{!}{%
    \begin{tabular}{c|c|c||c|c|c}
        \hline
        \hline
        Selection & Topology & Target & Data & Pre-fit/Data & Post-fit/Data \\
        \hline
        \multirow{6}{*}{\numu in \fhcalt} 
        &\multirow{2}{*}{0$\pi$}      
            & FGD1 & 33443 & 0.91 &	1.00 \\
            && FGD2 & 33156 & 0.91 &	1.00 \\
            \cline{2-6}
        &\multirow{2}{*}{$1\pi^+$}      
            & FGD1 & 7713 & 1.09 & 1.03\\
            && FGD2 & 6281 & 1.09 & 1.03\\
            \cline{2-6}
        &\multirow{2}{*}{Other}      
            & FGD1 & 8026 & 0.88 & 0.99\\
            && FGD2 & 7700 & 0.84 & 0.95\\

        \hline
        \multirow{6}{*}{\numub in \rhcalt}
            &\multirow{2}{*}{0$\pi$}      
            & FGD1 & 8388 & 0.97 & 1.00\\
            && FGD2 & 8334 & 0.94 & 0.98\\

            \cline{2-6}
        &\multirow{2}{*}{1$\pi^-$}      
            & FGD1 & 698 & 1.00 & 0.98\\
            && FGD2 & 650 & 0.96 & 0.98\\
            \cline{2-6}
        &\multirow{2}{*}{Other}      
            & FGD1 & 1472 & 0.88 & 1.00\\
            && FGD2 & 1335 & 0.89 & 1.03\\

        \hline
        \multirow{6}{*}{\numu in \rhcalt}
            &\multirow{2}{*}{0$\pi$}      
            & FGD1 & 3594 & 0.89 & 1.00 \\
            && FGD2 & 3433 & 0.92 & 1.03\\

            \cline{2-6}
        &\multirow{2}{*}{1$\pi^+$}      
            & FGD1 & 1111 & 1.04 & 1.04\\
            && FGD2 & 926 & 1.01 & 0.99 \\
            \cline{2-6}
        &\multirow{2}{*}{Other}      
            & FGD1 & 1344 & 0.80 & 0.96 \\
            && FGD2 & 1245 & 0.81 & 0.96 \\

    \hline
    \hline
    \end{tabular}
    }
    \caption{Number of events in each of the ND selections for data and the ratio to the prediction before and after the fit to data.}
    \label{tab:nd_event_rates}
\end{table}

\begin{figure*}[htbp]
    \centering
    \begin{subfigure}[t]{0.49\textwidth}
    \includegraphics[width=\textwidth, trim=0mm 0mm 0mm 7mm, clip]{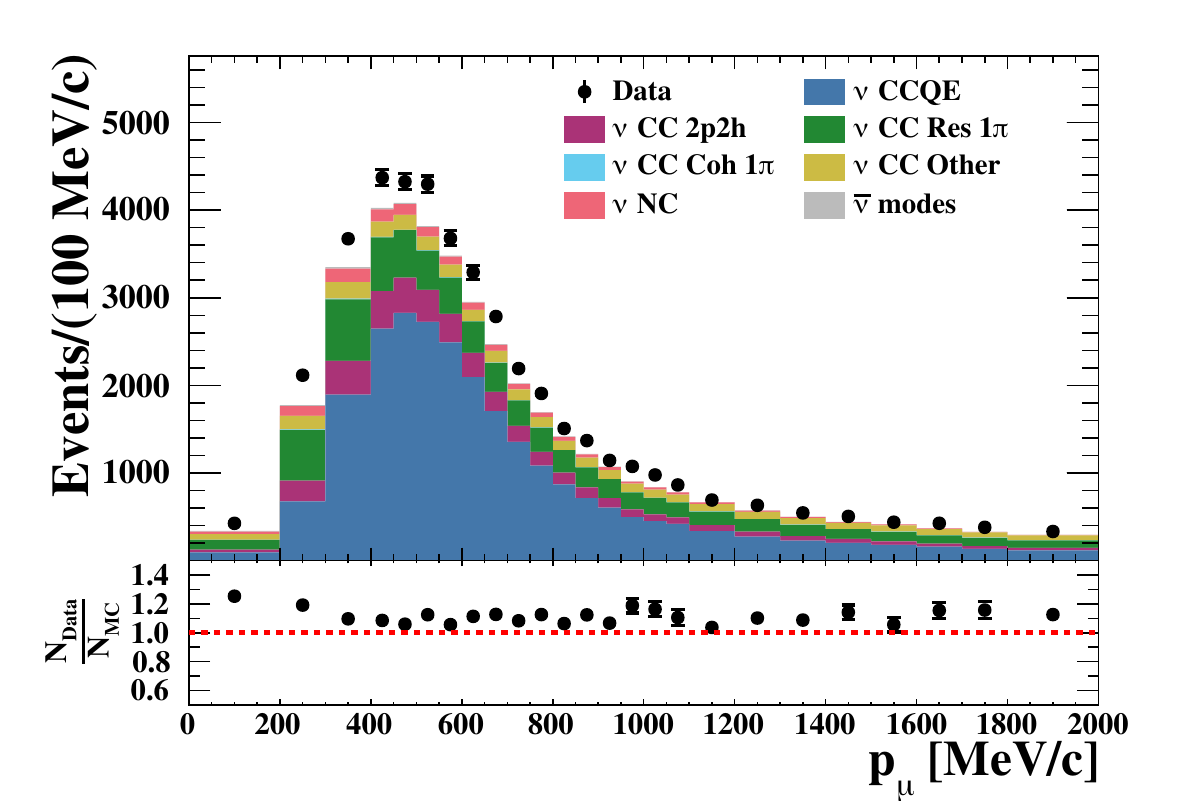}
    \includegraphics[width=\textwidth, trim=0mm 0mm 0mm 7mm, clip]{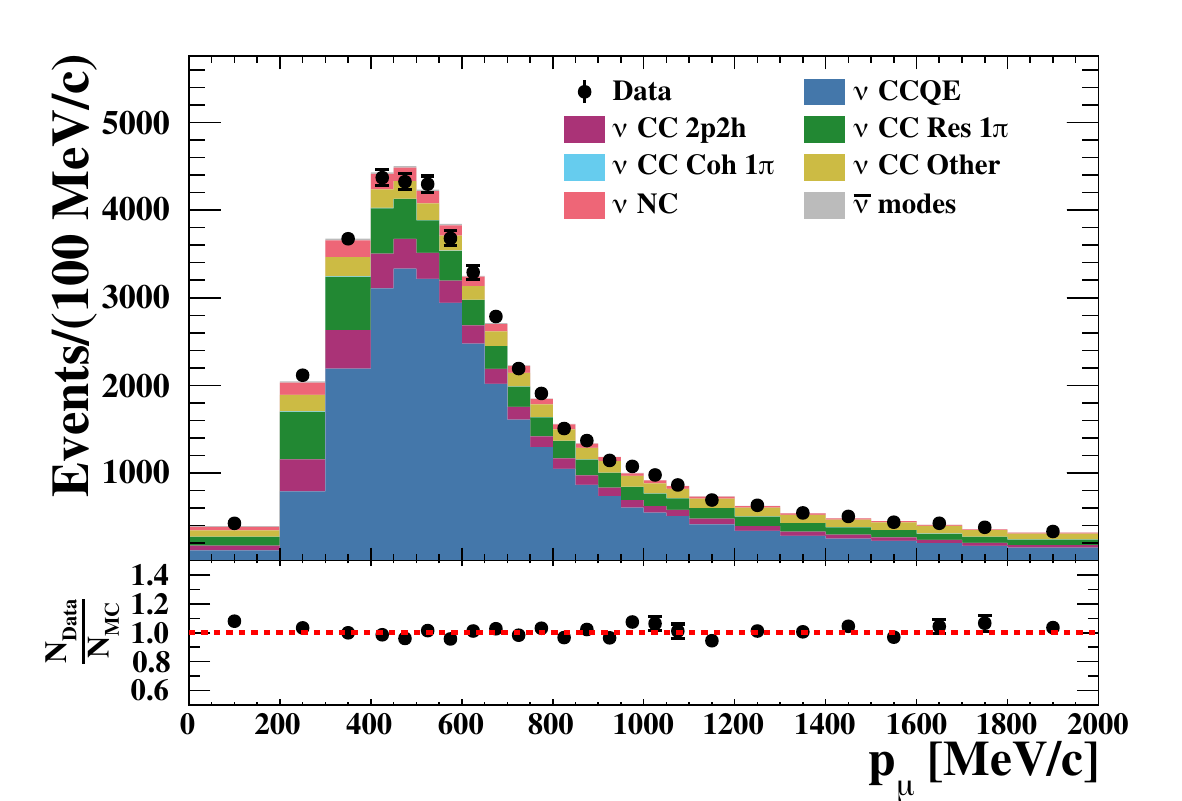}
    \caption{FGD1 \fhcalt \numu CC0$\pi$}
    \end{subfigure}
    \begin{subfigure}[t]{0.49\textwidth}
    \includegraphics[width=\textwidth, trim=0mm 0mm 0mm 7mm, clip]{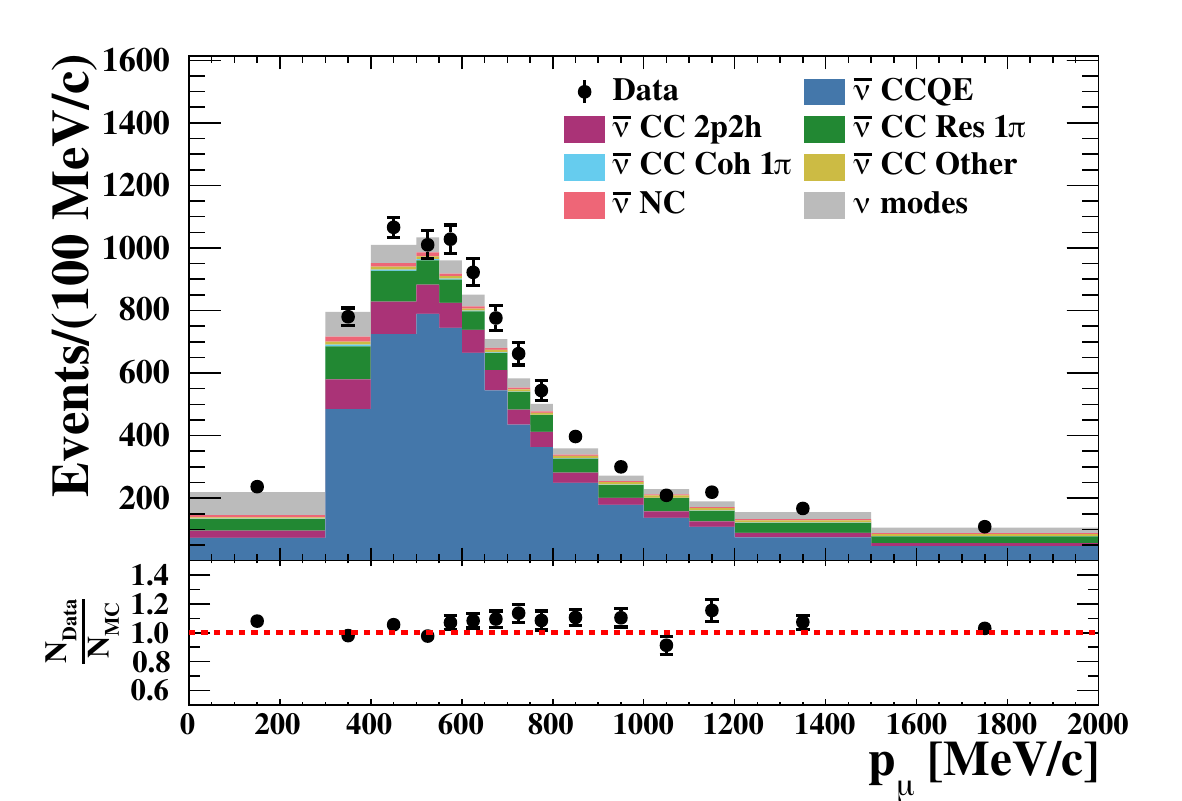}
    \includegraphics[width=\textwidth, trim=0mm 0mm 0mm 7mm, clip]{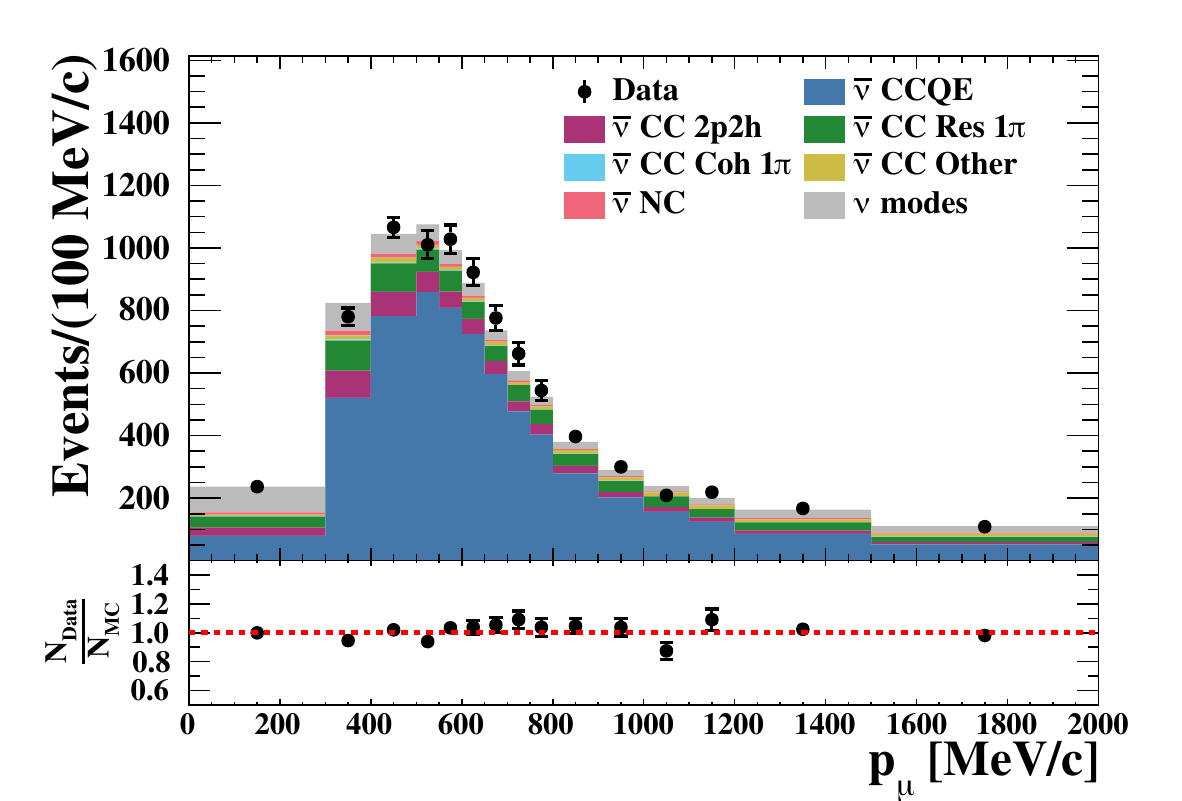}
    \caption{FGD2 \rhcalt \numub CC0$\pi$}
    \end{subfigure}
    \caption{Comparison of predicted pre-fit (top) and post-fit (bottom) event distributions for the ND FGD1 \fhcalt \numu \CCzeropi sample (left) and FGD2 \rhcalt \numub \CCzeropi sample (right). The data and prediction are shown in the reconstructed momentum of the muon candidate, and the simulation is broken down by interaction channel. The bottom insets show the ratio of data to simulation.}
    \label{fig:NDmode_predictions}
\end{figure*}

The observed and predicted \fhcalt \numu FGD1 \CCzeropi events projected onto \pmu are shown in \autoref{fig:NDmode_predictions} before and after the fit to data. Before the fit, there is a notable under-prediction which is largest at low \pmu. The fit increases the CCQE and 2p2h components and decreases the $1\pi$ components in the prediction to agree with the data. For comparison, the \rhcalt \numub FGD2 \CCzeropi selection is also shown in \autoref{fig:NDmode_predictions}, where there is agreement between the prediction and the data before the fit, which marginally improves after the fit. This showcases the ability of the systematic uncertainty treatment in the analysis to modify and constrain the modelling of neutrino and anti-neutrino interactions on carbon and oxygen separately, and the strength of having a sign-selecting ND.

\subsection{Assessing model compatibility with data}
\label{sec:ndfit_pvalue}
A $p$-value is calculated to assess the probability of the model given the data, and represents the probability that a model with a test statistic equal to or larger than the observed data is found. 
Simulated data sets, referred to as ``toys'', are created by varying the systematic uncertainties in the model according to their input covariances before the ND analysis, and statistical fluctuations are applied.
The model is fit to each toy and the $(-2\ln\mathcal{L})_\mathrm{min}$ is calculated. The $p$-value is defined as the fraction of the simulated data sets with \newline $(-2\ln\mathcal{L})_\mathrm{min}^\mathrm{Toy} \geq (-2\ln\mathcal{L})_\mathrm{min}^\mathrm{Data}$. An a priori criteria of $p>0.05$ is required of the ND analysis for the results to be used in the oscillation analysis. Using a total of 895 simulated data sets, $p=0.74$, demonstrating good agreement between the model and the data. 

\begin{table}[htbp]
    \centering
    \begin{tabular}{c | c | c || c}
    \hline\hline
        \multicolumn{3}{c||}{Likelihood}  & \multirow{2}{*}{$p$-value} \\
        \multicolumn{3}{c||}{contributor} & \\
        \hline
        \multirow{2}{*}{\numu in \fhcalt}
            & \multirow{2}{*}{0$\pi$} & FGD1 & 0.93 \\
            &                         & FGD2 & 0.93 \\
          
        \hline
        \multirow{2}{*}{\numub in \rhcalt}   
            & \multirow{2}{*}{0$\pi$} & FGD1 & 0.20 \\
            &                         & FGD2 & 0.15 \\
          
        \hline
        \multirow{2}{*}{\numu in \rhcalt}   
            & \multirow{2}{*}{0$\pi$} & FGD1 & 0.54 \\
            &                         & FGD2 & 0.45 \\
            
        \hline
        \multicolumn{3}{c||}{All samples}   & 0.82 \\
        \hline
        \multicolumn{3}{c||}{Neutrino flux} & 0.46 \\
        \multicolumn{3}{c||}{ND detector}   & 0.06 \\
        \multicolumn{3}{c||}{Cross section} & 0.01 \\
        \hline
        \multicolumn{3}{c||}{All samples, all syst.} & 0.74 \\
    \hline\hline
    \end{tabular}
    \caption{$p$-values comparing the variations of the model before the ND analysis and the model fit to the data, broken down by likelihood contributors, and showing the $p$-value for all samples, and the total $p$-value including all samples and all systematic uncertainties.}
    \label{tab:p_values}
\end{table}

Breaking down the $(-2\ln\mathcal{L})_\mathrm{min}$ contributions by the likelihoods from the selected samples and systematic uncertainties in \autoref{tab:p_values}, the selected samples are generally described well with $p=0.82$, with individual $p$-values for the \CCzeropi selections between $p=0.15-0.93$. 
Splitting the neutrino flux contributions into \fhcalt \numu, \fhcalt \numub, \rhcalt \numu and \rhcalt \numub, $p=0.74,0.74,0.31,0.37$ respectively, showing good compatibility. The cross-section systematics are the worst contributor with $p=0.01$, coming predominantly from parameters that are pulled away from their external constraints, e.g. $M_A^{QE}$, $M_A^{RES}$ and $C_5^A$. When instead varying the systematic uncertainty parameters with respect to their constraints after fitting to data, the cross-section model $p$-value improves to approximately $p=0.3$. This indicates that the cross-section model before the fit to data is unfavourable, but after the fit to data is satisfactory. 
The near-detector analysis constrains the product of the neutrino flux, ND detector, and neutrino interaction uncertainties, leading to large correlations between the systematic uncertainties, as demonstrated in \autoref{fig:corr_matrix}. Therefore, studying one group's $p$-value in isolation from the other is not exact. For this reason, the $p$-values from the uncertainty parameters do not have to follow the same strict criteria of $p>0.05$. However, the low $p$-value does highlight the need for continued effort in developing realistic neutrino interaction models and associated uncertainties.

%% file: SK.tex
\begin{figure*}[htbp]
\centering
\begin{subfigure}[b]{0.45\textwidth}
\includegraphics[width=\textwidth,trim=0mm 0mm 0mm 0mm, clip]{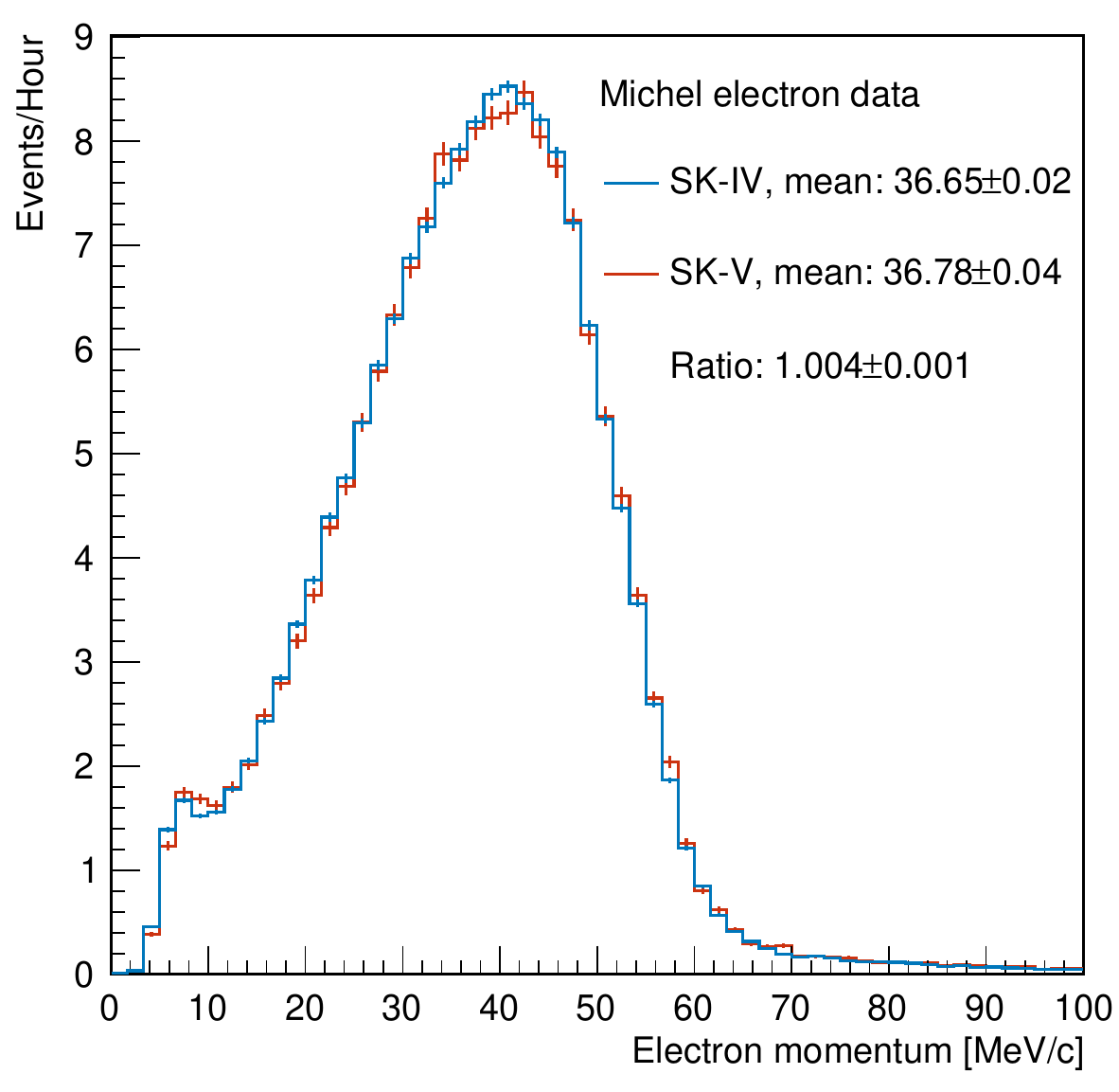}
\end{subfigure} 
\begin{subfigure}[b]{0.45\textwidth}
\includegraphics[width=\textwidth,trim=0mm 0mm 0mm 0mm, clip]{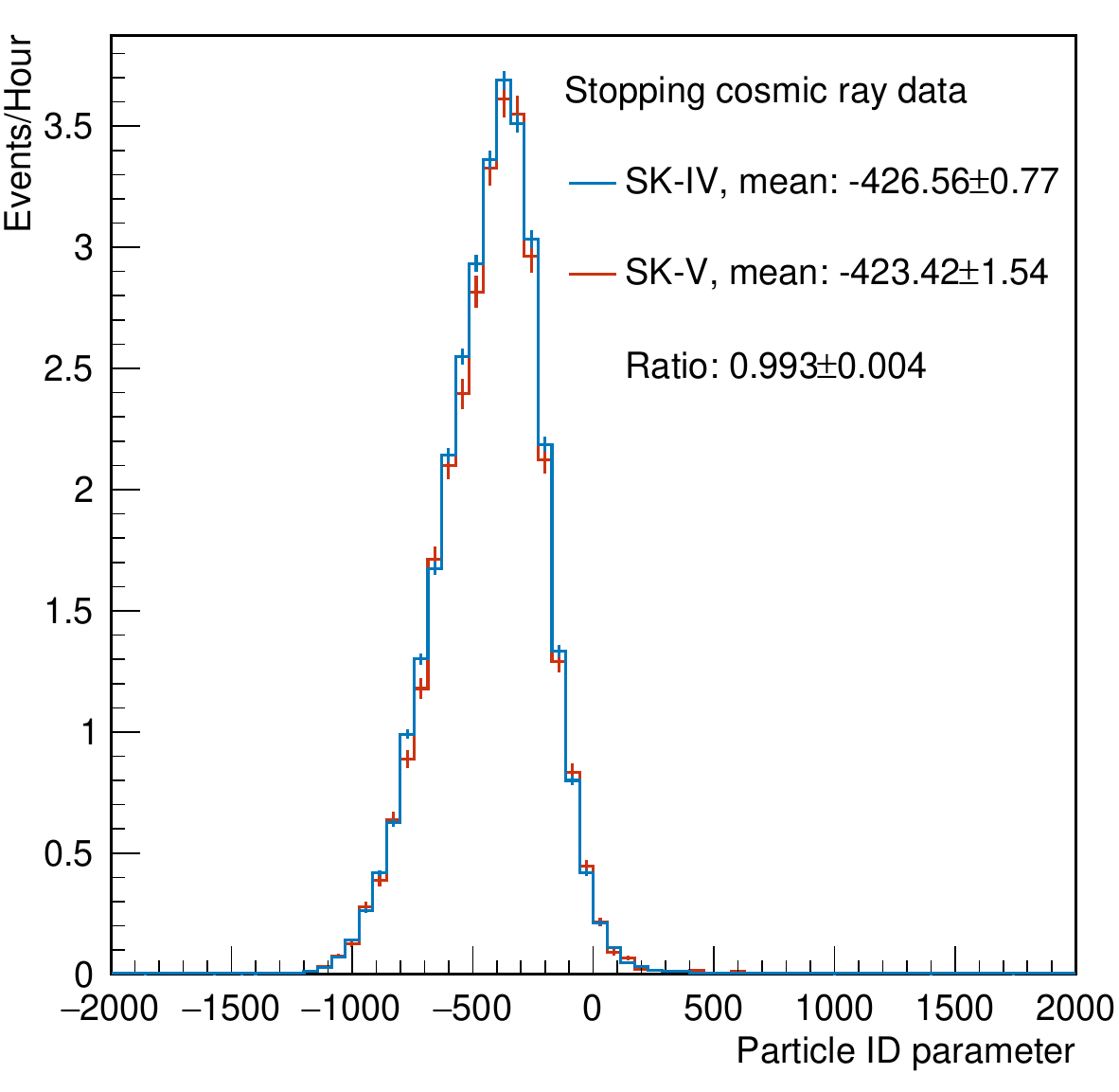}
\end{subfigure}
\caption{Reconstruction performance at the FD of stopping cosmic-ray muons and the Michel electrons from their decays. The left panel shows the reconstructed momentum distribution of those electrons for data taken during the SK-IV (blue) and SK-V (red) detector periods. The right is a similar comparison showing the parent muon's particle ID parameter, which separates events into electron-like (positive values) and muon-like (negative values) categories. The uncertainty on the data points is statistical.}
\label{fig:sk4sk5mucomp}
\end{figure*}

The FD event selection in this analysis is the same as used in previous T2K results~\cite{Abe:2021gky}; only the data have been updated, and the selection is briefly reviewed here. Similarly, the method of evaluating systematic uncertainties related to the FD is unchanged from previous analysis, where atmospheric events in SK are used to calculate the uncertainties using a MCMC-based approach.

The event reconstruction in SK uses both charge and timing information from hits in the PMTs, and particles are detected using their Cherenkov rings. The vertex position, momentum, and particle type of each ring is reconstructed~\cite{Super-Kamiokande:2019gzr}. Muons and electrons are differentiated by their ring profiles, where muons generally produce ``sharper'' rings due to less scattering, and electrons produces ``fuzzier'' rings due to their electromagnetic showers.
All samples in this analysis are based on observing one electron-like (\re) or muon-like (\rmu) primary Cherenkov ring, and a specific number of delayed triggers relative to the primary interaction, consistent with a Michel electron from an unseen charged pion's decay chain (referred to as decay electron, or ``\de''). 
Three samples are selected in the \fhcalt data: 
a CCQE-like \nue sample (\fhcalt \re with 0 \de), 
a CCQE-like \numu sample (\fhcalt \rmu with 0 or 1 \de), 
and a CC single pion-like \nue sample (\fhcalt \re with 1 \de).  
Similarly, there are two single-ring \rhcalt data samples:  
a CCQE-like \nueb sample (\rhcalt \re with 0 \de) and 
a CCQE-like \numub sample (\rhcalt \rmu with 0 or 1 \de). Unlike the ND, the FD is not magnetised and can therefore not determine the charge of the outgoing particles.

Since the start of T2K operations in 2009, the gain of the SK inner detector's PMTs has increased at a rate of at most a few percent per year. 
In previous T2K analyses, this effect was corrected during the reconstruction stage using a run-by-run global correction factor for all PMTs.
However, the gain drift differs based on the PMT production year, and the current analysis adopts a more detailed correction that accounts for these differences.
All T2K FD data in this analysis have been reprocessed and reconstructed using the updated correction.
The change to the gain correction results in a change in the observed charge available to the reconstruction algorithm relative to previous analyses, even when processing the same event. 
This may cause small shifts in an event's reconstructed parameters, including the number of rings, and each ring's particle type and momenta, which has caused some events to migrate into or out of the oscillation analysis samples with respect to the previous analyses.
For the reprocessed run $1-9$ data there are in total
1 more \fhcalt \re,
1 fewer \fhcalt \rede, 
1 more \rhcalt \re, 
and 3 fewer \rhcalt \rmu events compared to previous oscillation analysis. The migration of the events is summarised in \autoref{tbl:skmigration}.  
As the gain correction is applied to data and not to the simulation, the event migration has been cross-checked in both atmospheric neutrino and cosmic-ray muon data samples, which are used to evaluate FD detector uncertainties in the T2K analysis. In both studies, the level of migration was found to be consistent with that observed in the T2K beam data.

\begin{table}[htbp]
\resizebox{.5\textwidth}{!}{%
\begin{tabular}{l c |c|c|c|c}
\hline
\hline
\multicolumn{2}{c|}{Selection}         & Inward & Outward & Overlap & Net Change \\
\hline
\multirow{2}{*}{\rmu}   & \fhcalt     & 7 & 7 & 236 & 0 \\
                        & \rhcalt     & 3 & 6 & 134 & -3 \\
\hline
\multirow{2}{*}{\re}    & \fhcalt     & 4 & 3 & 72 & +1 \\
                        & \rhcalt     & 1 & 0 & 15 & +1 \\
\hline
\rede                   & \fhcalt     & 0 & 1 & 14 & -1 \\

\hline
\hline
\end{tabular}
}
\caption{Summary of event migrations at the FD after reprocessing data from the previous T2K analysis~\cite{Abe:2021gky, T2K:2019bcf}. ``Inward'' refers to newly added events that were not present in the previous analysis,  ``outward'' refers to events that were lost to the update, and ``overlap'' refers to the number of events that are common to the two analyses.}
\label{tbl:skmigration}
\end{table}

This analysis is the first to include data following the refurbishment of the FD in 2018, after the detector had been prepared for the gadolinium phase~\cite{Super-Kamiokande:2021the} but still using the ultrapure water without gadolinium, referred to as the SK-V period. Following this work, T2K's run 10 was under slightly different detector conditions than that of the previous data sets. This period had a larger background rate primarily at $\mathcal{O}(\text{MeV})$ energies, irrelevant to T2K's analysis.
During the run, the water's attenuation length, as measured by through-going cosmic-ray muons, was found to be stable above 90~m, consistent with data taken before the refurbishment, albeit slightly longer.
This suggests event reconstruction and detector uncertainties should similarly be consistent between the data periods, and several cross-checks were performed to confirm this.

\autoref{fig:sk4sk5mucomp} shows such a comparison between stopping cosmic-ray muon data and their Michel electrons taken during the run 9 and run 10 data periods at SK. The similarity of the distributions over both data sets highlights the stability of the detector and reconstruction algorithm following the refurbishment in 2018. 
Though only the reconstructed Michel momentum distribution and the parent muon's particle ID parameter are shown in the figure, distributions for other reconstructed parameters used in the T2K event selection showed similar high consistency. Kolmogorov-Smirnov tests of the expected events in run 10 confirmed this.
This was true for other calibration data as well as for atmospheric neutrino data, and small differences in these distributions were within current uncertainties.

\begin{figure}[htbp]
\centering
\includegraphics[width=0.45\textwidth,trim=0mm 0mm 0mm 10mm, clip]{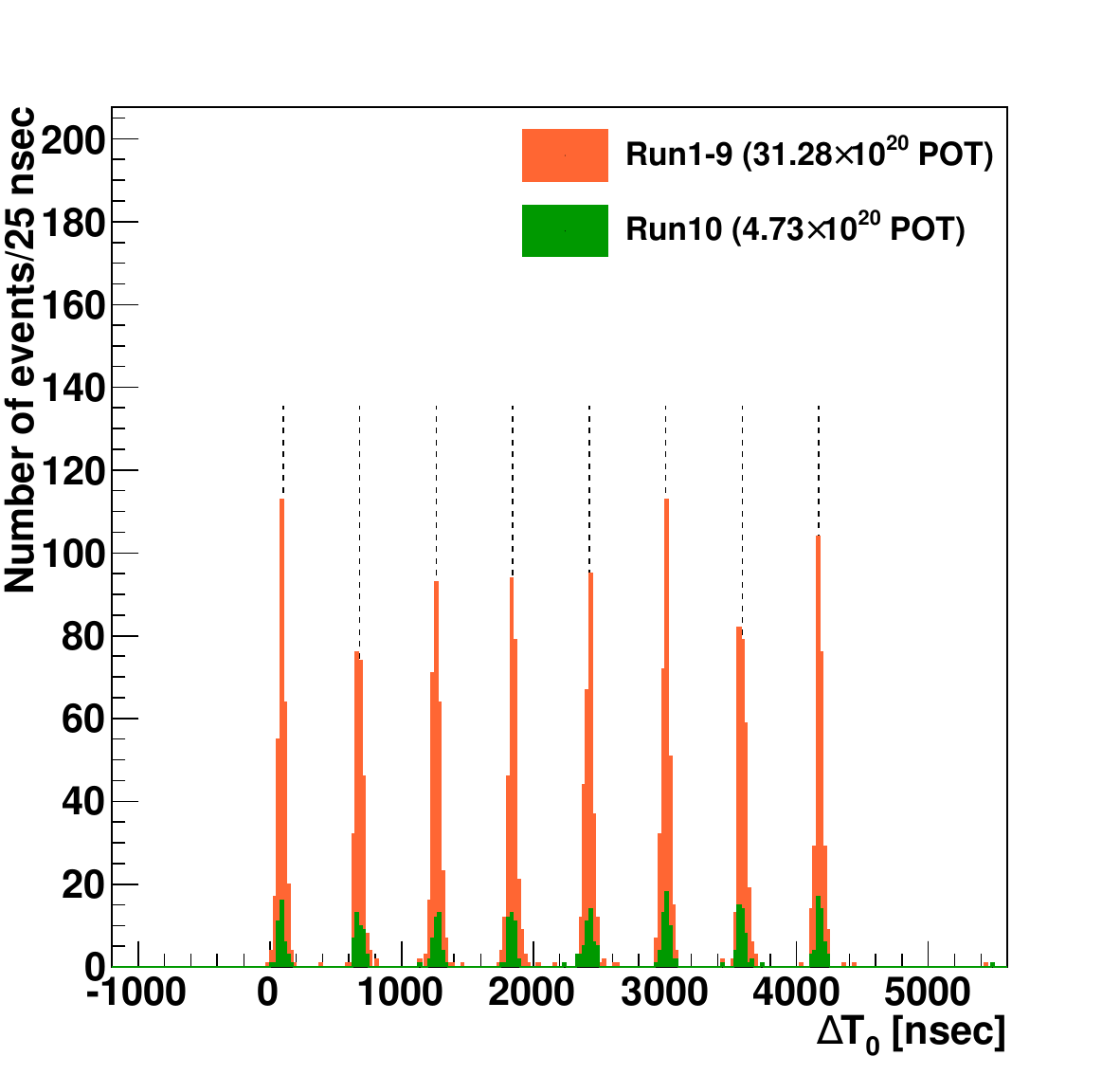}
\caption{Event timing at the FD for fully contained events collected during runs $1-9$ and run 10, overlaid with the central value of the expectation from the beam bunch timing structure.}
\label{fig:run10timing}
\end{figure}

Good detector stability was also found for the timing and selection of events observed in the T2K beam. 
The distribution of event times relative to the start of the spill at J-PARC is shown in \autoref{fig:run10timing} for events with minimal outer detector activity, labelled fully-contained events.
Events from run 10 showed a 34.2~ns RMS relative to their nearest expected bunch timing (dotted lines in the figure), consistent with that from previous runs.

Amongst the 354 selected fully-contained events in run 10, 75 were selected as \rmu, 18 as \re, and there were no new \rede events for the analysis described in the next section. The number of events in each selections is presented in \autoref{sec:oa_results}, \autoref{tab:oa:events:oscsystbestfit}.
\FloatBarrier

%% file: OA.tex
\begin{figure*}[htbp]
\centering
\begin{subfigure}[b]{0.49\textwidth}
\includegraphics[width=\textwidth]{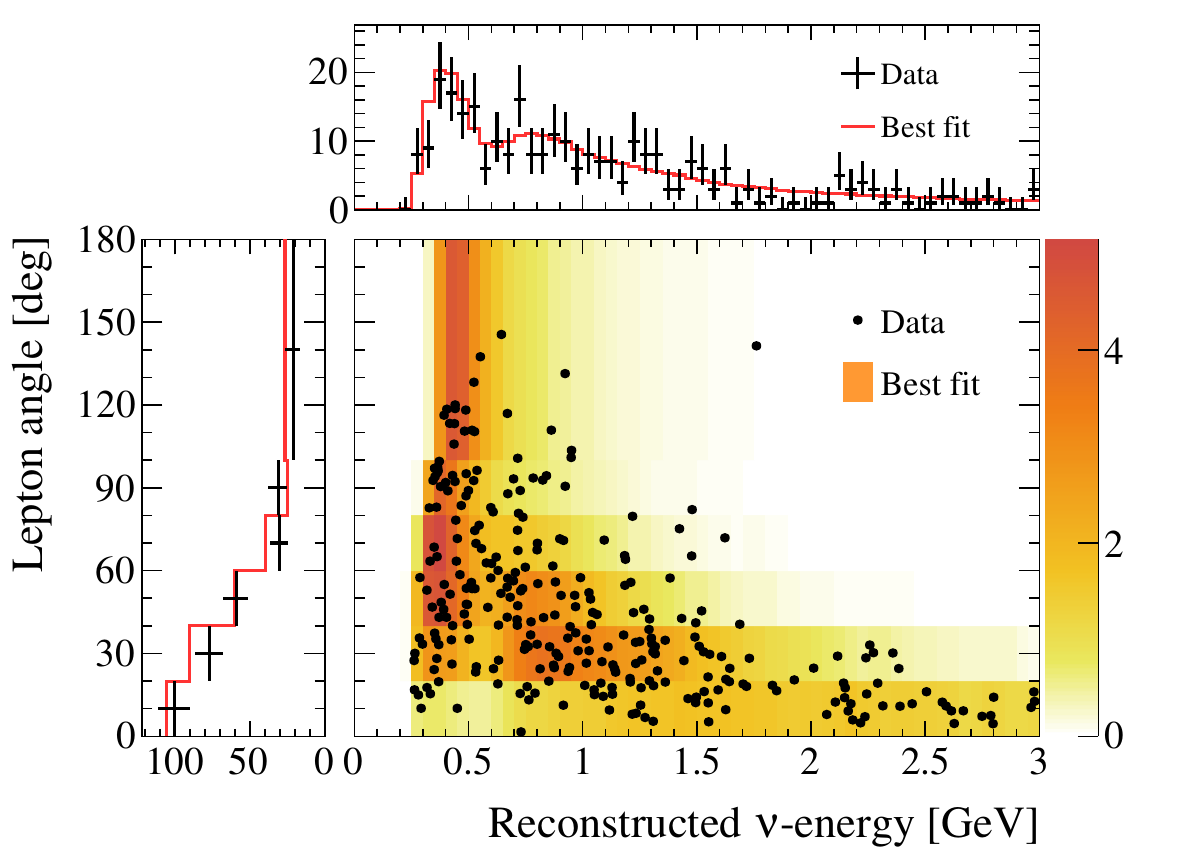}
\caption{\fhcalt \rmu}
\end{subfigure}
\begin{subfigure}[b]{0.49\textwidth}
\includegraphics[width=\textwidth]{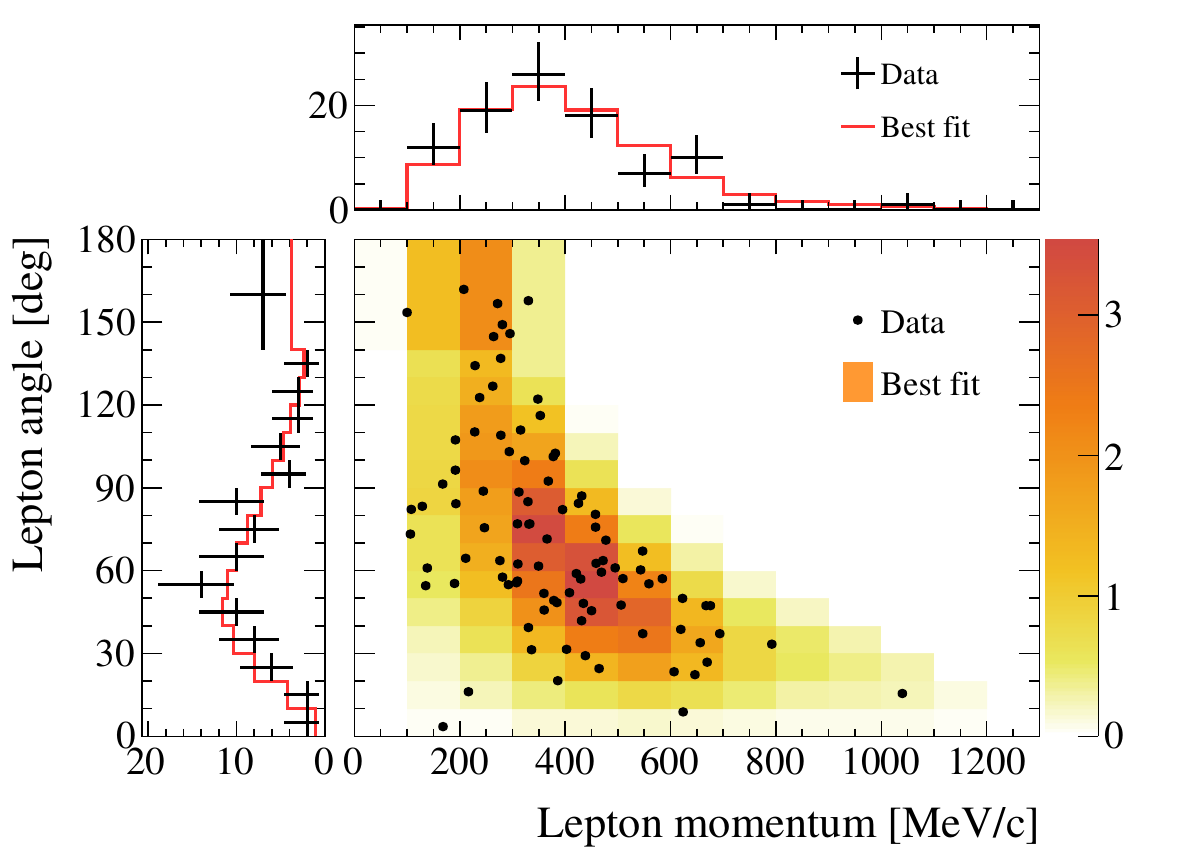}
\caption{\fhcalt \re}
\end{subfigure}
\\ \vspace{1em}
\begin{subfigure}[b]{0.49\textwidth}
\includegraphics[width=\textwidth]{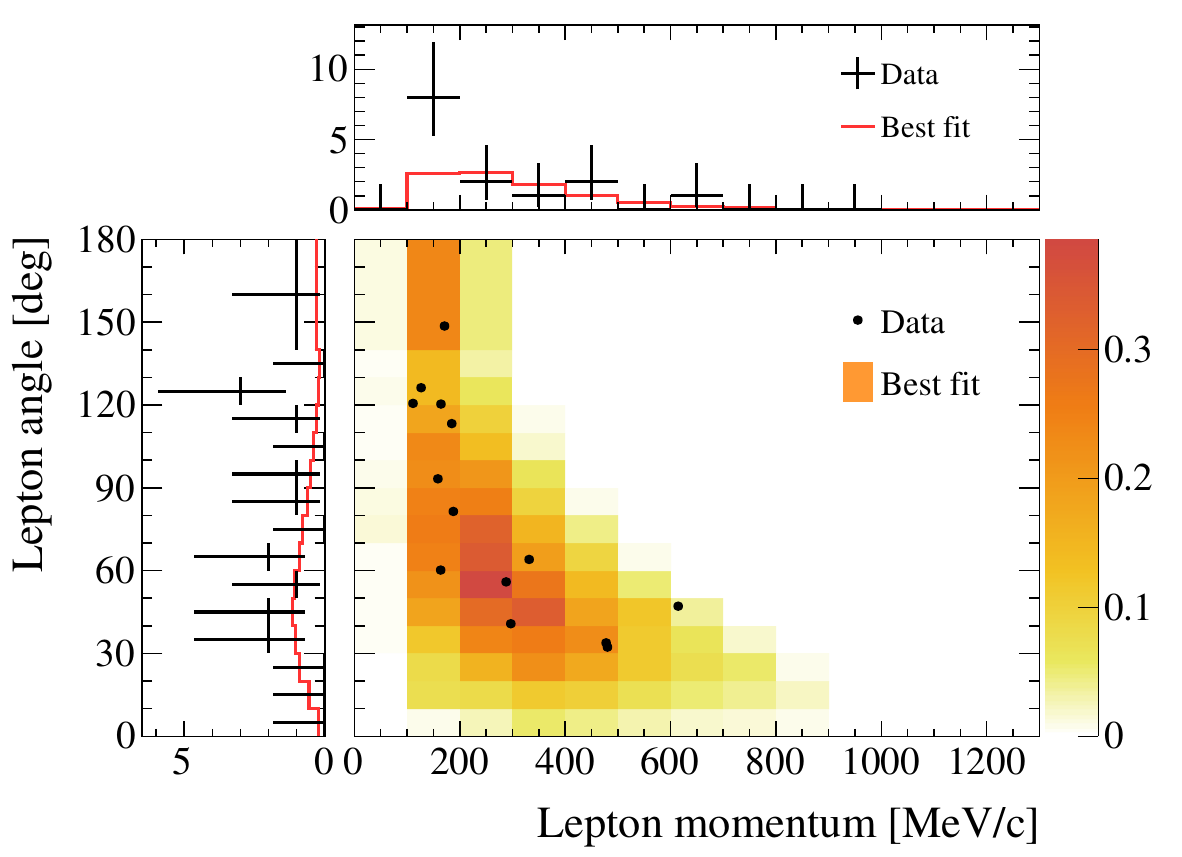} \caption{\fhcalt \rede}
\end{subfigure}
\\ \vspace{1em}
\begin{subfigure}[b]{0.49\textwidth}
\includegraphics[width=\textwidth]{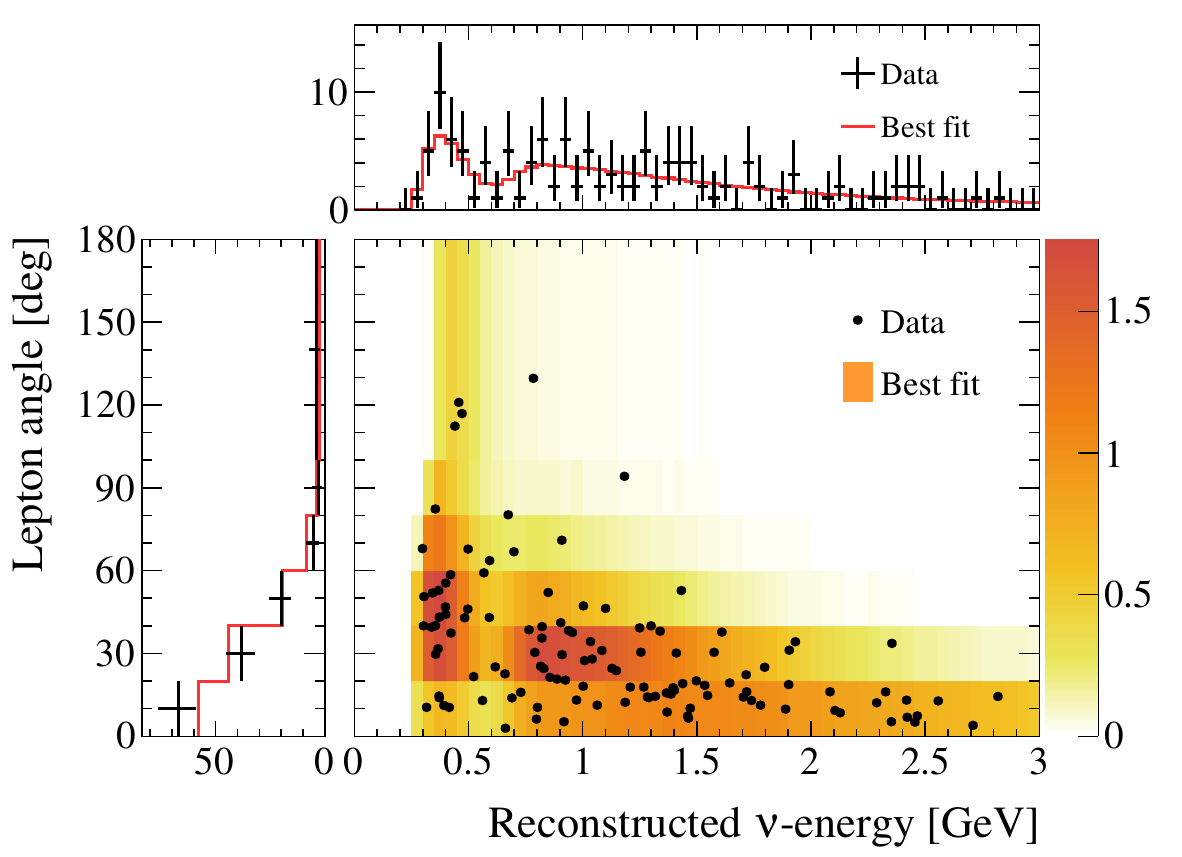}
\caption{\rhcalt \rmu}
\end{subfigure}
\begin{subfigure}[b]{0.49\textwidth}
\includegraphics[width=\textwidth]{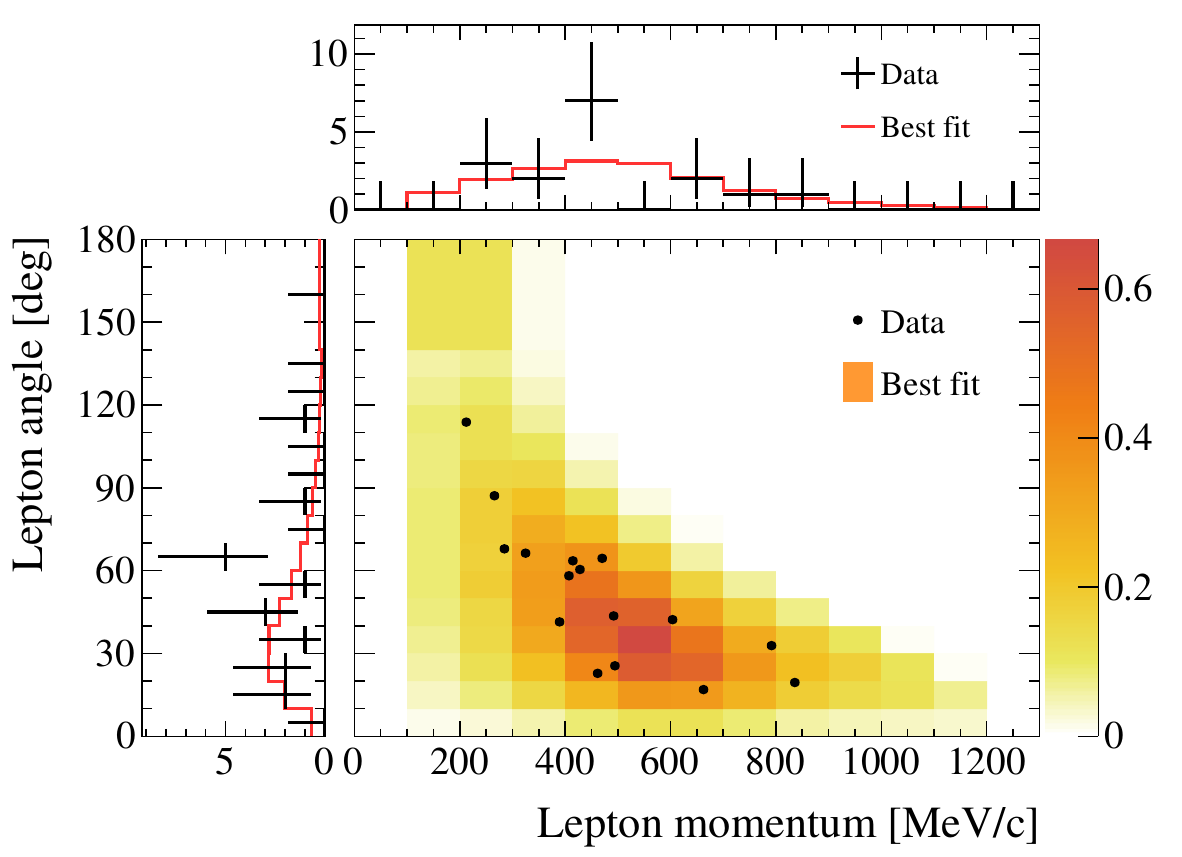}
\caption{\rhcalt \re}
\end{subfigure}
\caption{The events in the full data set for the five FD samples, shown in reconstructed lepton momentum and the angle between the neutrino beam and the lepton in the lab frame. The coloured background in the two-dimensional plot shows the expected number of events from the frequentist analysis, using the best-fit values for the oscillation and systematic uncertainty parameters, applying the reactor constraint on \ssqthonethree. The insets show the events projected onto each single dimension, and the red line is the expected number of events from the best-fit. The uncertainty represents the $1\sigma$ statistical uncertainty on the data.}
\label{fig:oa:dist}
\end{figure*}


This section presents the three-flavour oscillation analysis from the full data set presented in \autoref{fig:oa:dist}, including the constraints from the ND analysis in \autoref{sec:nd_fit}. The analyses at the FD are first introduced, followed by the constraints on the oscillation parameters from the Bayesian and frequentist data analyses in \autoref{sec:oa:bayesian} and \autoref{sec:oa:freq}, respectively. The comparison of the Bayesian and frequentist analyses are presented in \autoref{sec:oa_cross_fitter}, and the new result is put in the context of current world data in \autoref{sec:oa:comp}. The results presented in this section include the uncertainty inflation procedure from simulated data studies mentioned in \autoref{sec:interactionModel_fds}, whose results are discussed in detail later in \autoref{sec:fakeData} and \autoref{app:appendix_fakedata}.

\begin{table}[htbp]
\centering
\begin{tabular}{l c|cccc|c}
\hline
\hline
\multicolumn{2}{c|}{\multirow{2}{*}{Sample}} & \multicolumn{4}{c|}{True \deltacp (rad.)} & \multirow{2}{*}{Data}\\
\multicolumn{2}{c|}{}           & $-\pi/2 $ & $0$ & $ \pi/2$ & $  \pi$ & \\
\hline
\multirow{2}{*}{\rmu} & \fhcalt  & $346.61$  & $345.90$  & $346.57$  & $347.38$  & $318$     \\
                      & \rhcalt  & $135.80$  & $135.45$  & $135.81$  & $136.19$  & $137$     \\
\hline
\multirow{2}{*}{\re}  & \fhcalt  & $96.55$   & $81.59$   & $66.89$   & $81.85$   & $94$      \\
                      & \rhcalt  & $16.56$   & $18.81$   & $20.75$   & $18.49$   & $16$      \\
\hline
\rede                 & \fhcalt  & $9.30$    & $8.10$    & $6.59$    & $7.79$    & $14$      \\
\hline
\hline
\end{tabular}
\caption{Predictions for the number of events at the FD using oscillation parameters and systematic uncertainty parameters at their best-fit values whilst varying \deltacp.}
\label{tab:oa:events:oscsystbestfit}
\end{table}

\begin{figure}[htbp]
\centering
\includegraphics[width=0.95\columnwidth]{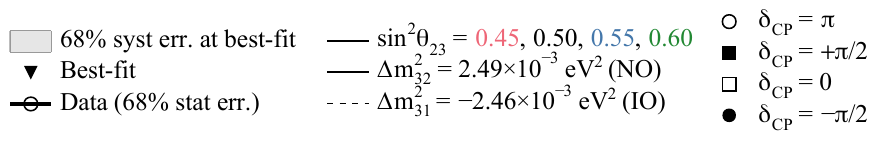} \\
\includegraphics[width=0.95\columnwidth,trim=0mm 0mm 0mm 8mm, clip]{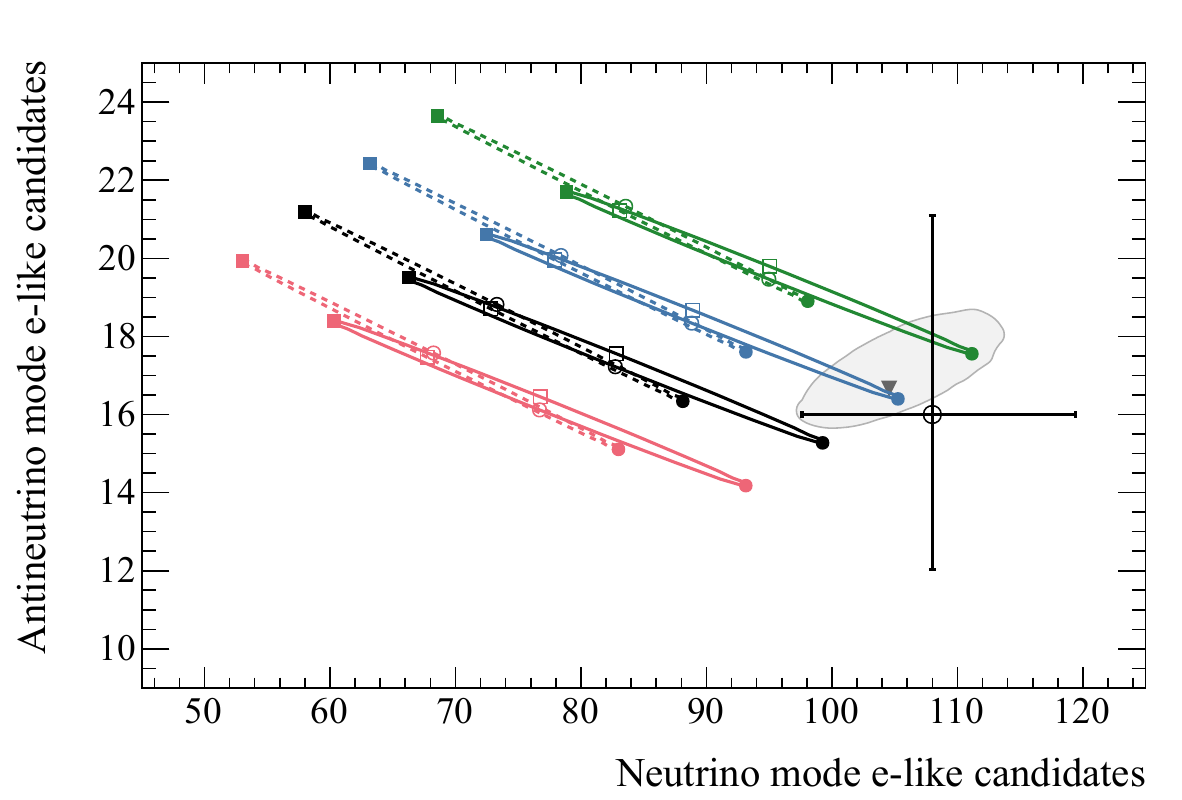} \\
\includegraphics[width=0.95\columnwidth]{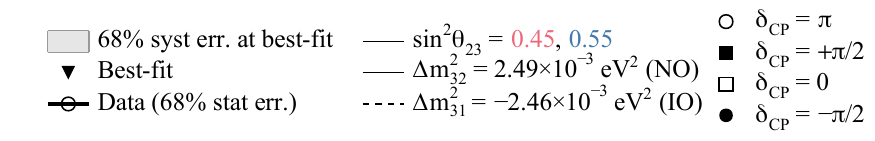} \\
\includegraphics[width=0.95\columnwidth,trim=0mm 0mm 0mm 8mm, clip]{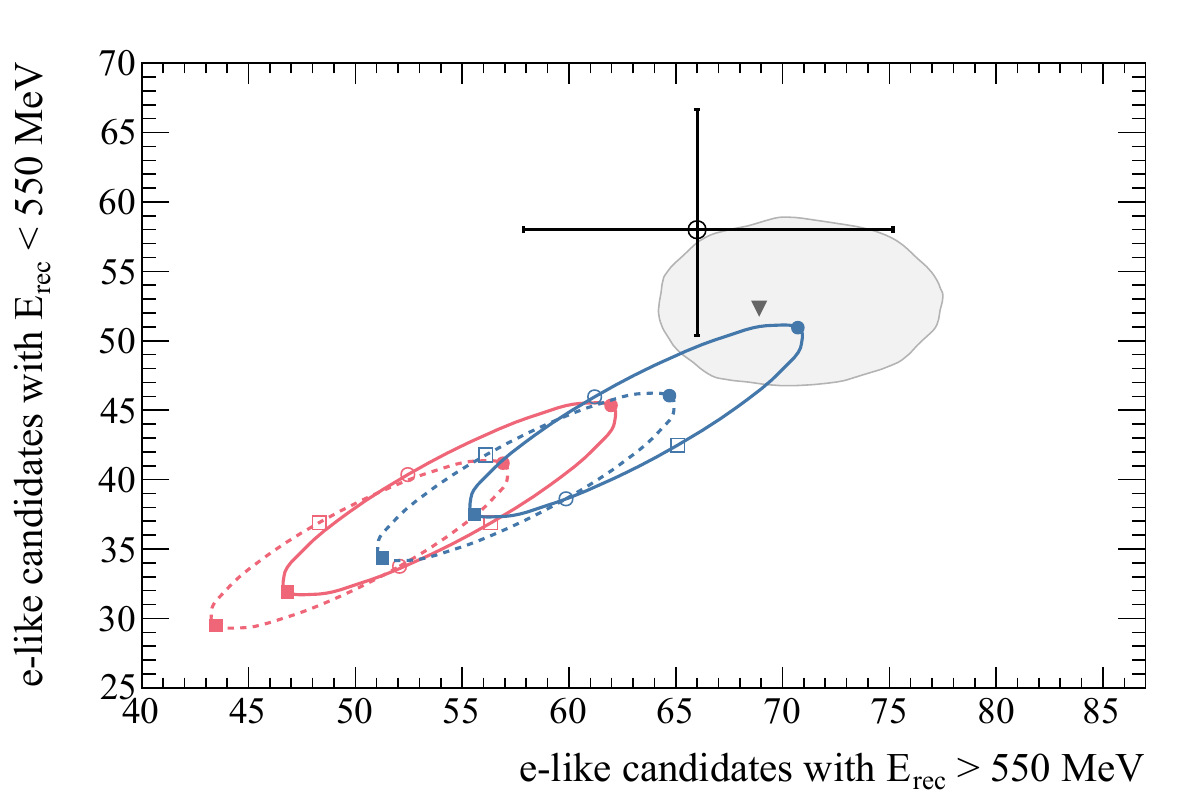}
\caption{The number of \fhcalt \re+ \rede versus \rhcalt \re events
(top, leading $\sin\deltacp$ dependence) and \fhcalt \re+ \rede+ \rhcalt \re events above and below $E_{rec}=550~\text{MeV}$ (bottom, leading $\cos\deltacp$ dependence), with the predicted number of events for various sets of oscillation parameters, as shown by the different coloured ellipses.
The values for the neutrino mass splitting are from the frequentist analysis of data, where $\dmsqtwothree=2.40\times10^{-3}~\textrm{eV}^2$ ($\dmsqthreeone=-2.46\times10^{-3}~\textrm{eV}^2$) is the best-fit point in the normal (inverted) ordering.
The uncertainties represent the 68\% confidence interval for the mean of a Poisson distribution given the observed data point. The underlaid contours contain the predicted number of events for 68\% of simulated experiments, varying the systematic uncertainty parameters around the best-fit values from the fit to ND data, and oscillation parameters set to the best-fit values from a fit to data. The overlaid triangle point shows the predicted number of events with both oscillation and systematic uncertainty parameters at their data best-fit values.
}
\label{fig:oa:bievent}
\end{figure}

The impact of \deltacp on the number of events in the selections is shown in \autoref{tab:oa:events:oscsystbestfit}, where there is a relatively small sensitivity in the \rhcalt \re selection, and most sensitivity comes from the \fhcalt \re selection, owing to the number of events in each sample.
To summarise the results, the number of observed electron neutrino events are plotted against the observed anti-neutrino events in \autoref{fig:oa:bievent}, where the data favours $\deltacp\sim-\pi/2$, $\dmsqtwothree>0$, and $\ssqthtwothree>0.50$; i.e. near maximal CP violation, the normal mass ordering, and the upper octant in the PMNS paradigm. The \re+\rede events in \fhcalt and the \re events in \rhcalt are sensitive to \sindcp, the neutrino mass ordering, and the octant of $\theta_{23}$, and their energy spectra has some sensitivity to \cosdcp, as illustrated in \autoref{fig:oa:bievent}.
Compared to T2K's previous analysis~\cite{Abe:2021gky,T2K:2019bcf}, the data are now closer to the best three-flavour fit prediction, resulting in a slightly weaker constraint on \deltacp. The weaker constraint is, however, more compatible with the expected sensitivity of the experiment, discussed later in \autoref{sec:oa:freq}.

\begin{table*}[htbp]
\centering
\begin{tabular}{c c | c c c | c | c}
\hline
\hline
\multicolumn{2}{c|}{\multirow{2}{*}{Sample}} & \multicolumn{3}{c|}{Uncertainty source (\%)} & \multirow{2}{*}{Flux$\otimes$Interaction (\%)} & \multirow{2}{*}{Total (\%)}\\
\multicolumn{2}{c|}{}                 & Flux      & Interaction & FD + SI + PN   &  \\
\hline
\multirow{2}{*}{\rmu}   & $\nu$     & 2.9 (5.0) & 3.1 (11.7)    & 2.1 (2.7)     & 2.2 (12.7)    & 3.0 (13.0) \\
                        & \nub      & 2.8 (4.7) & 3.0 (10.8)    & 1.9 (2.3)     & 3.4 (11.8)    & 4.0 (12.0) \\
\hline
\multirow{2}{*}{\re}    & $\nu$     & 2.8 (4.8) & 3.2 (12.6)    & 3.1 (3.2)     & 3.6 (13.5)    & 4.7 (13.8) \\
                        & \nub      & 2.9 (4.7) & 3.1 (11.1)    & 3.9 (4.2)     & 4.3 (12.1)    & 5.9 (12.7) \\
\hline
\rede                   & $\nu$     & 2.8 (4.9) & 4.2 (12.1)    & 13.4 (13.4)   & 5.0 (13.1)    & 14.3 (18.7) \\

\hline
\hline
\end{tabular}
\caption{Uncertainties on the number of events in each FD sample broken down by source after (before) the fit to ND data. ``FD+SI+PN'' combines the uncertainties from the FD detector, secondary particle interactions (SI), and photo-nuclear (PN) effects. 
``Flux$\otimes$Interaction'' denotes the combined effect from the ND constrained flux and interaction parameters, and the unconstrained interaction parameters. 
The change in the ``FD+SI+PN'' uncertainties before and after the ND fit is an indirect effect due to the change of interaction mode fractions in the samples after the ND fit.}
	\label{tab:oa:percenterrors:postND}
\end{table*}

The systematic uncertainties on the predicted number of events before and after the fit to ND data is given in \autoref{tab:oa:percenterrors:postND}. After the fit, the total uncertainty is reduced by a factor 2--5 depending on the sample, with the impact from flux and interaction uncertainties reduced by more than 60\%. 
After the ND fit, the interaction uncertainties are of similar size to the FD detector, pion secondary interaction, and photo-nuclear systematic uncertainties for all samples except the \rede, which is dominated by FD detector uncertainties. 
The FD detector uncertainties characterise the performance of SK and its reconstruction, the pion secondary interaction uncertainties were discussed in \autoref{sec:int_fsi} and are informed by external $\pi-A$ scattering data, and the photo-nuclear uncertainty comes from when Cherenkov photons are absorbed by the nuclei in the FD, causing particles to be mis-reconstructed or entirely missed due to the lack of any Cherenkov rings.
Although the impact from uncertainties in the flux and interaction model are similar for the selections at about 3\% when considered separately, they significantly correlate with each other after the fit to ND data, which causes the combined uncertainty from the ND-constrained interaction parameters and the neutrino flux to be smaller than the sum of their squares.

\begin{figure*}[htbp]
\centering
\begin{subfigure}[b]{0.49\textwidth}
\includegraphics[width=\textwidth,trim=0mm 0mm 0mm 10mm, clip]{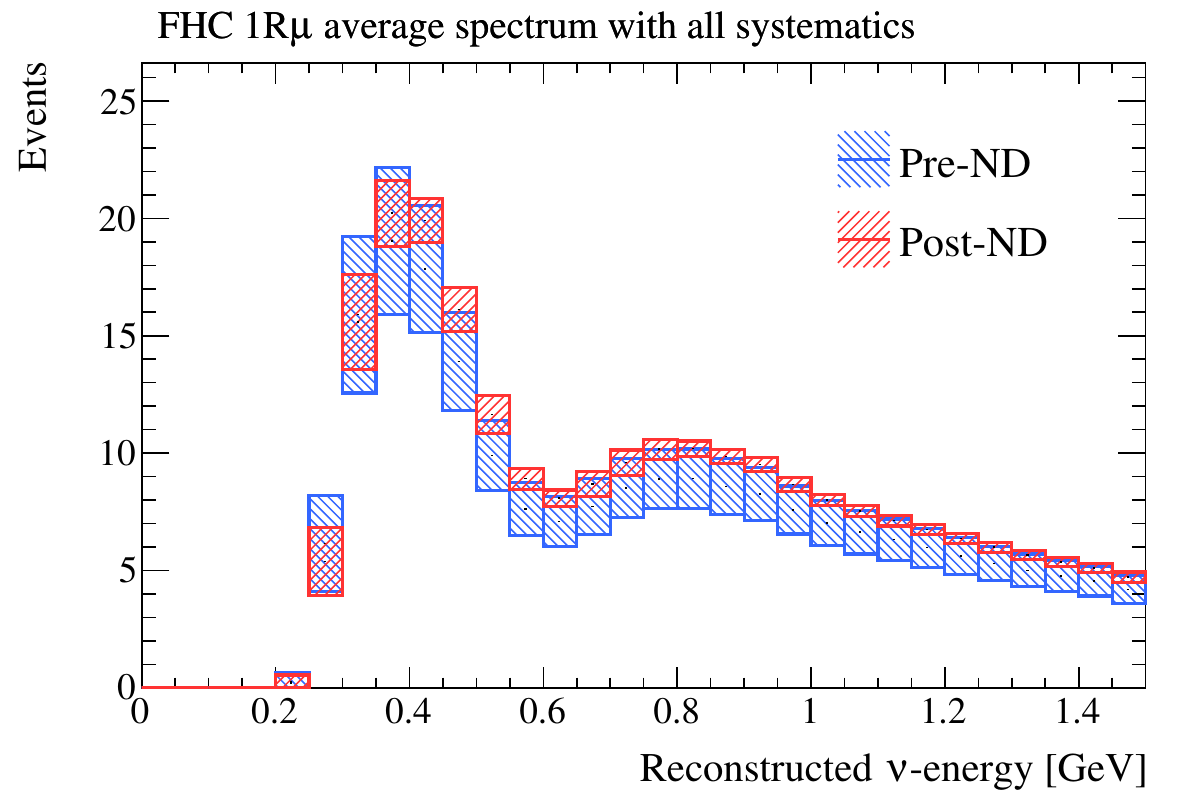}
\caption{\fhcalt \rmu}
\end{subfigure}
\begin{subfigure}[b]{0.49\textwidth}
\includegraphics[width=\textwidth,trim=0mm 0mm 0mm 10mm, clip]{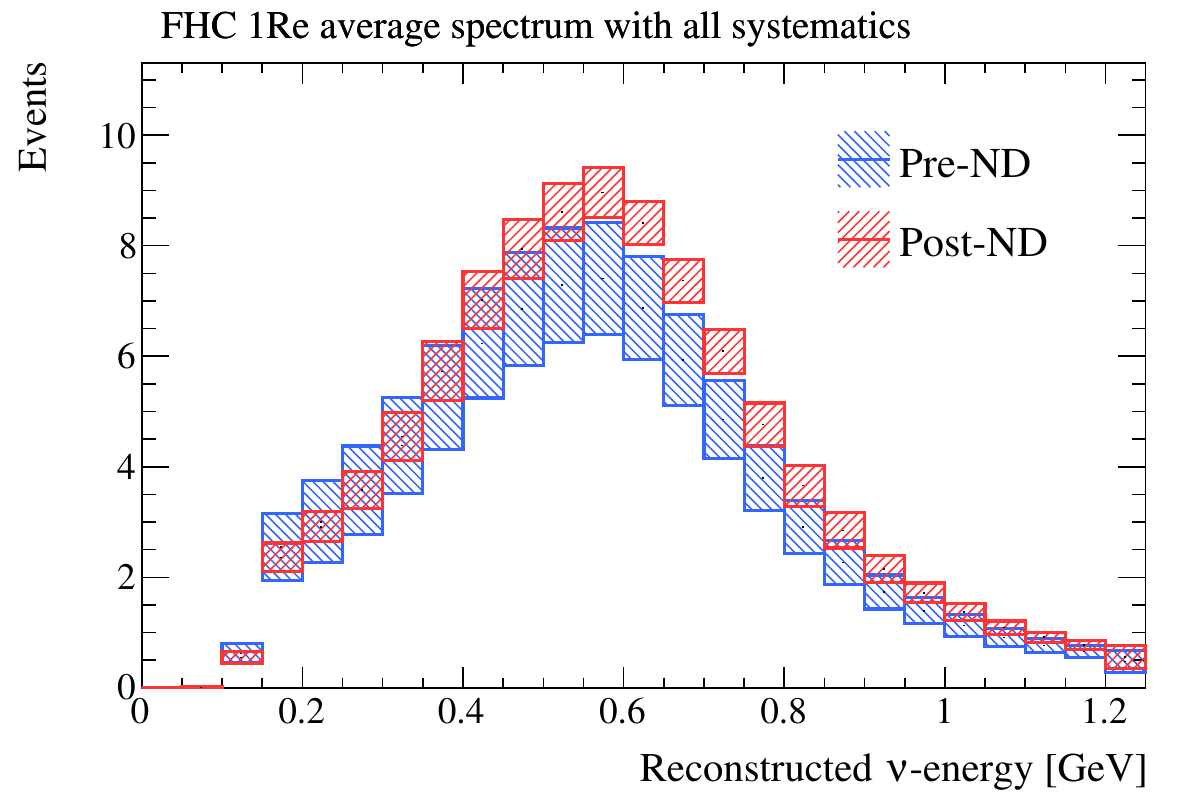}
\caption{\fhcalt\re}
\end{subfigure}
\\ \vspace{1em}
\begin{subfigure}[b]{0.49\textwidth}
\includegraphics[width=\textwidth,trim=0mm 0mm 0mm 10mm, clip]{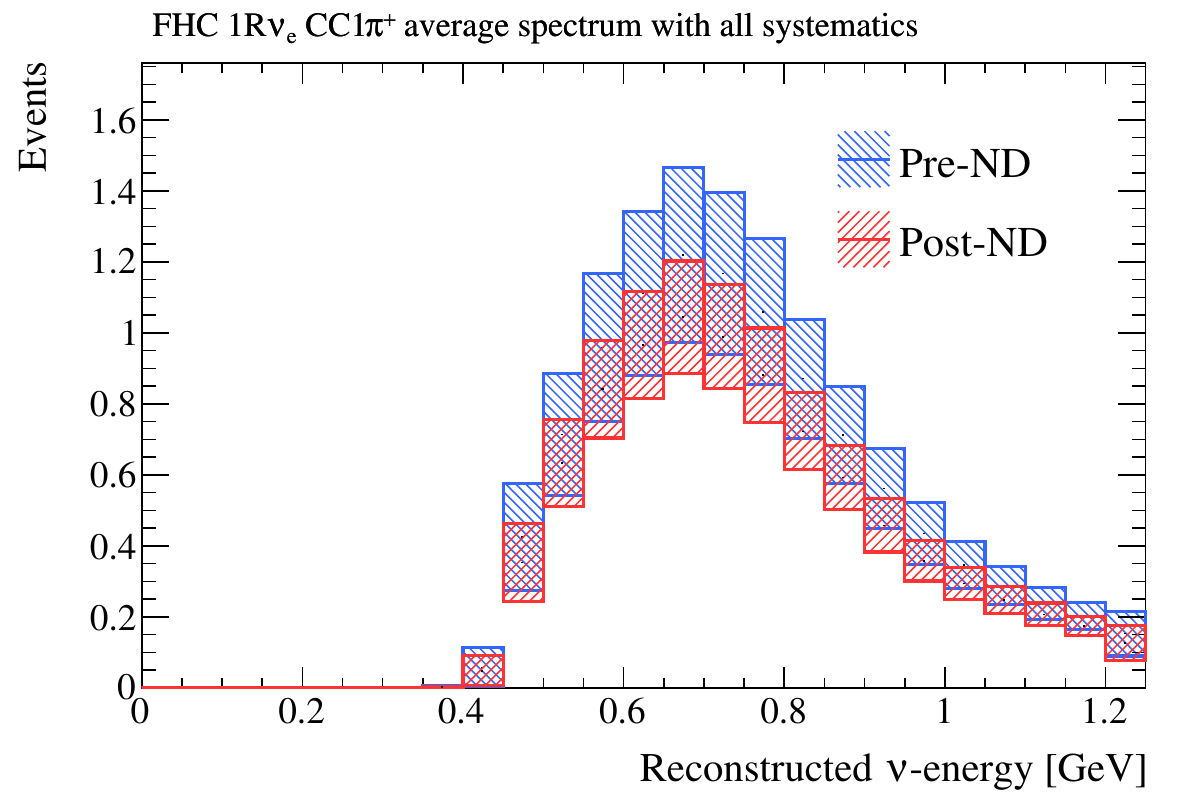} \caption{\fhcalt \rede}
\end{subfigure}
\\ \vspace{1em}
\begin{subfigure}[b]{0.49\textwidth}
\includegraphics[width=\textwidth,trim=0mm 0mm 0mm 10mm, clip]{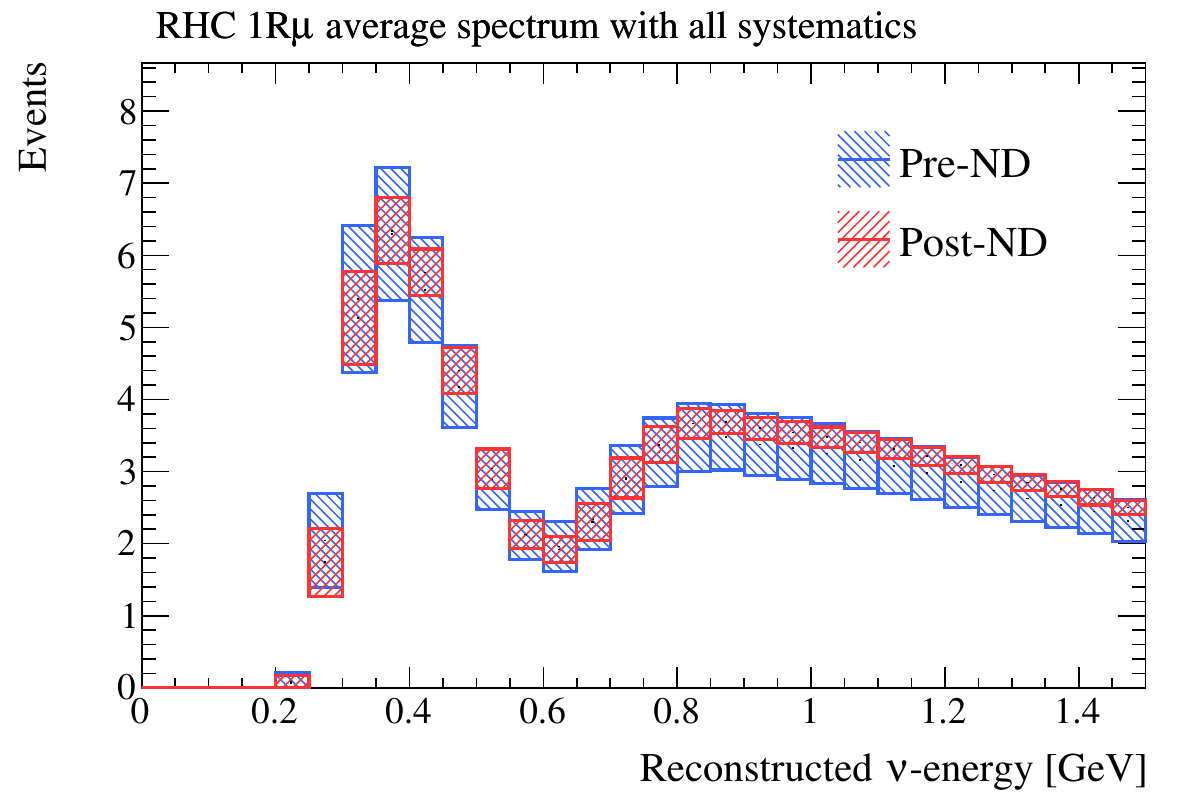}
\caption{\rhcalt \rmu}
\end{subfigure}
\begin{subfigure}[b]{0.49\textwidth}
\includegraphics[width=\textwidth,trim=0mm 0mm 0mm 10mm, clip]{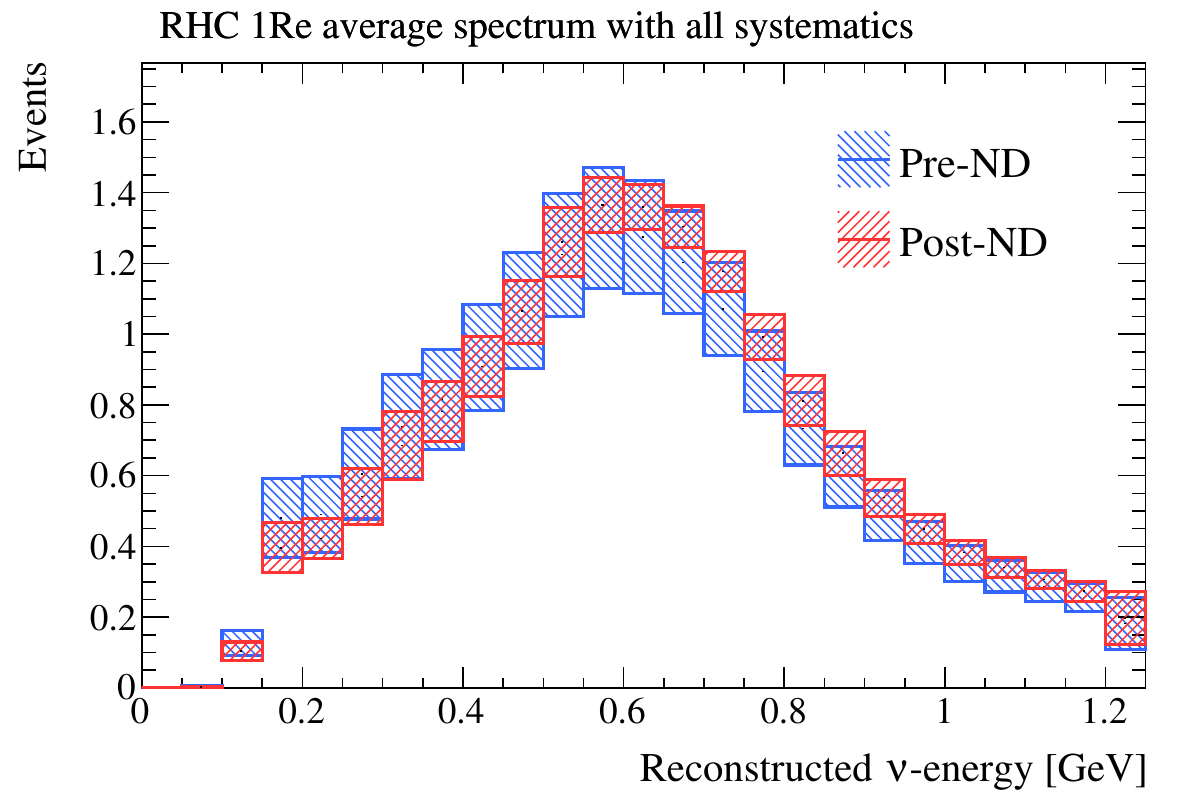}
\caption{\rhcalt \re}
\end{subfigure}
\caption{Total uncertainty on the reconstructed neutrino energy spectrum in the FD selections before and after the ND analysis of data. The oscillation parameters are set to values near the T2K best-fit point, specified in \autoref{app:appendix_fakedata}, \autoref{tab:asimov_pars}.}
\label{fig:oa:errorbands}
\end{figure*}

These constraints are used to build the predictions for the FD energy spectra including all uncertainties, as shown in \autoref{fig:oa:errorbands}. The five lower-$Q^2$ parameters have no external constraints, and the expected sensitivity from a FD-only fit (excluding the ND) is used as the uncertainty. This is \emph{solely} for the purpose of providing a representative uncertainty on the events when an ND fit is not used, and this uncertainty is not used elsewhere in the analysis.

The degrees of freedom from the oscillation parameters are of the form $\sin^{2}\theta_{ij}$, $\Delta m^{2}_{ij}$, and \deltacp. T2K is not sensitive to the ``solar'' oscillation parameters \ssqthonetwo and \dmsqonetwo, therefore constraints from the world averages reported in PDG 2019~\cite{Tanabashi:2018oca} are imposed\footnote{$\ssqthonetwo=0.307(\pm0.013)$, $\dmsqonetwo=7.53(\pm0.18)\times10^{-5}~\mathrm{eV^2}$}, where the frequentist analysis fixes the parameters and the Bayesian analysis accounts for their uncertainties.
An additional constraint may be imposed on \ssqthonethree from the world average reported in PDG 2019~\cite{Tanabashi:2018oca}, referred to as the ``reactor constraint''\footnote{$\ssqthonethree=2.18(\pm0.07)\times10^{-2}$}. The reactor constraint has a significant effect on the sensitivity to other oscillation parameters of interest, notably \deltacp. Accordingly, results are presented with and without this constraint applied. The reactor constraint is applied as a Gaussian penalty to the test statistic for both the frequentist and Bayesian analyses.

\subsection{Bayesian results}
\label{sec:oa:bayesian}
\begin{figure*}[htbp]
\centering
\includegraphics[width=\textwidth]
{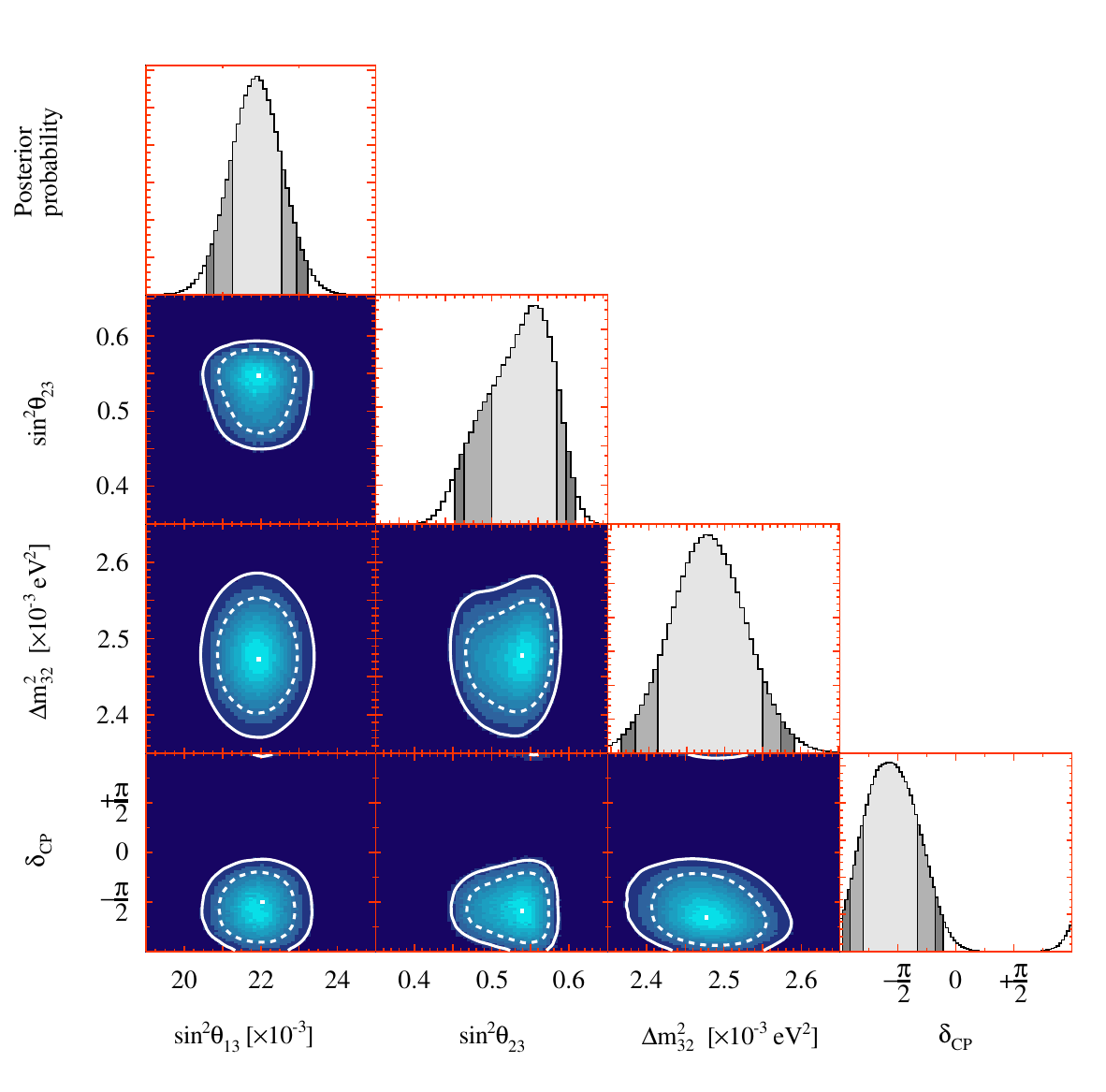}
\caption{Marginalised posterior probability densities from the Bayesian analysis for oscillation parameters of interest from a fit to data with the reactor constraint on \ssqthonethree applied. The two-dimensional posteriors have 68\% (dashed) and 90\% (solid) credible levels indicated and the point with highest posterior probability. The one-dimensional posteriors have 68\%, 90\%, and 95\% credible intervals indicated in different shades of grey. All credible regions are calculated from marginalising over both mass orderings, although panels displaying \dmsqtwothree show only the portions of the distributions in the normal mass ordering ($\dmsqtwothree>0$).}
\label{fig:oa:mach3:trianglewRC}
\end{figure*}

The Bayesian results presented in this section are obtained by sampling the posterior distributions through MCMC~\cite{metropolis,hastings} analysis, using the ND and FD selections simultaneously. The MCMC analysis presented in \autoref{sec:nd_fit} is utilised for the ND. 
The $e$-like samples use both the reconstructed angle between the outgoing lepton and the mean neutrino direction, and the reconstructed neutrino energy assuming a CCQE interaction and a struck nucleon at rest (\autoref{eq:enurec}). For the \rede selection---which is dominated by $1e^{-}1\pi^+$ final states---the nucleon mass is replaced by the $\Delta(1232)$ mass. The $\mu$-like samples only use the reconstructed neutrino energy assuming a CCQE interaction.
The posterior probability at the FD first includes the product of Poisson probabilities for observing the number of events in the data given the model prediction per bin across all samples. A Gaussian multivariate distribution is used to include the effect of external constraints on the systematic uncertainty parameters. The general form of the likelihood is the same as the ND analysis, presented in ~\autoref{eq:barlowbeeston_total}, but excludes the statistical uncertainty on the simulation for the FD.

Credible regions are extracted from lower dimensional marginalised posterior distributions for parameters of interest by adding up the highest probability density region until a certain fraction of the distribution is captured.
Flat priors are used over the entire ranges of \ssqthtwothree, \dmsqtwothree, \deltacp (or \sindcp), and Gaussian priors are applied on \dmsqonetwo and \ssqthonetwo. For \ssqthonethree either a flat or a Gaussian prior is applied via the aforementioned reactor constraint. The priors for normal and inverted orderings are the same, namely 50\%.

\autoref{fig:oa:mach3:trianglewRC} shows several marginalised posterior distributions for oscillation parameters of interest. Two-dimensional distributions for every combination of the four oscillation parameters of interest are shown with the $68\%$ and $90\%$ credible intervals in dashed and solid lines, respectively. Each two-dimensional posterior distribution also shows the point of highest probability density. Marginalised one-dimensional posterior probability distributions are also given for each of the four oscillation parameters with $68\%$, $90\%$, and $95\%$ credible intervals in different shades of grey.

\subsubsection{Atmospheric oscillation parameters}
\begin{figure}[htbp]
\centering
\includegraphics[width=0.95\columnwidth,trim=0mm 0mm 0mm 5mm, clip]{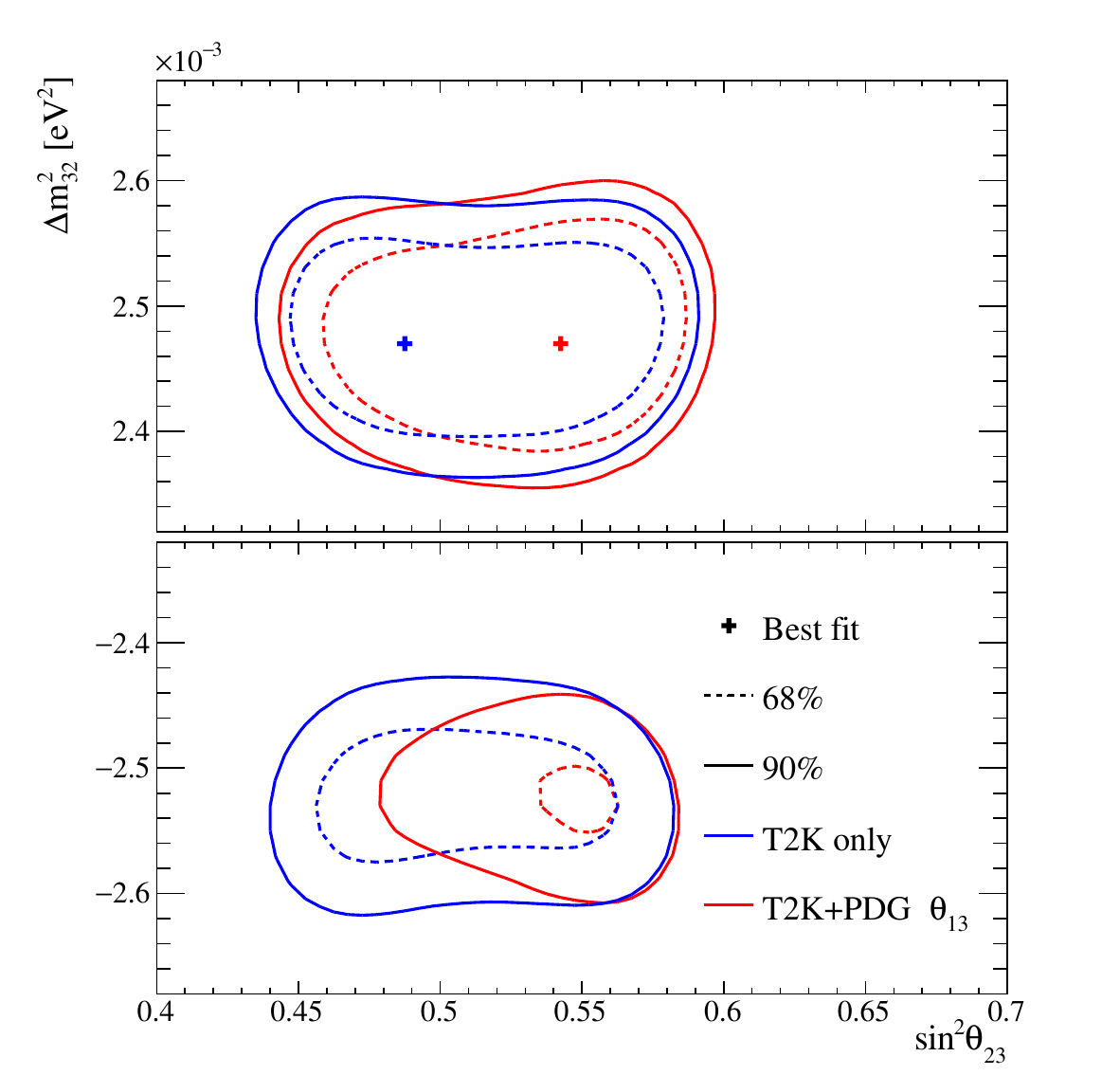}
\caption{68\% and 90\% credible intervals from the marginalised $\ssqthtwothree-\dmsqtwothree$ posterior distribution with (red) and without (blue) the reactor constraint applied. The top (bottom) shows the proportion of probability density in the normal (inverted) mass ordering.}
\label{fig:oa:mach3:disappRCcomp}
\end{figure}

\begin{figure}[htbp]
\centering
\includegraphics[width=0.95\columnwidth,trim=0mm 0mm 0mm 12mm, clip]{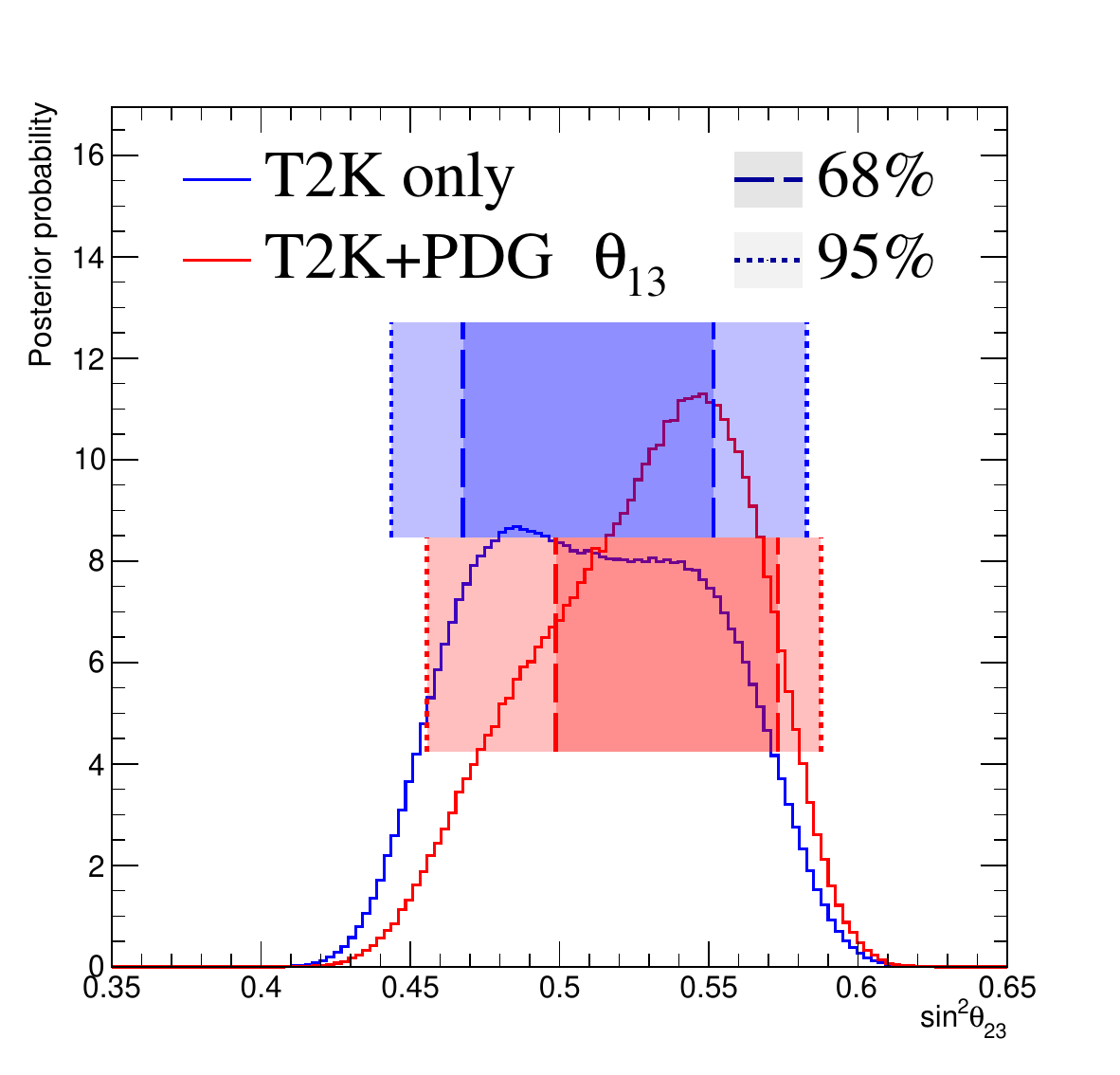}
\caption{The marginalised posterior probability density of \ssqthtwothree with (red) and without (blue) the reactor constraint on \ssqthonethree applied. The shaded areas show the 68\% and 95\% regions of highest posterior density, equivalent to the 1$\sigma$ and 2$\sigma$ credible intervals.}
\label{fig:oa:mach3:th23RCcomp}
\end{figure}

The effects of applying the reactor constraint on the $\ssqthtwothree-\dmsqtwothree$ contours is shown in \autoref{fig:oa:mach3:disappRCcomp}. Applying the constraint increases the probability density in the upper octant and the normal neutrino mass ordering. 
The marginalised posterior probability distribution of \ssqthtwothree with and without the reactor constraint is shown in \autoref{fig:oa:mach3:th23RCcomp}. The posterior probabilities are largely overlapping, with a preference for the upper octant when using the reactor constraint, and there is barely any octant preference without the reactor constraint.

\begin{table}[htbp]
\centering
\resizebox{.5\textwidth}{!}{%
\begin{tabular}{r l|c c|c}
\hline
\hline
 & & \multicolumn{2}{c|}{\ssqthtwothree} & \multirow{2}{*}{Sum} \\
 & & $< 0.5$ & $> 0.5$ & \\
\hline
\multirow{2}{*}{\dmsqtwothree} & $>0$ (NO) & 0.195 (0.260) & 0.613 (0.387) & 0.808 (0.647) \\
 & $<0$ (IO) & 0.035 (0.152) & 0.157 (0.201) & 0.192 (0.353) \\
\hline
\multicolumn{2}{c|}{Sum} &0.230 (0.412) & 0.770 (0.588) & 1.000\\
\hline
\hline
\end{tabular}
}
\caption{Fractions of posterior probability in different combinations of the mass ordering and $\theta_{23}$ octant from fit to T2K data with (without) the reactor constraint on \ssqthonethree. NO (IO) refers to the normal (inverted) neutrino mass ordering.}
\label{table:BayesTable}
\end{table}

The results for the atmospheric parameters are summarised in \autoref{table:BayesTable}, showing the proportion of the posterior probability that lies in the different mass orderings and $\theta_{23}$ octant, with and without the reactor constraint.
A flat prior distribution on both \dmsqtwothree and \ssqthtwothree is equivalent to comparing the likelihood that T2K's data is described by the different choices of hypotheses.
The analysis with (without) the reactor constraint sees a Bayes factor (BF) of 3.35 (1.43) for the upper over the lower $\theta_{23}$ octant; 4.21 (1.83) for the normal over inverted mass ordering; and a combined factor of 1.58 (0.63) for upper $\theta_{23}$ octant and normal ordering. When calculating the BFs, the alternate hypothesis is any other combination of octant and mass ordering. Interpreting the largest BFs with the Jeffreys' scale, there is substantial evidence for the normal ordering when marginalising over the octant, and substantial evidence for the upper octant when marginalising over the mass ordering. In the more recent interpretation of BFs by Kass and Raftery~\cite{10.2307/2291091}, these both correspond to positive evidence. Importantly, the Jeffreys and Kass--Raftery definitions of ``evidence'' do not equate to the criteria often used in particle physics. For instance, a probability of 95.4\% (``$2\sigma$'') is equivalent to a BF of 20.7, which is deemed as ``decisive'' on the Jeffreys' scale, and as ``strong'' on the Kass--Raftery scale.

\subsubsection{The CP-violating phase \deltacp, and \ssqthonethree}
\begin{figure}[htbp]
\centering
\includegraphics[width=0.95\columnwidth,trim=5mm 20mm 0mm 12mm, clip]{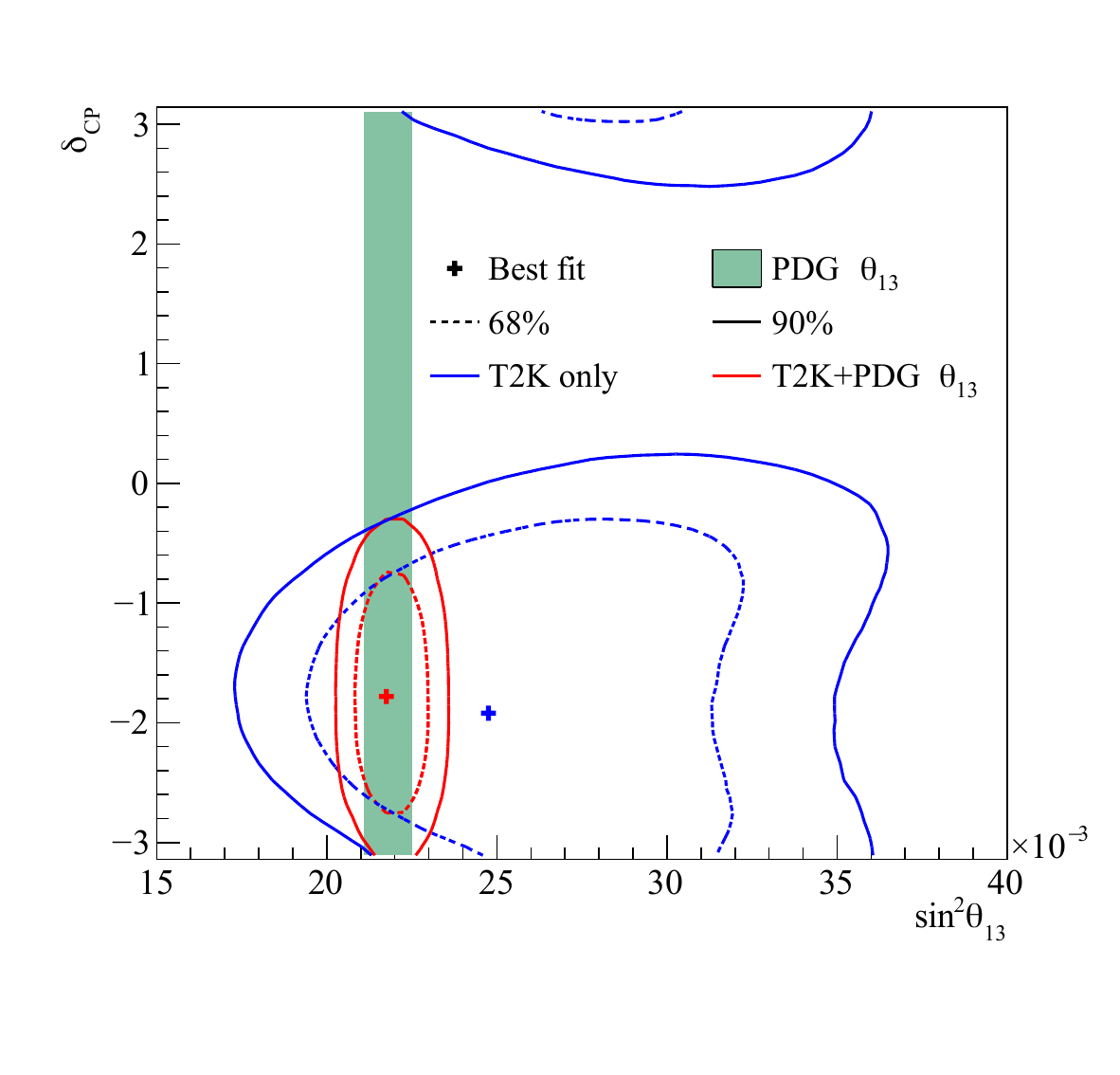}
\caption{68\% and 90\% credible intervals from the marginalised $\ssqthonethree-\deltacp$ posterior distribution with (red) and without (blue) the reactor constraint (green band) applied, marginalised over both mass orderings.}
\label{fig:oa:mach3:appRCcomp}
\end{figure}

\begin{figure}[htbp]
\centering
\includegraphics[width=0.95\columnwidth,trim=0mm 0mm 0mm 12mm, clip]{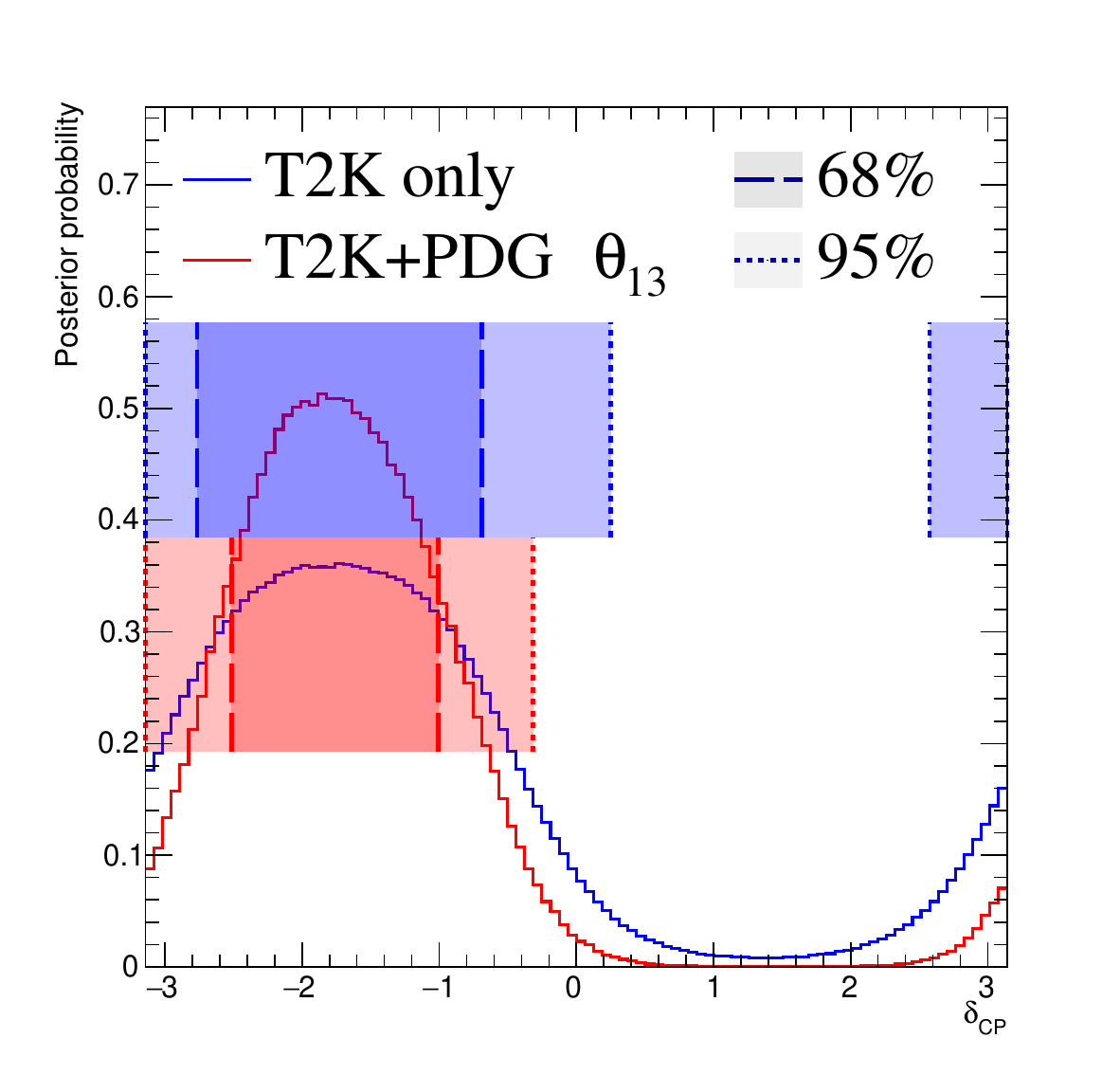}
\caption{The marginalised posterior probability density of \deltacp with (red) and without (blue) the reactor constraint applied. The shaded areas show the 68\% and 95\% regions of highest posterior density, equivalent to the 1$\sigma$ and 2$\sigma$ credible intervals.}
\label{fig:oa:mach3:dcpRCcomp}
\end{figure}

A comparison of $\ssqthonethree-\deltacp$ contours with and without the reactor constraint is shown in \autoref{fig:oa:mach3:appRCcomp}.
The regions are in good agreement, with a majority of the 1$\sigma$ regions overlapping, comparable with the reactor constraint. A comparison of the \deltacp posterior distributions is shown in \autoref{fig:oa:mach3:dcpRCcomp}, showing the impact of the reactor constraint on T2K's \deltacp result. The external constraint breaks the partially degenerate effects of \ssqthonethree and \deltacp on the \nue appearance, leading to the \fhcalt \re and \rede selections having a larger sensitivity to \deltacp.

\subsubsection{The Jarlskog invariant}
\begin{figure}[htbp]
\centering
\includegraphics[width=0.95\columnwidth,trim=0mm 0mm 0mm 12mm, clip]{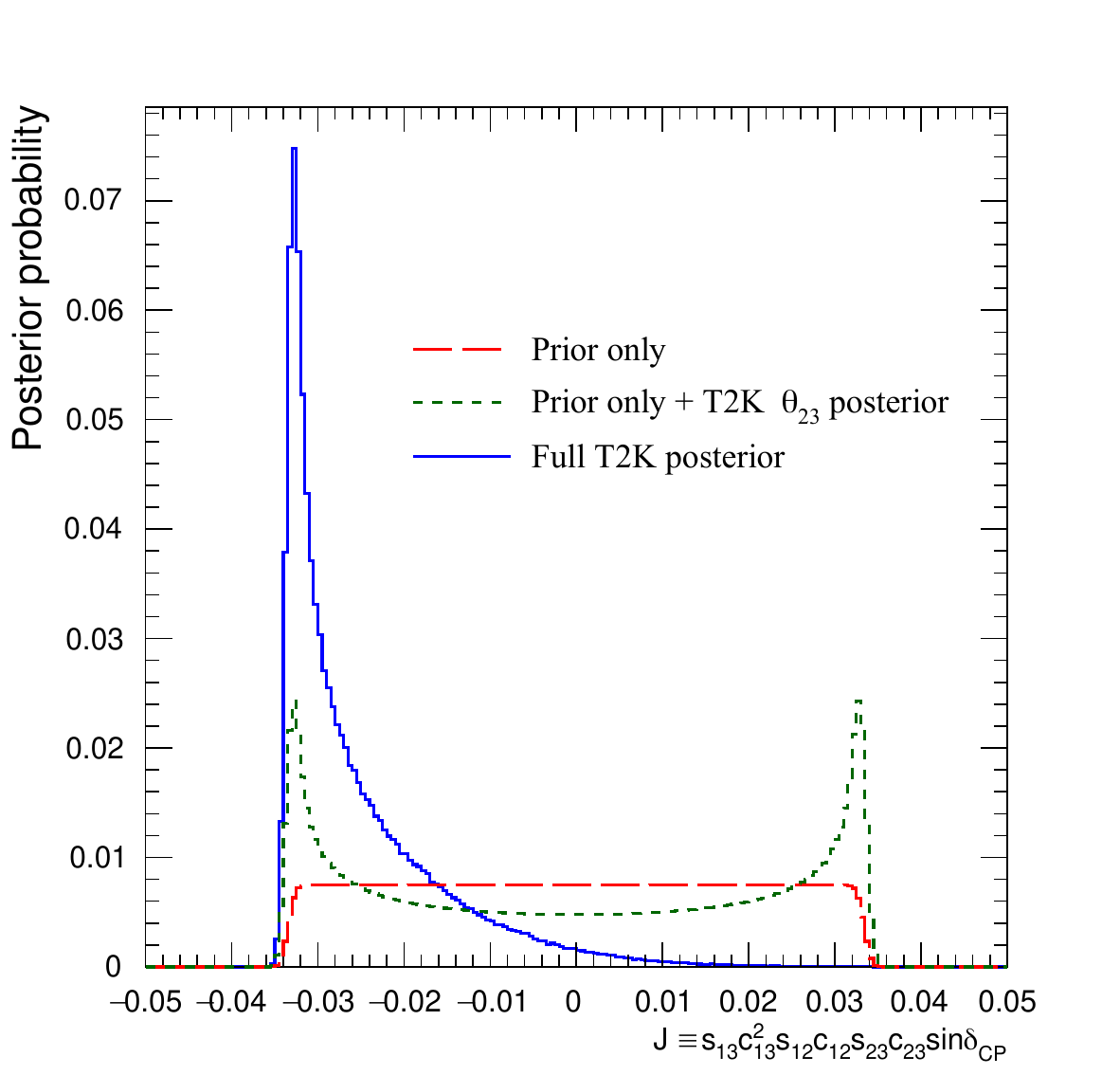}
\caption{Posterior probability distributions for the Jarlskog invariant using a prior distribution from the 2019 PDG reactor constraint on $\theta_{13}$\cite{Tanabashi:2018oca} (red), prior from all parameters except sampling $\theta_{23}$ from the T2K posterior (green), and the full T2K posterior (blue). All three posterior probabilities used a prior probability distribution flat in \deltacp.}
\label{fig:oa:mach3:jarl1D_incremental}
\end{figure}

\begin{figure}[htbp]
\centering
\includegraphics[width=0.95\columnwidth,trim=0mm 0mm 0mm 12mm, clip]{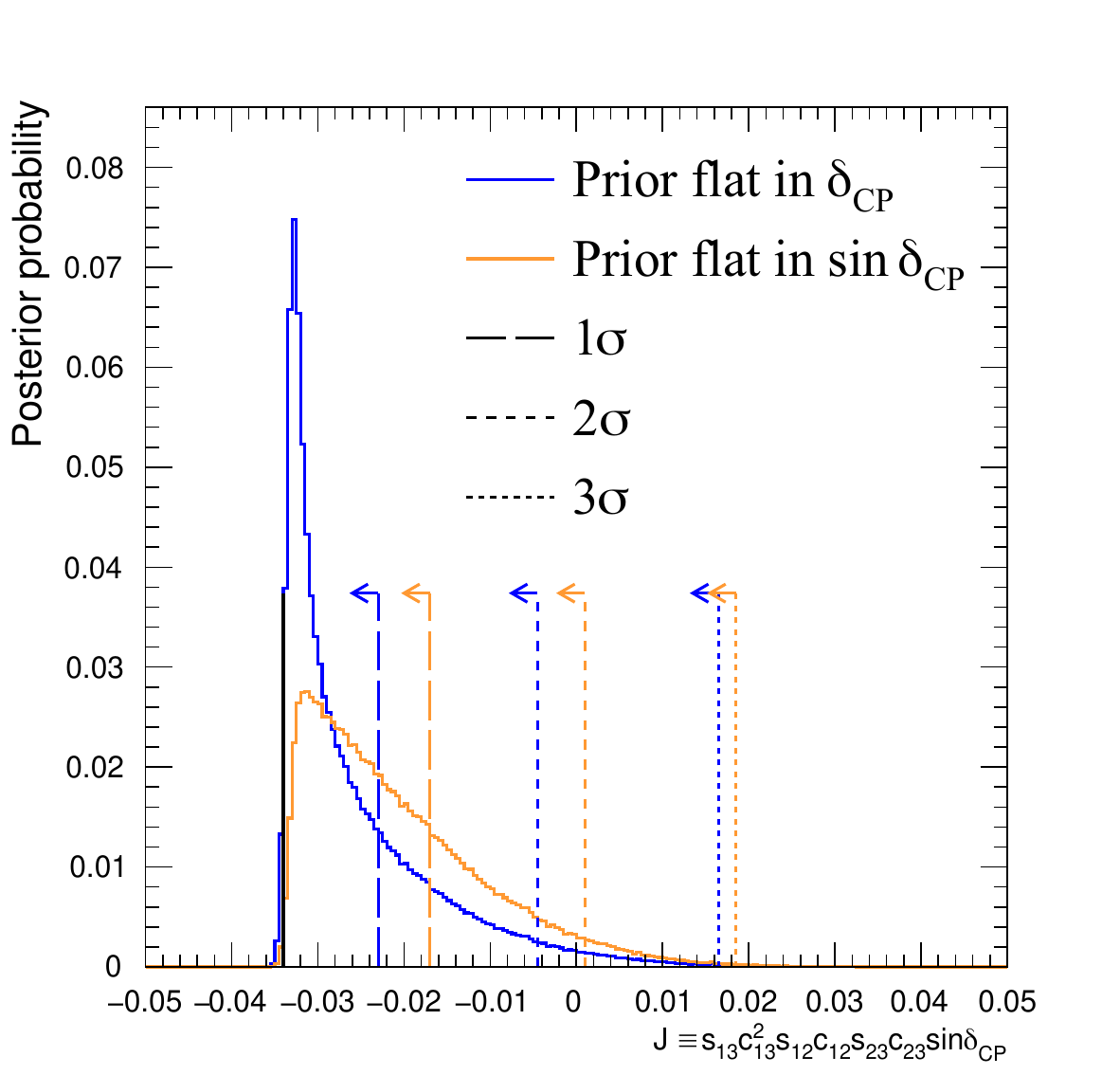}
\caption{Posterior probability distributions for the Jarlskog invariant taken from posterior distributions with priors that are either flat in \deltacp (blue) or flat in \sindcp (orange). 1$\sigma$, 2$\sigma$, and 3$\sigma$ credible intervals are shown as the region between the vertical black solid line and the specified vertical dashed lines.}
\label{fig:oa:mach3:jarl1D}
\end{figure}

The sampled posterior probability density is in part a function of the PMNS mixing angles and $\deltacp$, which means the probability distribution for the Jarlskog invariant~\cite{PhysRevLett.55.1039, JARLSKOG2005323},
\begin{equation}
J=\sin\theta_{13} \cos^2\theta_{13} \sin\theta_{12} \cos\theta_{12} \sin\theta_{23} \cos\theta_{23} \sin\deltacp
\label{eq:jarlskog}
\end{equation}
can be extracted directly from the steps in the MCMC.
The posterior distribution for $J$ is presented in \autoref{fig:oa:mach3:jarl1D_incremental}, which favours a near-maximal negative $J$. 
The prior probability distribution is largely flat in the range $J=[-0.035,0.035]$, with the fall-off beyond that coming from external $\theta_{12}$ and $\theta_{13}$ constraints.
The preference for \ssqthtwothree values near maximal mixing has the effect of picking out the more extreme values of $J$.
When sampling the full posterior probability, which incorporates the \deltacp constraint, a preference for negative values of $J$ emerges.
The blue curve in \autoref{fig:oa:mach3:jarl1D_incremental} is recreated in \autoref{fig:oa:mach3:jarl1D} showing the $1\sigma$, $2\sigma$, and $3\sigma$ credible intervals.
Two-dimensional credible regions for the Jarlskog invariant against both \ssqthtwothree and \deltacp are included in \autoref{app:appendix_2DJarlskog}.

Although this analysis does not rule out CP-conserving values of \deltacp at $2\sigma$, it does rule out $J=0$ at the $2\sigma$ level and excludes the $J>0.17$ region at $>3\sigma$ with a flat prior probability in \deltacp.
The dependence of $J$ on the choice of a prior flat in \deltacp or flat in \sindcp is shown in \autoref{fig:oa:mach3:jarl1D}.
The prior flat in \sindcp flattens out the Jarlskog distribution, which in turn slightly expands the 2$\sigma$ credible interval to where $J=0$ is just included.
These conclusions agree with previous studies on the impact of the \deltacp prior at T2K~\cite{Abe:2021gky,T2K:2019bcf}.

\begin{table}[htbp]
\centering
\begin{tabular}{l c |c c}
\hline
\hline
\multicolumn{2}{c|}{\multirow{2}{*}{Selection}}  & \multicolumn{2}{c}{$p$-value} \\
\multicolumn{2}{c|}{}        & Shape     & Rate    \\
\hline
\multirow{2}{*}{\rmu} & \fhcalt  & 0.48   & 0.18 \\
                      & \rhcalt & 0.85   & 0.74 \\
\hline
\multirow{2}{*}{\re} & \fhcalt   & 0.19  & 0.49 \\
                    & \rhcalt   & 0.61   & 0.39 \\
\hline
\rede  & \fhcalt & 0.86   & 0.22 \\
\hline
\multicolumn{2}{c|}{All} & 0.73   & 0.30 \\
\hline
\hline
\end{tabular} \\
\caption{Breakdown of posterior predictive $p$-values by sample, quoted separately using a shape or rate based calculation, demonstrating good compatibility between the model and the data.}
\label{table:GOF_wRC}
\end{table}

\subsubsection{Goodness-of-fit analysis}
\begin{figure*}[htbp]
\begin{center}
\begin{subfigure}[b]{0.33\textwidth}
\includegraphics[width=\textwidth,trim=0mm 0mm 0mm 9mm, clip]{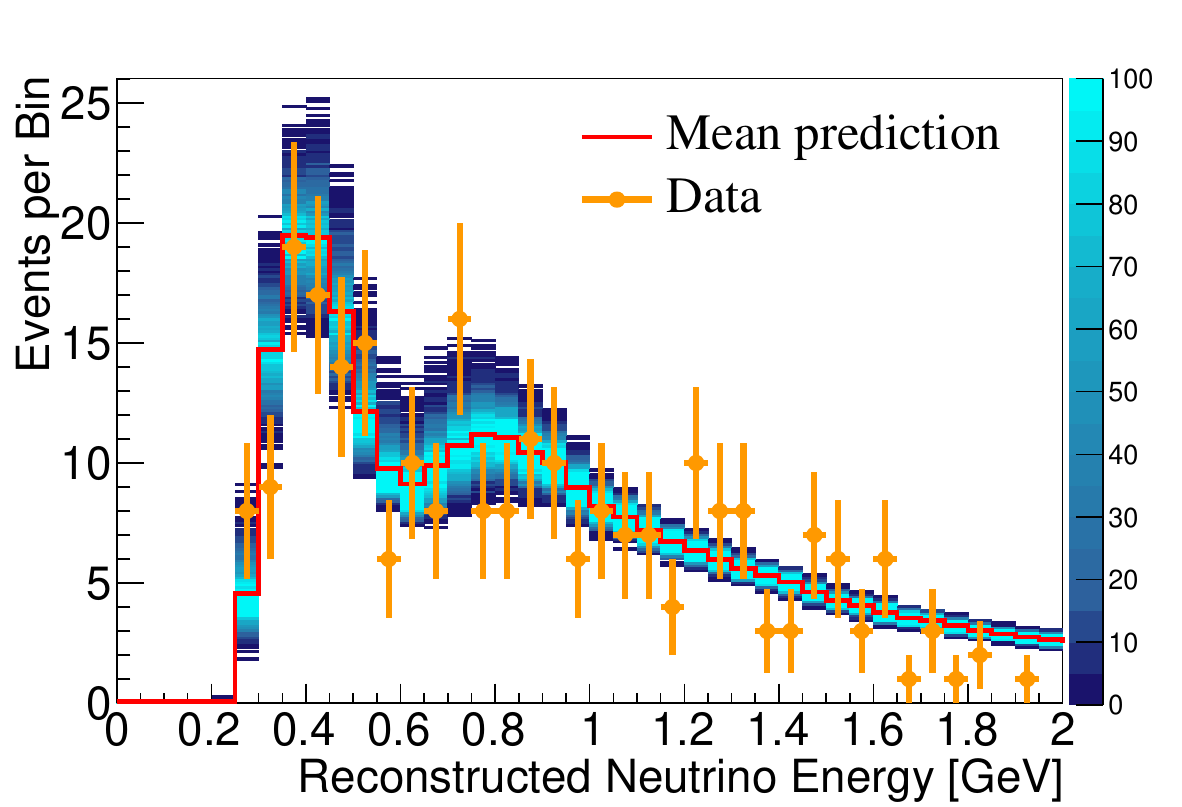}
\caption{\fhcalt \rmu}
\end{subfigure}
\begin{subfigure}[b]{0.33\textwidth}
\includegraphics[width=\textwidth,trim=0mm 0mm 0mm 9mm, clip]{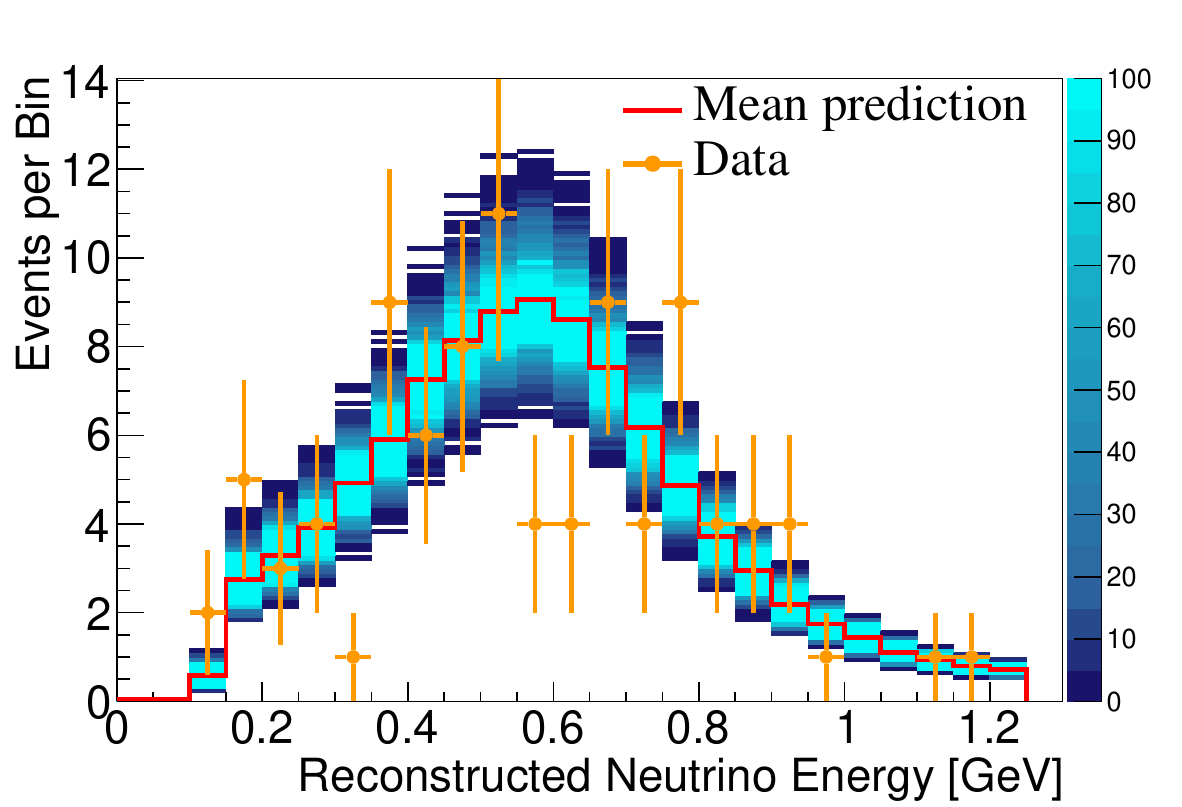}
\caption{\fhcalt \re}
\end{subfigure}
\begin{subfigure}[b]{0.33\textwidth}
\includegraphics[width=\textwidth,trim=0mm 0mm 0mm 9mm, clip]{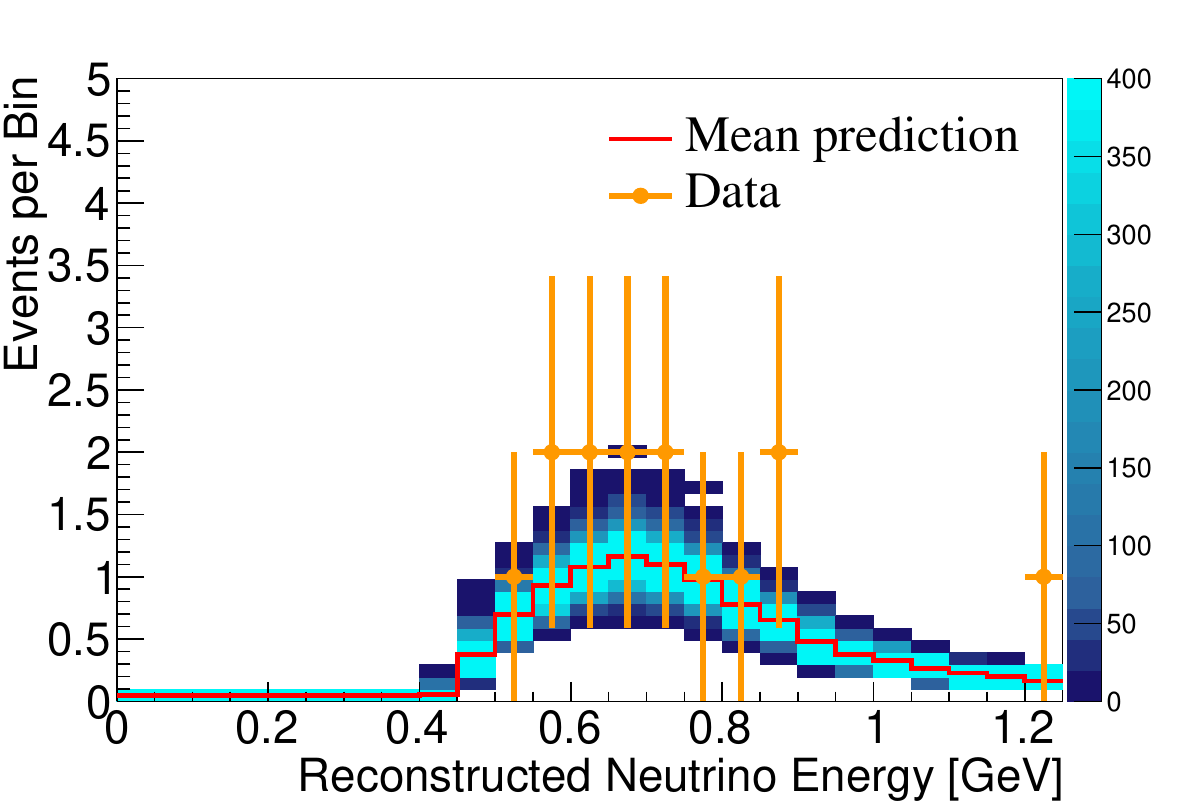} \caption{\fhcalt \rede}
\end{subfigure}
\hspace{0.33\textwidth}

\begin{subfigure}[b]{0.33\textwidth}
\includegraphics[width=\textwidth,trim=0mm 0mm 0mm 9mm, clip]{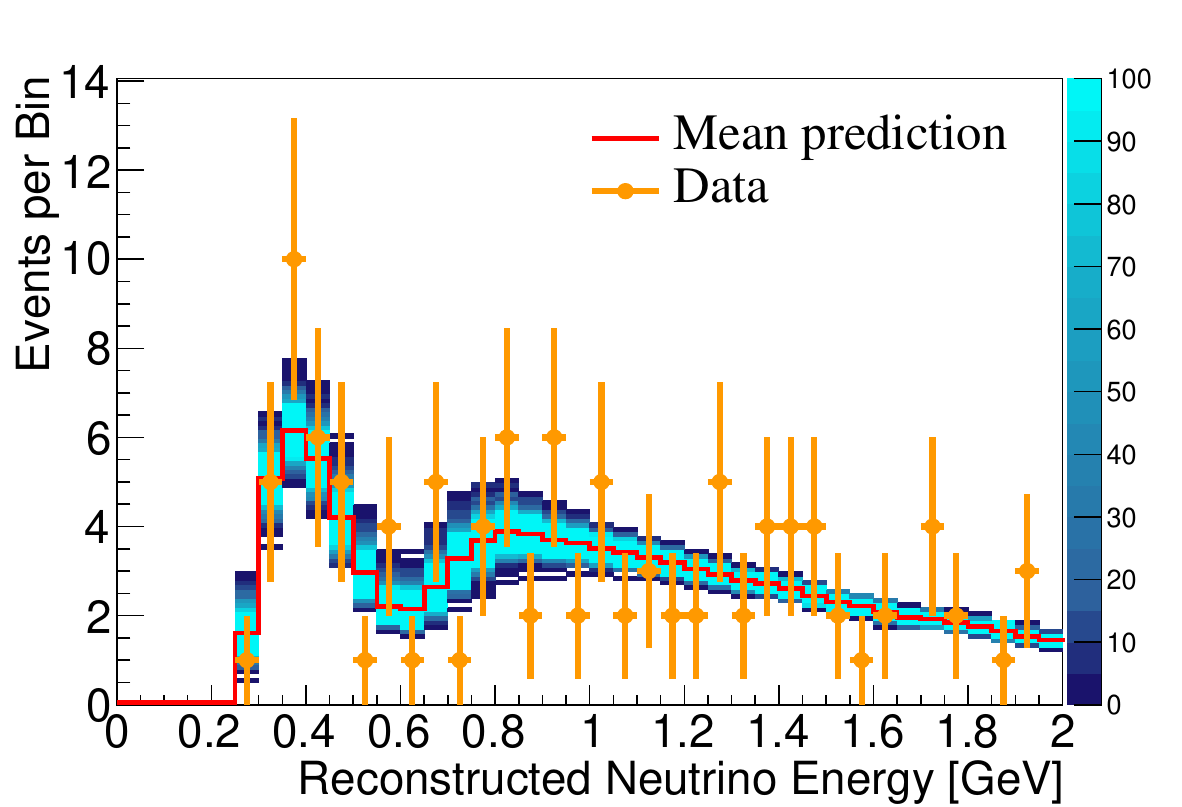}
\caption{\rhcalt \rmu}
\end{subfigure}
\begin{subfigure}[b]{0.33\textwidth}
\includegraphics[width=\textwidth,trim=0mm 0mm 0mm 9mm, clip]{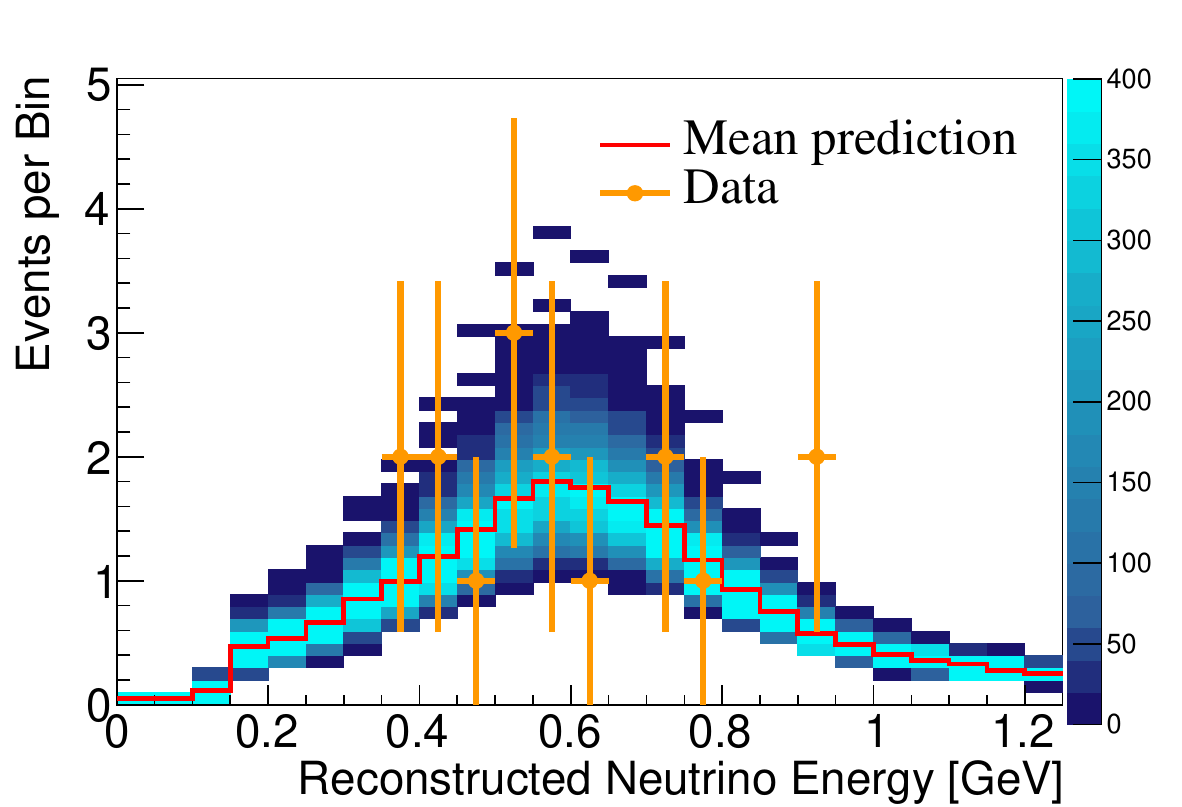}
\caption{\rhcalt \re}
\end{subfigure}
\end{center}
\caption{The reconstructed neutrino energy distributions of each FD sample. Data with Poisson uncertainties are shown in orange and the distributions of the predictions are shown in the coloured background, with the mean of those distributions overlaid in red. The $z$-axis represents the number of MCMC samples that had a prediction in a specific bin, and its intensity is directly proportional to the probability. The predictions are built by sampling both the nuisance and oscillation parameters from the posterior probability distribution in the Bayesian analysis. \ssqthonethree is constrained from T2K data alone with no reactor constraint applied. }
\label{fig:oa:mach3:postpred}
\end{figure*}

Predictions for the five samples at the FD are formed in \autoref{fig:oa:mach3:postpred}, using the posterior probability distributions for the systematic uncertainties and oscillation parameters from the fit to data.
By eye, the predictions agree well with the data, which are plotted as orange data points with statistical uncertainties applied. To quantify the model agreement with the data, the posterior predictive $p$-values~\cite{Gelman_Post} are calculated.
These $p$-values can be calculated using either the total number of events per sample (rate-based) or the events per bin of each sample (shape-based). It can also be split by sample, or calculated as a total $p$-value.
When including all samples, the shape-based and rate-based approach give $p=0.73$ and $p=0.30$ respectively. The $p$-values from both shape- and rate-based calculations broken down by sample and in total are tabulated in \autoref{table:GOF_wRC}. Good $p$-values are demonstrated for all cases.

\subsection{Frequentist results}
\label{sec:oa:freq}
As in previous T2K analyses, the frequentist results are obtained
using the marginal likelihood $\mathcal{L}_\mathrm{marg}(\theta)$ $=$ $\int \mathrm d\eta\,p(\eta)$ $\mathcal{L}(\theta,\eta)$ as the test statistic. Here, $\mathcal{L}(\theta,\eta)$ is the binned Poisson likelihood for the parameter of interest, $\theta$, and the nuisance parameters, $\eta$. The statistical treatment of nuisance parameters in the fit is thus identical to the Bayesian analysis and assumes a prior probability distribution $p(\eta)$. The numerical integration is performed by varying systematic uncertainties with a Gaussian covariance matrix from the ND analysis in \autoref{sec:nd_fit} as a constraint, and varying the other oscillation parameters with a flat prior probability distribution on \ssqthtwothree, \deltacp, \dmsqtwothree, and \ssqthtwoonethree, or a Gaussian prior on \ssqthtwoonethree.
Confidence intervals and regions are constructed with two different methods.
For critical parameters with known boundary effects, the Feldman--Cousins (FC) method~\cite{Feldman:1997qc} is utilised to calculate the coverage. This is performed for the result using the reactor constraint, on the one-dimensional confidence intervals in \deltacp and \ssqthtwothree, and their joint confidence region.
For generating the ensemble of experiments for FC evaluation, the nuisance oscillation parameters are varied from the posterior distribution obtained by fitting a representative simulated data set, sometimes referred to as ``Asimov data''. This simulated data set is generated at the global best-fit point using the reactor constraint. Since the FC method is computationally intensive, the remaining confidence regions are constructed using constant $\Delta \chi^2(\theta) = \chi^2(\theta) - \min_{\theta'} \chi^2(\theta')$ values via Wilks's theorem~\cite{10.1214/aoms/1177732360}, where $\chi^2 = -2\ln \mathcal{L}_\mathrm{marg}$. Whether $\Delta \chi^2$ is computed with respect to the minimum over both mass orderings, or the minimum in each mass ordering separately, is indicated in each of the results from the frequentist analysis.
The frequentist analysis bins the $e$-like FD samples in reconstructed lepton angle and reconstructed lepton momentum, and the $\mu$-like samples in reconstructed lepton angle and the reconstructed neutrino energy, defined in the same way as in the Bayesian analysis presented in \autoref{sec:oa:bayesian}. In previous analyses, the $\mu$-like samples were binned only in reconstructed neutrino energy, and adding the lepton angle information increases the $1\sigma$ expected sensitivity to \dmsqtwothree by $\mathcal{O}(1\%)$.

\begin{table*}[htbp]
\centering
\bgroup
\def\arraystretch{1.3}
\begin{tabular}{l | c c | c c}
\hline\hline
\multirow{2}{*}{Parameter}  & \multicolumn{2}{c|}{With reactor constraint} & \multicolumn{2}{c}{Without reactor constraint}\\
        & Normal ordering    & Inverted ordering & Normal ordering    & Inverted ordering \\
\hline
\deltacp  (rad.)         & $-1.97_{-0.62}^{+0.97}$ & $-1.44_{-0.59}^{+0.56}$ & $-2.22_{-0.81}^{+1.25}$   & $-1.29_{-0.83}^{+0.72}$ \\
$\ssqthonethree/10^{-3}$  & --- & --- & $28.0_{-6.5}^{+2.8}$ & $31.0_{-6.9}^{+3.0}$ \\
$\ssqthtwothree $ & $0.561_{-0.038}^{+0.019}$    & $0.563_{-0.032}^{+0.017}$ & $0.467_{-0.018}^{+0.106}$     & $0.466_{-0.019}^{+0.103}$ \\
$\dmsqtwothree/10^{-3}\,(\mathrm{eV^2})$ & $2.494_{-0.058}^{+0.041}$ & --- & $2.495_{-0.058}^{+0.041}$ & --- \\
$|\Delta m^2_{31}|/10^{-3}\,(\mathrm{eV^2})$ & --- & $2.463_{-0.056}^{+0.042}$ & --- & $2.463_{-0.055}^{+0.043}$ \\
\hline\hline
\end{tabular}
\egroup
\caption{Results for the oscillation parameters from the fit to data with and without the reactor constraint in the frequentist analysis, with the confidence intervals estimated using the constant $\Delta \chi^2$ method.}
\label{tab:oa:freq:BestFit}
\end{table*}

Global best-fit values are given in \autoref{tab:oa:freq:BestFit}. As noted in the Bayesian section, the results with and without the reactor constraint are compatible, with the former resulting in stronger constraints on \deltacp and \ssqthtwothree. All the following results are from the fit to data using the reactor constraint.

\subsubsection{The CP-violating phase \deltacp, and mass ordering}
\label{sec:cp_phase_mo_results}
\begin{figure}[htbp]
\centering
\includegraphics[width=0.95\columnwidth]{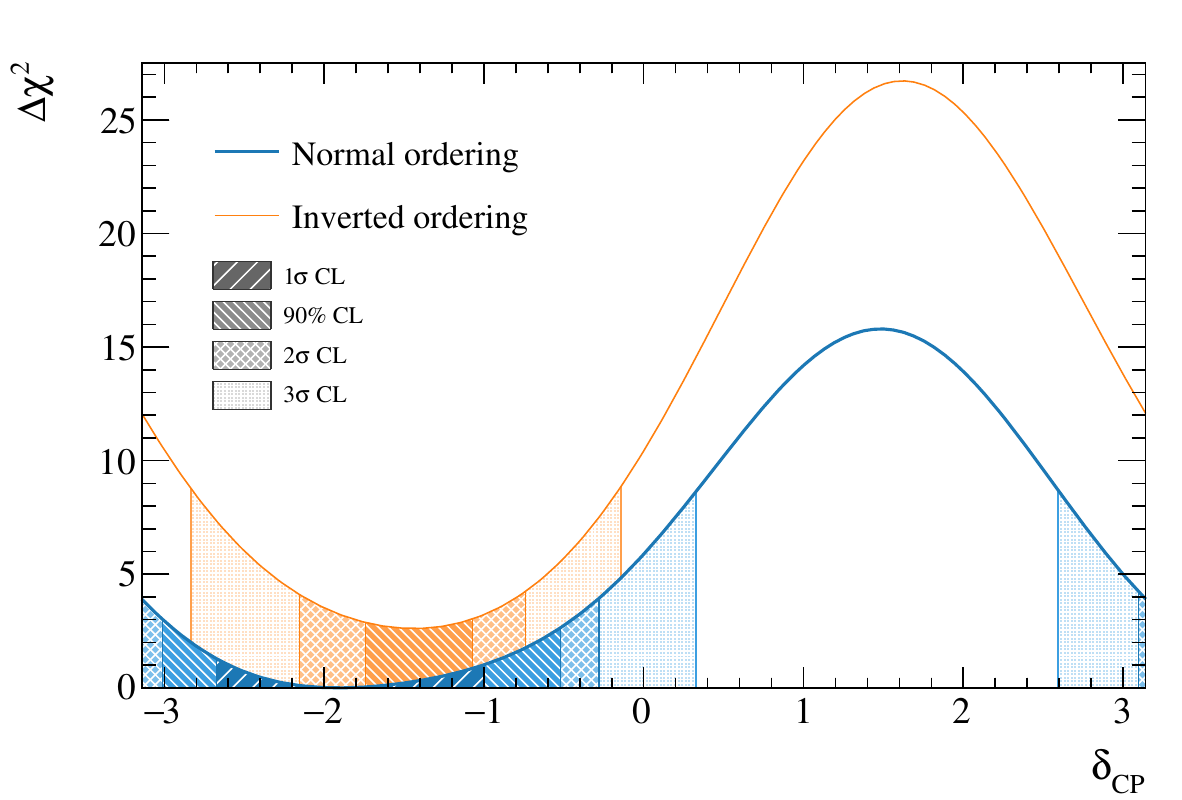}
\caption{The $\Delta\chi^2$ distribution in \deltacp from fitting to the data with the reactor constraint applied. The confidence intervals in the shaded regions are calculated using the FC method.}
\label{fig:oa:FC:dcp}
\end{figure}

\autoref{fig:oa:FC:dcp} shows the $\Delta\chi^2$ distributions for \deltacp in both mass orderings with FC-adjusted confidence intervals, which are also summarised in \autoref{tab:CI:FC:dcpth23}. A large region of $\sin\deltacp > 0$ is excluded at $>3\sigma$ confidence level (CL), whereas the CP-conserving values $\deltacp=0,\pi$ are excluded at 90\% CL. In particular, $\deltacp = \pi$ is just inside the $2\sigma$ interval.

\begin{table*}[htbp]
\centering
\bgroup
\def\arraystretch{1.1}
\begin{tabular}{l | c c | c c}
\hline\hline
Confidence  & \multicolumn{2}{c|}{\deltacp (rad.)} & \multicolumn{2}{c}{\ssqthtwothree} \\
Level & Normal ordering & Inverted ordering & Normal ordering & Inverted ordering \\
\hline
$1\sigma$ & $[-2.67, -1.00]$ & --- & $[0.529, 0.582]$ & --- \\
$90\%$ & $[-3.01, -0.52]$ & $[-1.74,-1.07]$ & $[0.444, 0.593]$ & $[0.536, 0.584]$  \\
$2\sigma$ & $[-\pi, -0.28] \cup{} [3.10, \pi]$ & $[-2.16, -0.74]$ & $[0.436, 0.598]$ & $[0.512, 0.592]$ \\
$3\sigma$ & $[-\pi, 0.33] \cup{} [2.59, \pi]$ & $[-2.83, -0.14]$ & N/A & N/A \\
\hline\hline
\end{tabular}
\egroup
\caption{FC-corrected confidence intervals for \deltacp and \ssqthtwothree from the fit to data in the frequentist analysis, using the reactor constraint on \ssqthtwoonethree. The $3\sigma$ FC correction was not computed for \ssqthtwothree.}
\label{tab:CI:FC:dcpth23}
\end{table*}

\begin{figure}[htbp]
\centering
\includegraphics[width=0.95\columnwidth]{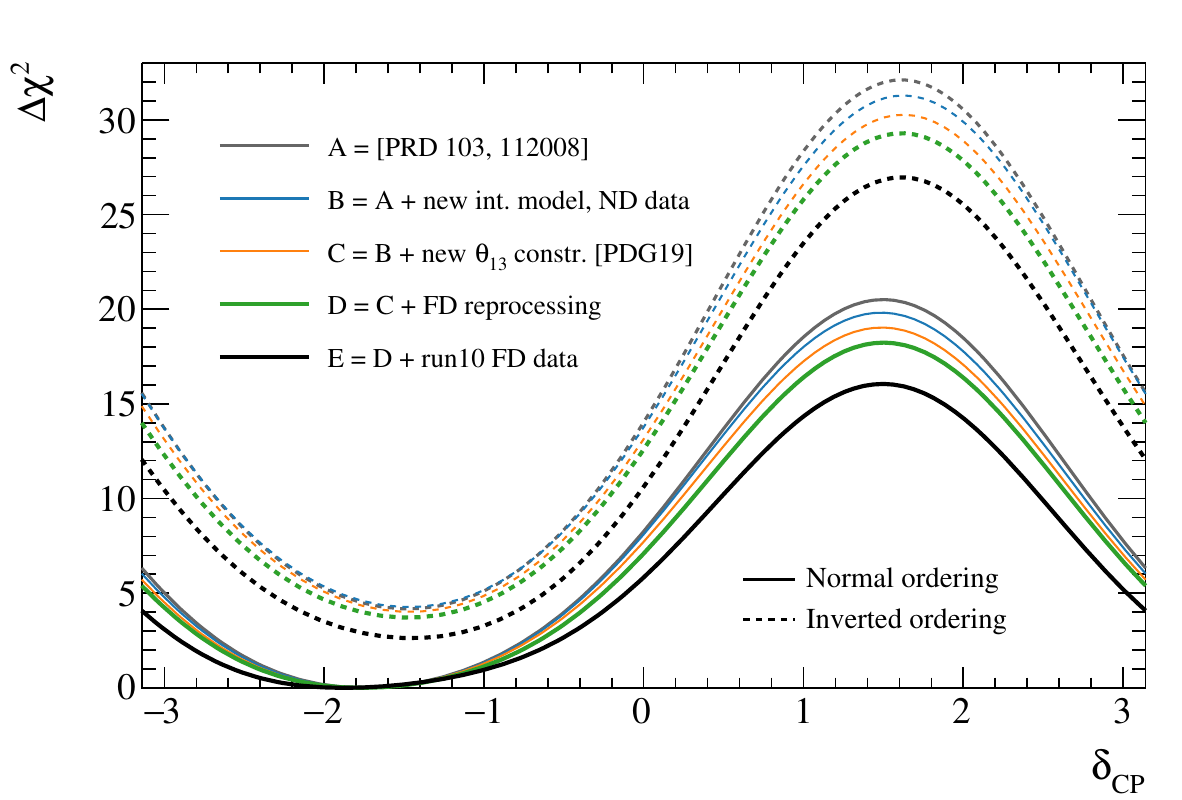}
\caption{The $\Delta\chi^2$ distribution in \deltacp for incremental modifications from the previous analysis~\cite{Abe:2021gky}
to this result, for normal and inverted mass orderings. ``E'' corresponds to this analysis, except that unlike the main frequentist result, the $\mu$-like samples do not use the scattering angle information for better compatibility with the previous T2K analysis.}
\label{fig:oa:evolution:dcp}
\end{figure}

\begin{figure*}[htbp]
\centering
\begin{subfigure}[b]{\columnwidth}
\includegraphics[width=0.95\textwidth]{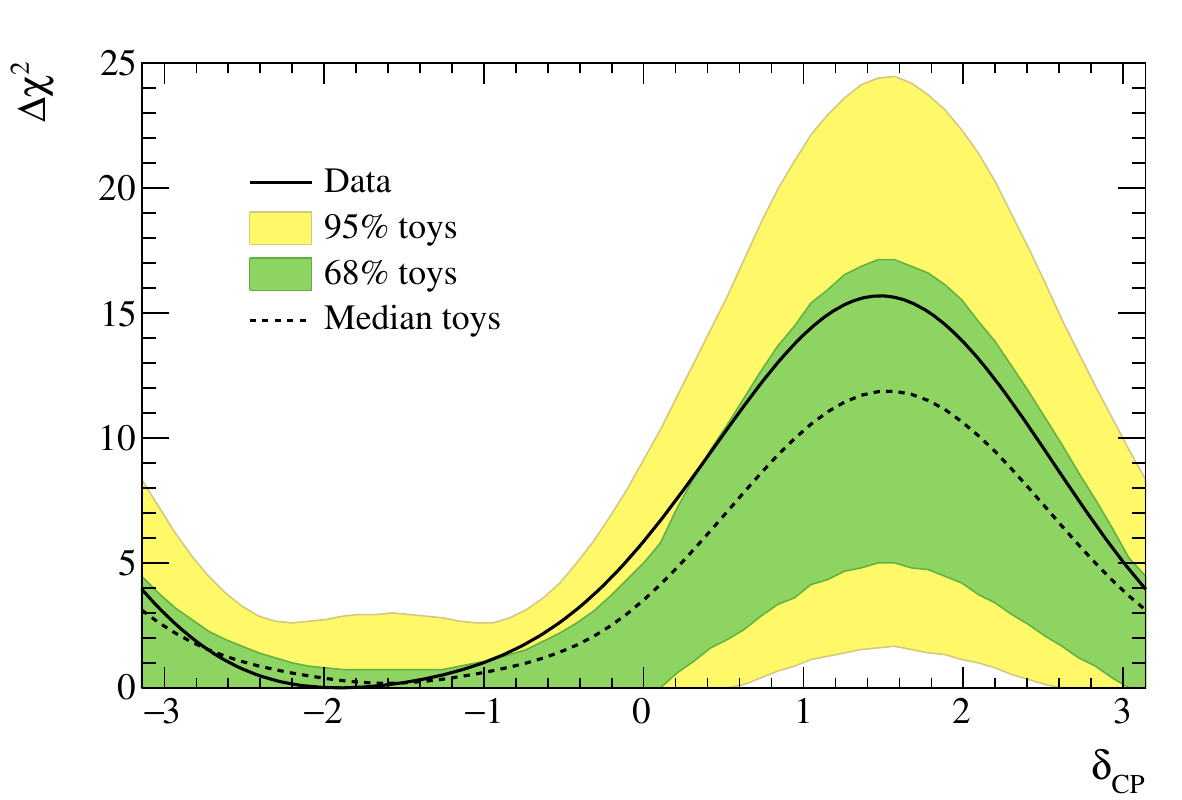}
\end{subfigure}
\begin{subfigure}[b]{\columnwidth}
\includegraphics[width=0.95\textwidth]{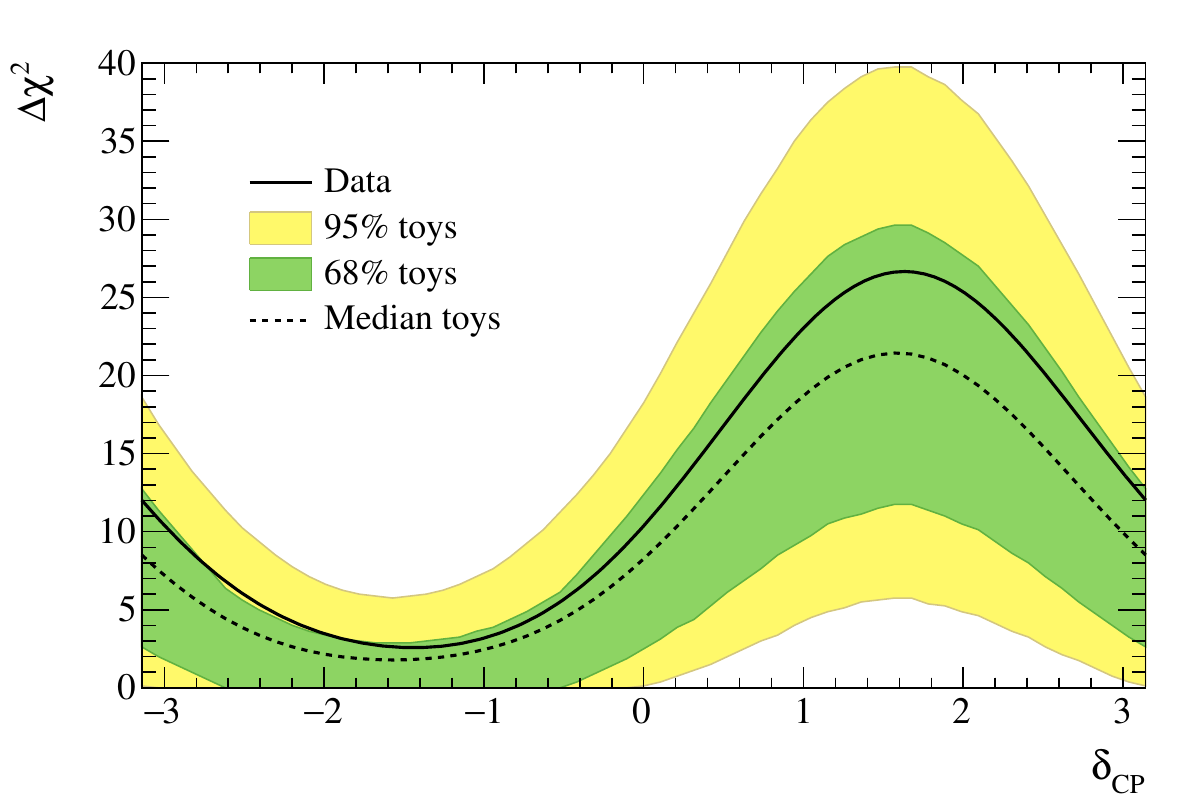}
\end{subfigure}
\caption{The $\Delta\chi^2$ distribution in \deltacp from fitting to the data assuming normal (left) and inverted (right) neutrino mass ordering, with the reactor constraint applied. The distribution is overlaid with the expectations from an ensemble of toy simulated experiments created with true normal ordering and $\deltacp = -\pi/2$, showing the $\Delta\chi^2$ for 68\% and 95\% of the toys, and their median.}
\label{fig:oa:brazil:dcp}
\end{figure*}

As was also seen in the Bayesian analysis, the constraint on \deltacp is weaker compared to T2K's previous analysis~\cite{Abe:2021gky,T2K:2019bcf}. \autoref{fig:oa:evolution:dcp} shows the impact on the $\Delta\chi^2$ of \deltacp after each update introduced in this analysis, all of which weaken the \deltacp constraint, with the largest contribution being the addition of the latest data in run 10 at the FD. The data is now more consistent with the expectation, shown in \autoref{fig:oa:brazil:dcp} for both normal and inverted ordering. In most of the \deltacp parameter space, T2K is below the upper limit of the 68\% expectation band of ensemble experiments at maximal CP~violation. The inverted mass ordering is disfavoured at more than $1\sigma$ for all \deltacp values, mostly consistent with the expected sensitivity at $\sin\deltacp = -1$. Replacing each sample's event distribution in data by the expectation of the model shows that the stronger \deltacp constraint observed in data compared to the expectation comes from the \fhcalt \rede sample.

\subsubsection{Atmospheric oscillation parameters}
\begin{figure}[htbp]
\centering
\includegraphics[width=0.95\columnwidth]{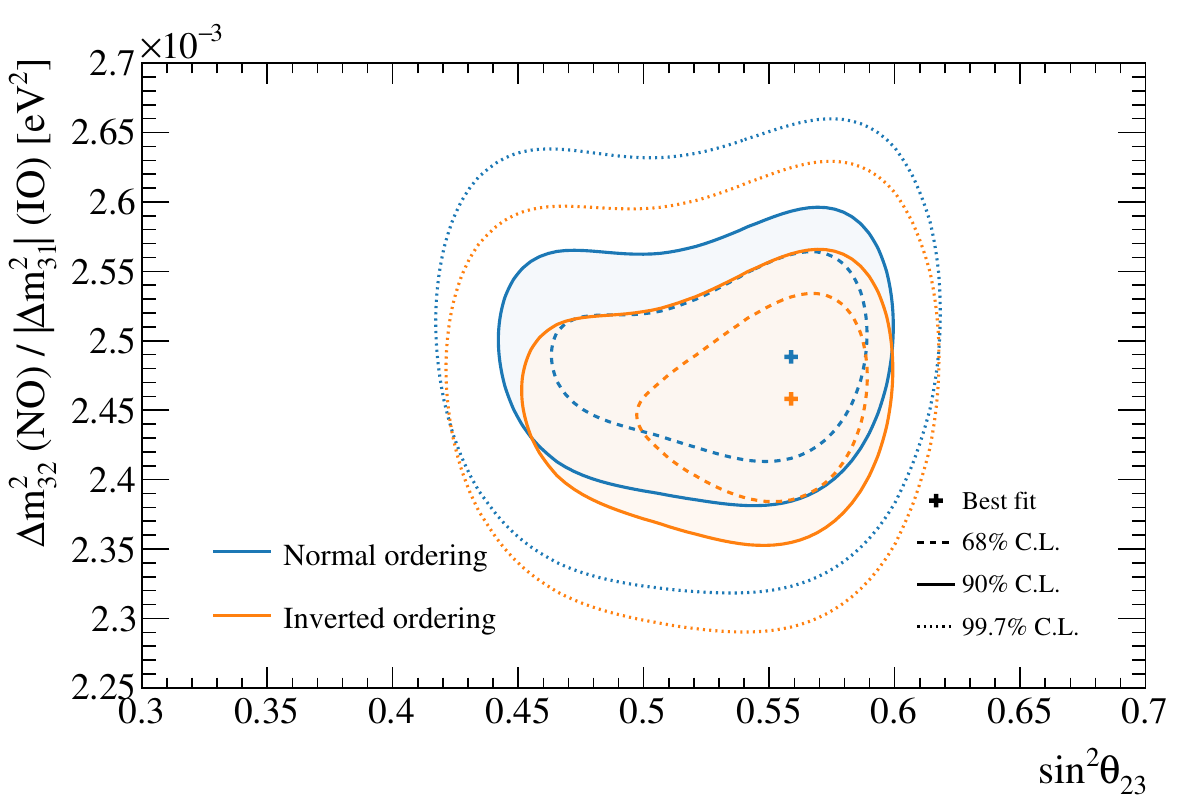}
\caption{Confidence regions in $\ssqthtwothree-\dmsqtwothree$ ($|\Delta m^2_{31}|$ in inverted mass ordering) for the data fit with the reactor constraint applied, obtained with the constant $\Delta \chi^2$ method, where in each mass ordering hypothesis a fixed mass ordering is assumed.}
\label{fig:oa:RC:th23dm2}
\end{figure}

\begin{figure}[htbp]
\centering
\includegraphics[width=0.95\columnwidth]{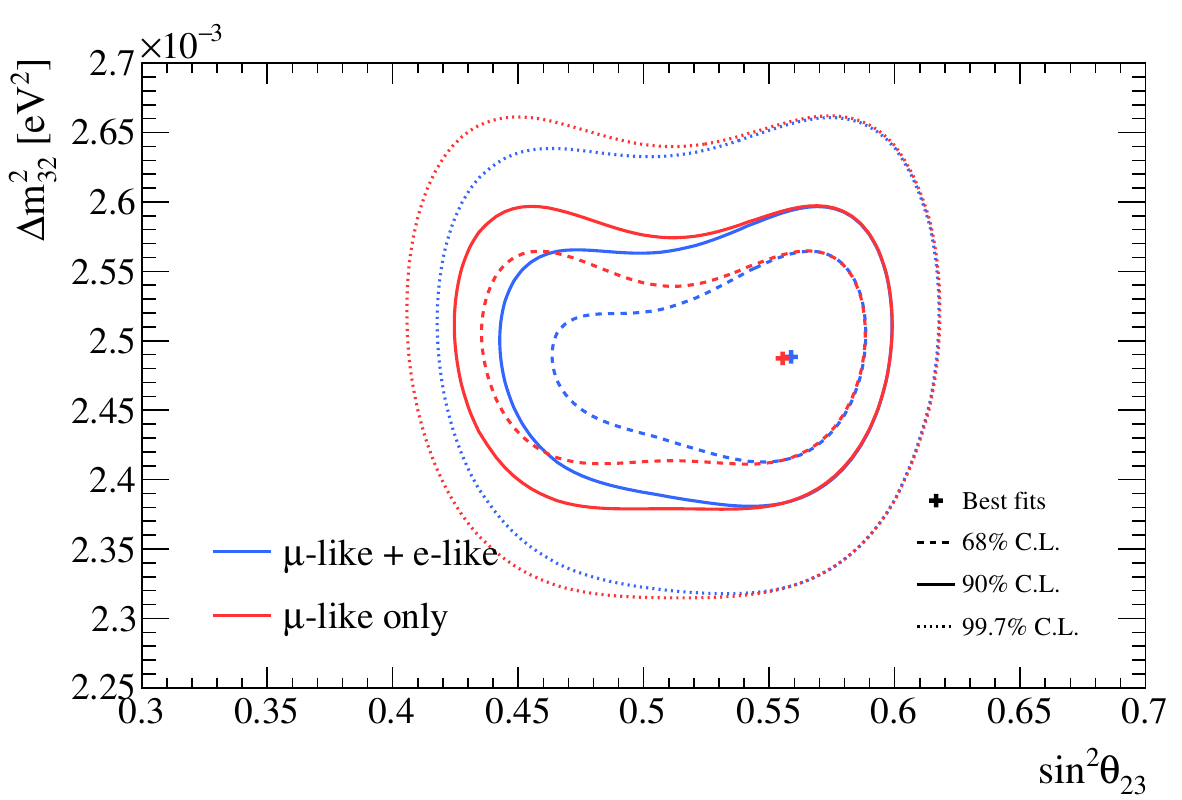}
\caption{Comparison of confidence regions in $\ssqthtwothree - \dmsqtwothree$ for normal ordering, between a full fit and a fit using only the $\mu$-like samples. The intervals are calculated with the constant $\Delta \chi^2$ method, and applying the reactor constraint.}
\label{fig:oa:muonly:th23dm2_NH}
\end{figure}

The $\ssqthtwothree-\dmsqtwothree$ confidence intervals are presented in \autoref{fig:oa:RC:th23dm2}. The contours are compatible for both mass orderings, with a slight shape change compared to the previous T2K analysis, to now marginally prefer the upper octant. \autoref{fig:oa:muonly:th23dm2_NH} shows the contour from a fit using only the \rmu samples, which shows that the constraint is dominated by the \rmu samples, with the \re samples providing the sensitivity to the octant of $\theta_{23}$.

\begin{figure}[htbp]
\centering
\includegraphics[width=0.95\columnwidth]{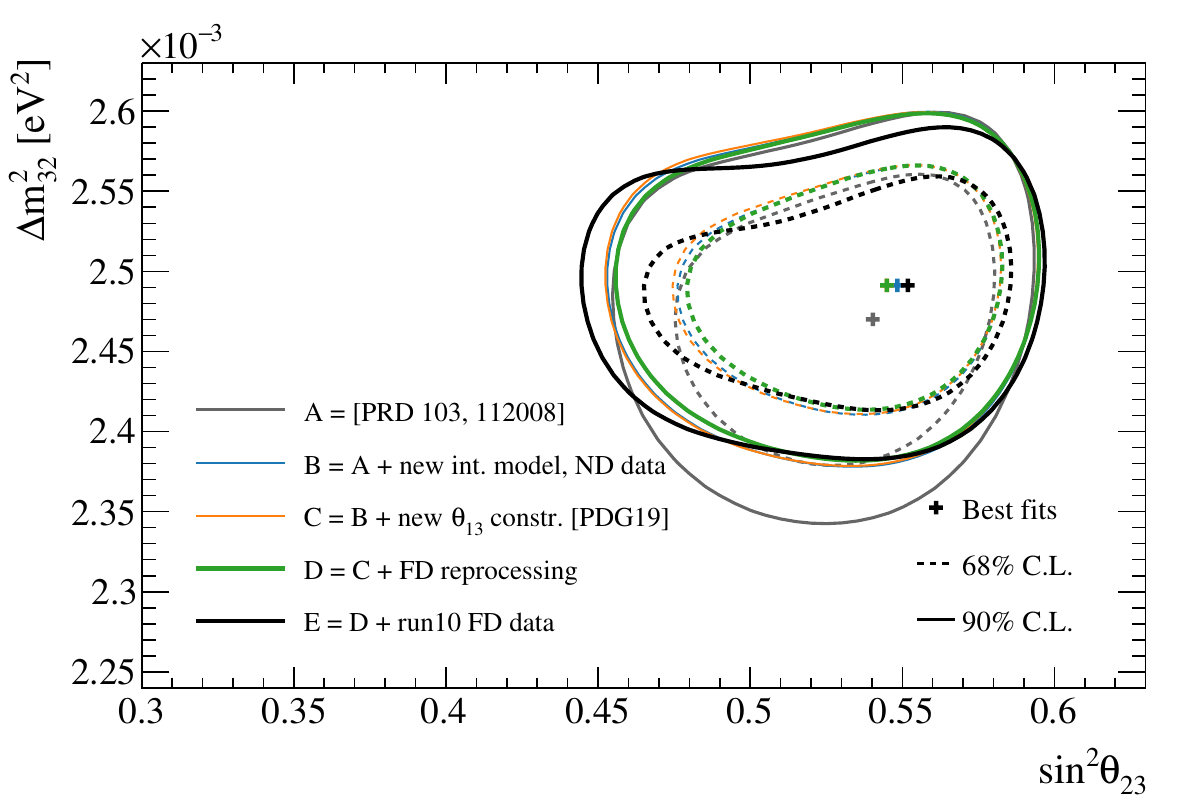}
\caption{The $\Delta\chi^2$ distribution in \dmsqtwothree and \ssqthtwothree for normal ordering, showing incremental modifications of the previous analysis~\cite{Abe:2021gky,T2K:2019bcf} to this result. 
``E'' corresponds to this analysis, except that unlike the main frequentist result, the $\mu$-like samples do not use the scattering angle information for better compatibility with the previous analysis. The best-fit point for ``C'' in orange is the same as ``D'' in green.}
\label{fig:oa:evolution:th23dm2}
\end{figure}

The evolution of the $\ssqthtwothree-\dmsqtwothree$ contour from the fit to data after introducing each update in the analysis is shown in \autoref{fig:oa:evolution:th23dm2}. The most significant impact on the \dmsqtwothree constraint comes from changing the cross-section model and updating the ND constraint. For \dmsqtwothree, improvements in the removal energy uncertainty have significantly reduced the uncertainty \emph{before} the smearing based on simulated data studies has been applied. Thanks to increased robustness of the uncertainty model, the size of the smearing has also been reduced by a factor of 2.8, discussed in detail in \autoref{sec:fakeData}. For \ssqthtwothree, there is a slight shift in shape from the latest data in the FD. The new data favour a slightly larger \ssqthtwothree, and so pushes less against the boundary of maximal mixing. This results in a slightly weaker constraint in the lower octant and similar constraint in the upper octant compared to the previous analysis.

\begin{figure}[htbp]
\centering
\includegraphics[width=0.95\columnwidth]{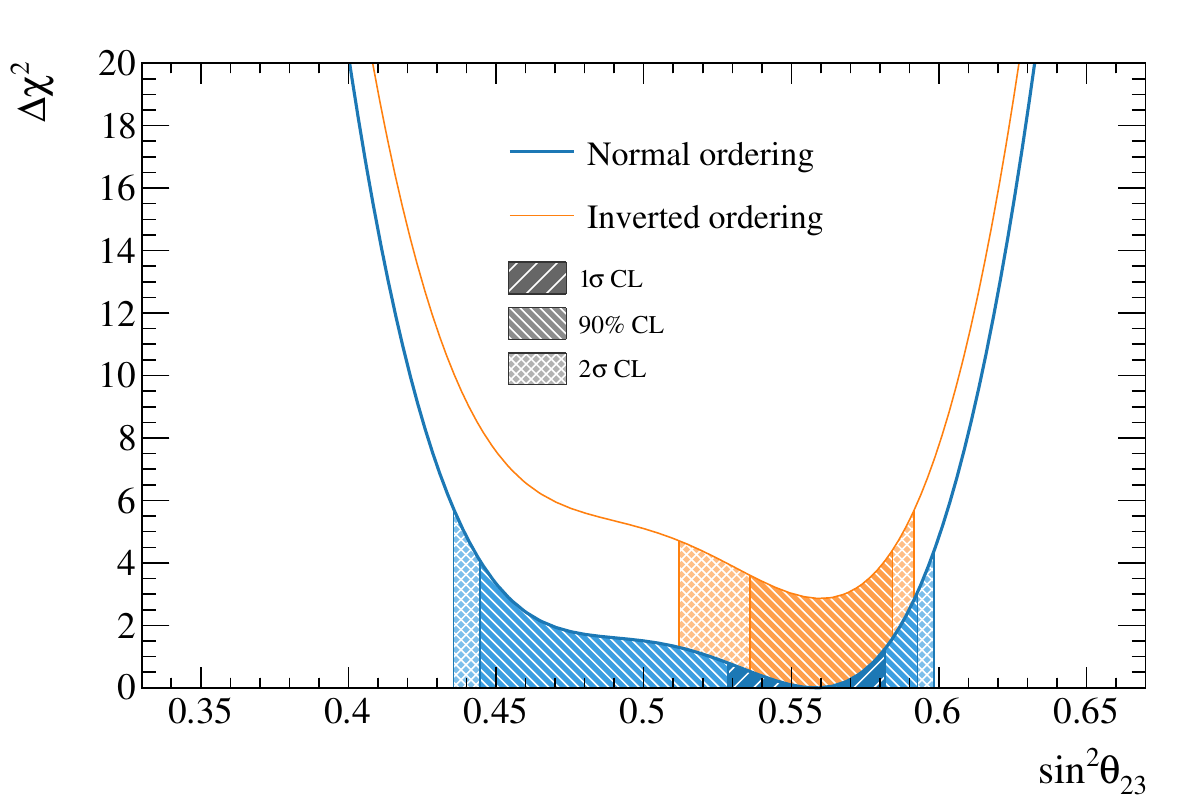}
\caption{The $\Delta\chi^2$ distribution in \ssqthtwothree for fitting to the data with the reactor constraint applied. The confidence intervals in the shaded regions are calculated using the FC method.}
\label{fig:oa:FC:th23}
\end{figure}

\begin{figure}[htbp]
\centering
\includegraphics[width=0.95\columnwidth]{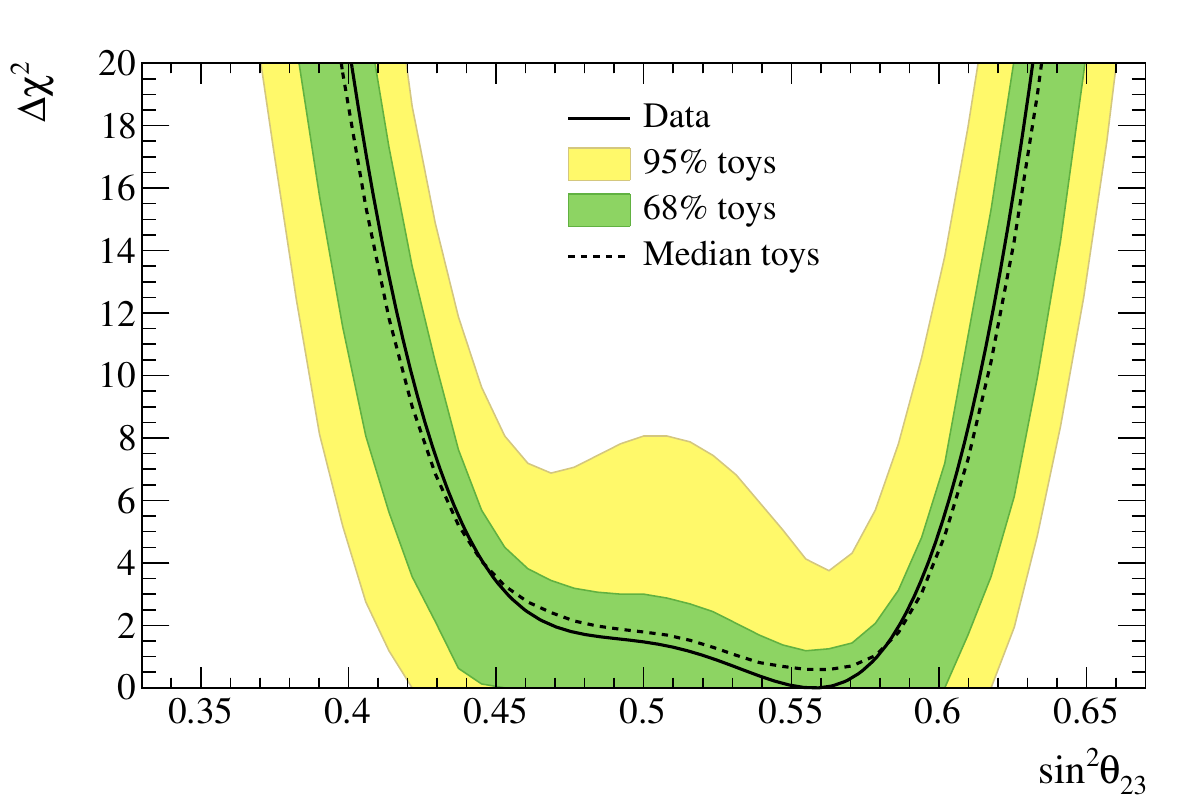}
\caption{The $\Delta\chi^2$ distribution in \ssqthtwothree for fitting to the data with the reactor constraint applied, overlaid with the expected distributions from an ensemble of simulated experiments created with true NO and $\ssqthtwothree = 0.56$.}
\label{fig:oa:brazil:th23:NH}
\end{figure}

The $\Delta \chi^2$ distribution for \ssqthtwothree is shown in \autoref{fig:oa:FC:th23} with the confidence intervals summarised in \autoref{tab:CI:FC:dcpth23}. The new data at the FD has reduced compatibility with maximal mixing, which is now outside the FC-corrected $1\sigma$ confidence interval. Whilst the upper octant is favoured at $1\sigma$ CL, the data is still compatible with both octant hypotheses at 90\% CL. These results are compatible with the sensitivity, shown in \autoref{fig:oa:brazil:th23:NH}.

\subsection{Cross-fitter comparisons}
\label{sec:oa_cross_fitter}
\begin{figure*}[htbp]
\centering
\begin{subfigure}[b]{0.49\textwidth}
\includegraphics[width=\textwidth]{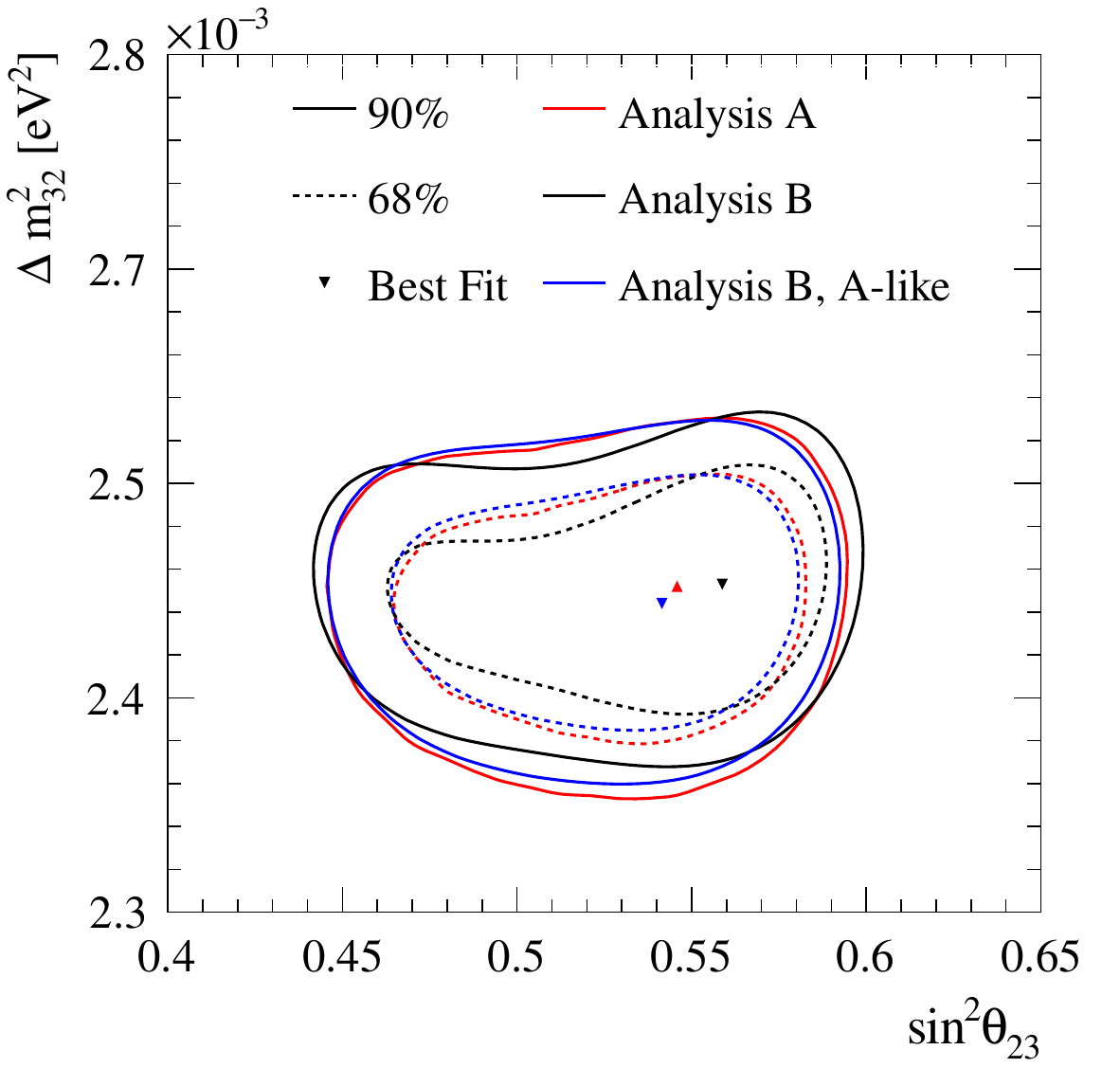}
\label{fig:oa:M3PTcomp:disapp}
\end{subfigure}
\begin{subfigure}[b]{0.49\textwidth}
\includegraphics[width=\textwidth]{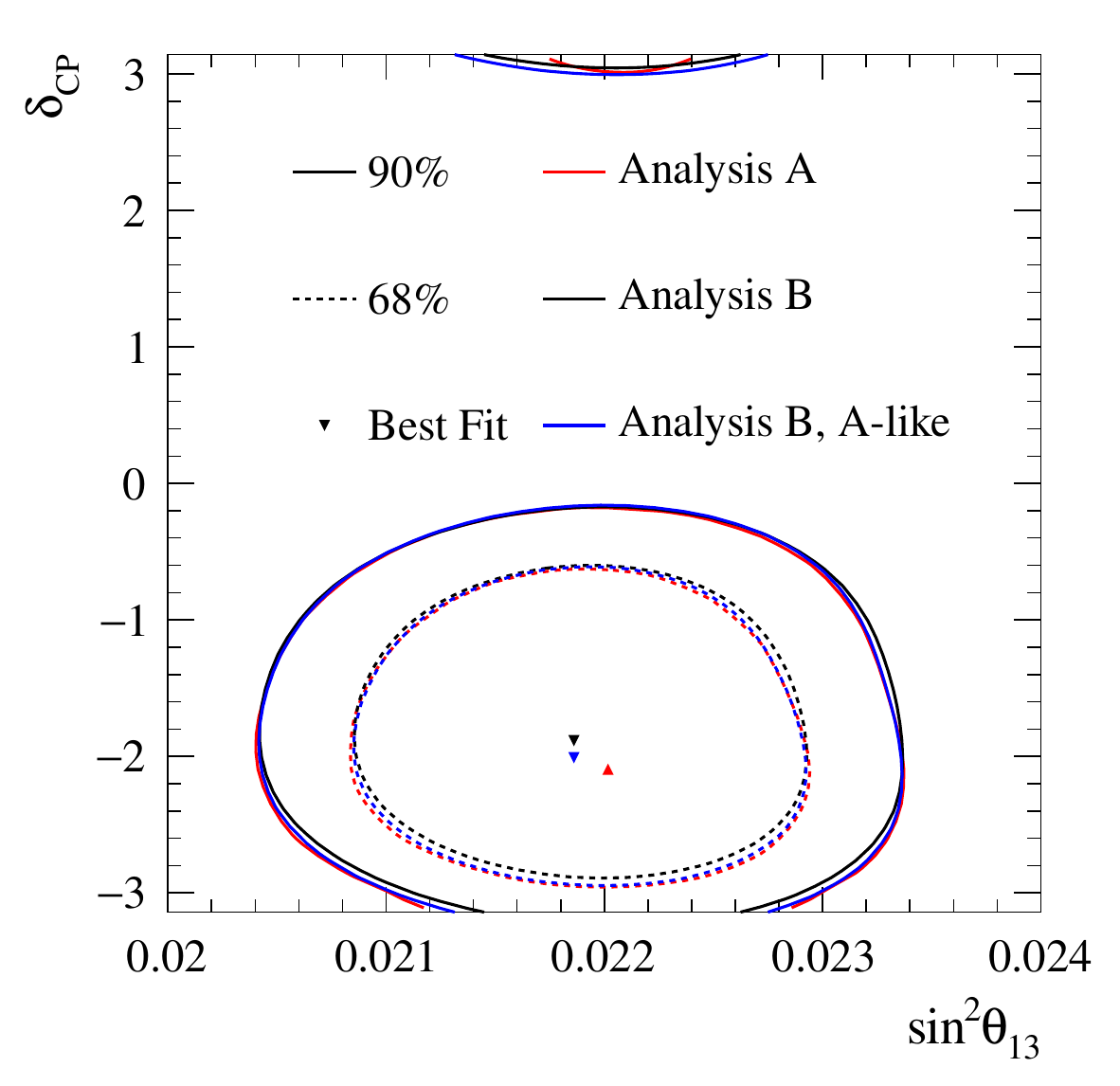}
\label{fig:oa:M3PTcomp:app}
\end{subfigure}
\caption{Comparison of the 68\% and 90\% confidence intervals from fits to data from the Bayesian analysis (``Analysis A'') and the frequentist analysis (``Analysis B''), discussed in \autoref{sec:oa_cross_fitter}. ``Analysis B, A-like'' configures the frequentist analysis in the same way as the Bayesian analysis, using the same binning at the FD and the same MCMC-based ND analysis. The contours are extracted from fits that fix the neutrino mass ordering to the normal ordering and apply the reactor constraint on \ssqthonethree.}
\label{fig:oa:M3PTcomp}
\end{figure*}

To compare the consistency of the Bayesian analysis described in \autoref{sec:oa:bayesian} and the frequentist analysis described in \autoref{sec:oa:freq}, the Bayesian posterior distributions were recast into frequentist $\Delta\chi^{2}$ distributions comparable to the frequentist analysis.
\autoref{fig:oa:M3PTcomp} shows comparisons of $\ssqthtwothree-\dmsqtwothree$ and $\ssqthonethree-\deltacp$ contours from fits to data from both frameworks.
The minor differences between the resulting contours can be attributed to two distinct analysis choices: the way in which the constraints from the near-detector analysis on the systematic uncertainties are applied, and the choice of kinematic variables in which the far detector samples are binned.
The Bayesian analysis uses a ND analysis with irregular rectangular binning for the ND samples to better adapt to differences in $p_\mu-\cos\theta_\mu$ phase space density, whereas the frequentist analysis' ND constraint uses regular binning.
Both analyses use the lepton scattering angle for the $e$-like samples, but differ in the use of reconstructed energy (Bayesian) or reconstructed lepton momentum (frequentist) for the $\mu$-like samples. Additionally, the frequentist analysis also uses the reconstructed lepton scattering angle for the $\mu$-like samples to disentangle systematic uncertainties related to energy scale and mis-reconstructed backgrounds at the FD.
Other differences, like the non-Gaussian nature of parameters included in the ND constraint or event-by-event vs. binned oscillation probability calculation, had little effect.
When the frequentist analysis is configured to impose the constraints from the ND MCMC analysis and bin the FD samples similarly to the Bayesian analysis, these minor differences abate, shown in \autoref{fig:oa:M3PTcomp}.
The uncertainty models of both analyses were validated against each other for consistency and were found to agree.

\subsection{Comparisons with other experiments}
\label{sec:oa:comp}
\begin{figure}[htbp]
\centering
\includegraphics[width=0.95\columnwidth]{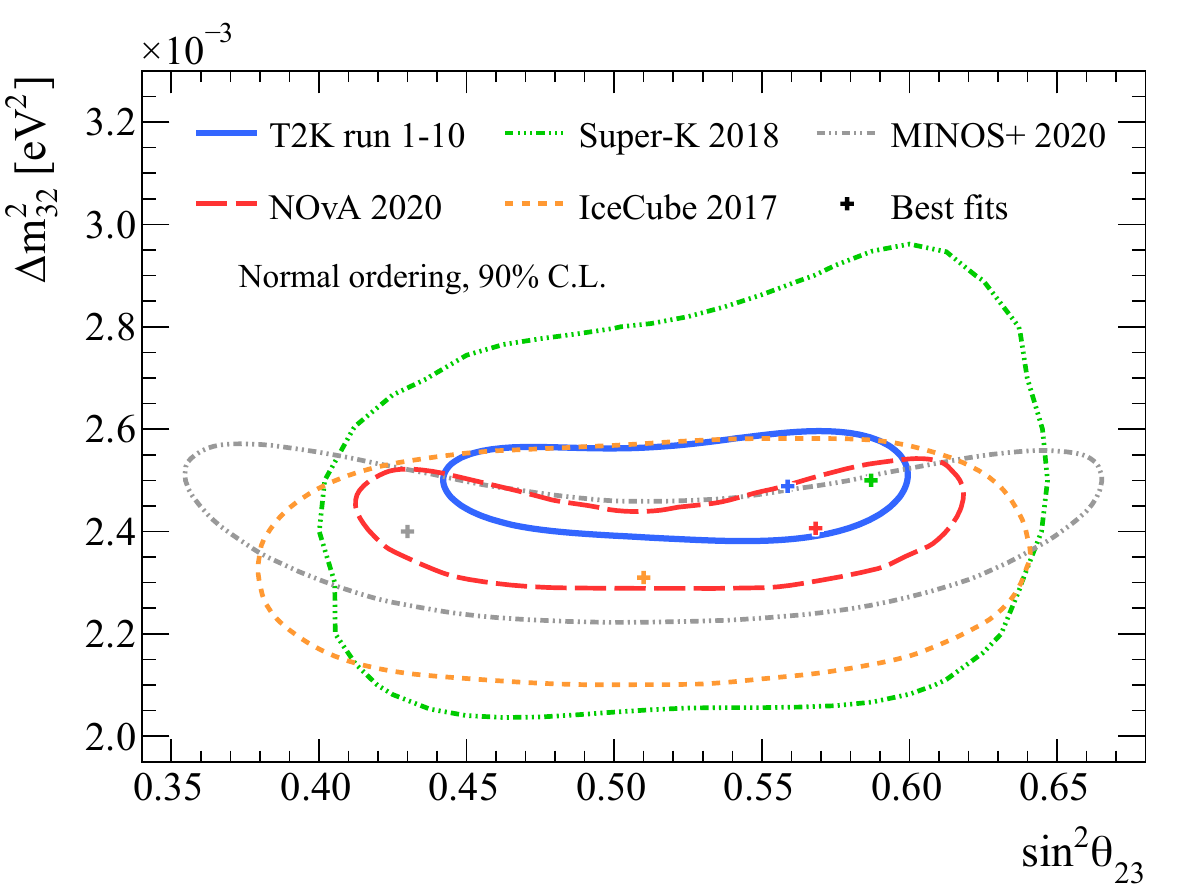}
\caption{Comparison of the 90\% confidence regions in $\ssqthtwothree-\dmsqtwothree$ for normal ordering with \nova~\cite{NOvA:2021nfi}, Super-K~\cite{Super-Kamiokande:2017yvm}, IceCube~\cite{icecube_2018_10_21234_B4105H}, and MINOS+~\cite{MINOS:2020llm}. The \nova and IceCube constraints are obtained with the FC method, but with different treatment of the mass ordering: \nova takes the minimum over both mass orderings, whereas the IceCube contours assume normal ordering. The T2K, Super-K, and MINOS+ contours are computed with the constant $\Delta\chi^2$ method, assuming normal ordering.}
\label{fig:oa:compexp:th23dm2_NH}
\end{figure}

\begin{figure*}[htbp]
\centering
\includegraphics[width=\textwidth,trim=0 0 50 0,clip]{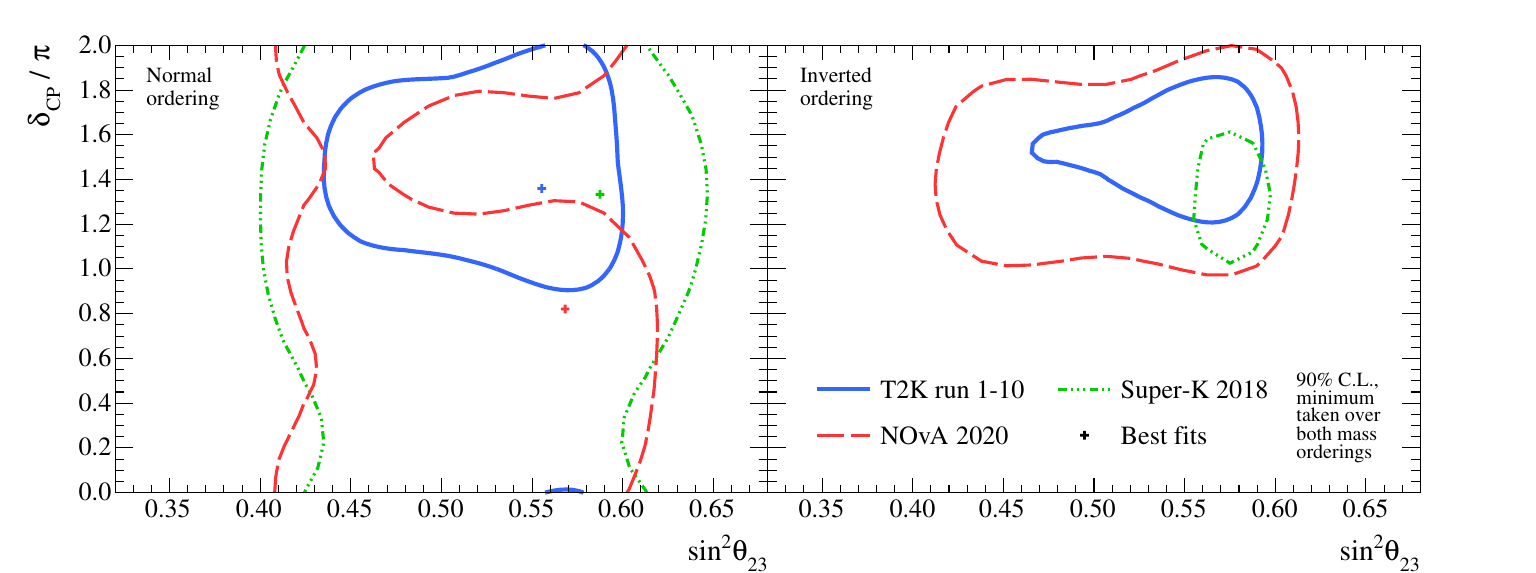}
\caption{Comparison of 90\% confidence regions in $\ssqthtwothree - \deltacp$ over both mass orderings with \nova~\cite{NOvA:2021nfi} and Super-K~\cite{Super-Kamiokande:2017yvm}. The T2K and \nova confidence regions have been computed using the FC method, whereas the Super-K results are obtained with the constant $\Delta\chi^2$ method.}
\label{fig:oa:compexp:th23dcp}
\end{figure*}

In the global context of neutrino oscillation experiments, these results provide leading constraints on both the atmospheric oscillation parameters, \dmsqtwothree and \ssqthtwothree, and the CP-violating phase, \deltacp. 
Whereas the other experiments profile over parameters to calculate the $\Delta \chi^2$, T2K instead calculates the marginal likelihood for the $\Delta\chi^2$.
\autoref{fig:oa:compexp:th23dm2_NH} shows the 90\% confidence regions in $\ssqthtwothree-\dmsqtwothree$ for the normal ordering from the frequentist analysis, compared to \nova, SK and IceCube. There is general agreement between the experiments, with T2K providing the strongest constraints on both parameters. 
\autoref{fig:oa:compexp:th23dcp} compares the 90\% confidence regions in $\ssqthtwothree-\deltacp$ for both orderings to \nova and SK. The confidence intervals on \ssqthtwothree significantly overlap, as do the intervals for \deltacp. In the normal ordering, T2K excludes large regions of the \nova constraint at 90\% confidence interval, and \nova excludes parts of T2K's 90\% confidence interval. In the inverted ordering, the experiments consistently favour the $\pi<\deltacp<2\pi$ region, with a weak preference for the upper octant. Importantly, there is no significant tension between the experiments, and more data is needed to elucidate the matter. Furthermore, the joint oscillation analyses with the \nova and SK collaborations will help address this.

%% file: Fakedata.tex
Simulated data studies with the frequentist analysis were used to investigate the impact of alternative model predictions and data-driven tunes, discussed in \autoref{sec:interactionModel_fds}, on the oscillation parameter constraints. The oscillation analysis in \autoref{sec:oa_results} had these uncertainty inflation strategies applied, and this section summarises the procedure, with details provided in \autoref{app:appendix_fakedata}. 

\subsection{Methodology}
In the simulated data studies, the prediction from an alternative model is treated as the data at the ND, and is fit with the usual systematic uncertainty model. The parameters are fit to simulated data at the ND and are propagated to the FD, and the reconstructed energy spectrum and oscillation parameter constraints are compared to an ``Asimov'' data set. 
In an Asimov analysis, the parameters for the systematic uncertainties are set to specific values and the predicted spectra at each detector is treated as the data, giving an expectation of the sensitivity if no statistical fluctuations were present.
For the oscillation parameters, two separate Asimov points were tested: one close to T2K's best-fit point, and one with $\deltacp=0$ and non-maximal \ssqthtwothree, detailed in \autoref{app:appendix_fakedata}, \autoref{tab:asimov_pars}. This section only presents results with the Asimov data near T2K's best-fit point. The PDG 2019 reactor constraint on \ssqthonethree is applied in the following studies, but had little impact on the overall conclusions.
Although simulated data sets can result in both weaker and stronger constraints on the oscillation parameters than the expected sensitivity, they are only used to \emph{inflate} the uncertainties in this analysis.

The simulated data set procedure mainly identifies two types of effects:
\begin{itemize}[leftmargin=*]
    \item \textbf{Systematic uncertainty model shortcomings:} If the systematic uncertainty model is robust, or if the effect of the alternative model is small, the oscillation parameter contours obtained with the simulated data sets will not see a bias with respect to the expected sensitivity. The bias is quantified as the percentage change of the middle of the $1\sigma$ confidence interval of an oscillation parameter, relative to the $1\sigma$ from systematic uncertainties in the expected sensitivity analysis. An example is discussed in \autoref{subsec:nonqe_fds_results}.
    
    \item \textbf{ND to FD extrapolation issues:} Some alternate models may not produce a significant bias on the oscillation parameters, often due to the low sensitivity of the samples they affect. Issues in the extrapolation process can be exposed by comparing three distributions: i) the predicted spectrum at the FD from fitting the Asimov data set at the ND, ii) the predicted spectrum at the FD from fitting to the simulated data at the ND, iii) and the predicted spectrum at the FD when applying the alternative model directly. Even though the bias on the oscillation parameters at T2K statistics may be small, simulated data studies may guide which of the systematic uncertainties are important to address in future T2K analyses and upcoming high-statistics experiments, such as Hyper-Kamiokande~\cite{Hyper-Kamiokande:2018ofw} and DUNE~\cite{DUNE:2020ypp}. An example is discussed in \autoref{sec:minerva_fds}.
\end{itemize}

All the individual biases are summed in quadrature and are used to inflate the confidence interval for \dmsqtwothree, due to its simple Gaussian probability density. For \deltacp, the effect of systematic uncertainties is much smaller and the probability density is non-Gaussian, so a different method is applied. Each simulated data set is studied to see if it impacts any major claims in the analysis; in this analysis the 90\% confidence interval of \deltacp. This is done by calculating the difference in the $\Delta\chi^2$ distribution for \deltacp for the Asimov data and the simulated data, and adding the difference to the $\Delta\chi^2$ distribution for \deltacp  from the data, where the 90\% confidence interval was calculated using the FC method mentioned in \autoref{sec:oa:freq}.

Simulated data sets can drastically increase or decrease the number of events at both detectors. In such cases, comparing the constraints on the oscillation parameters to the expected sensitivity conflates the effects of propagating mis-modelling from the ND analysis with the impact of increased or decreased statistics from the simulated data set. For instance, an alternative model that increases the number of \nueany at the FD near the oscillation maximum will likely lead to a stronger constraint on \deltacp due to the measurement being dominated by statistical uncertainties in those samples.
To gauge this effect, the three predictions from the ND to FD extrapolation studies, outlined earlier, are used. If the model from the ND simulated data analysis predicts the spectrum of the alternative model well at the ND, and correctly predicts the spectrum at the FD compared to when directly applying the alternative model, a ``scaled Asimov'' approach is utilised. 
In these cases, two changes to the procedure are made: a) the propagated ND constraint is the expected sensitivity to the systematic parameters if the real data is as predicted by the pre-fit model, and b) the variation to the model that was used to build the simulated data set is also applied to the simulation that is being fit at the FD. This removes most of the statistical effect and better captures the features of propagating a mis-modelling in the ND analysis. For this analysis, the scaled approach was only used for the 2p2h Martini simulated data set.

The simulated data studies all concerned the interaction model and were detailed in \autoref{sec:interactionModel_fds}. Details of the simulated data study procedure and two examples are provided in \autoref{app:appendix_fakedata}.

\subsection{Results}
\autoref{tab:bias_table_fakedata} summarises the observed biases on the oscillation parameters, showing the simulated data set with the highest impact from each category. The full results are shown in \autoref{app:appendix_fakedata}, \autoref{tab:bias_table_fakedata_full}. The impact of the simulated data studies on \ssqthtwothree and \deltacp was found to be small compared to the impact of statistical and systematic uncertainties. The largest bias on \dmsqtwothree relative to the systematic uncertainty was found to be 57.8\% from the pion SI simulated data set, and 20.8\% relative to the overall uncertainty.
Selected simulated data studies were added in quadrature\footnote{The non-CCQE, data-driven low pion momentum, low $Q^2$ pion suppression from \minerva, pion SI, the CCQE form factor with the largest impact (z-expansion, upper variation), Martini 2p2h, and the removal energy simulated data studies were selected.} to avoid double counting similar physics effects, leading to an overall smearing on \dmsqtwothree of $1.35\times10^{-5}~\mathrm{eV^2}$. For comparison, the overall uncertainty on \dmsqtwothree from the expected sensitivity study, before the simulated data procedure, was $5.7\times10^{-5}~\mathrm{eV^2}$ and is dominated by the uncertainty from statistics, which was $5.3\times10^{-5}~\mathrm{eV^2}$.
Generally, the simulated data studies had a smaller impact on \deltacp, due to its uncertainty being dominated by the statistics in the electron-like selections at the FD.
\begin{table}[htbp]
\centering
\begin{tabular}{ll ccc}
\hline \hline
Simulated data set                  & Relative to & \ssqthtwothree & \dmsqtwothree & \deltacp \\
\hline
\multirow{2}{*}{CCQE z-exp high}    & Total       & 0.3\%  & 2.1\%  & 0.4\%   \\ %
                                    & Syst.       & 0.7\%  & 5.7\%  & 1.7\%  \vspace{2mm} \\
                                    
\multirow{2}{*}{CCQE removal energy}& Total       & 0.0\%  & 4.8\%  & 1.3\%   \\ %
                                    & Syst.       & 0.0\%  & 13.4\% & 5.2\%  \vspace{2mm} \\
\multirow{2}{*}{Non-CCQE}           & Total       & 8.7\%  & 11.8\%    & 1.7\%   \\ %
                                    & Syst.       & 21.3\% & 32.7\%   & 6.9\%  \vspace{2mm} \\  %
\multirow{2}{*}{2p2h Martini}       & Total       & 0.7\%  & 2.7\%  & 0.4\%   \\  %
                                    & Syst.       & 1.6\%  & 7.3\%  & 1.6\%  \vspace{2mm} \\
\multirow{2}{*}{\minerva pion tune} & Total       & 2.9\%  & 2.5\%  & 0.9\%   \\ %
                                    & Syst.       & 7.2\%  & 6.8\%  & 3.5\%  \vspace{2mm} \\
\multirow{2}{*}{Data-driven pion}   & Total       & 4.7\%  & 6.5\%  & 1.0\%   \\ %
                                    & Syst.       & 11.6\% & 17.9\% & 3.9\%  \vspace{2mm} \\
\multirow{2}{*}{Pion SI}            & Total       & 0.7\%  & 20.8 \% & 1.0\% \\
                                    & Syst.       & 1.9\%  & 57.8 \% & 4.6\% \\ 
\hline\hline
\end{tabular}
\caption{Biases on the main oscillation parameters for each simulated data set, calculated as the shift in the middle of the $1\sigma$ confidence interval relative to the overall uncertainty from systematic sources (``Syst.'') and the total (``Total'') to one decimal place.}
\label{tab:bias_table_fakedata}
\end{table}

In the previous T2K analyses~\cite{Abe:2021gky,T2K:2019bcf}, the simulated data study for the nucleon removal energy had a significant impact, especially on \dmsqtwothree, and an additional uncertainty was introduced. In this analysis, the updated nucleon removal energy uncertainty, described in \autoref{sec:interactionModel}, has caused it to no longer be the dominant source of systematic uncertainty.

\autoref{tab:dcp_intervals} shows the changes to the 90\% confidence interval for \deltacp for each of the simulated data studies. The non-CCQE and the data-driven pion momentum modification simulated data sets had the largest impact, shifting the 90\% CL by 0.09 and 0.07 respectively. The change to the 90\% confidence limits does not alter the conclusions on the exclusion of CP-conserving values presented in \autoref{sec:oa_results}.
\begin{table}[htbp]
\centering 
 \begin{tabular}{l|c c} 
 \hline
 \hline
\multirow{2}{*}{Simulated data set} & \multicolumn{2}{c}{Change to 90\% C.L. of \deltacp}   \\ 
                                    & -3.01 & -0.52 \\ 
\hline
CCQE z-exp high                 & 0.05 & 0.04 \\
CCQE removal energy             & 0.00 & 0.02 \\
\textbf{Non-CCQE}               & 0.06 & \textbf{0.09} \\
2p2h Martini                    & 0.04 & 0.04 \\
\minerva pion tune              & 0.05 & 0.04 \\
\textbf{Data-driven pion}       & \textbf{0.07} & 0.04 \\

Pion SI                         & 0.00 & 0.01 \\ 
\hline
\hline
 \end{tabular}
 \caption{Shifts of the 90\% confidence interval boundaries of \deltacp, in radians, as a result of the simulated data studies. The values in the top row correspond to the results of the data fit, assuming normal ordering. The values for each simulated data set are added to (subtracted from) the right (left) \deltacp interval edge from the data fit. Only the absolute size of the shift is taken into account. The simulated data sets with the largest impact are typed in bold.}
 \label{tab:dcp_intervals}
\end{table}

%% file: Conclusions.tex
The T2K collaboration has measured the three-flavour PMNS neutrino oscillation parameters \dmsqtwothree, \ssqthonethree, \ssqthtwothree, \deltacp, the Jarlskog invariant $J$, and the mass ordering, using the statistics at the FD equivalent to $3.6\times10^{21}$ POT. T2K continues to favour neutrino oscillations with near-maximal CP violation, in the upper octant of \ssqthtwothree, in the normal mass ordering, with a \ssqthonethree consistent with the measurements by reactor experiments. 

The analysis included $4.72\times10^{20}$ POT more neutrino data at the FD, and $5.73(4.48)\times10^{20}$ POT more (anti-)neutrino data at the ND. For the first time, a neutrino flux constraint using charged pion data from a T2K replica target at NA61/SHINE was used, which approximately halves the flux uncertainty before the ND analysis. An updated neutrino interaction model with a refined initial-state and removal-energy model with associated uncertainties was also employed, amongst others. High statistics ND data was used to constrain the neutrino flux and interaction model uncertainties at the FD, which also constrains the wrong-sign background of the neutrino beam with the magnetised ND. Biases from unmodelled systematic uncertainties were studied through simulated data studies, which acted to inflate the \dmsqtwothree and \deltacp confidence intervals. These results present the strongest constraints on several neutrino oscillation parameters, and are more consistent with the expected sensitivities to the oscillation parameters compared to T2K's previous analysis. 

The results are limited by statistics, and T2K will continue to take data as J-PARC upgrades~\cite{T2K:2019eao,Oyama:2020kev} the neutrino beam for the Hyper-Kamiokande experiment~\cite{Hyper-Kamiokande:2018ofw}. In preparation, the T2K beamline has recently undergone a long shutdown, and will be operating the magnetic horns at 320~kA current, with beam power in excess of 700~kW in the near future~\cite{T2K:2019eao,Oyama:2020kev}.
Upcoming analyses at the FD will expand selections to include multiple Cherenkov rings, increasing statistics by approximately 30\%. 
The FD has also begun collecting data with gadolinium doped in the ultra-pure water~\cite{Super-Kamiokande:2021the}, drastically increasing the efficiency in tagging interactions producing neutrons.
At the ND, selections are being developed to improve the understanding of nuclear effects in neutrino interactions, such as 2p2h, in-medium corrections, and the initial state, thus addressing the larger systematic uncertainties in this analysis. 
Furthermore, the ND280 upgrade~\cite{T2K:2019bbb,Attie:2021yeh,Blondel:2020hml} will be ready to take data in 2023, providing significantly improved reconstruction capabilities for low momentum protons and pions with full angular coverage, which will allow for detailed study of nuclear effects, in addition to measurements of neutron kinematics.
Moreover, the T2K collaboration is actively working with the NOvA and SK collaborations on combined neutrino oscillation analyses, taking advantage of synergies in experiment design to lift current degeneracies and increase statistics.

%% file: Acknowledgements.tex
\begin{acknowledgements}
We thank the J-PARC staff for superb accelerator performance. We thank the CERN NA61/SHINE Collaboration for providing valuable particle production data. We acknowledge the support of MEXT,   JSPS KAKENHI (JP16H06288, JP18K03682, JP18H03701, JP18H05537, JP19J01119, JP19J22440, JP19J22258, JP20H00162, JP20H00149, JP20J20304) and bilateral programs(JPJSBP120204806, JPJSBP120209601), Japan; NSERC, the NRC, and CFI, Canada; the CEA and CNRS/IN2P3, France; the DFG (RO 3625/2), Germany; the INFN, Italy; the Ministry of Education and Science(DIR/WK/2017/05) and the National Science Centre (UMO-2018/30/E/ST2/00441 ), Poland; the RSF19-12-00325, RSF22-12-00358 and the Ministry of Science and Higher Education (075-15-2020-778), Russia; MICINN (SEV-2016-0588, PID2019-107564GB-I00, PGC2018-099388-BI00, PID2020-114687GB-I00) Government of Andalucia (FQM160, SOMM17/6105/UGR) and the University of Tokyo ICRR's Inter-University Research Program FY2023 Ref. J1, and ERDF funds and CERCA program, Spain; the SNSF and SERI (200021\_185012, 200020\_188533, 20FL21\_186178I), Switzerland; the STFC and UKRI, UK; and the DOE, USA. We also thank CERN for the UA1/NOMAD magnet, DESY for the HERA-B magnet mover system, the BC DRI Group, Prairie DRI Group, ACENET, SciNet, and CalculQuebec consortia in the Digital Research Alliance of Canada, GridPP and the Emerald High Performance Computing facility in the United Kingdom, and the CNRS/IN2P3 Computing Center in France. In addition, the participation of individual researchers and institutions has been further supported by funds from the ERC (FP7), “la Caixa” Foundation (ID 100010434, fellowship code LCF/BQ/IN17/11620050), the European Union’s Horizon 2020 Research and Innovation Programme under the Marie Sklodowska-Curie grant agreement numbers 713673 and 754496, and H2020 grant numbers RISE-GA822070-JENNIFER2 2020 and RISE-GA872549-SK2HK; the JSPS, Japan; the Royal Society, UK; French ANR grant number ANR-19-CE31-0001; the SNF Eccellenza grant number PCEFP2\_203261; and the DOE Early Career programme, USA. For the purposes of open access, the authors have applied a Creative Commons Attribution licence to any Author Accepted Manuscript version arising.
\end{acknowledgements}

%% file: Appendix.tex
\appendix
\section{Data release}
\label{sec:data_release}
\input{Appendices/DataRelease.tex}

\section{Detailed simulated data studies}
\label{app:appendix_fakedata}
\input{Appendices/FakeDataDetails.tex}

\section{Two-dimensional Jarlskog credible regions}
\label{app:appendix_2DJarlskog}
\input{Appendices/Jarlskog2D.tex}

\section{Additional critical values for \ssqthtwothree and \deltacp}
\input{Appendices/CriticalValues.tex}

\section{Systematic uncertainty parameters from the analysis of ND data}
\label{app:appendix_NDFitValues}
\input{Appendices/NDTables.tex}

%% file: Appendices/DataRelease.tex
A digital data release has been prepared for this publication in Ref.~\cite{t2k_collaboration_2023_7741399}, containing all the results from the Bayesian and frequentist analyses presented in \autoref{sec:oa_results}. They are presented in \texttt{ROOT} format, containing \texttt{TGraph} and \texttt{TH1D} objects. Detailed confidence and credible intervals are provided for the parameters of interest, with and without the constraint on \ssqthonethree from reactor experiments applied, in both neutrino mass orderings. A \texttt{README} is provided outlining the details and conventions of the data release. The neutrino flux release is separately provided in Ref.~\cite{megan_friend_2021_5734307}.

%% file: Appendices/FakeDataDetails.tex
This section expands on the summary of the simulated data studies provided in \autoref{sec:fakeData}, providing details of two different studies and their effects. 
All studies are done with parameters listed in \autoref{tab:asimov_pars}, where one set is close to the T2K best-fit point, and the other has $\deltacp=0$ and non-maximal $\ssqthtwothree=0.45$ with the parameter set A being presented in this paper.

First, the non-CCQE study is described in more detail as it produced the largest bias on oscillation parameters. Then the \minerva single-pion suppression study is detailed---having a small impact at T2K statistics, but will likely have larger impact in higher statistics experiments. The section finishes with a summary of the impact of all the simulated data studies.

\begin{table}
\centering
\begin{tabular}{l|c c}
\hline \hline
\multirow{2}{*}{Parameter}              & \multicolumn{2}{c}{Parameter set} \\ 
                                        & A         & B \\
     \hline
     \deltacp                           & -1.601    & 0.0 \\
     \ssqthtwothree                     & 0.528     & 0.45 \\
     \ssqthonethree                     & \multicolumn{2}{c}{0.0218} \\
     \ssqthonetwo                       & \multicolumn{2}{c}{0.307} \\
     $\dmsqtwothree/10^{-3} (\text{eV}^2)$  & \multicolumn{2}{c}{2.509} \\
     $\dmsqonetwo/10^{-5} (\text{eV}^2)$    & \multicolumn{2}{c}{7.530} \\
\hline \hline
\end{tabular}
\caption{The oscillation parameters used in the expected sensitivity fits, which are used to evaluate the relative size of the biases for the simulated data studies.}
\label{tab:asimov_pars}
\end{table}

\subsection{Non-CCQE simulated data set}
\label{subsec:nonqe_fds_results}
\begin{figure}[htbp]
    \centering
    \includegraphics[width=0.95\columnwidth]{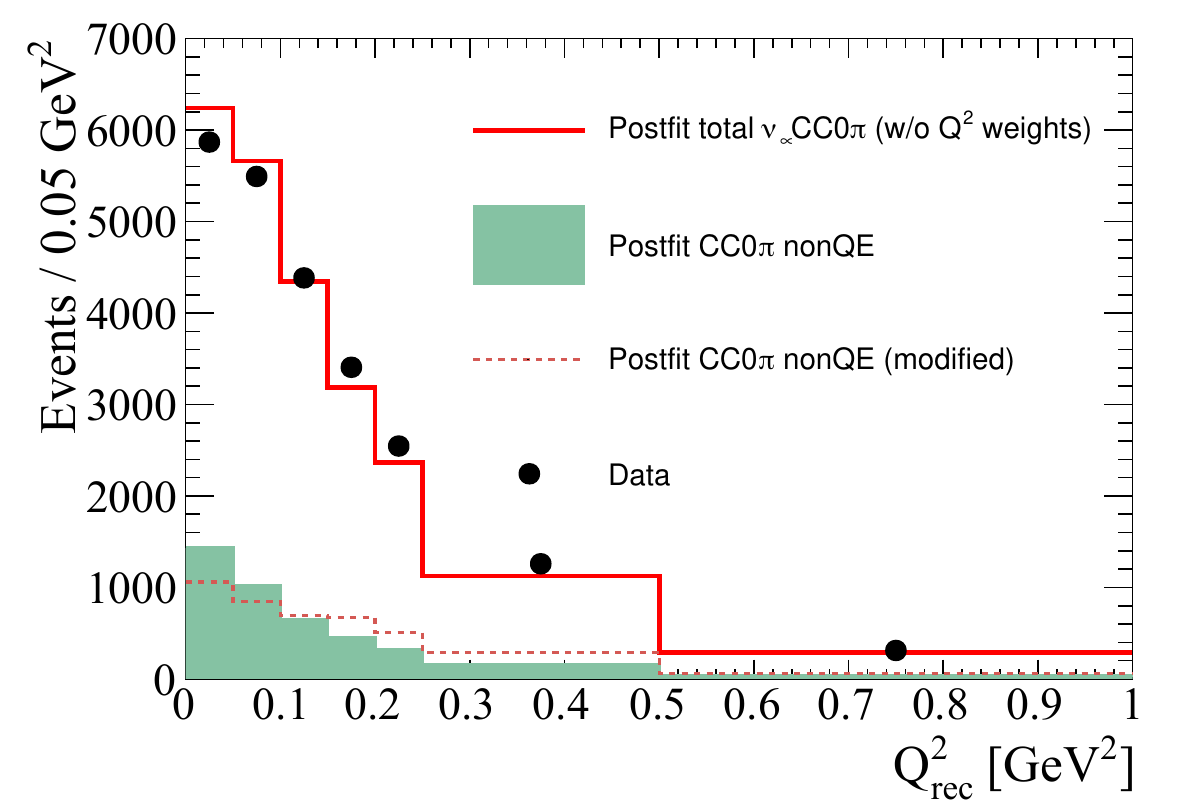}
    \caption{The ND prediction after the fit to data for the FGD1 \numu \CCzeropi sample is shown with the data, where the non-CCQE contribution is shown in shaded green. The dotted red line shows the non-CCQE contribution needed for the overall prediction to match the data. The weight is the ratio of the dotted red line to the shaded green.}
  \label{fig:nonqe_weights}
\end{figure}

As described in \autoref{sec:interactionModel}, the $Q^2$ normalisation parameters included in this analysis are introduced to provide \emph{ad hoc} freedom in the NEUT CCQE model, based on known discrepancies with data~\cite{Abe:2020jbf,Abe:2020uub, Ruterbories:2018gub,Rodrigues:2015hik}. The robustness of this parametrisation is tested and its impact on oscillation parameter measurements by assuming that the underestimation of data is entirely due to non-CCQE interactions. The simulated data set to replicate this effect is devised by
\begin{itemize}
    \item The prediction after the ND fit to data is weighted to restore the CCQE $Q^2$ parameters to their nominal values (i.e. unity).
    \item The reconstructed $Q^2_{rec}$ distributions are built for data and the simulation from the previous step using \autoref{eq:q2rec}, with bins matching the ranges of the $Q^2$ parameters. Since the study concerns \CCzeropi interactions, this is only done for the \CCzeropi samples.
    \item The simulated non-CCQE events with true \CCzeropi topology are varied until the prediction matches the data .
    \item The change in the non-CCQE true \CCzeropi events is then applied as a set of weights to the nominal prediction as a function of true $Q^{2}_{rec}$, resulting in the simulated data set.
\end{itemize}
This process is applied separately to \numu and \numub samples to extract different weights for neutrino and anti-neutrino events, where the effect was smaller for the \numub samples. The weight extraction process and the resulting weights for neutrino events are shown in \autoref{fig:nonqe_weights}.

\begin{figure}[htbp]
    \centering
    \begin{subfigure}[h]{0.95\columnwidth}
    \includegraphics[width=\textwidth,trim=0mm 5mm 0mm 10mm, clip]{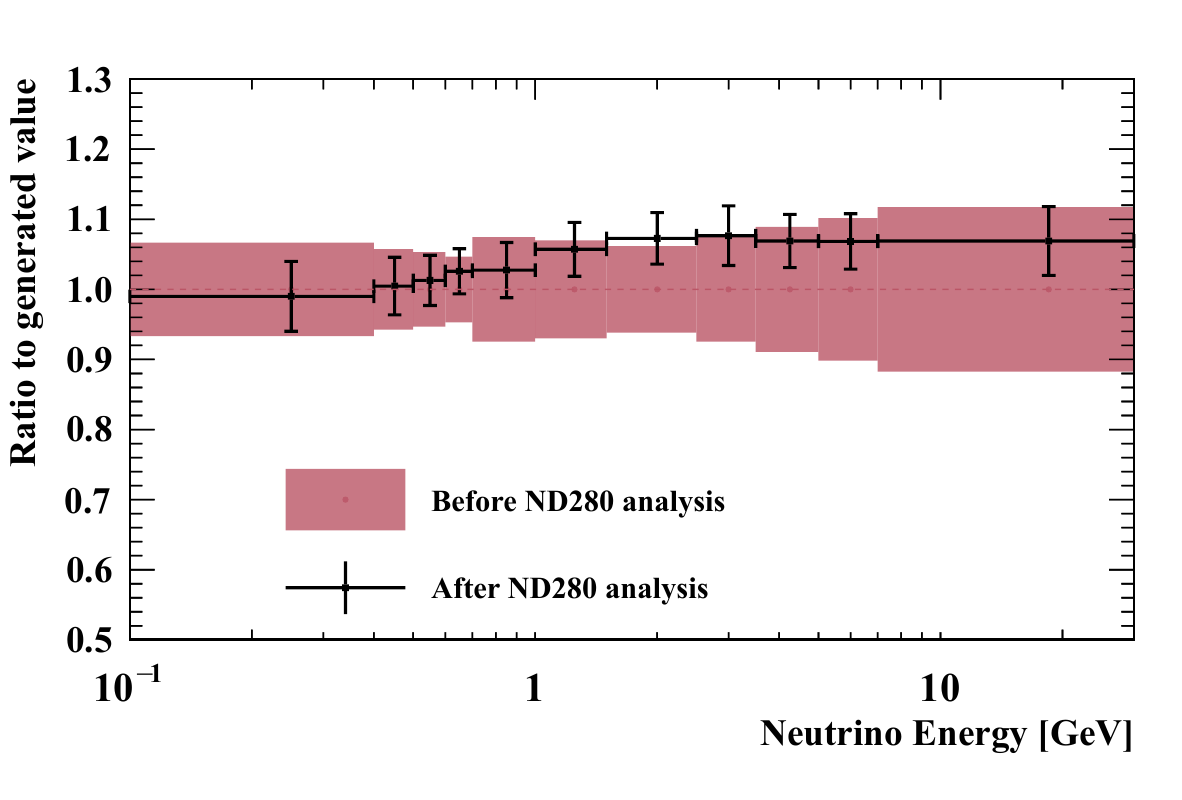}
    \caption{FD \fhcalt \numu flux parameters}
    \end{subfigure}
    \begin{subfigure}[h]{0.95\columnwidth}
    \includegraphics[width=\textwidth,trim=0mm 0mm 0mm 10mm, clip]{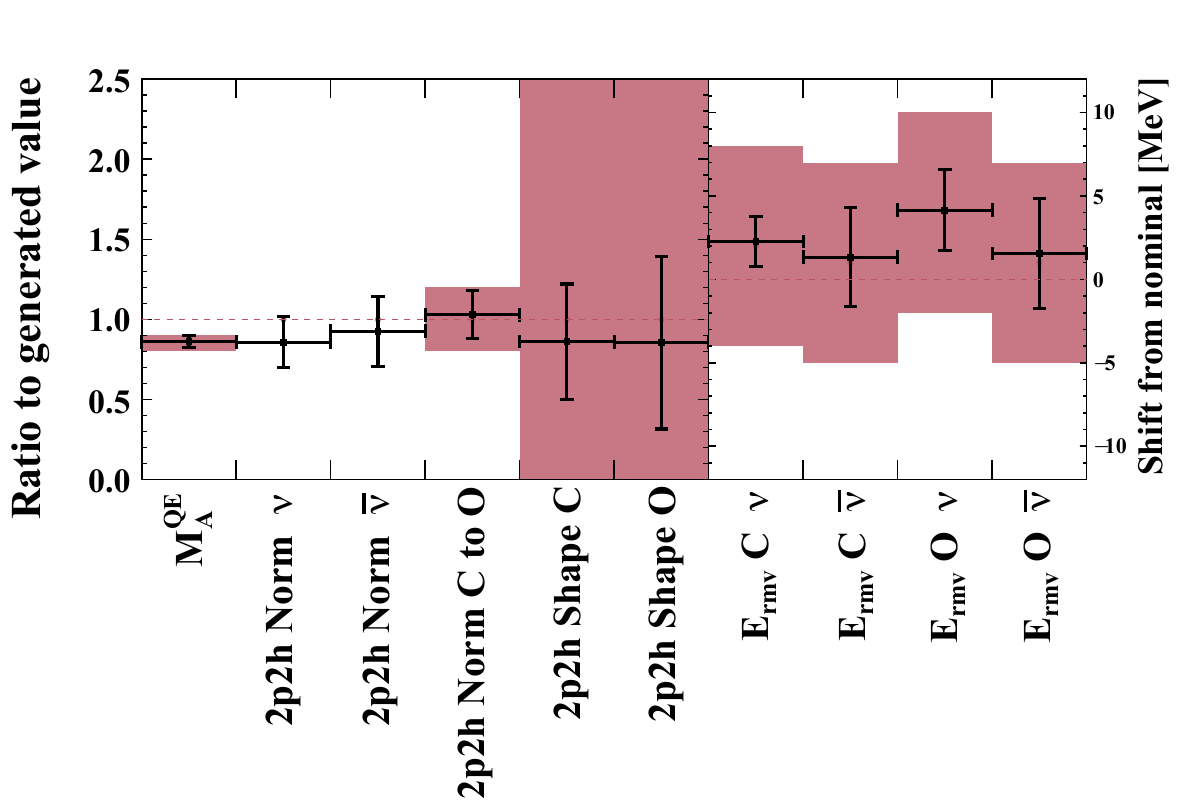}
    \caption{\CCzeropi cross-section parameters}
    \end{subfigure}
    \begin{subfigure}[h]{0.95\columnwidth}
    \includegraphics[width=\textwidth,trim=0mm 5mm 0mm 10mm, clip]{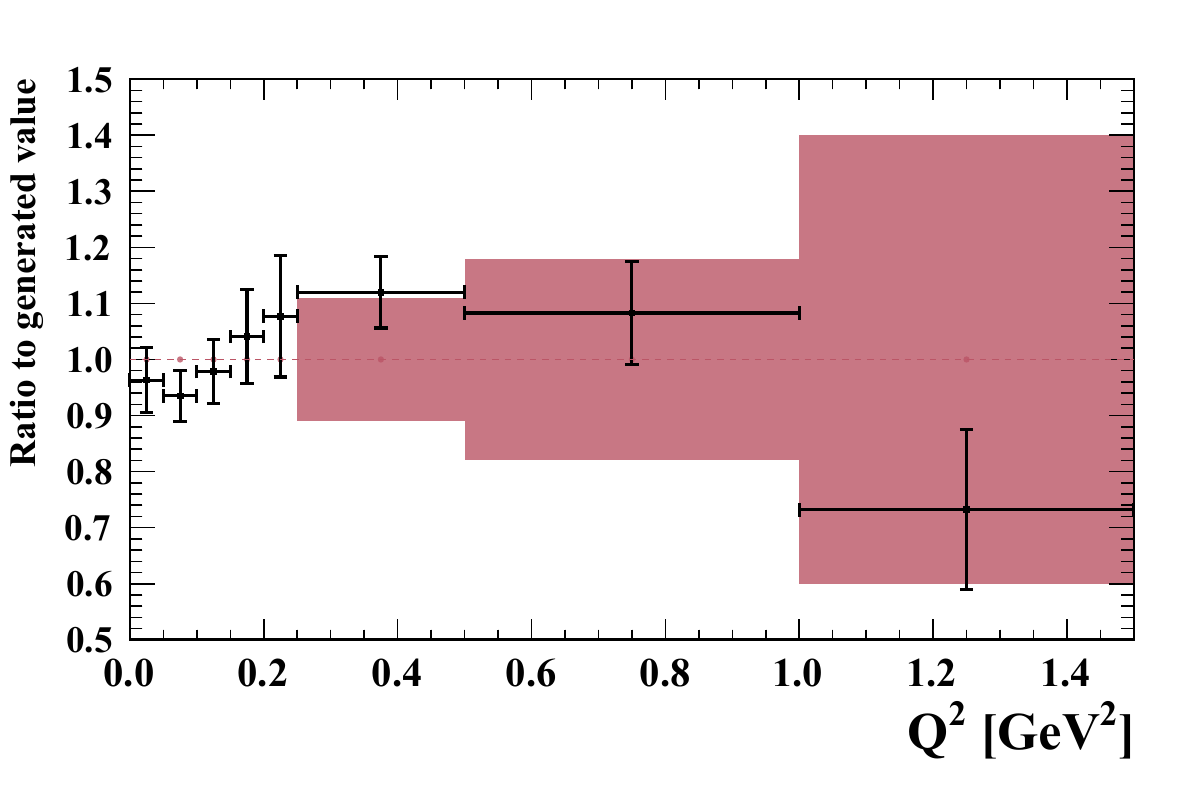}
    \caption{$Q^2$-dependent CCQE parameters}
    \end{subfigure}
    \caption{
    Constraints on the \fhcalt \numu flux (top) and \CCzeropi cross-section parameters (middle, bottom) from the fit to the non-CCQE simulated data set (black points, black lines), overlaid on the input uncertainty (red band). The parameters are presented as a ratio to the generated value in NEUT, except for the removal energy parameters which show the shifts in units of MeV. 
    }
    \label{fig:nonQE_ndfit}
\end{figure}

The resulting distribution was used as the simulated data in the ND fit. As shown in \autoref{fig:nonQE_ndfit}, the neutrino flux, 2p2h $\nu$ normalisation, and the $Q^2$ parameters shift to accommodate the model variation, notably present in the higher $E_\nu$ region. This is largely expected, since non-CCQE events are often from higher energy neutrinos due to the interaction cross section increasing with $E_\nu$, whereas CCQE plateaus. 

\begin{figure}[htbp]
    \centering
    \includegraphics[width=0.95\columnwidth,trim=0mm 0mm 20mm 5mm, clip]{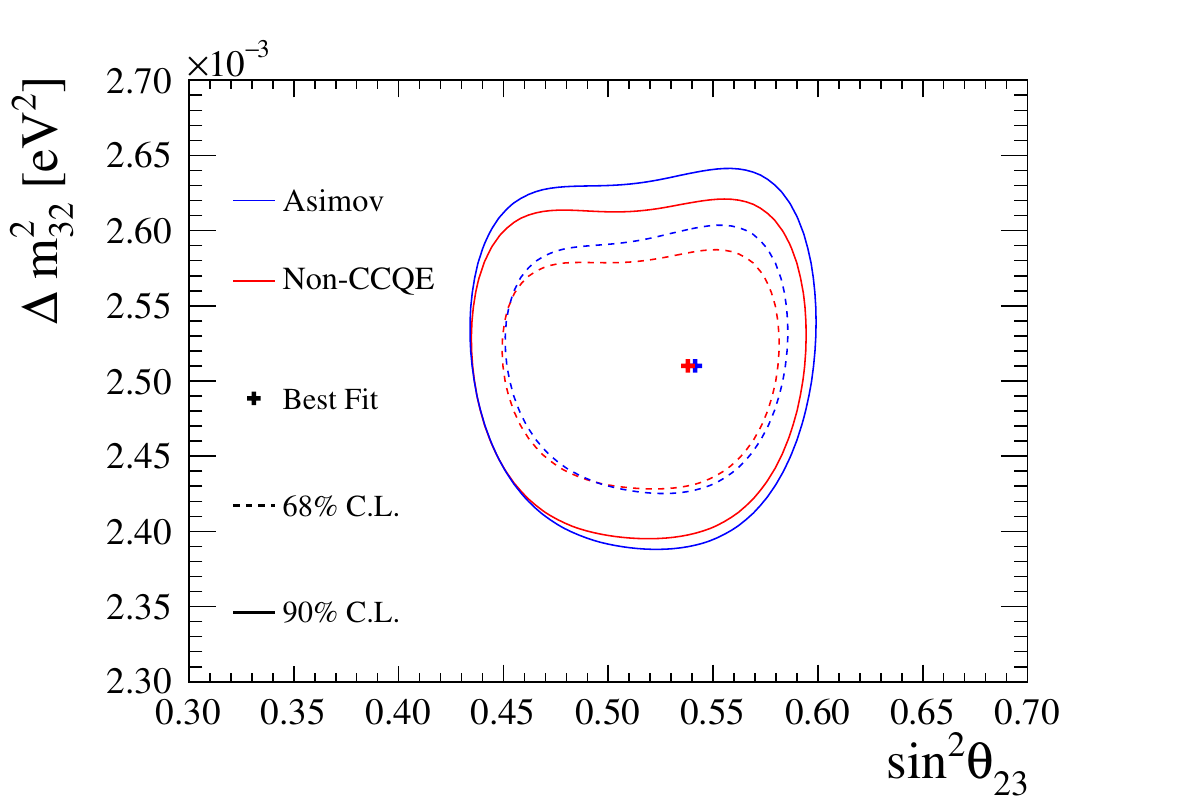}
    \includegraphics[width=0.95\columnwidth,trim=0mm 0mm 20mm 5mm, clip]{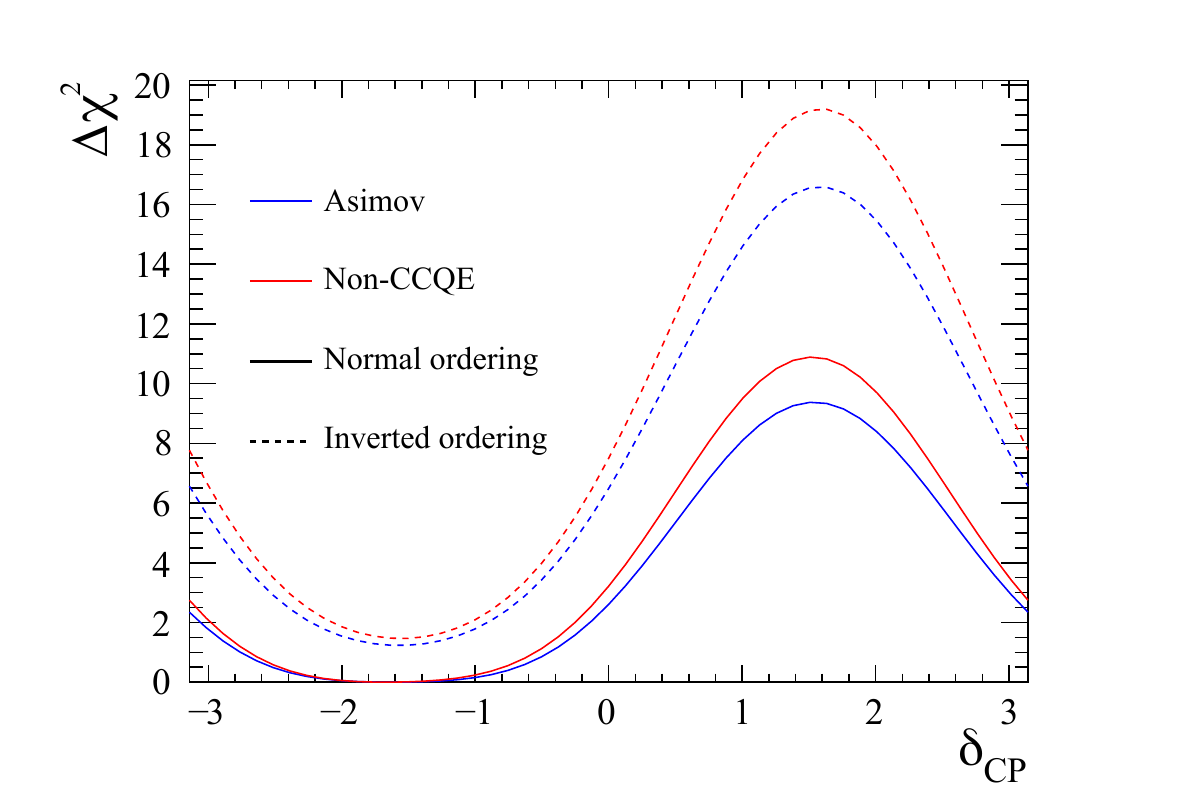}
    \caption{Comparison of confidence regions in $\ssqthtwothree-\dmsqtwothree$ (top) for the expected sensitivity fit (blue) and the non-CCQE simulated data study (red), and the $\Delta \chi^2$ as a function of \deltacp (bottom).}
    \label{fig:nonqe_ptheta_results}
\end{figure}

The same set of weights were used to create a corresponding simulated data set for the FD samples. The ND constraints from this simulated data study were propagated to the FD simulation, and used to extract oscillation parameters. \autoref{fig:nonqe_ptheta_results} shows the oscillation parameter contours for an expected sensitivity fit and the ``non-CCQE'' simulated data set. Quantitatively, the simulated data set changes the interval width of 32.7\% on \dmsqtwothree and 21.3\% on \ssqthtwothree with respect to the size of the systematic uncertainty, $\sigma_{syst}$. This is equivalent to an uncertainty inflation of $0.69\times10^{-5}~\text{eV}^2$. This simulated data set has an increased sensitivity to \deltacp, and no significant bias in the best-fit value is observed.

This study attempts to address the differences between neutrinos and anti-neutrinos by applying different weights for the two. Another study may also take into account the differences between interactions on carbon and oxygen targets, which some models predict should have different behaviour at low $Q^2$. However, the ability for the ND analysis to distinguish behaviour at low $Q^2$ on carbon and oxygen is limited by statistics, and such a detailed study is currently not warranted.

\subsection{\minerva single-pion suppression}
\label{sec:minerva_fds}
Studies of \minerva's neutrino-induced single-pion production (SPP) data found that the GENIE event generator over-predicted the cross section~\cite{Stowell:2019zsh}. The effect was particularly pronounced at low 4-momentum transfer, where a suppression of the cross section by $\sim60\%$ was needed to match the data. The data-driven tune from the paper was applied to T2K SPP events to gauge the impact of this low-$Q^2$ modification on the oscillation analysis. 
Importantly, the SPP model used in the GENIE event generator by \minerva differs from that in NEUT for this analysis, especially in the low-$Q^2$ region, and T2K SPP data~\cite{T2K:2019yqu} sees little need for a suppression with recent NEUT versions~\cite{Avanzini:2021qlx}. 
Furthermore, \minerva sees the largest suppression in CC$1\pi^0$ selections, which are barely selected in T2K's FD. The CC$1\pi^+$ selection, which has a larger overlap with selections in T2K's FD, has a significantly weaker suppression in the publication. The low-$Q^2$ suppression should therefore be considered as a conservative variation, designed to build an extreme simulated data study to analyse ``worst-case scenario'' mismodelling. 

\begin{figure}[htbp]
    \centering
    \includegraphics[width=0.95\columnwidth]{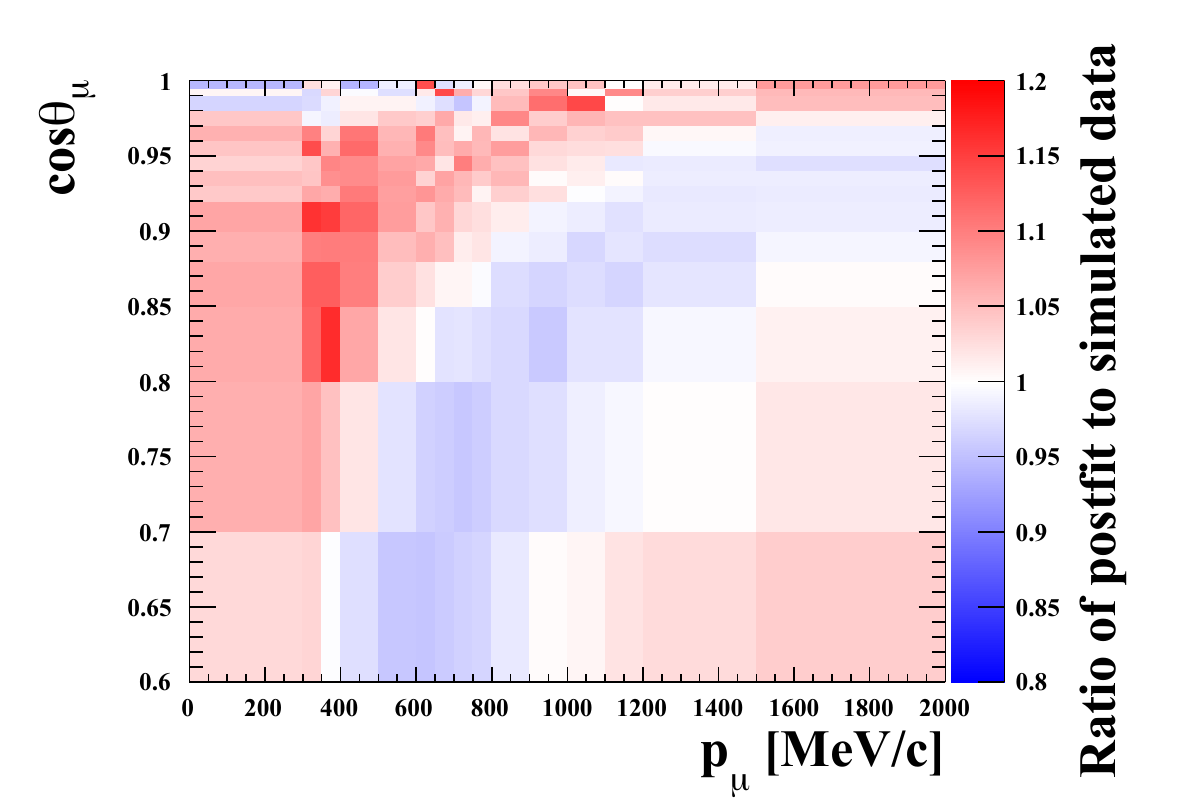}
    \caption{Ratio between the prediction after the fit to the simulated data from the \minerva pion suppression to the simulated data itself, for the ND FGD1 \CConepi sample.}
    \label{fig:nd_postfit_ratio_minerva}
\end{figure}

After the fit to the simulated data, the majority of affected parameters are related to SPP, with little change to the neutrino flux parameters and the \CCzeropi interaction parameters. This is because they are largely constrained by the dominant \CCzeropi selections at the ND, which are barely affected by the simulated data. $C_5^A$ was pulled down, which acts to decrease the overall SPP cross section, and $M_A^{RES}$ is pulled up, which increases the high $Q^2$ SPP cross section. Since CC coherent events primarily occupy the low-$Q^2$ region, they too are suppressed by the CC coherent normalisation parameter, which is pulled lower. The 2p2h normalisation and highest $Q^2$ CCQE normalisations are also pulled low, as such events occupy a similar space in $\pmu-\cosmu$ to SPP events in the \CCzeropi selection. The ratio of the prediction versus the simulated data at the ND is shown in \autoref{fig:nd_postfit_ratio_minerva}, where the \CConepi sample is over-predicted by 20\% in certain $\pmu-\cosmu$ regions. Hence, the current uncertainty model at T2K can not account for a sizeable low-$Q^2$ suppression of SPP events. Importantly, this has a low impact on the dominant ND \CCzeropi selections, since SPP is a sub-leading contributor.
 
\begin{figure}[htbp]
    \centering
    \includegraphics[width=0.95\columnwidth,trim=10mm 0mm 20mm 10mm, clip]{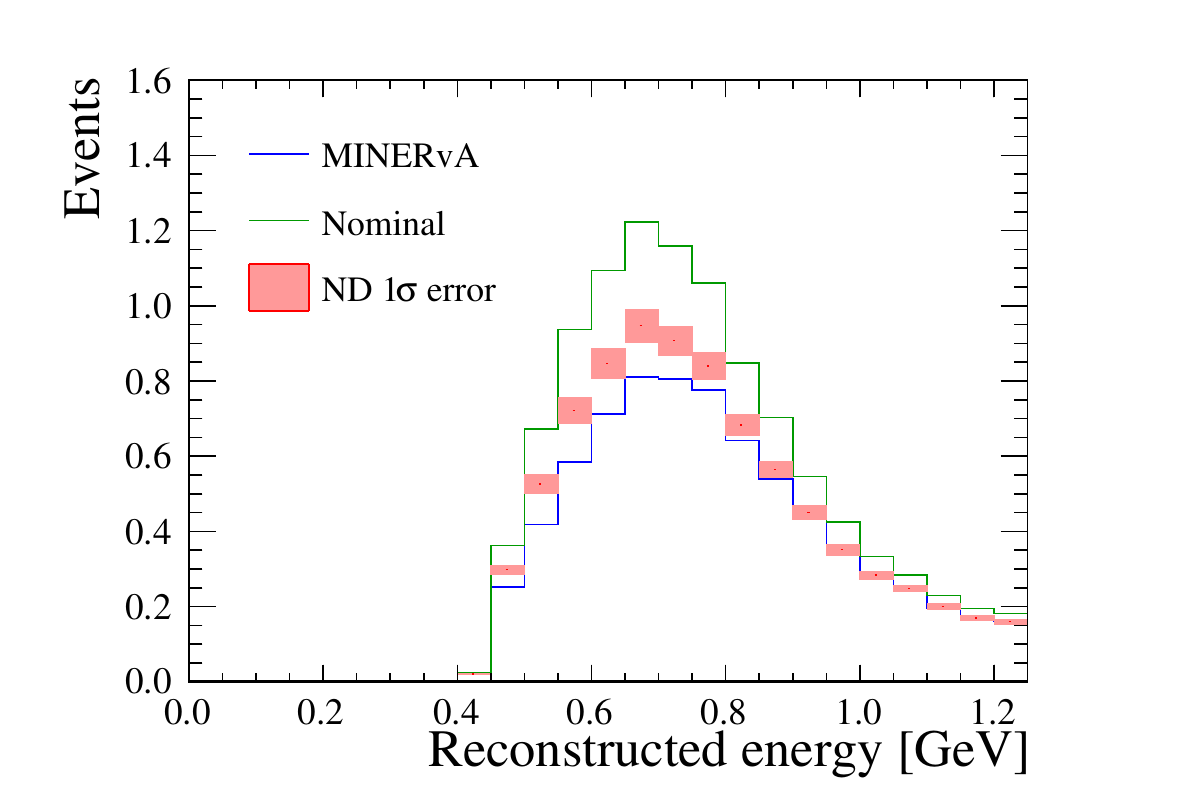}
    \caption{The predictions for the \rede selection at the FD using the unchanged model (green) and including the \minerva pion-suppression (blue). The red bands show the propagated ND prediction from the fit to the simulated data set, with the 1$\sigma$ uncertainty bands.}
    \label{fig:minerva_fds_near-vs-far}
\end{figure}

The prediction for the SPP-dominated \rede selection is shown in \autoref{fig:minerva_fds_near-vs-far} and echoes the ND prediction; the prediction from fitting the ND model to the simulated data can not replicate the alternative model. However, the impact on the oscillation parameters is relatively small as the \rede sample has low statistics and occupies higher $E_\nu$ compared to the dominant \fhcalt \re selection, which provides most of the \nue appearance at T2K. This issue may present significant biases for experiments with higher statistics of SPP, such as \nova, DUNE and Hyper-Kamiokande. It highlights the need for high-quality single-pion neutrino data, and the development of sufficiently robust and flexible models to describe them.

\subsection{Summary of biases}
As detailed in \autoref{sec:fakeData}, a selection of the largest contributors to the bias were used to inflate the confidence intervals for the two oscillation parameters, $\Delta m^2$ and \deltacp. The down-selection was used to avoid double-counting similar physics effects---for instance the CCQE axial form factor described by either the 3-component or z-expansion models---by including the simulated data set with the largest bias. This section gives the details of the biases for all the simulated data sets on the oscillation parameters.

\autoref{tab:bias_table_fakedata_full} presents the full bias table for the three oscillation parameters \ssqthtwothree, \dmsqtwothree, and \deltacp. As with previous T2K analyses, \deltacp is significantly less sensitive to systematic uncertainties, owing to the relative dominance of statistical uncertainty in the single ring electron-like samples at the FD. The shifts in the 90\% confidence interval of \deltacp for each simulated data study is shown in \autoref{tab:dcp_intervals_full}, where the shifts are dominated by the three largest simulated data studies: the pion SI, the non-CCQE, and data-driven pion tune.

\begin{table}[htbp]
\centering
\begin{tabular}{ll ccc}
\hline \hline
Simulated data set                  & Relative to & \ssqthtwothree & \dmsqtwothree & \deltacp \\
\hline
\multirow{2}{*}{CCQE 3-comp nom.}       & Total       & 1.0\%  & 0.4\%  & 0.8\%   \\
                                    & Syst.       & 2.5\%  & 1.1\%  & 3.1\%  \vspace{2mm} \\
\multirow{2}{*}{CCQE 3-comp high}       & Total       & 1.3\%  & 0.7\%  & 0.3\%   \\
                                    & Syst.       & 3.2\%  & 1.8\%  & 1.1\%   \vspace{2mm}\\ 
\multirow{2}{*}{CCQE 3-comp low}      & Total       & 0.7\%  & 0.2\%  & 0.2\%   \\
                                    & Syst.       & 1.7\%  & 0.6\%  & 0.8\%   \vspace{2mm}\\ 
\multirow{2}{*}{CCQE z-exp nom.}        & Total       & 2.5\%  & 0.2\%  & 0.6\%   \\
                                    & Syst.       & 6.1\%  & 0.6\%  & 2.2\%   \vspace{2mm}\\
\multirow{2}{*}{CCQE z-exp high}       & Total       & 0.3\%  & 2.1\%  & 0.4\%   \\
                                    & Syst.       & 0.7\%  & 5.7\%  & 1.7\%   \vspace{2mm}\\
\multirow{2}{*}{CCQE z-exp low}       & Total       & 3.1\%  & 0.2\%  & 0.1\%   \\
                                    & Syst.       & 7.5\%  & 0.6\%  & 0.6\%   \vspace{2mm}\\
\multirow{2}{*}{CCQE removal energy}     & Total       & 0.0\%  & 4.8\%  & 1.3\%   \\
                                    & Syst.       & 0.0\%  & 13.4\% & 5.2\%   \vspace{2mm}\\
\multirow{2}{*}{Non-CCQE}           & Total       & 8.7\%  & 11.8\%    & 1.7\%   \\
                                    & Syst.       & 21.3\% & 32.7\%   & 6.9\%   \vspace{2mm}\\
\multirow{2}{*}{2p2h Martini}       & Total       & 0.7\%  & 2.7\%  & 0.4\%   \\
                                    & Syst.       & 1.6\%  & 7.3\%  & 1.6\%   \vspace{2mm}\\
\multirow{2}{*}{\minerva pion tune} & Total       & 2.9\%  & 2.5\%  & 1.0\%   \\
                                    & Syst.       & 7.2\%  & 6.8\%  & 3.5\%   \vspace{2mm}\\
\multirow{2}{*}{Data-driven pion}   & Total       & 4.7\%  & 6.5\%  & 1.0\%   \\
                                    & Syst.       & 11.6\% & 17.9\% & 3.9\%   \vspace{2mm}\\
\multirow{2}{*}{Pion SI}            & Total       & 0.7\%  & 20.8\% & 1.0\% \\
                                    & Syst.       & 1.9\%  & 57.8\% & 4.6\% \\ 
\hline\hline
\end{tabular}
\caption{Biases on the main oscillation parameters for each simulated data set, calculated as the shift in the middle of the $1\sigma$ confidence interval relative to the overall uncertainty from systematic sources (``Syst.'') and the total (``Total'') to one decimal place.}
\label{tab:bias_table_fakedata_full}
\end{table}

\begin{table}[htbp]
\centering 
 \begin{tabular}{l|c c} 
 \hline
 \hline
\multirow{2}{*}{Simulated data set} & \multicolumn{2}{c}{Change to 90\% C.L. of \deltacp}   \\ 
                                    & -3.01 & -0.52 \\ 
\hline
CCQE 3-comp nom.                & 0.04 & 0.02 \\
CCQE 3-comp high                & 0.05 & 0.03  \\
CCQE 3-comp low                 & 0.04 & 0.03 \\
CCQE z-exp nom.                 & 0.01 & 0.01 \\ 
CCQE z-exp high                 & 0.05 & 0.04 \\
CCQE z-exp low                  & 0.00 & 0.00 \\
CCQE removal energy             & 0.00 & 0.02 \\
\textbf{Non-CCQE}   & 0.06 & \textbf{0.09} \\
2p2h Martini                    & 0.04 & 0.04 \\ 
\minerva pion tune              & 0.05 & 0.04 \\
\textbf{Data-driven pion} & \textbf{0.07} & 0.04 \\

Pion SI                         & 0.00 & 0.01 \\ 
\hline
\hline
 \end{tabular}
 \caption{Shifts of the 90\% confidence interval boundaries of \deltacp, in radians, as a result of the simulated data studies. The values in the top row correspond to the results of the data fit, assuming normal ordering. The values for each simulated data set are added to (subtracted from) the right (left) \deltacp interval edge from the data fit. Only the absolute size of the shift is taken into account. The simulated data sets with the largest impact are typed in bold.}
 \label{tab:dcp_intervals_full}
\end{table}

%% file: Appendices/Jarlskog2D.tex
\begin{figure}[htbp]
\centering
\includegraphics[width=0.95\columnwidth,trim=0mm 0mm 0mm 12mm, clip]{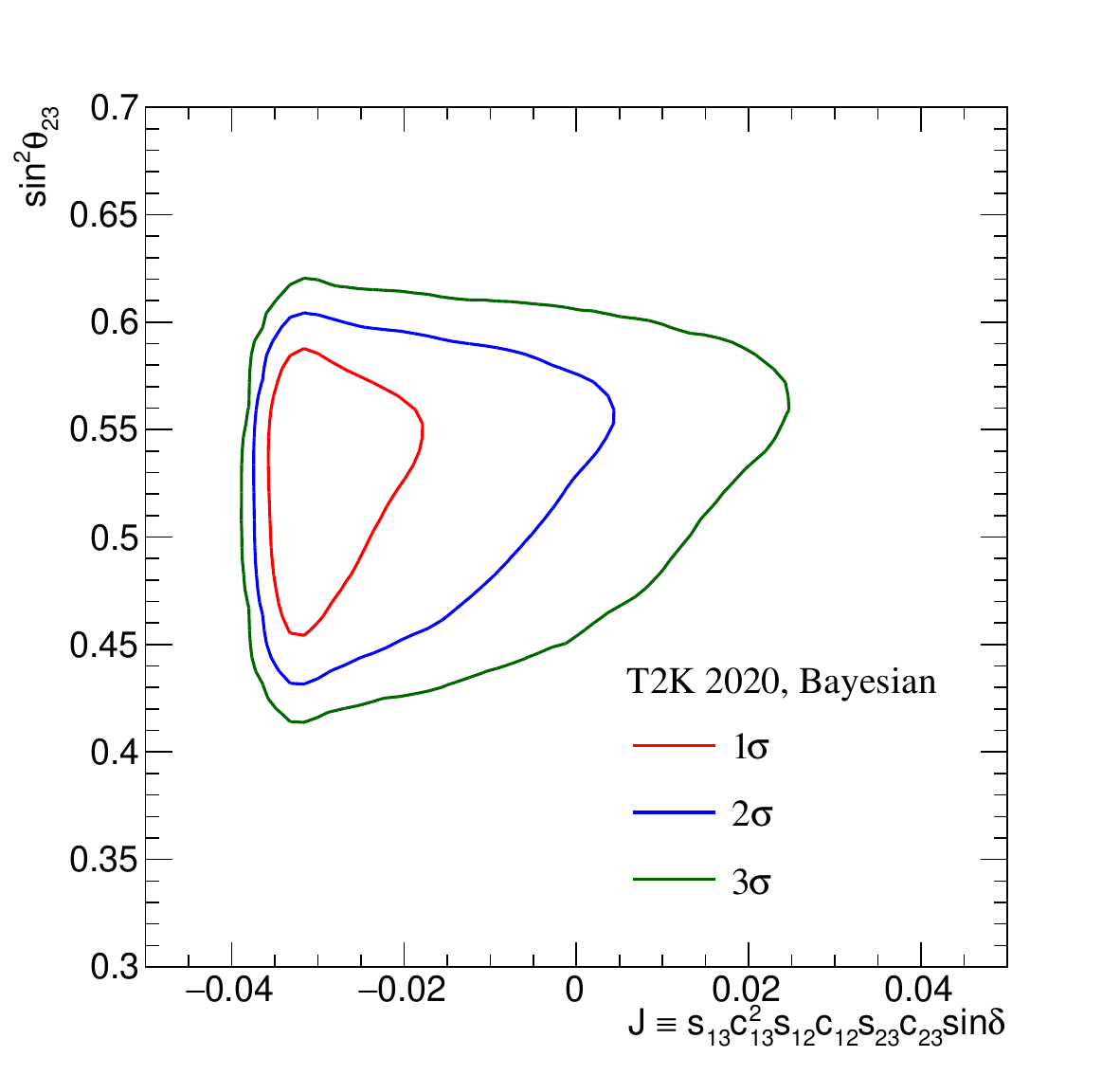}
\caption{1$\sigma$, 2$\sigma$, and 3$\sigma$ credible regions in $J-\ssqthtwothree$ space extracted from the Bayesian analysis discussed in \autoref{sec:oa:bayesian}.}
\label{fig:oa:mach3:th23jarl}
\end{figure}
\begin{figure}[htbp]
\centering
\includegraphics[width=0.95\columnwidth,trim=0mm 0mm 0mm 12mm, clip]{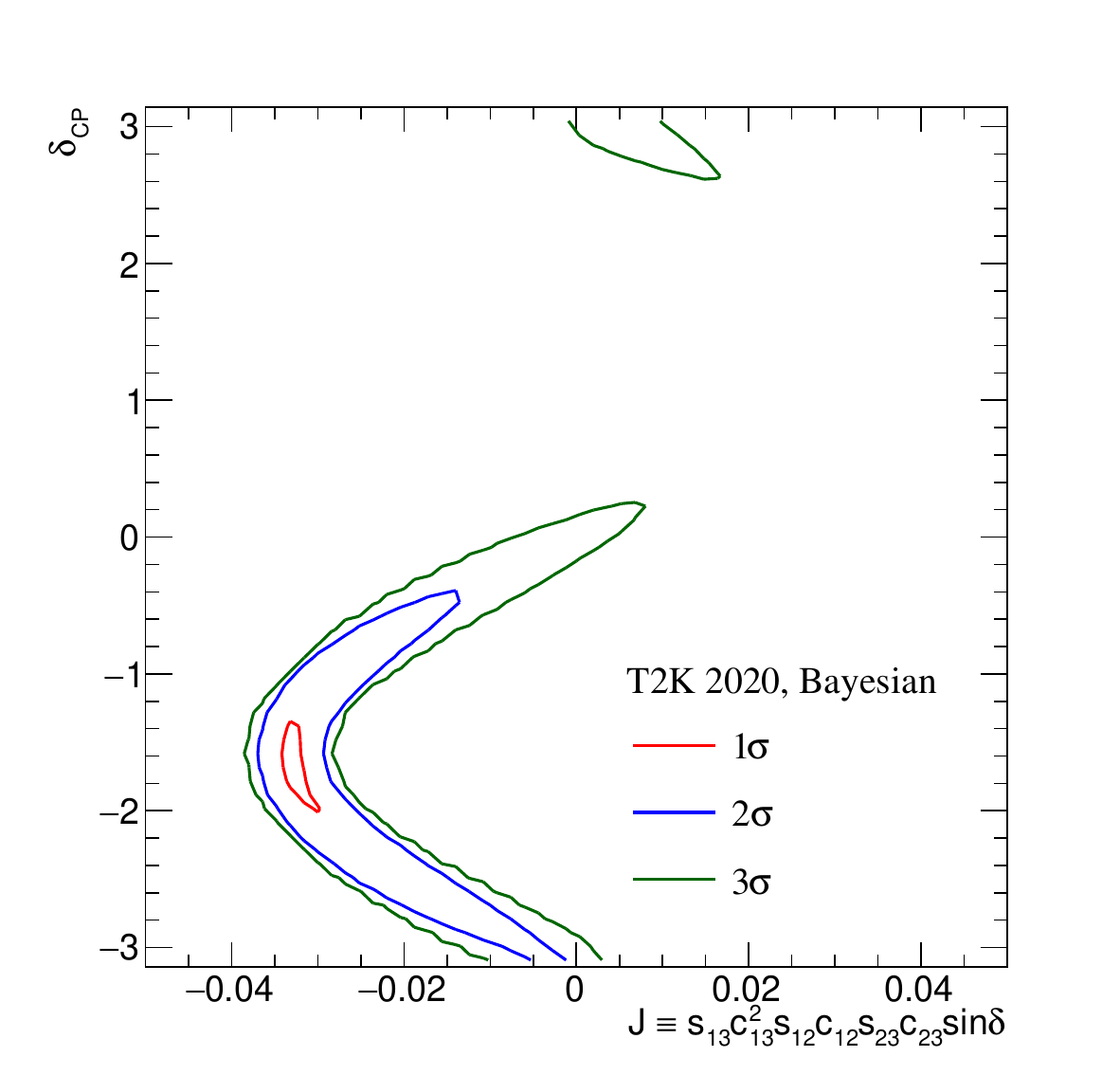}
\caption{1$\sigma$, 2$\sigma$, and 3$\sigma$ credible regions in $J-\deltacp$ space extracted from the Bayesian analysis discussed in \autoref{sec:oa:bayesian}.}
\label{fig:oa:mach3:dcpjarl}
\end{figure}

The measurements of \deltacp and \ssqthtwothree are the main constraints on the Jarlskog invariant from T2K's analysis.
Both of these parameter measurements have the effect of preferring the more extreme values of the Jarlskog invariant that are otherwise allowed, as shown earlier in \autoref{fig:oa:mach3:jarl1D_incremental}.
T2K's preference for values of $\theta_{23}\sim45\degree$ maximises the factor of $\sin{\theta_{23}}\cos{\theta_{23}}$. Its best-fit value being slightly above that mark allows for smaller values of $J$ as illustrated in \autoref{fig:oa:mach3:th23jarl}.
Similarly, T2K's preference for values of $\deltacp\approx-90\degree$ maximises the \sindcp factor and selects negative values of $J$.
\autoref{fig:oa:mach3:dcpjarl} illustrates the influence of the \deltacp posterior with values of \deltacp further away from maximal CP-violation mapping to smaller preferred magnitudes of $J$.

%% file: Appendices/CriticalValues.tex
\begin{figure*}[htbp]
\centering
\begin{subfigure}[h]{0.95\columnwidth}
\includegraphics[width=\textwidth]{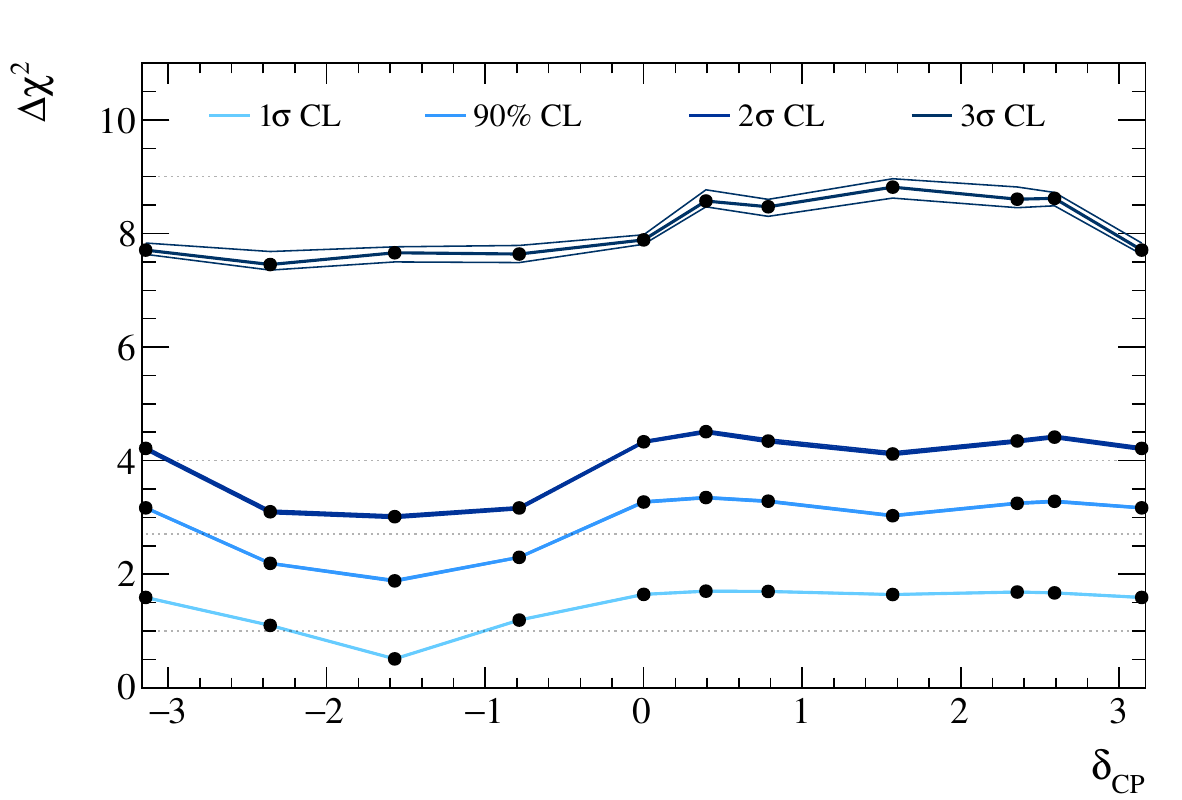}
\caption{Normal ordering, \deltacp.}
\end{subfigure}
\begin{subfigure}[h]{0.95\columnwidth}
\includegraphics[width=\textwidth]{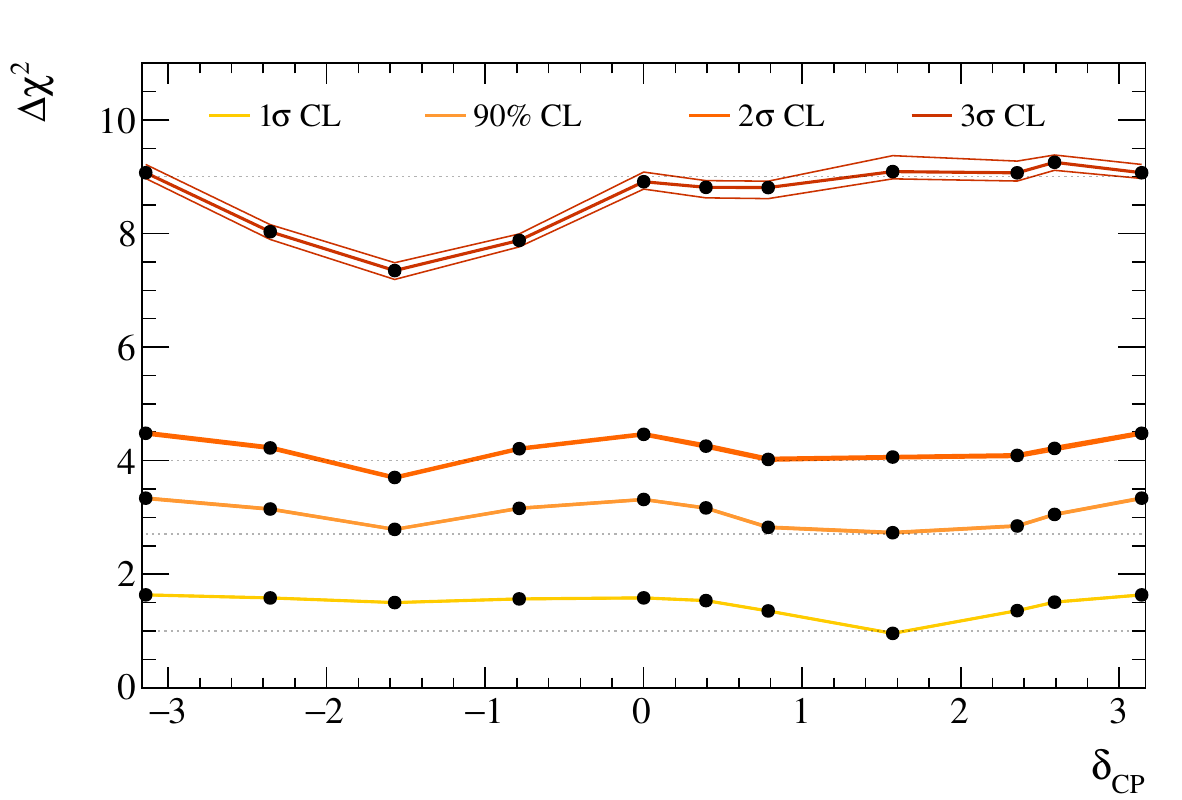}
\caption{Inverted ordering, \deltacp.}
\end{subfigure}
\begin{subfigure}[h]{0.95\columnwidth}
\includegraphics[width=\textwidth]{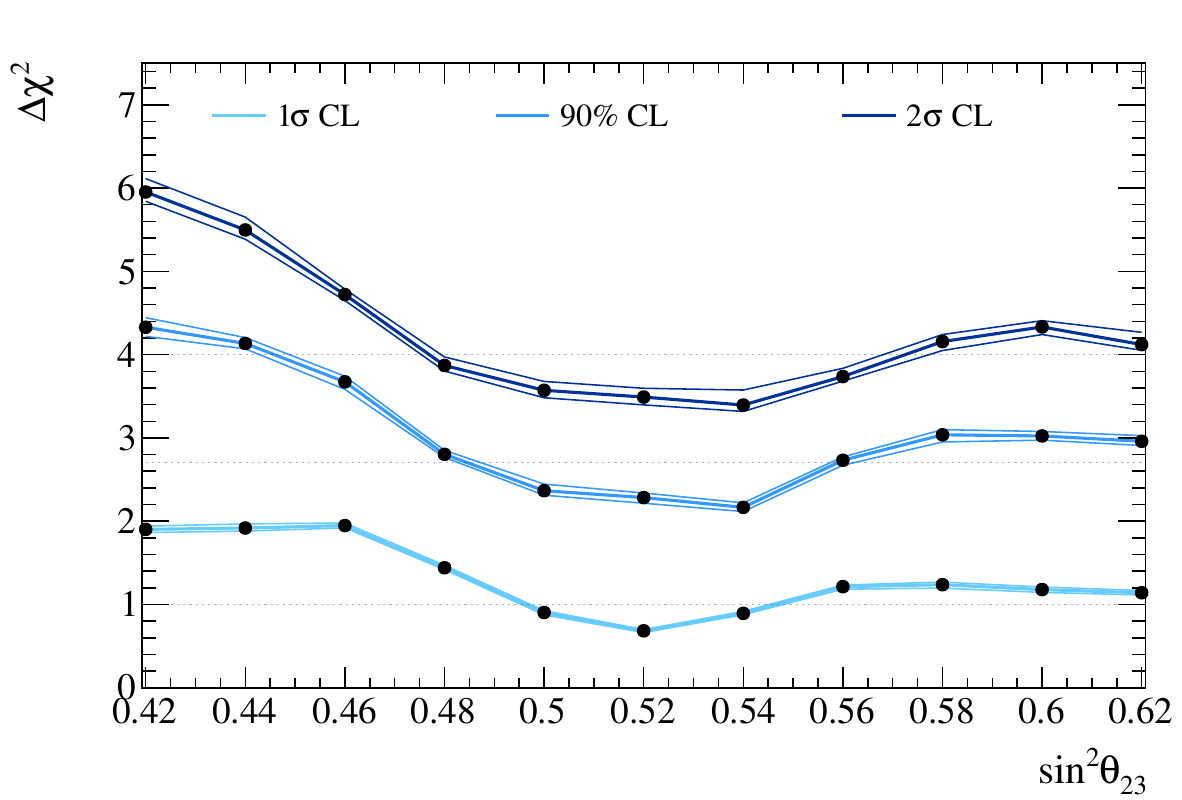}
\caption{Normal ordering, \ssqthtwothree.}
\end{subfigure}
\begin{subfigure}[h]{0.95\columnwidth}
\includegraphics[width=\textwidth]{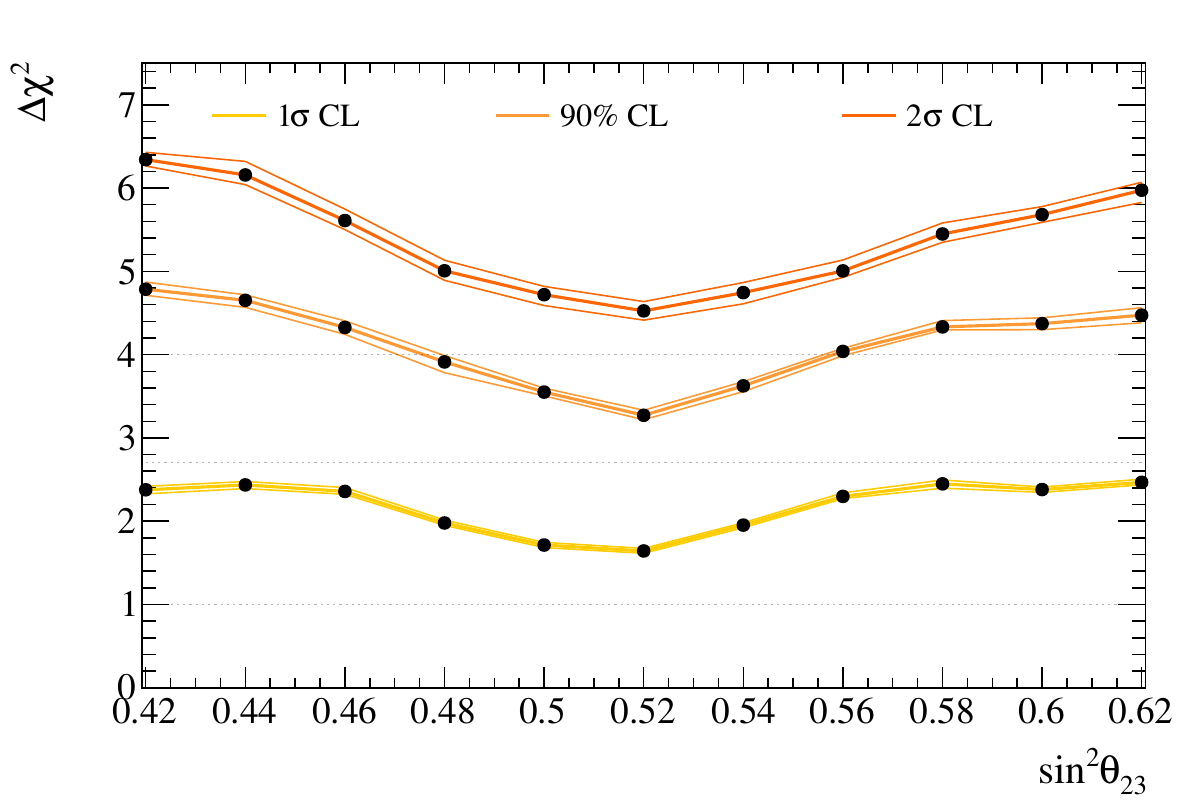}
\caption{Inverted ordering, \ssqthtwothree.}
\end{subfigure}
\caption{FC-corrected critical $\Delta \chi^2$ values for 1D-fits in \deltacp and \ssqthtwothree, computed at true values indicated with black dots and linearly interpolated in between. The error bands show the $1\sigma$ toy-statistical uncertainty (binomial confidence interval). The dotted lines show the asymptotic values from Wilks's theorem for reference.}
\label{fig:oa:appendix:dchi2c:1D}
\end{figure*}
\begin{figure*}[htbp]
\centering
\begin{subfigure}[h]{0.95\columnwidth}
\includegraphics[width=\textwidth]{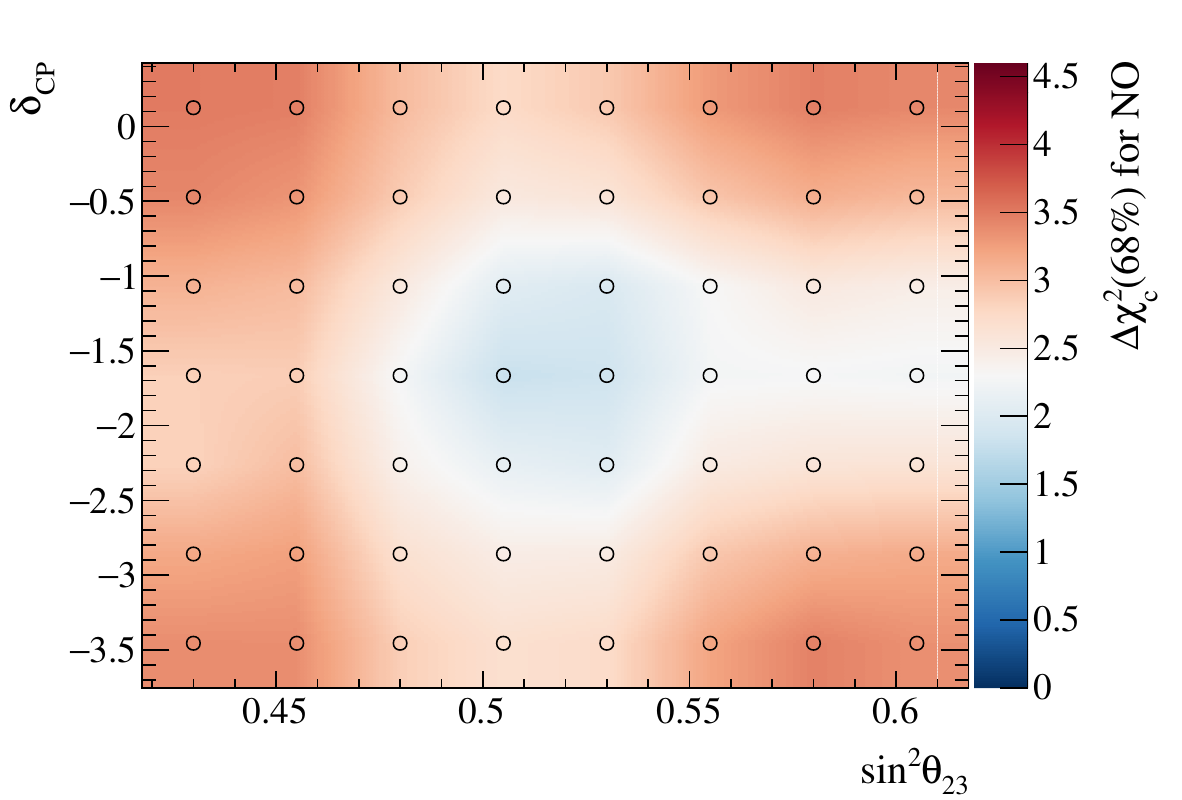}
\caption{Normal ordering, 68\%.}
\end{subfigure}
\begin{subfigure}[h]{0.95\columnwidth}
\includegraphics[width=\textwidth]{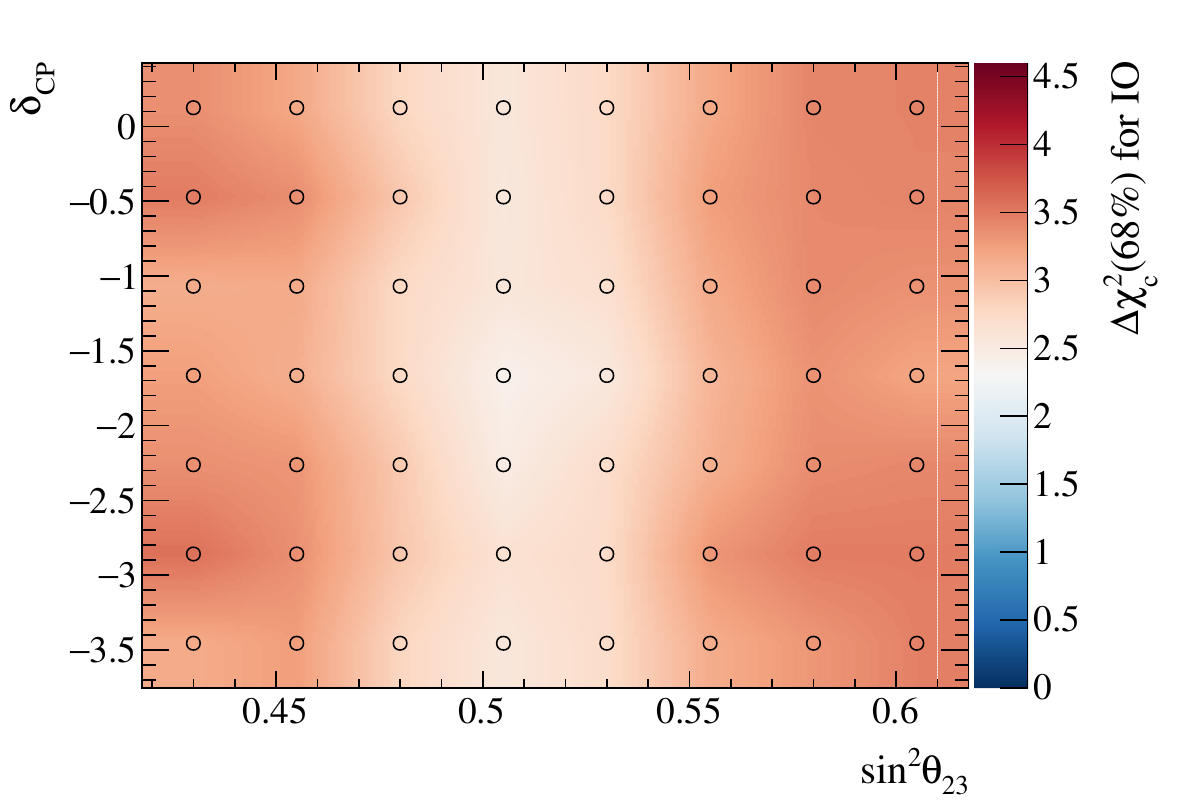}
\caption{Inverted ordering, 68\%.}
\end{subfigure}
\begin{subfigure}[h]{0.95\columnwidth}
\includegraphics[width=\textwidth]{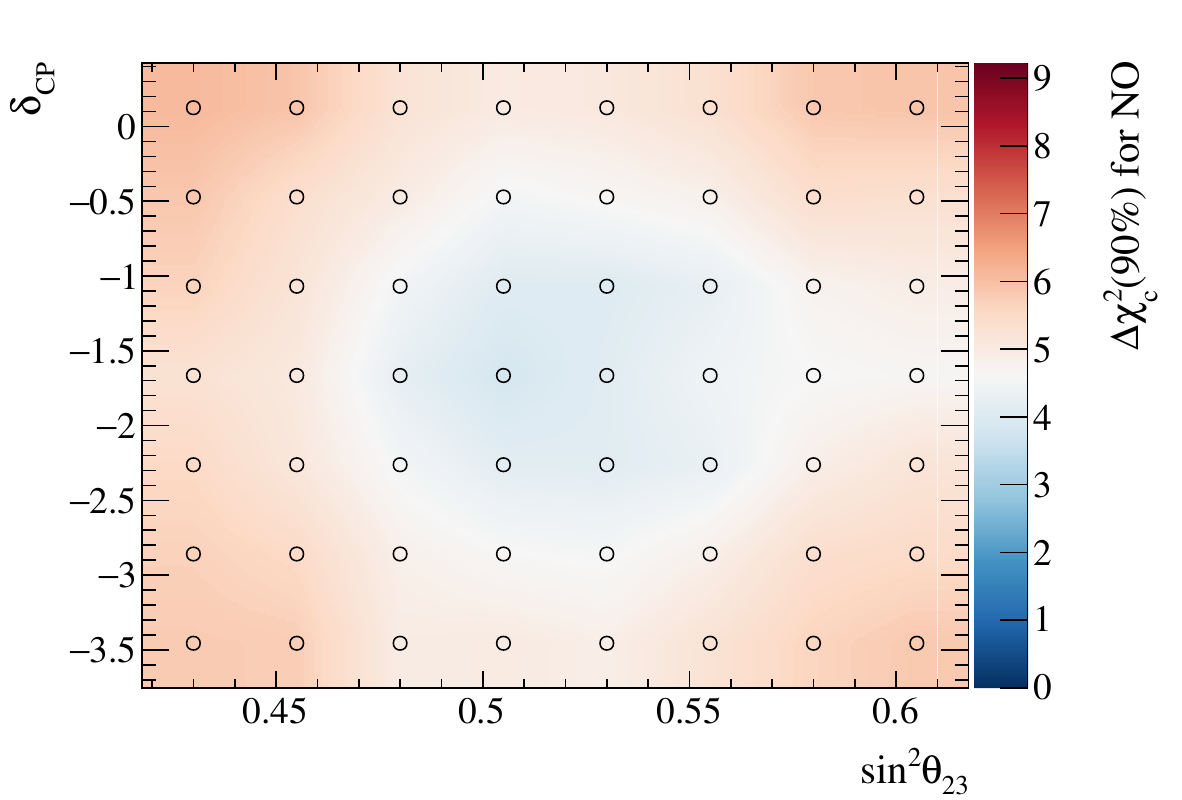}
\caption{Normal ordering, 90\%.}
\end{subfigure}
\begin{subfigure}[h]{0.95\columnwidth}
\includegraphics[width=\textwidth]{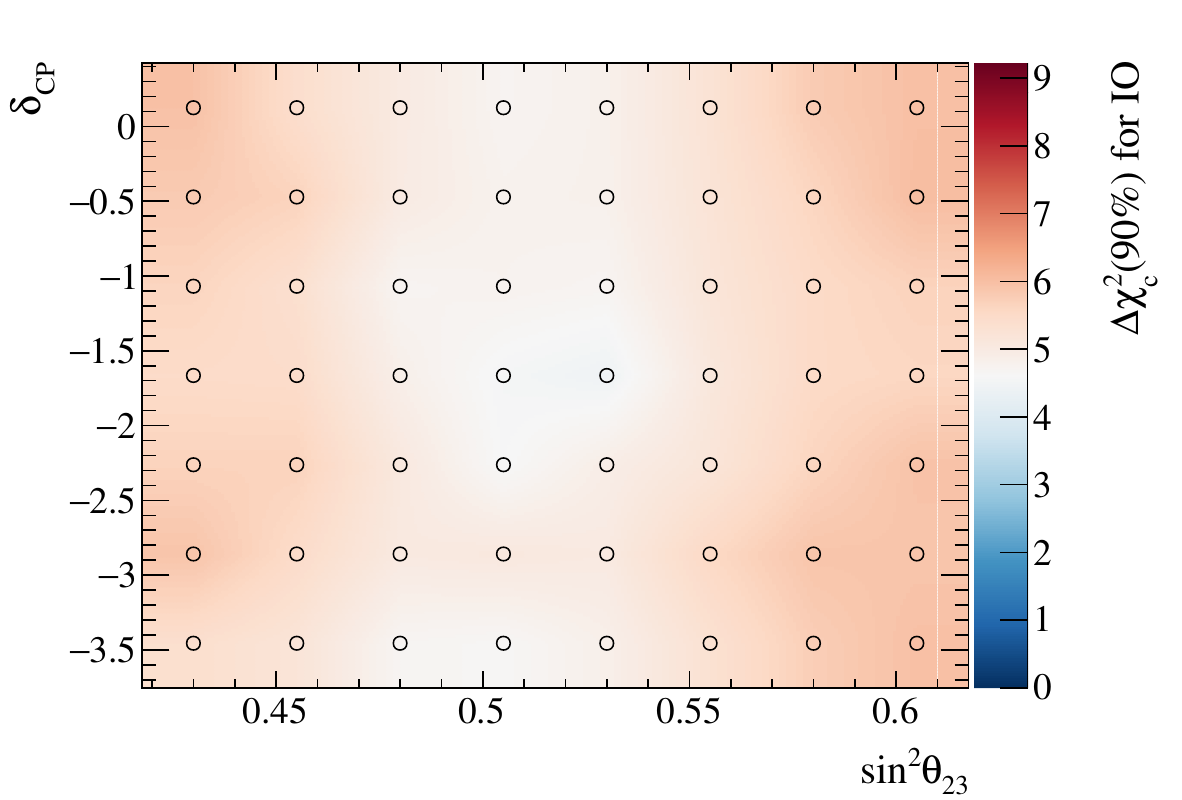}
\caption{Inverted ordering, 90\%.}
\end{subfigure}
\caption{68\% and 90\% FC-corrected critical $\Delta \chi^2$ values for 2D-fits of $\ssqthtwothree-\deltacp$ over both mass orderings, computed at the points indicated by circles and bi-linearly interpolated for the regions in between the points. The colour scheme is chosen to show the asymptotic values from Wilks's theorem in white.}
\label{fig:oa:appendix:dchi2c:2D}
\end{figure*}

Critical values for the Feldman--Cousins confidence intervals and regions are shown in \autoref{fig:oa:appendix:dchi2c:1D} for one-dimensional fits in \deltacp and \ssqthtwothree, and in \autoref{fig:oa:appendix:dchi2c:2D} for two-dimensional fits of $\ssqthtwothree-\deltacp$. In both cases, the upper limit of the $1\sigma$ toy-statistical uncertainty interval on the critical values is conservatively used for the confidence interval computation.
Compared to the asymptotic values from Wilks's theorem, the physical boundaries at $\deltacp = \pm \pi/2$ and $\ssqthtwothree \approx (2 \cos^2\theta_{13})^{-1} \approx 0.513$ tend to pull the critical values down, whereas degeneracies of \deltacp, \ssqthtwothree around these physical boundaries, and mass ordering, pull the critical values higher. Further differences are also caused by the assumed distribution of true oscillation parameter values for the nuisance oscillation parameters, i.e. those not plotted here. These are thrown from a distribution close to their posterior distributions from the fit to data, and the MC prediction with true oscillation probabilities set to the T2K best-fit point is used as the ``data''. This is chosen instead of the actual data to avoid including the effects of statistical fluctuations in data.

%% file: Appendices/NDTables.tex
The parameters from the uncertainty model after the fit to ND data are presented here. The carbon-only uncertainties are not propagated to the FD, as no carbon is present there. The values before and after the fit to ND data, and the uncertainty from the diagonal entry of the covariance matrix, are shown. The values are tabulated in \autoref{tab:skfhcflux} and \autoref{tab:skrhcflux} for the flux parameters, and \autoref{tab:xsecpostfit} for the cross-section parameters. The fit results are from the gradient-descent fitter, and the uncertainty is evaluated by scanning around the best-fit point. These parameters are only valid in the context of the exact model used in this analysis, and the context shouldn't be interpreted as an actual global constraint on the parameters.

\begin{table}[htbp]
\centering
\begin{tabular}{l | c | c }
\hline
\hline
Flux parameter & Pre-fit & Post-fit \\
\hline
FD $\numu$ [0.0, 0.4] & $ 1.00 \pm 0.07 $ & $ 1.11 \pm 0.05$ \\
FD $\numu$ [0.4, 0.5] & $ 1.00 \pm 0.06 $ & $ 1.10 \pm 0.04 $ \\
FD $\numu$ [0.5, 0.6] & $ 1.00 \pm 0.05 $ & $ 1.08 \pm 0.03 $ \\
FD $\numu$ [0.6, 0.7] & $ 1.00 \pm 0.05 $ & $ 1.07 \pm 0.03 $ \\
FD $\numu$ [0.7, 1.0] & $ 1.00 \pm 0.07 $ & $ 1.06 \pm 0.04 $ \\
FD $\numu$ [1.0, 1.5] & $ 1.00 \pm 0.07 $ & $ 1.03 \pm 0.04 $ \\
FD $\numu$ [1.5, 2.5] & $ 1.00 \pm 0.06 $ & $ 1.03 \pm 0.04 $ \\
FD $\numu$ [2.5, 3.5] & $ 1.00 \pm 0.07 $ & $ 1.01 \pm 0.04 $ \\
FD $\numu$ [3.5, 5.0] & $ 1.00 \pm 0.09 $ & $ 0.98 \pm 0.04 $ \\
FD $\numu$ [5.0, 7.0] & $ 1.00 \pm 0.10 $ & $ 0.92 \pm 0.04 $ \\
FD $\numu$ [7.0, 30.0] & $ 1.00 \pm 0.12 $ & $ 0.91 \pm 0.04 $ \\
\hline

FD $\numub$ [0.0, 0.7] & $ 1.00 \pm 0.09$ & $ 1.06 \pm 0.08 $ \\
FD $\numub$ [0.7, 1.0] & $ 1.00 \pm 0.06 $ & $ 1.04 \pm 0.05 $ \\
FD $\numub$ [1.0, 1.5] & $ 1.00 \pm 0.07 $ & $ 1.04 \pm 0.06 $ \\
FD $\numub$ [1.5, 2.5] & $ 1.00 \pm 0.08 $ & $ 1.05 \pm 0.07 $ \\
FD $\numub$ [2.5, 30.0] & $ 1.00 \pm 0.08 $ & $ 1.04 \pm 0.06 $ \\
\hline

FD $\nue$ [0.0, 0.5] & $ 1.00 \pm 0.06 $ & $ 1.09 \pm 0.04 $ \\
FD $\nue$ [0.5, 0.7] & $ 1.00 \pm 0.05 $ & $ 1.08 \pm 0.04 $ \\
FD $\nue$ [0.7, 0.8] & $ 1.00 \pm 0.05 $ & $ 1.06 \pm 0.04 $ \\
FD $\nue$ [0.8, 1.5] & $ 1.00 \pm 0.06 $ & $ 1.04 \pm 0.04 $ \\
FD $\nue$ [1.5, 2.5] & $ 1.00 \pm 0.08 $ & $ 1.00 \pm 0.04 $ \\
FD $\nue$ [2.5, 4.0] & $ 1.00 \pm 0.09 $ & $ 0.98 \pm 0.04 $ \\
FD $\nue$ [4.0, 30.0] & $ 1.00 \pm 0.09 $ & $ 0.98 \pm 0.05 $ \\
\hline

FD $\nueb$ [0.0, 2.5] & $ 1.00 \pm 0.10 $ & $ 1.02 \pm 0.09$ \\
FD $\nueb$ [2.5, 30.0]  & $ 1.00 \pm 0.13 $ & $ 1.09\pm 0.11 $ \\
\hline
\hline
\end{tabular}
\caption{FD \fhcalt flux parameters before and after the fit to ND data, including the uncertainty from the diagonal of the covariance matrix. The values in brackets show the range of \Enu in units of GeV for each parameter.}
\label{tab:skfhcflux}
\end{table}

\begin{table}[htbp]
\centering
\begin{tabular}{l | c | c}
\hline
\hline
Flux parameter & Pre-fit & Post-fit \\
\hline
FD $\numu$ [0.0, 0.7] & $ 1.00 \pm 0.09 $ & $ 1.11 \pm 0.06 $ \\
FD $\numu$ [0.7, 1.0] & $ 1.00 \pm 0.06 $ & $ 1.07 \pm 0.05 $ \\
FD $\numu$ [1.0, 1.5] & $ 1.00 \pm 0.06 $ & $ 1.07 \pm 0.04 $ \\
FD $\numu$ [1.5, 2.5] & $ 1.00 \pm 0.07 $ & $ 1.07 \pm 0.04 $ \\
FD $\numu$ [2.5, 30.0]  & $ 1.00 \pm 0.07 $ & $ 1.02 \pm 0.04 $ \\
\hline

FD $\numub$ [0.0, 0.4]  & $ 1.00 \pm 0.07 $ & $ 1.09 \pm 0.05 $ \\
FD $\numub$ [0.4, 0.5]  & $ 1.00 \pm 0.06$ & $ 1.09 \pm 0.04 $ \\
FD $\numub$ [0.5, 0.6]  & $ 1.00 \pm 0.06 $ & $ 1.07 \pm 0.04 $ \\
FD $\numub$ [0.6, 0.7]  & $ 1.00 \pm 0.05 $ & $ 1.06 \pm 0.03 $ \\
FD $\numub$ [0.7, 1.0]  & $ 1.00 \pm 0.08 $ & $ 1.09 \pm 0.04 $ \\
FD $\numub$ [1.0, 1.5]  & $ 1.00 \pm 0.08 $ & $ 1.06 \pm 0.04 $ \\
FD $\numub$ [1.5, 2.5]  & $ 1.00 \pm 0.06 $ & $ 1.01 \pm 0.04 $ \\
FD $\numub$ [2.5, 3.5]  & $ 1.00 \pm 0.07 $ & $ 1.01 \pm 0.05 $ \\
FD $\numub$ [3.5, 5.0]  & $ 1.00 \pm 0.09 $ & $ 0.95 \pm 0.06 $ \\
FD $\numub$ [5.0, 7.0]  & $ 1.00 \pm 0.09 $ & $ 0.95 \pm 0.06 $ \\
FD $\numub$ [7.0, 30.0]  & $ 1.00 \pm 0.12 $ & $ 0.93 \pm 0.09 $ \\
\hline

FD $\nue$ [0.0, 2.5] & $ 1.00 \pm 0.09 $ & $ 1.03 \pm 0.07 $ \\
FD $\nue$ [2.5, 30.0] & $ 1.00 \pm 0.08 $ & $ 1.03 \pm 0.07 $ \\
\hline

FD $\nueb$ [0.0, 0.5] & $ 1.00 \pm 0.06 $ & $ 1.08 \pm 0.04 $ \\
FD $\nueb$ [0.5, 0.7] & $ 1.00 \pm 0.05 $ & $ 1.07 \pm 0.04 $ \\
FD $\nueb$ [0.7, 0.8] & $ 1.00 \pm 0.06 $ & $ 1.06 \pm 0.04 $ \\
FD $\nueb$ [0.8, 1.5] & $ 1.00 \pm 0.06 $ & $ 1.04 \pm 0.04 $ \\
FD $\nueb$ [1.5, 2.5] & $ 1.00 \pm 0.08 $ & $ 1.01 \pm 0.06 $ \\
FD $\nueb$ [2.5, 4.0] & $ 1.00 \pm 0.09 $ & $ 1.01 \pm 0.07 $ \\
FD $\nueb$ [4.0, 30.0] & $ 1.00 \pm 0.15 $ & $ 1.09 \pm 0.13 $ \\
\hline
\hline
\end{tabular}
\caption{FD \rhcalt flux parameters before and after the fit to ND data, including the uncertainty from the diagonal of the covariance matrix. The values in brackets show the range of \Enu in units of GeV for each parameter.}
\label{tab:skrhcflux}
\end{table}

\begin{table*}[htbp]
\centering
\begin{tabular}{l|c|c|l}
\hline
\hline
Parameter & Pre-fit & Post-fit & Comment \\
\hline
$M_A^{QE}~(\text{GeV}/c^2)$ & $ 1.03 \pm 0.06 $ & $ 1.17 \pm 0.04 $ & \\
\hline

$Q^2 <0.05~\text{GeV}^2$       &$ 1.00 \pm \infty $ & $ 0.78 \pm 0.05 $ & \multirow{8}{*}{\parbox[t]{3.5cm}{Norm. on true CCQE events in true $Q^2$.}}\\
$0.05<Q^2 <0.10~\text{GeV}^2$  &$ 1.00 \pm \infty $ & $ 0.89 \pm 0.04 $ &\\
$0.10<Q^2 <0.15~\text{GeV}^2$  &$ 1.00 \pm \infty $ & $ 1.03 \pm 0.05 $ &\\
$0.15<Q^2 <0.20~\text{GeV}^2$  &$ 1.00 \pm \infty $ & $ 1.03 \pm 0.08 $ &\\
$0.20<Q^2 <0.25~\text{GeV}^2$  &$ 1.00 \pm \infty $ & $ 1.09 \pm 0.10 $ &\\
$0.25<Q^2 <0.50~\text{GeV}^2$  &$ 1.00 \pm 0.11 $ & $ 1.26 \pm 0.06 $ &\\
$0.50<Q^2 <1.00~\text{GeV}^2$  &$ 1.00 \pm 0.18 $ & $ 1.14 \pm 0.08 $ &\\
$Q^2 >1.00~\text{GeV}^2$       &$ 1.00 \pm 0.40 $ & $ 1.26 \pm 0.14 $ &\\
\hline

$\Delta E_{rmv}^C \nu$ (MeV) & $ 2.00 \pm 6.00 $ & $ -2.38 \pm 1.75 $  &\\
$\Delta E_{rmv}^C \nub$ (MeV) & $ 0.00 \pm 6.00 $ & $ 1.64 \pm 1.93 $  &\\
$\Delta E_{rmv}^O \nu$ (MeV) & $ 4.00 \pm 6.00 $ & $ 2.55 \pm 3.08 $  &\\
$\Delta E_{rmv}^O \nub$ (MeV) & $ 0.00 \pm 6.00 $ & $ -1.26 \pm 3.19 $  &\\
\hline

2p2h norm. $\nu$    &$ 1.00 \pm \infty $ & $ 1.06 \pm 0.15 $  &\\
2p2h norm. \nub     &$ 1.00 \pm \infty $ & $ 0.72 \pm 0.16 $  &\\
2p2h norm. C$\rightarrow$O  & $ 1.00 \pm 0.20 $ & $ 1.05 \pm 0.15 $  &\\
\hline

2p2h shape C  & $ 0.00 \pm 3.00 $ & $ 0.97 \pm 0.46 $  & \multirow{2}{*}{\parbox[t]{3.5cm}{-1 is non-$\Delta$-like, 0 is Nieves \etal~\cite{Nieves:2011yp}, +1 is $\Delta$-like.}}\\
2p2h shape O  & $ 0.00 \pm 3.00 $ & $ 0.00 \pm 0.17 $  & \\
\hline
2p2h low-\Enu $\nu$ & $ 1.00 \pm 1.00 $ & $ 1.00 \pm 1.00 $ & \multirow{4}{*}{\parbox[t]{3.5cm}{+1 is Nieves-like~\cite{Nieves:2011yp}, 0 is Martini-like~\cite{Martini:2009uj}. Not fit at ND.}} \\
2p2h high-\Enu $\nu$ & $ 1.00 \pm 1.00 $ & $ 1.00 \pm 1.00 $ & \\
2p2h low-\Enu \nub  & $ 1.00 \pm 1.00 $ & $ 1.00 \pm 1.00 $ \\
2p2h high-\Enu \nub   & $ 1.00 \pm 1.00 $ & $ 1.00 \pm 1.00 $ \\
\hline

$C^A_5$  & $ 0.96 \pm 0.15 $ & $ 0.98 \pm 0.06 $ \\
$M_A^{RES}~(\text{GeV}/c^2)$ & $ 1.07 \pm 0.15 $ & $ 0.79 \pm 0.05 $ \\\hline
$I_{1/2}$ non-res norm. low-$p_\pi$ $\overline{\nu}_\mu$  & $ 0.96 \pm 0.96 $ & $ 0.96 \pm 0.96 $ & Not fit at ND. \\\hline
$I_{1/2}$ non-res norm. & $ 0.96 \pm 0.40 $ & $ 0.87 \pm 0.23 $ & \\
CC coh. C norm.  & $ 1.00 \pm 0.30 $ & $ 0.61 \pm 0.22 $ \\
CC coh. O norm.  & $ 1.00 \pm 0.30 $ & $ 0.61 \pm 0.22 $ \\
\hline

Coulomb corr. $\nu$ & $ 1.00 \pm 0.02 $ & $ 1.00 \pm 0.02 $ \\
Coulomb corr. \nub & $ 1.00 \pm 0.01 $ & $ 1.00 \pm 0.01$ \\
\hline
\nue/\numu norm.  & $ 1.00 \pm 0.03 $ & $ 1.00 \pm 0.03 $ & \multirow{2}{*}{\parbox[t]{3.5cm}{No ND selection, poorly constrained.}}\\
\nueb/\numub norm.  & $ 1.00 \pm 0.03 $ & $ 1.00 \pm 0.03 $ \\
\hline

CC Bodek-Yang on/off DIS  & $ 0.00 \pm 1.00 $ & $ 1.04 \pm 0.19 $ & \multirow{2}{*}{\parbox[t]{3.5cm}{+1 is B-Y supp. off, 0 is B-Y supp. on.~\cite{Bodek:2003wc,Bodek:2005de}}} \\
CC Bodek-Yang on/off multi-$\pi$  & $ 0.00 \pm 1.00 $ & $ -0.03 \pm 0.18 $ & \\
\hline
CC multiplicity multi-$\pi$  & $ 0.00 \pm 1.00 $ & $ 0.14 \pm 0.71 $ & \parbox[t]{3.5cm}{+1 is AGKY-like~\cite{Yang:2009zx}, 0 is NEUT-like.} \\
\hline
CC misc. norm.  & $ 1.00 \pm 1.00 $ & $ 2.28 \pm 0.43 $ \\
CC DIS+multi-$\pi$ norm. $\nu$ & $ 1.00 \pm 0.04 $ & $ 1.06 \pm 0.03 $ \\
CC DIS+multi-$\pi$ norm. \nub & $ 1.00 \pm 0.07 $ & $ 0.94 \pm 0.06 $ \\
\hline

NC coh. norm. & $ 1.00 \pm 0.30 $ & $ 1.02 \pm 0.30 $ & \multirow{2}{*}{\parbox[t]{3.5cm}{No ND selection, poorly constrained.}}\\
NC $1\gamma$ norm.  & $ 1.00 \pm 1.00 $ & $ 1.00 \pm 1.00 $ \\
\hline
NC other ND norm.  & $ 1.00 \pm 0.30 $ & $ 1.66 \pm 0.13 $ & Not propagated to FD. \\
NC other FD norm.  & $ 1.00 \pm 0.30 $ & $ 1.00 \pm 0.30 $ & Not fit at ND.\\
\hline

Pion FSI Quasi-Elastic  & $ 1.00 \pm 0.29 $ & $ 0.83 \pm 0.09 $ & \multirow{5}{*}{\parbox[t]{3.5cm}{Scaling of pion scattering probabilities relative to the constraint from external data~\cite{PinzonGuerra:2018rju}}}.\\
Pion FSI Quasi-Elastic $p_\pi>500~\text{MeV}/c$  &$ 1.00 \pm 0.47 $ & $ 0.75\pm 0.16 $ \\
Pion FSI Inelastic  &$ 1.00 \pm 1.10 $ & $ 1.71 \pm 0.31 $ \\
Pion FSI Absorption  &$ 1.00 \pm 0.31 $ & $ 1.19 \pm 0.12 $ \\
Pion FSI Charge Exchange  &$ 1.00 \pm 0.44 $ & $ 0.78 \pm 0.34 $ \\
\hline
\hline
\end{tabular}
\caption{Cross-section parameters before and after the fit to ND data, including the uncertainty from the diagonal of the covariance matrix. The parameters are detailed in \autoref{sec:interactionModel}. Parameters without external constraints are labelled with an uncertainty $\pm \infty$.}
\label{tab:xsecpostfit}
\end{table*}